\documentclass[a4paper]{article}

\usepackage{fullpage}
\usepackage{graphicx}
\usepackage{amsmath,amssymb}
\usepackage{bbm}
\usepackage{bm}

\numberwithin{equation}{section}
\numberwithin{figure}{section}
\numberwithin{table}{section}

\renewcommand{\Im}{\ensuremath{\mathop\mathrm{Im}}}
\renewcommand{\Re}{\ensuremath{\mathop\mathrm{Re}}}
\newcommand{\ads}{\ensuremath{\mathrm{AdS}}}				
\newcommand{\SU}{\ensuremath{\mathrm{SU}}}
\newcommand{\PSU}{\ensuremath{\mathrm{PSU}}}
\newcommand{\SO}{\ensuremath{\mathrm{SO}}}
\newcommand{\SL}{\ensuremath{\mathrm{SL}}}
\renewcommand{\O}{\ensuremath{\mathrm{O}}}
\newcommand{\U}{\ensuremath{\mathrm{U}}}
\newcommand{\mbold}[1]{\mbox{\boldmath$#1$}}					
\newcommand{\dd}{\mathrm{d}}									
\newcommand{\ds}{\mathrm{d}s}									
\renewcommand{\S}{\ensuremath{\mathrm{S}}}
\newcommand{\adsfivesfive}{\ensuremath{\ads_5 \times \S^{5}}}	
\newcommand{\bra}[1]{\left<#1\right|}							
\newcommand{\ket}[1]{\left|#1\right>}							
\newcommand{\vev}[1]{\left<#1\right>}							
\newcommand{\smatrix}[3]{\left<#1\!\left|\vphantom{#1 #2 #3}#2\right|#3\right>}
\newcommand{\tr}{\mathop\mathrm{Tr}}							
\newcommand{\str}{\mathop\mathrm{sTr}}							
\newcommand{\commute}[2]{\ensuremath{\left[#1\mathop{,}#2\right]}}		
\newcommand{\defeq}{\mathrel{\mathord{:}\mathpunct{=}}}					
\newcommand{\hodge}{\raisebox{0.25ex}[0cm][0cm]{\ensuremath{\mathord{\star}}}}	
\newcommand{\gym}{\ensuremath{g_\text{\tiny\scalebox{.9}{YM}}}}
\newcommand{\gs}{\ensuremath{g_\text{\scriptsize s}}}
\newcommand{\rh}{\ensuremath{r_{\!\circ}}}
\newcommand{\GR}{\ensuremath{G^{\scriptscriptstyle R}}}
\newcommand{\SDBI}{\ensuremath{S_\text{\tiny DBI}}}
\newcommand{\ii}{\ensuremath{i}}
\newcommand{\ee}{\ensuremath{e}}
\renewcommand{\L}{\ensuremath{\mathcal L}}
\newcommand{\N}{\ensuremath{\mathcal N}}
\newcommand{\G}{\ensuremath{\mathcal G}}
\newcommand{\R}{\ensuremath{\mathfrak R}}
\newcommand{\wn}{\ensuremath{\mathfrak w}}
\newcommand{\qn}{\ensuremath{\mathfrak q}}
\newcommand{\mn}{\ensuremath{\mathfrak m}}
\newcommand{\del}{\partial}
\newcommand{\db}[1]{D$#1$-brane}
\newcommand{\D}[1]{D$#1$}
\newcommand{\threevec}[1]{\text{\boldmath$\mathrm{#1}$\unboldmath}}		
\newcommand{\fourvec}[1]{\ensuremath{\vec #1}}		
\newcommand{\T}{{\mathcal T}}			
\newcommand{\OT}{{\mathcal O}_{\T}}
\newcommand{\OF}{{\mathcal O}_{F^2}}
\newcommand{\order}[1]{O\!\left(#1\right)}

\title{In-medium effects in the holographic quark-gluon plasma}
\author{Felix Rust\\[1em] \small Max-Planck-Institut f\"ur Physik (Werner-Heisenberg-Institut),\\ \small F\"ohringer Ring 6, 80805 M\"unchen, \small rust@mpp.mpg.de}
\date{}

\begin{document}

\maketitle

\begin{abstract}

In this article we use the gauge/gravity duality to investigate various
properties of strongly coupled gauge theories, which we interpret as models for
the quark-gluon plasma (QGP). In particular, we use variants of the \D3/\D7
setup as an implementation of the top-down approach of connecting string theory
with phenomenologically relevant gauge theories.

We focus on the effects of finite temperature and finite density on fundamental
matter in the holographic quark-gluon plasma, which we model as the $\N=2$
hypermultiplet in addition to the $\N=4$ gauge multiplet of supersymmetric
Yang-Mills theory.

We use a setup in which we can describe the holographic plasma at finite
temperature and either baryon or isospin density and investigate the properties
of the system from three different viewpoints.

(i) We study meson spectra. Our observations at finite temperature and particle
density are in qualitative agreement with phenomenological models and
experimental observations. They agree with previous publications in the
according limits.

(ii) We study the temperature and density dependence of transport properties of
fundamental matter in the QGP. In particular, we obtain diffusion coefficients.
Furthermore, in a kinetic model we estimate the effects of the coupling strength
on meson diffusion and therewith equilibration processes in the QGP.

(iii) We observe the effects of finite temperature and density on the phase
structure of fundamental matter in the holographic QGP. We trace out the phase
transition lines of different phases in the phase diagram.

\end{abstract}

\section{Introduction}

The entire content of matter and radiation in the universe is a manifestation of
the energy unleashed in some unknown process which we commonly refer to as the
\emph{big bang}. This event is thought of as the moment of the creation of
matter, space and time --- the universe. From that moment on energy existed in
various manifestations. At an order of magnitude of $\text{10}^{-\text{33}}$
seconds after the big bang, quarks formed. Today we experimentally detect these
particles together with leptons and the force mediating gauge bosons as the
fundamental constituents of all visible matter. The interaction of these
particles is described incredibly accurately within two different theories.
Processes taking place at energy levels below the TeV scale involving
electromagnetic, weak and strong interaction are described accurately by the
\emph{standard model} of particle physics, although the strong force is hard to
exploit theoretically at low energies for mathematical reasons. The fourth of
the known forces, gravity, is described within the separate framework of
\emph{general relativity}.

While many aspects of the particles that make up our world are well understood,
others remain a mystery. Among the latter is the behavior of matter under
conditions that must have existed shortly after the big bang. The earliest
period of the universe that can either be described by theoretical models and
numerical simulations or probed by experiments is ranging from about
$\text{10}^{-\text{33}}$ seconds after the big bang when quarks and gluons
emerged until approximately $\text{10}^{-\text{6}}$ seconds after the big bang
when hadronization of quarks set in.  During this
early phase matter existed in conditions of extremely high density and
temperature. Under these conditions quarks are not confined and do not form
hadrons. Instead they are moving independently and interact with each other
predominantly via the exchange of gluons, which mediate the force of strong
interactions. As typical for plasmas, the freely moving quarks allow for (color)
charge screening. Matter in this phase is therefore referred to as the
\emph{quark-gluon plasma} (QGP).

From the beginning on, the universe expanded and cooled down. After
$\text{10}^{-\text{6}}$ seconds at a critical value of the temperature of
approximately 160--190 MeV  the energy dependent coupling constant of the
strong force rose to a value that let confinement set in. Eventually quarks
combined to bound states and formed the hadronic matter that is now composing
the galaxies visible in the universe. Only very particular regions of the
present universe come into consideration for providing conditions extreme enough
to contain matter in the phase of the quark-gluon plasma. Such barren places are
the cores of neutron stars. These are remnants of supernova explosions of stars
of about 20--30 solar masses. Due to the extremely high gravitational pressure,
the hadronic matter existing on the planets surface in deeper layers is squeezed
together to such an extent that electrons and protons combine to neutrons
(thereby emitting neutrinos). From the surface towards the core of these objects
the pressure increases. In the inner layers even neutrons are not stable
anymore. Instead the quarks and gluons may interact individually to appear as
the QGP. Other temporary habitats of the quark-gluon plasma seem to exist on
earth: The experiments conducted at heavy ion colliders are dedicated to monitor
the processes occurring at collisions of heavy nuclei at energies high enough to
produce a fireball of extremely hot and dense matter. Such experiments are
hosted at the Super Proton Synchrotron (SPS), the Relativistic Heavy Ion
Collider (RHIC) accelerating gold nuclei, and in future also the Large Hadron
Collider (LHC) which can be used to accelerate lead nuclei, as well as future
SIS experiments at the Facility for Antiproton and Ion Research (FAIR). The
state of matter observed at RHIC is a strongly coupled system composed of
deconfined quarks and gluons, the strongly coupled quark-gluon plasma (sQGP).

To get an impression of the processes occurring during the first moments after
the creation of the universe\,---\,including the interactions of quarks and
gluons which lead to the genesis of the hadronic matter that composes our
world\,---\,it is necessary to understand the properties of the quark-gluon
plasma. This will eventually allow for deeper insight into the process of
hadronization and the phase transition from the quark-gluon plasma to the
hadronic phase. Further knowledge about the nature of matter may also allow for
progress in finding the answer to questions about the nature of dark matter and
dark energy, the majority of the energy content of our universe\,---\,by far
greater than the contributions visible matter can account for. 

Still a manageable theoretical description of the interaction of quarks and
gluons in the strongly coupled systems observed at experiments is not
straightforward, although the standard model contains a theory of quarks and
gluons, known as quantum chromodynamics (QCD). It is the strong coupling that
impedes the solution of the equations of motion of QCD at low energies.
Analytical answers from QCD are obtained from perturbation expansions in the
coupling constant, which do not converge at strong coupling. Therefore today
there is no analytic description of the formation of bound states of quarks or
the interaction of quarks and gluons in the sQGP from first principles. One
successful alternative to obtain results at strong coupling is lattice gauge
theory, which tries to simulate the dynamics of QCD numerically on a number of
discrete points in spacetime. While this approach gave answers to numerous
questions, by nature it cannot produce analytical results which would lead to
conceptual insights. It moreover approximates spacetime as a coarse grid and so
far has to incorporate some simplifications of QCD\@.

A completely different approach to the quark-gluon plasma can be pursued from a
point of view that also motivates the work presented in this work. A possible
alternative to the description of strongly coupled quarks and gluons may be
given in terms of \emph{string theory}. This theory assumes strings to be the
fundamental degrees of freedom, from which all matter is composed. The different
elementary particles we know are thought to arise as the different oscillation
modes of the strings. Initially, around 1970, it aimed to explain the relation
between spin $J$ and mass $m$ of the resonances found in then performed
collision experiments, $J=\alpha_0 + \alpha' m^2$, with $\alpha'$ known as the
``Regge slope''. The idea was to describe the force between quarks as if a
string of tension $1/\alpha'$ holds the particles together. Despite
modeling the Regge behavior, the theory failed to describe the observed cross
sections correctly and was successfully displaced by QCD wherever applicable.
Nevertheless, the understanding and interpretation of string theory evolved to a
great extent, especially the possibility to describe quantized gravity attracted
interest. Today it is the most promising candidate for a unified description of
all known forces of nature within one single theory. In this sense it can be
thought of as a generalization of the successful standard model by including a
description of gravity. The world of strings appears as a stunningly complex
system that may give answers to such fundamental questions as the origin of the
number of spacetime dimensions we live in, and allow for a formulation of
quantum gravity. Still much of the theory has to be understood and almost no
predictions lie within the reach of experimental verification.

However, during the past dozen years evidence mounted that indeed there are
connections between string theory and gauge theories, like QCD. In the mid
1990s, during the so called second string theory revolution, it was discovered
that string theory not only features strings as degrees of freedom. In addition,
there are higher dimensional objects, called \emph{branes} as an allusion to
membranes. Branes and strings interact with each other. In this way branes
influence the degrees of freedom introduced by the string oscillations. As the
understanding of string theory grew, it was discovered that certain limits of
string theory contain the degrees of freedom of particular non-Abelian gauge
theories. These insights heralded a new era of applications of string theory to
problems in gauge theory. It began in 1997 with Juan Maldacena's discovery of
analogies between the classical limit of so called type~IIB string theory,
including branes, and the $\N=4$ supersymmetric Yang-Mills quantum gauge field
theory. It is possible to establish a one to one mapping between the degrees of
freedom of both theories. Maldacena therefore speculated that both of them are
different descriptions of the same physical reality. The formulation of this
conjecture is known as the \emph{\ads/CFT correspondence}
\cite{Maldacena:1997re}. As we will discuss later, this correspondence between
type~IIB string theory in Anti-de~Sitter space (\ads) and the $\N=4$
supersymmetric non-Abelian conformal field theory (CFT) is especially suitable
to describe the strongly coupled regime of gauge theories. An astonishing
feature of the \ads/CFT correspondence is that it conjectures the equivalence of
a \emph{classical} theory of gravity and a \emph{quantum} field theory. In some
aspects this field theory resembles the properties of QCD\@. Moreover, it
relates the strongly coupled regime of the quantum field theory to the weakly
coupled regime of the related gravity theory. One therefore can obtain strong
coupling results of field theory processes by means of well established
perturbative methods on the gravity side. Finally, the \ads/CFT correspondence
allows to interpret the quantum field theory to be the four dimensional
representation of processes in string theory, which is defined in ten spacetime
dimensions. Because of these properties the correspondence is more generally
also referred to as \emph{gauge/gravity duality} and is said to realize the
\emph{holographic principle}. So far there is no mathematically rigorous proof
for the correspondence to hold. Nevertheless, in all cases that allowed for a
direct comparison of results from both theories, perfect matching was found.

The \ads/CFT correspondence is considered as one of the most important
achievements in theoretical physics of the last decades. However, by now the
string theory limit which exactly corresponds to QCD is not known. Albeit the
direct way ahead towards a comprehensive analytical description of strongly
coupled QCD is not foreseeable, numerous cornerstones where already passed and
some junctions and connections to the related physical disciplines where found.
Examples are deeper insights into the connection of black hole physics to
thermodynamics, the relation to finite temperature physics, and the discovery of
quantities like the famous ratio $\eta/s$ of shear viscosity to entropy density
that are universal for large classes of theories. The motivation to use the
\ads/CFT correspondence to explore the strongly coupled quark-gluon plasma
therefore is twofold. On the one hand side there is the attempt to provide a
description of strongly coupled quarks and gluons as a supplement to QCD\@. In
this way string theory might contribute to further understanding of gauge
theories. On the other hand a phenomenologically relevant application of string
theory can be used as a benchmark to evaluate the capabilities of string theory
in describing nature. In this way string theory might benefit from the
exploration of new regimes of QCD, so far described predominantly by quantum
field and lattice gauge theories. The ability to produce the sQGP in collision
experiments for the first time may allow to check predictions from string
theory. There is well-founded hope that the quark-gluon plasma can provide a
link between string theory and experiment.

We will make use of the \ads/CFT correspondence to investigate
strongly coupled systems. The models we use for this purpose will be various
modifications of the gauge/gravity duality that allow for the description of
quantum field theories that feature certain aspects known from QCD\@. Which
aspects and which parameter regions we can cover with this approach will be
pointed out in the introductory section on \ads/CFT and in those sections where
we introduce the models. It is interesting in its own to see how far the
correspondence may be extended and which facets of quarks and gluons can be
modeled at all. However, it is even more fascinating to see that already today
some properties of phenomenological relevance can be captured by a so called
holographic description of the sQGP via the \ads/CFT correspondence. Such results
allow for comparison with lattice gauge theory and effective field theories. A
vast number of attempts to apply the correspondence to the dynamics of quarks
and gluons has been under investigation during the past years. The questions
pursued in this work are the following.
\begin{itemize}
	\item Can quarks and gluons combine to form hadrons inside the quark-gluon
		plasma? How does the spectrum of bound states of quarks, esp.\ of
		mesons, look like inside the sQGP?
		
	\item How do these spectra and the lifetime of mesons depend on temperature
		and quark density?
		
	\item How do quarks and their bound states move through the plasma?
	
	\item What effects has the strong coupling?	
\end{itemize}
The answers we obtain are by part of qualitative nature, or can be expected to
receive corrections, which can be calculated as soon as progress in the field
allows to relax some limiting assumptions. Nevertheless, it is amazing to see
that the gauge/gravity duality can give answers to these questions in terms of a
minimal number of input parameters. We do not follow the so called ``bottom-up''
approach, also known as \ads/QCD. There the goal would be to find gravity duals
to phenomenological gauge theories which incorporate certain desired aspects of
QCD. Instead we pursue the ``top-down'' approach. This means that we are aware
of the fact that the \ads/CFT correspondence is a phenomenon discovered in string
theory. We try to construct models which are consistent solutions of string
theory and observe the consequences on the gauge theory side. From this point of
view, the results obtained by means of the \ads/CFT correspondence can be
interpreted as a sign of the predictive power of string theory.

\medskip

This article is organized as follows. Section~\ref{chap:adscft} gives a brief
introduction to the gauge/gravity duality, the extensions which are of relevance
for the derivation of our results, and a short discussion of the application to
QCD and the quark-gluon plasma. However, we will not try to give an introduction
neither to string theory nor to quantum field theory, supersymmetry or general
relativity. Nevertheless, these theories are the basis of this work. Especially
string theory is on the one hand the basis of this work, on the other hand too
rich to provide a broad background in detail here. Therefore, we will provide
the necessary theoretical arguments and details wherever needed in a hopefully
adequate manner. The remaining sections deal with the answers of the questions
mentioned above. Each of them contains a brief introductory section and one or
more technical sections which lead to results that will be discussed at the end
of each section. In particular, section~\ref{chap:specFuncs} deals with meson
spectra at finite temperature and particle density, and discusses the influence
of these parameters on the spectra. In section~\ref{chap:transport} transport
coefficients of quarks and mesons in the plasma are calculated. Comparison with
weak coupling results enables us to estimate the effects of strong coupling. In
section~\ref{chap:phaseDiag} we examine the lifetime and stability of mesons at
different temperatures and particle densities and in this way get new insights
into the structure of the phase diagram of the dual field theory. A summary and
discussion of the results is finally given in section~\ref{chap:conclusion}.
Appendices clarify conventions and notational issues and present some
calculations in a more detailed form than the main text allows for.

\section{The AdS/CFT correspondence and extensions}
\label{chap:adscft}

\vspace{-.2\baselineskip}
Conventional holograms are able to encode truly three-dimensional information on
a two-dimensional surface. Analogously, in particle physics and quantum gravity,
the equivalence of information contained in a theory defined in some lower
dimensional space and a different theory on a higher dimensional domain, is
referred to as the \emph{holographic principle}. One of the first observations
of such kind of holography was the discovery that the information captured
inside the horizon radius of a black hole, i.\,e.\  the entropy given by the number
of possible microstates, can be described in terms of the horizon surface area
alone \cite{Bekenstein:1973ur}. This observation suggests the existence of a
holographic realization of quantum gravity.

Another observation of a holographically realized connection between
gravitational physics and quantum mechanics was made by Juan Maldacena at the
end of the so-called second string theory revolution. He then conjectured the
equivalence of a supergravity theory in Anti-de~Sitter spacetime (\ads) and a
certain type of conformal field theory (CFT) \cite{Maldacena:1997re}. This
discovery triggered an enormous amount of efforts to establish the long sought
connection between quantum gauge field theories and gravity, which did not abate
so far. The fact that the theory on the \ads side of the correspondence
can be expressed as a low energy limit of string theory, which naturally
incorporates gravity, is widely interpreted as a support of the claim of string
theory to offer a formalism which allows for a unified description of all known
fundamental forces of nature.

In this section we briefly review the Maldacena conjecture and some of the
extensions invented during the last decade. Instead of giving an exhaustive
review, we merely draw a hopefully concise and consistent sketch of the whole
picture. In doing so we emphasize those features of the correspondence that are
most important for the developments in the subsequent sections. Classical
reviews which deal with the subject in depth are
refs.~\cite{Aharony:1999ti,D'Hoker:2002aw}. For the sake of clarity, we
restrict explicit calculations to a minimum here.

\subsection{The original AdS/CFT correspondence}

The \ads/CFT correspondence as it was formulated by Maldacena in 1997 relates two
particular theories which we introduce subsequently. Afterwards we rephrase the
conjecture of Maldacena, before we comment on the relaxation of the underlying
assumptions, other generalizations and the applicability to QCD and the
quark-gluon plasma in the following sections.

	\subsubsection[$\N=4$ \, Super-Yang-Mills theory]{\boldmath$\N=4$\unboldmath{} \, Super-Yang-Mills theory}
	\label{sec:N4theory}
		
		One of the two theories related by the \ads/CFT correspondence is $\N=4$
		super-Yang-Mills theory in four spacetime dimensions of Minkowski
		topology. It is a supersymmetric quantum field theory with
		$\SU(\N=4)_\text{R}$ R-symmetry, which rotates the four supercharges
		into each other. All fields are arranged in one supersymmetry multiplet.
		The on-shell field content is given by six real spacetime scalars $X^i$
		with $i=1,2,\ldots,6$, one spacetime vector field $A$ and four two
		component spin~$1/2$ left Weyl fermions $\lambda^a$ with
		$a=1,2,3,4$. Under R-symmetry transformations the six scalars transform
		as an antisymmetric $\bm{\underline 6}$ of rank two. The Weyl fermions
		represent a $\bm{\underline 4}$ and the vector field is a singlet.
		
		With respect to \emph{gauge symmetries}, all the fields constitute one
		single multiplet, called the $\N=4$ gauge multiplet. They transform
		under the adjoint representation of the gauge symmetry group $\SU(N)$,
		where the integer $N$ is left as a parameter for now. Later we will
		interpret it as the number of color degrees of freedom.
		
		The according gauge indices labeling the elements of the gauge symmetry
		generators $T^k$ with $k=1,2,\ldots,N^2-1$ are suppressed in our
		notation, e.\,g.\  the notation $X^i$ for a spacetime scalar is the short
		form of $X^{(i)k}\,T^k$ where $T^k$ as a matrix has elements
		${(T^k)^m}_n$ labeled by $m,n=1,2,\ldots,N$. One would write
		out the elements of $X^i$ as ${X^{im}}_n$, where $i$ labels the
		index which transforms under the $\bm{\underline 6}$ of the R-symmetry
		and $m,n$ are the indices transforming under $\SU(N)$ gauge symmetries.
		We label spacetime directions by $\mu$ and $\nu$. With this convention
		and the field strength tensor $F=\dd A+A\wedge A$ the unique Lagrangian
		\cite{D'Hoker:2002aw} reads as
		\begin{equation}
		\label{eq:N4Lagrangian}
		\begin{aligned}
			\L = \; \tr\Bigg[ & -\frac{1}{2\gym^2} F_{\mu\nu}F^{\mu\nu} + \frac{\theta_I}{8\pi^2}F_{\mu\nu}\,\hodge F^{\mu\nu} - i  \bar{\lambda}^a \bar\sigma^\mu D_\mu \lambda_a 
			 -  D_\mu X^i \, D^\mu X^i +  \gym\,C^{ab}_i \lambda_a \commute{X^i}{\lambda_b}\\
			& +  \gym\,C_{iab} \bar{\lambda}^a \big[X^i,\bar{\lambda}^b\big] + \frac{\gym^2}{2}  \big[X^i,X^j\big]^2 \Bigg],
		\end{aligned}
		\end{equation}
		where the trace is performed over the suppressed gauge indices and $D$
		is the gauge covariant derivative. The symbol $\theta_I$ denotes the
		real valued instanton angle, the $C_i^{ab}$ and $C_{iab}$ are related to
		the structure constants of the R-symmetry group, and there is one
		dimensionless coupling constant $\gym$ in this Lagrangian. The energy
		dimensions of the operators and constants are given by
		\begin{equation}
			[A]=[X^i]=1, \qquad [\lambda^a]=\frac{3}{2},\qquad [\gym]=[\theta_I]=0,
		\end{equation}
		so the action $S=\int\dd^4x\:\L$ is scale invariant. In fact $\N=4$
		theory is invariant under transformations generated by the conformal
		symmetry group $\SO(4,2)\cong\SU(2,2)$ composed by Poincar\'e
		transformations, scaling and so-called superconformal transformations as
		well as under the above mentioned R-symmetry group $\SU(4)_\text{R}$.
		These transformations compose the global symmetry group denoted by
		$\PSU(2,2\mathop{|}4)$. Note that these symmetries are realized also in
		the quantized theory and not broken by anomalies. Moreover, the
		Lagrangian of $\N=4$ super-Yang-Mills theory is \emph{unique}. In
		contrast to other supersymmetric theories, which allow for different
		choices of the potential for the superfields, the form of the action is
		completely determined by the demand for renormalizability.
		
		There even is a further symmetry. With respect to the complex
		combination of the coupling $\gym$ and the instanton angle $\theta_I$
		given by
		\begin{equation}
			\tau = \frac{\theta_I}{2\pi} + \ii\,\frac{4\pi}{\gym^2}
		\end{equation}
		the action is invariant under $\tau\mapsto\tau+1$. Generalizing this
		symmetry, the Montonen-Olive conjecture states that the theory is
		invariant under $\SL(2,\mathbbm{Z})$ transformations acting on the
		complex coupling $\tau$, this symmetry is denoted as S-duality. It
		includes a transformation $\tau\mapsto -1/\tau$ which
		indicates that the theory describes a duality between strongly and
		weakly coupled regimes. We will not make use of this duality, though.

		\subsubsection*{The large $N$ limit and the connection to string theory}
		\label{sec:largeNlimit}
		
\begin{figure}
				\centering
				\includegraphics[width=.7\linewidth]{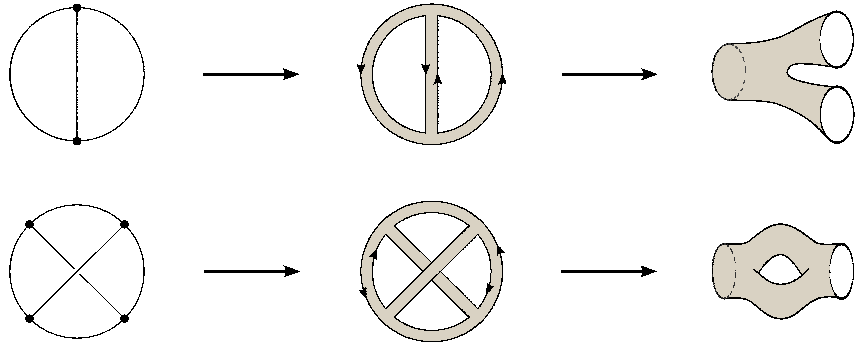}
				\caption[Feynman, double line, and string diagrams]{
					Feynman diagrams (left) can be translated to double line
					diagrams (middle), which in turn can be interpreted as
					Riemann surfaces of well defined topology (shaded).
					These surfaces (deformed to the shape on the right) can
					be interpreted as stringy Feynman diagrams.
				}
				\label{fig:doubleLine}
			\end{figure}

			The $\N=4$ super Yang-Mills theory introduced here is strongly
			related to string theory, which becomes visible in the limit of
			asymptotically many colors, $N\to\infty$, while the effective
			coupling $\lambda=\gym^2N$ is kept fix. This is the so-called
			\emph{'t~Hooft limit} \cite{Hooft:1973jz}. The motivation to
			consider the large $N$ limit is to find a parameter which allows for
			perturbative calculations in strongly coupled gauge theories, namely
			1/N. To illustrate this in a simplified way for the
			theory given by \eqref{eq:N4Lagrangian}, we note that one can
			schematically write the interaction terms of this Lagrangian as
			\begin{equation}
				\L \sim \tr \left[ \partial\Phi_i\,\partial\Phi_i 
					+ \gym c^{ijk}\Phi_i\Phi_j\Phi_k
					+ \gym^2 d^{ijkl}\Phi_i\Phi_j\Phi_k\Phi_l\right],
			\end{equation}
			where $\Phi_i$ are any of the bosonic fields $X_i$ or $A$ (and the
			fermions are related to them by supersymmetry). Note that the
			three-point vertices are proportional to $\gym$ while the four-point
			vertices are proportional to $\gym^2$. After the introduction of
			$\tilde\Phi_i=\gym\Phi_i$ the Lagrangian acquires the form
			\begin{equation}
			\label{eq:newPhiLagr}
				\L \sim \frac{1}{\gym^2} \tr \left[ \partial\tilde\Phi_i\,\partial\tilde\Phi_i 
					+ c^{ijk}\tilde\Phi_i\tilde\Phi_j\tilde\Phi_k
					+ d^{ijkl}\tilde\Phi_i\tilde\Phi_j\tilde\Phi_k\tilde\Phi_l\right].
			\end{equation}
			Remember that all fields $\Phi_i$ of $\N=4$ SYM theory transform in
			the adjoint representation of the gauge group $\SU(N)$. So $\Phi_i$
			can be written in a matrix notation, where the elements of the
			matrix are denoted by ${(\Phi_i)^a}_b$, with
			$a,b=1,2,\ldots,N$ transforming in the fundamental and
			antifundamental representation, respectively. In a Feynman diagram a
			propagator for some particle $\Phi_i$ then corresponds to a double
			line, with one line corresponding to the upper and one to the lower
			index $a$, $b$ of the gauge group, see figure~\ref{fig:doubleLine}.
			
			In this double line notation we can now order diagrams in an
			expansion parametrized by $1/N$ to see that the
			contributions to gauge invariant processes may be ordered according
			to the Euler characteristic of the Feynman diagram, i.\,e.\  they are
			ordered according to topology of the diagram. As an example consider
			the diagrams in fig.~\ref{fig:doubleLine}. For the amplitude
			corresponding to some Feynman diagram, a propagator introduces a
			factor of $\gym^2=\lambda/N$ while the above Lagrangian
			\eqref{eq:newPhiLagr} shows that vertices pick up a factor of
			$1/\gym^2=N/\lambda$. From the double line
			notation it is clear that each closed line therein represents a loop
			which introduces a factor $N$. Moreover, think of the diagrams as
			describing polyhedrons which are characterized by vertices, edges
			(propagators), and faces which are the regions separated by the
			edges. We observe that the factors of $N$ for a diagram with $V$
			vertices, $E$ edges and $F$ faces (i.\,e.\  loops of lines in the double
			line diagrams) appear in powers of
			\begin{equation}
				N^{V-E+F} \lambda^{E-V} = N^{\chi} \lambda^{E-V} = N^{2-2g} \lambda^{E-V}.
			\end{equation}
			The number $V-E+F=\chi=2-2g$ is the Euler character of the
			polyhedron described by a Feynman diagram. The genus of the
			corresponding Riemann surface is given by $g$. From this dependence
			on $N$ we see that diagrams with smallest $g$, i.\,e.\  planar diagrams
			with $g=0$, contribute with highest order, while diagrams with
			topologies of higher genus $g$ are suppressed by factors of $N^{2g}$
			relative to the planar ones. In this way any process in the field
			theory can be decomposed into diagrams ordered by their genus $g$ in
			the double line notation. The amplitude $\mathcal M$ of a given
			process may then be obtained by a sum of the contributions from all
			relevant Feynman diagrams,
			\begin{equation}
				\mathcal M = \sum_{g=0}^\infty N^{2-2g}\,f_g(\lambda).
			\end{equation}
			This type of expansion is exactly the same as the one obtained by
			performing an expansion of diagrams describing the interaction of
			closed oriented strings, the type II string theories, upon
			recognizing the parameter
			$1/N=\gym^2/\lambda$ as being proportional
			to \gs, the string coupling constant \cite{Aharony:1999ti}. The hope
			is, that the topologies of Feynman diagrams reflect the contribution
			of the string theory diagrams with the same topology. From the
			standard examples shown in figure~\ref{fig:doubleLine} we see that
			the genus $g$ represents the number of loops in the associated
			string theory diagram.
			
			The large $N$ limit corresponds to weakly coupled string theory,
			as $\gs\propto\lambda/N$. In this limit we only have to
			consider the leading diagrams with genus $g=0$. These are the gauge
			theory processes described by planar diagrams, corresponding to tree
			level diagrams in string theory.
			
			These arguments are of heuristic nature. For instance, there are
			effects like instantons in a gauge theory, which can not be treated
			in a $1/N$ expansion. Such effects therefore should match
			the according non-perturbative effects in string theory.

	\subsubsection{Type IIB supergravity}
		
		The preceding section introduced the quantum field theory, which
		represents one of the two theories connected by the \ads/CFT
		correspondence. The second theory is type~IIB supergravity. Supergravity
		theories are supersymmetric gauge field theories containing a spin $2$
		field identified with the graviton, the quantum field of gravitation.
		Supergravity thereby is an attempt to combine supersymmetric field
		theory with general relativity.
		
		Even though Supergravity is an interesting field to study on its own
		right, it can be embedded in a larger and more general framework. In
		fact supergravity is a certain limit of string theory. A brief comment
		on this perception will follow below. As string theory revealed that a
		consistent description of the forces and matter of nature requires ten
		spacetime dimensions, we will be interested in a formulation of
		supergravity in ten-dimensional backgrounds. There are different
		supergravity theories in ten spacetime dimensions, which can be
		constructed from compactifications of a \emph{unique} causal unitary
		$11$-dimensional supergravity theory \cite{D'Hoker:2002aw}.
		
		The one formulation we will make use of throughout this monograph, and
		which is the one most intimately connected with the \ads/CFT
		correspondence is the so-called type~IIB supergravity in ten spacetime
		dimensions. This theory is a $\N=2$ supersymmetric theory with a field
		content given by the bosonic fields $G$ (symmetric rank $2$, the
		metric), $C$ (scalar, axion), $\Phi$ (scalar, dilaton), $B$ (rank $2$
		antisymmetric, Kalb-Ramond field), $A_2$ (rank $2$ antisymmetric), $A_4$
		(self dual rank $4$ antisymmetric). The fermions of the theory satisfy
		Majorana-Weyl conditions and are given by two $\psi^I$ ($I=1,2$, spin
		$3/2$ gravitinos of same chirality) and two fields $\lambda^I$
		($I=1,2$, spin $/1/2$ dilatinos of same chirality, which is
		opposite to the chirality of the gravitinos). This theory is chiral in
		the sense that it is parity violating \cite{D'Hoker:2002aw}.
		
		The action of type~IIB supergravity may be written down in terms of the
		field strengths
		\begin{xalignat}{2}
			F_1 &= \dd C, & H_3 &=\dd B, \\
			F_3 &= \dd A_2, & \tilde F_3 &= F_3-CH_3,\\
			F_5 &= \dd A_4, & \tilde F_5 &= F_5 - \frac{1}{2}\, A_2\wedge H_3 + \frac{1}{2}\,B\wedge F_3,
		\end{xalignat}
		and then reads
		\begin{equation}
		\label{eq:IIBLagr}
			\begin{aligned}
				S_\text{IIB}	=\; & +\frac{1}{2\kappa^2} \int\!\!\dd^{10}x\; \sqrt{|\det G|}\left( e^{-2\Phi} \left( 2 \mathcal R + 8 \partial_\mu\Phi\,\partial^\mu\Phi - |H_3|^2\right)
						  -  |F_1|^2 - |\tilde F_3|^2 - |\tilde F_5|^2\right)\\
						&  -\frac{1}{2\kappa^2} \int  A_4\wedge H_3\wedge F_3 
						  + \text{fermions},
			\end{aligned}
		\end{equation}
		where $\kappa$ is the Newton constant and $\mathcal R$ is the Ricci
		scalar. Additionally, at the level of the equations of motion one has to
		impose the self-duality constraint
		\begin{equation}
			\hodge\tilde F_5 = \tilde F_5.
		\end{equation}
		
		\subsubsection*{Type~IIB supergravity as a string theory limit}
		
		Starting from the Polyakov action to describe string world sheets,
		tachyonic string modes were discovered in the derived spectrum. These
		tachyons indicate an instability of the theory. To arrive at a stable
		and causal theory, one should remove these tachyonic excitations from
		the spectrum. To do so, one may modify the Polyakov action by
		introducing supersymmetry, and truncate the spectrum of physical states
		in a consistent way, a procedure called GSO-Projection (after the
		inventors Gliozzi, Scherk and Olive). This projection exactly leaves a
		spacetime supersymmetric spectrum.
		
		This procedure not only removes the tachyonic ground state from the
		closed string spectrum but additionally demands a number of $10$
		spacetime dimensions to preserve causality. The lowest remaining modes
		after the GSO projection represent the ground state of the remaining
		theory. For instance, in the Neveu-Schwarz sector of the string
		excitations this ground state happens to be described by massless
		excitations of strings. The so-called level matching condition demands
		this state to be generated from a vacuum $\ket{0}$ by the action of two
		creation operators, a left- and a right-moving one,
		$\alpha^\mu_{-1}\tilde\alpha^\nu_{-1}\ket{0}$. These excitations may be
		described by a tensor valued field $M$ with components $M^{\mu\nu}$.
		This field in turn decomposes into a symmetric part with components
		$G^{\mu\nu}$ (describing the degrees of freedom of the graviton),
		antisymmetric components $B^{\mu\nu}$ (the $B$-field) and the scalar
		$\Phi$ (the dilaton) that determines the trace of $M$.
		
		Computing the masses of string excitations generated by more than two
		creation operators acting on the vacuum, unveils that these excitations
		describe fields which represent particles of finite positive mass
		proportional to $1/\alpha'$. In a low energy theory compared
		to the energy scale of inverse string length, or equivalently on length
		scales that do not resolve the stringy nature of the fundamental theory
		one may approximate the strings by pointlike particles, effectively
		described as $\alpha'\to 0$. In this limit, however, all massive modes
		gain infinite masses and will not effect the low energy dynamics. The
		low energy theory may therefore only contain particles described by the
		massless supersymmetry multiplet to which the fields $B$, $G$ and $\Phi$
		belong. The action remaining for these fields exactly describes the
		supergravity action. We are interested in the sector of closed string
		excitations with same chirality for the left and right moving
		excitations, which is called type~IIB string theory and leads to
		type~IIB supergravity in the low energy limit.
		
		In the action \eqref{eq:IIBLagr} above, the fields $G$, $B$ and $\Phi$
		can be found in the first line, they originate from the Neveu-Schwarz
		sector (NS-NS) of the string theory fields, while the second and third
		lines contain the Ramond sector (R-R) contributions.

		\subsubsection*{Extremal p-brane solutions and D-branes}
			
			Solutions to the supergravity equations of motion with non-trivial
			charges of $(p+1)$-forms $A_{p+1}$ are called \emph{p-branes}. These
			solutions exhibit Poincar\'e-invariance in $(p+1)$ dimensions, their
			name thus stems from the number $p$ of spatial dimensions included
			in this symmetry group. In this sense these solutions are higher
			dimensional generalizations of membranes, which one would denote as
			$2$-branes in this context.
			
			Note that the flux $f$ of the field strength $F_{p+2}=\dd A_{p+1}$
			through some surface $\Sigma$ is conserved since $\dd f =
			\int_\Sigma \dd F_{p+2} =0$, as $F_{p+2}$ is an exact and thus
			closed form. Moreover, the electric coupling of the $p$-form to the
			$p$-brane with worldvolume $\Sigma_{p+1}$ of spacetime dimension
			$p+1$ can be described by the diffeomorphism invariant action
			\begin{equation}
				S_{p} = T_{p}\int\limits_{\Sigma_{p+1}} \!\!\!\!A_{p+1}.
			\end{equation}
			The proportionality constant $T_{p}$ denotes the tension of the
			$p$-brane. It has the interpretation of the energy or mass per unit
			area of the worldvolume,
			\begin{equation}
			\label{eq:braneTension}
				T_p = \frac{2\pi}{\gs \, (2\pi\ell_s)^{p+1}}.
			\end{equation}
			
			In type~IIB supergravity, there are $0$-forms, $2$-forms and
			$4$-forms, allowing for the following $p$-brane solutions. The
			$0$-forms allow for $(-1)$-brane solutions, so-called
			\D{(-1)}-instantons. Then there are $1$-branes, charged under and
			thus coupling to the according solutions of the $B$-field. The two
			dimensional $1$-branes are identified with the worldsheet of the
			fundamental strings of the underlying string theory. They are called
			F$1$-strings. The $1$-branes which couple to the $A_2$ field are
			called \D1-strings, and the $3$-brane solutions according to the
			$A_4$ field are called \db3s.
			
			The magnetic analogon to the electric couplings are given by the
			Hodge dual field strengths. The magnetic dual field strength to
			$F_n$ in a ten-dimensional background, is the $(10-n)$-form $\hodge
			F_n$, which has a $(9-n)$-form field as its potential. This in turn
			couples to a $(8-n)$-brane. In this way the type~IIB field strengths
			$F_1$ and $F_3$ allow for magnetic couplings to \db7s and \db5s.
			
			The naming of branes as D$p$-branes we just saw, arises from string
			theory. As we can see, the D$p$-branes are coupling to fields in the
			Ramond sector. The letter D is short for the \emph{Dirichlet
			boundary conditions} such a brane imposes on the dynamics of the
			endpoints of open strings. Namely, in string theory D-branes are
			identified with the surfaces on which open strings end
			\cite{Polchinski:1995mt,Johnson:2003gi}. The endpoints of these
			strings then have a well defined position in the direction
			perpendicular to the brane, namely the position of the brane. Such a
			specification of a certain value for a actually dynamical quantity
			is known as the imposition of Dirichlet boundary conditions. It is
			believed that the $p$-brane solutions in the supergravity limit of
			string theory may be identified with D$p$-branes in full string
			theory.
			
			The fact that $p$-branes are $(p+1)$-dimensional Poincar\'e
			invariant imposes restrictions on the metric. For instance, some
			$d$-dimensional spacetime which supports a $p$-brane will include a
			Poincar\'e invariant subspace with symmetry group
			$\mathbbm{R}^{p+1}\times \SO(1,p)$. Additionally, one can always
			find solutions which are maximally rotationally invariant in the
			$(d-p-1)$-dimensional space transverse to the brane. Thus, in
			particular a ten-dimensional spacetime supporting \db3s has an
			isometry group of $\mathbbm{R}^4\times \SO(1,3)\times \SO(6)$.
			
			Analog to Reissner-Nordstr\"om black holes in general relativity,
			the possible solutions of supergravity backgrounds can be
			parametrized by the mass $M$ of the $p$-brane solution and its RR
			charge $N$, which are functions of two parameters $r_\pm$, which can
			be interpreted as horizons of the solution. The case $r_+<r_-$
			exhibits a naked singularity and is therefore regarded as
			unphysical. In the limit of $r_+=r_-$ the brane is said to be an
			\emph{extremal $p$-brane}, while it is a \emph{non-extremal} black
			brane for $r_+>r_-$, with an event horizon. For details refer to
			ref.~\cite{Aharony:1999ti}. We restrict our attention to the case of
			extremal $p$-branes.
			
			The most general form of an extremal $p$-brane metric can be written
			in terms of a function $H$ as \cite{D'Hoker:2002aw}
			\begin{equation}
			\label{eq:d-braneMetric}
				\ds^2 = {H(\breve{y})}^{-\frac{1}{2}}\, \eta_{\mu\nu} \dd x^\mu \dd x^\nu + {H(\breve{y})}^\frac{1}{2} \, \dd \breve{y}^2.
			\end{equation}
			Here, the coordinates of the vector $\breve{y}$
			parametrize the space transverse to the brane, and $\eta$ is the
			$(p+1)$-dimensional Minkowski metric.
			
			Supported by the insight that D-branes are dynamical objects of the
			theory \cite{Polchinski:1995mt}, one can adopt the point of view
			that the above geometry is generated by a stack of $N\in\mathbbm N$
			branes placed in an initially flat $d$-dimensional Minkowski
			spacetime at a positions $\breve{y}_i$, with
			$i=1,2,\ldots,N$. Asymptotically far away from the stack one can
			therefore expect the whole spacetime to become flat again. String
			theory calculations then restrict the function $H(\breve{y})$ to
			\begin{equation}
				\label{eq:Hfunc}
				H(\breve{y}) = 1 + \sum_{i=1}^{N}\frac{\gs \, (4\pi)^{(5-p)/2}\, \Gamma\left(\text{$\frac{7-p}{2}$}\right){\alpha'}^{(d-p-3)/2}}{\left| \breve{y} - \breve{y}_i \right|^{d-p-3}},
			\end{equation}
			where $\gs$ is the string coupling constant and $\alpha'$
			parametrizes the string tension.
			
			Of special interest for this article are \db3s and \db7s. For the
			introduction of the \ads/CFT correspondence, it is useful to look at
			\db3s first. \db7s will become an important ingredient for
			generalizations of the correspondence.

		\subsubsection*{\D3-branes and Anti-de~Sitter space}
		\label{sec:D3andAdS}

			There are several aspects which make \db3s especially interesting.
			First of all, \db3s by definition introduce four-dimensional
			Poincar\'e symmetry, the resulting ten-dimensional geometry for
			$p=3$ is regular. Moreover, the solution for the axion and dilaton
			fields ($C$ with $F_1=\dd C$ and $\Phi$ in \eqref{eq:IIBLagr}) can
			be shown to be constants. In addition, the field strength $F_5$ is
			self-dual. For our considerations the metric will be a central
			quantity. Especially the case of $N$ coincident \db3s located at a
			position $y_\text{\tiny \D3}$ in a spacetime of dimension $d=10$
			will be important. From \eqref{eq:Hfunc} we see that the function
			$H(\vec y)$ in this case is given by
			\begin{equation}
				H(\breve{y}) = 1+ \frac{4\pi \gs N{\alpha'}^2}{\left|\breve{y} - \breve{y}_\text{\tiny \D3}\right|^4}\,.
			\end{equation}
			We introduce the quantity $R$ simply as an abbreviation,
			\begin{equation}
				\label{eq:defineR}
				R^4=4\pi \gs N{\alpha'}^2.
			\end{equation}
			However, a few lines below we will see that this parameter has a
			crucial geometric interpretation. With $R$ we write
			$H(\breve{y})$ as
			\begin{equation}
				H(\breve{y}) = 1+\frac{R^4}{\left|\breve{y} - \breve{y}_\text{\tiny \D3}\right|^4}\,.
			\end{equation}
			By a coordinate shift we may always denote the position
			$\breve{y}_\text{\tiny \D3}$ as the origin of
			the $y$ coordinates and set it to zero. The distance from the brane
			will be denoted by $r=|\breve{y}|$. The metric
			\eqref{eq:d-braneMetric} generated by a stack of \db3s therefore may
			be written as
			\begin{equation}
			\label{eq:d3metric}
				\ds^2 = \left(1+\frac{R^4}{r^4}\right)^{-\frac{1}{2}}\dd x_\mu\dd x^\mu + \left(1+\frac{R^4}{r^4}\right)^\frac{1}{2}\left(\dd r^2+ r^2 \dd\Omega_5^2\right).
			\end{equation}
			Far away from the stack of branes, at large $r\gg R$, where the
			influence of the branes on spacetime will not be sensible, the
			metric is asymptotically flat ten-dimensional Minkowski spacetime.
			However, in the limit of $r\to 0$ the metric appears to be
			singular. This limit is therefore known as the \emph{near horizon
			limit}. In fact spacetime is not singular in this limit but
			develops constant (negative) curvature. Because space is flat at
			large $r$, but has constant curvature at $r\to 0$ this limit is also
			referred to as the \emph{throat region}. In the near horizon limit
			at small $r\ll R$ the metric asymptotically becomes
			\begin{equation}
			\label{eq:AdS5S5metricInr}
				\ds^2 = \frac{r^2}{R^2}\,\dd x_\mu \dd x^\mu  + \frac{R^2}{r^2}\dd r^2  + R^2\dd\Omega_5^2.
			\end{equation}
			This is the product space $\adsfivesfive$, where the first two terms
			describe what is known as five-dimensional Anti-de~Sitter space, or
			$\ads_5$ for short. The parameter $R$ is called the radius of
			\ads space. The last term represents the familiar five-dimensional
			sphere, of radius $R$ as well. The geometry of Anti-de~Sitter space
			is crucial for the gauge/gravity duality. To discuss some
			properties we introduce the coordinate $z=R^2/r$ and write the
			metric as
			\begin{equation}
			\label{eq:AdS5S5metric}
				\ds^2 = \frac{R^2}{z^2}\left(\dd x_\mu \dd x^\mu + \dd z^2 \right) + R^2\dd\Omega_5^2.
			\end{equation}
			The metric \eqref{eq:AdS5S5metric} can be derived as the induced
			metric of a five-dimensional hypersurface which is embedded into a
			six-dimensional spacetime with metric
			\begin{equation}
			\label{eq:embedMetric}
				\dd s^2_6 = -\dd X_0^2 + \sum_{i=1}^4 \dd X_i^2 \pm \dd X_5^2,
			\end{equation}
			where the $X_i$ parametrize the six-dimensional space and the choice
			of the ambiguous sign depends on whether we aim for a metric on
			$\ads_5$ with Euclidean or Minkowski signature. Originally, the
			\ads/CFT correspondence was conjectured for Euclidean signature. The
			hypersurface which defines $\ads_5$ obeys
			\begin{equation}
			\label{eq:adsEmbed}
				X_\mu X^\mu = - R^2.
			\end{equation}
			We now parametrize this surface with so-called Poincar\'e
			coordinates $z\geq0$ and $x^\mu\in\mathbbm{R}$ with $\mu=0,1,2,3$,
			such that $x_\mu x^\mu = \pm (x^0)^2 + (x^1)^2 + (x^2)^2 + (x^3)^2$
			with the sign corresponding to the one in \eqref{eq:embedMetric} and
			\begin{equation}
				\begin{aligned}
					X_0 &= \frac{R^2 + z^2 + x_\mu x^\mu}{2z},\\
					X_4 &= \frac{R^2 - z^2 - x_\mu x^\mu}{2z},\\
					X_i &= R\,\frac{x^i}{z}, \qquad i=1,2,3,\\
					X_5 &= R\,\frac{x^0}{z}.
				\end{aligned}
			\end{equation}
			The hypersurface parametrized by $z$ and the $x^\mu$ fulfills
			\eqref{eq:adsEmbed} and therefore represents $\ads_5$. It has the
			induced metric
			\begin{equation}
			\label{eq:metricAdSonly}
				\ds^2 = \frac{R^2}{z^2}\left(\dd x_\mu \dd x^\mu + \dd z^2 \right)
			\end{equation}
			which appears as the first factor of the product spacetime
			\eqref{eq:AdS5S5metric}. Note that for the Minkowski signature
			background the restriction $z\geq 0$ leaves only one of the two
			separate hyperboloids described by \eqref{eq:adsEmbed}. The other
			half is parametrized by $z\leq 0$ and is a clone of the part we use.
			  The spacetime
			coordinates parametrized by $\vec x$ suggest to be related to four
			dimensional Euclidean or Minkowski spacetime, depending on the
			choice of sign in \eqref{eq:embedMetric}. The coordinate $z$ on the
			other hand is called the \emph{radial coordinate} of \ads\ space.
			When we establish the \ads/CFT dictionary we will pay special
			attention to the behavior of fields near the so-called
			\emph{conformal boundary} of \ads\ space. It is defined as the
			projective boundary which lies at $z\to0$ in the coordinates at
			hand. In the embedding space introduced above the boundary would be
			infinitely far away from the origin of the coordinate system. The
			metric \eqref{eq:metricAdSonly}, however, is diverging at the
			boundary, except we rescale it \cite{Witten:1998qj}. A scale factor
			$f(z)$ with a first order root of $f$ at $z=0$ will exactly cancel
			the divergence after rescalings
			\begin{equation}
				\ds^2 \mapsto f^2(z)\,\ds^2.
			\end{equation}
			As we are free to choose the function $f(z)$ as long as we do not
			introduce new roots in $f(z)$ or change the order of the root at
			$z=0$, we can choose between a family of rescaling functions $f$,
			which are related by some arbitrary function $w(z)$ as
			\begin{equation}
				f(z)\mapsto f(z)\, e^{w(z)}.
			\end{equation}
			This freedom therefore expresses the fact that the boundary of
			Anti-de~Sitter space is only well defined up to \emph{conformal
			rescalings}. Then the boundary at $z=0$ represents four-dimensional
			Euclidean or Minkowski spacetime, defined up to conformal
			rescalings.
			
			For later reference we point out the isometry group of \adsfivesfive
			here. The Lorentzian version with negative sign in
			\eqref{eq:embedMetric} clearly displays an $\SO(2,4)$ rotational
			invariance in the $\ads_5$ subspace, while the isometry group of the
			five sphere is $\SO(6)$.
			
			We interpreted supergravity as a limit of string theory in the last
			paragraphs. Moreover, we will work in the near horizon limit from
			now on. Consequently, we will work in a background spacetime with
			the topology of \adsfivesfive. The non-vanishing curvature of
			$\ads_5$ spacetime can be characterized by the Ricci scalar
			\begin{equation}
				\mathcal R = \frac{20}{R^2}.
			\end{equation}
			String theory, however, is not solved in curved backgrounds so far.
			To allow for a good approximation of type~IIB string theory by
			working in the supergravity limit, one should therefore arrange
			spacetime curvature to be small. A large \ads\ radius leads to small
			curvature. Note that by \eqref{eq:defineR} the relation of $R$ to
			the string scale $\ell_s=\sqrt{\alpha'}$ depends on two parameters
			of the theory. These are $N$ and the string coupling constant $\gs
			=e^\Phi$, which can be tuned by specifying a value of the arbitrary
			constant dilaton field $\Phi$. We thus see that the supergravity
			approximation seems to be valid only for $\gs N\ggg 1$. This
			guaranties $R\gg\ell_s$, such that the radius of the string theory
			background is large compared to the string length $\ell_s$. In this
			way $\gs N\ggg 1$ ensures that the strings do not resolve the curved
			nature of the background, and type~IIB string theory can be trusted
			as a good approximation to string theory on \adsfivesfive.
			
	\subsubsection{The Maldacena conjecture}
	\label{sec:theconjecture}
		
		In a famous publication from the year 1997, Juan Maldacena pointed out
		that there exists a connection between certain quantum field theories
		and classical supergravity theories \cite{Maldacena:1997re}. In
		particular, the degrees of freedom found in type~IIB supergravity on
		\adsfivesfive contain the large coupling limit of the $\N=4$ SYM theory
		in four dimensions.
		
		As a generalization, consider full string theory instead of the
		supergravity limit, and relax the limit of large coupling on the quantum
		field theory side. Maldacena then conjectured the equivalence of two
		theories, formulated as the \emph{\ads/CFT correspondence}. We
		summarize it as follows:
		
		\begin{quotation}
			Computations of observables, states, correlation functions and their
			dynamics yield the same result in the following two theories, which
			may therefore be regarded as physically equivalent.
			
			On the one side (\ads) there is $10$-dimensional type~IIB string
			theory on the spacetime \adsfivesfive. The $5$-form flux through the
			$\S^5$ given by the integer $N$, and the equal radii $R$ of $\ads_5$
			and $\S^5$ are related to the string coupling constant $\gs$ by
			$R^4=4\pi \gs N{\alpha'}^2$.\label{parameterrelations}
			
			On the other side (conformal field theory, CFT) of the
			correspondence there is a conformally symmetric four-dimensional
			$\N=4$ super-Yang-Mills theory with gauge group $\SU(N)$ and
			Yang-Mills coupling $\gym$, related to the string coupling by
			$\gym=2\pi\gs ^2$.
			
			This equivalence is conjectured to hold for any value of $N$ and
			$\gym$.
		\end{quotation}
		
		\noindent It is a remarkable feature of this correspondence that it
		relates a theory containing gravity to a quantum field theory, which
		otherwise lacks any description of gravity. In the supergravity limit, a
		$10$-dimensional \emph{classical} theory of gravity matches a
		four-dimensional \emph{quantum theory}. In fact a dictionary between
		operators of the quantum field theory and the supergravity fields can be
		established. We will comment on this below. However, the correspondence
		in its strong form, given above, is very general and thus allows hardly
		any applications. For instance, so far there is no formulation of string
		theory on curved spaces, such as \adsfivesfive. Nevertheless, there are
		interesting non-trivial limits in which explicit computations can be
		performed.
		
		The \emph{'t~Hooft limit} is defined as considering a \emph{fixed} value
		of the 't~Hooft coupling $\lambda=\gym^2N$ while $N \rightarrow \infty$.
		This yields a simplification of Feynman diagrams of the field theory. As
		we saw in section~\ref{sec:largeNlimit}, in this limit only planar
		diagrams contribute to physical processes. Note that a fixed value of
		$\lambda$ in the large $N$ limit implies weak coupling on the string
		theory side as the coupling constants are related by $2\pi\gs=\gym^2$.
		So on the string theory side this results in the limit of a classical
		string theory (no string loops) on \adsfivesfive.
		
		The \emph{Maldacena limit} implements a further restriction. Starting
		from the 't~Hooft limit, we let $\lambda\to\infty$. This of course
		prohibits perturbative computations on the field theory side, since here
		$\lambda$ is the effective coupling parameter. On the string theory
		side, though, this limit results in $\alpha'/R^2\to0$. So the curvature
		of the string theory background becomes small compared to the string
		length, which allows for consistent applications of the classical
		supergravity limit of string theory, which does not resolve the stringy
		nature of the fundamental building blocks of matter.
		
		Thus, working in the Maldacena limit not only allows to describe a
		quantum field theory in terms of a classical theory of gravity. It also
		allows to investigate the strongly coupled regime of the quantum field
		theory by performing calculations in the weakly coupled regime of the
		dual theory, where perturbative methods are applicable.
		
		The conjecture is not an ad hoc statement, but rather results from
		string theory arguments. Consider a stack of $N$ coincident \db3s which
		interact with open strings. In the low energy limit $\alpha'\to 0$ we
		have to consider infinitely short strings since $\ell_s^2=\alpha'$.
		These strings may end on any of the $N$ branes on the stack. As all the
		branes are coincident we can not distinguish between them, which implies
		an $\U(N)\cong\U(1)\times\SU(N)$ symmetry of the theory, where the
		$\U(1)$ factor gives the position of the brane and does
		not play a role here. It can be shown that the \db3s' solutions exhibit
		$\N=4$ supersymmetry.  Therefore in the low energy
		limit this theory describes precisely the conformal $\N=4$ $\SU(N)$
		gauge theory. Our special interest is the behavior of the strongly
		coupled regime of this theory, which is not accessible by perturbation
		theory. Instead of first taking the low energy limit and then the large
		coupling limit, we look at what happens if we proceed in reverse order.
		Starting from the stack of branes we are now interested in the strong
		coupling limit. From section~\ref{sec:D3andAdS} we know that the near
		horizon geometry in this case will have the topology of \adsfivesfive
		with radius $R^4=4\pi\gs N{\alpha'}^2$, so we are forced to consider
		string theory on curved backgrounds. We also mentioned that the low
		energy limit of string theory is captured by supergravity. If we adopt
		the attitude that the physics of our system should be the same
		regardless of the order in which we impose the limits, then in the
		Maldacena limit we have to consider strongly coupled gauge theory and
		supergravity as two descriptions of the same physical setup.

		The quantum field theory may be interpreted as a description of the
		dynamics of open strings ending on the \db3s. In the low energy limit
		$\alpha'\to 0$ the degrees of freedom (strings) are confined to the
		domain of the \db3s. In the \adsfivesfive geometry of the string theory
		background \eqref{eq:AdS5S5metric} this domain is parametrized by the
		coordinates along the boundary of $\ads_5$. So we can say that the
		\ads/CFT correspondence describes how a four-dimensional field theory
		defined on the boundary of five-dimensional \ads\ space encodes the
		information of a higher dimensional theory. In analogy to conventional
		holograms which encode three-dimensional information on a lower
		dimensional hyperspace (namely a two-dimensional surface), the \ads/CFT
		correspondence is said to realize the \emph{holographic principle}.

	\subsubsection{An AdS/CFT dictionary}
	\label{sec:dictionary}
		
		So far we recognized that the gauge/gravity duality allows for the
		reformulation of some problem defined in a gauge theory in terms of a
		gravity theory. In order to obtain quantitative answers, it is necessary
		to identify the corresponding quantities in both theories. The
		supergravity theory is formulated in terms of classical fields on a ten
		dimensional background, while the $\N=4$ SYM theory describes the
		dynamics of operators acting on quantum states in four spacetime
		dimensions. The relations between the parameters of the theories were
		introduced with the correspondence in section \ref{parameterrelations}. For
		the coupling constants $\gym$, $\gs$ and $\lambda$, as well as the \ads
		radius $R$, the string tension $\alpha'$ and the number of colors $N$,
		they are
		\begin{equation}
			R^4=4\pi\gs N\alpha'^2,\qquad 2\pi\gs=\gym^2,\qquad\lambda=\gym^2 N.
		\end{equation}
		
		Observables, however, are expressed in terms of correlation functions of
		gauge invariant operators of the quantum field theory. It is possible to
		translate correlation functions of the field theory to expressions in
		terms of supergravity fields. A precise prescription of how to
		accomplish this was given in two seminal papers from 1998 by Edward
		Witten \cite{Witten:1998qj}, and Gubser, Klebanov, Polyakov
		\cite{Gubser:1998bc}. As a result it is possible to establish a complete
		dictionary, which translates quantities from on side of the
		correspondence to the other.
		
		Since the domain on which the field theory is defined can be identified
		with the boundary of $\ads_5$ space, one can imagine supergravity fields
		$\phi$ in $\ads_5$ to interact with some conformally invariant operator
		$\mathcal O$ on the boundary. We denote the boundary value of the
		supergravity field by $\phi_0=\lim_{\partial\ads_5}\phi$. A coupling
		would look like
		\begin{equation}
		\label{eq:boundaryInterAction}
			S_\text{int} = \int\limits_{\partial\ads_5}\!\!\!\!\dd^4x\; \phi_0(\vec x)\, \mathcal O(\vec x).
		\end{equation}
		In this sense the boundary value $\phi_0$ of the supergravity field acts
		as the source of the operator $\mathcal O$ in the field theory. Such an
		interaction term appears in the generating functional for correlation
		functions, which we write schematically as
		\begin{equation}
			\left<\exp \int\limits_{\partial\ads_5}\!\!\! \phi_0\, \mathcal O\right>_{\!\!\text{\raisebox{.75em}{CFT}}}.
		\end{equation}
		Witten's proposal was to identify the generating functional for
		correlation functions of operators $\mathcal O$ with the partition
		function $Z_\text{sugra}$ of the supergravity theory, which is given by
		\begin{equation}
			Z_\text{sugra}[\phi_0] = \exp\left(-S_\text{sugra}[\phi]\right) \Big|_{\phi=\phi_0},
		\end{equation}
		where $S_\text{sugra}$ is the supergravity action. So the ansatz for the
		generating functional of correlation functions of operators of the field
		theory can be written as
		\begin{equation}
			\left<\exp \int\limits_{\partial\ads_5}\!\!\! \phi_0\, \mathcal O\right>_{\!\!\text{\raisebox{.75em}{CFT}}}
			= \exp\left(-S_\text{sugra}[\phi]\right) \Big|_{\phi=\phi_0}.
		\end{equation}
		Correlation functions for $\mathcal O$ can then be obtained in the usual
		way by evaluating the functional derivative of the generating functional
		with respect to the source $\phi_0$ of the operator. Explicit
		calculations will be performed in later sections. As a general example,
		some two point function would be obtained by solving the supergravity
		equations of motion, plugging these solutions into the action
		$S_\text{sugra}$, then expressing the result in terms of solution
		$\phi_0$ on the boundary, and eventually evaluating
		\begin{equation}
		\label{eq:2pointPrescription}
			\left< \mathcal O(x)\,\mathcal O(y) \right> = 
			\left. \frac{\delta}{\delta\phi_0(x)}\frac{\delta}{\delta\phi_0(y)} \, \exp \left(-S_\text{sugra}[\phi]\right)\right|_{\phi_0=0}.
		\end{equation}
		
		The remaining question is which operators are dual to which fields. The
		fact that there exists such a dictionary relies heavily on the
		symmetries of the two related theories. The symmetry of a theory
		reflects the transformation behavior of the field content and by the
		Noether theorem accounts for the conserved quantities (charges). We can
		expect that two equivalent theories share the same amount of degrees of
		freedom, which must be reflected in their symmetries.
		
		To ensure a gauge invariant field theory action, including the source
		term \eqref{eq:boundaryInterAction}, we have to restrict our attention
		to operators $\mathcal O$ which are gauge invariant. The local $\SU(N)$
		gauge symmetry of the quantum field theory in fact has no counterpart on
		the supergravity side in the Maldacena limit. The parameter $N$ is
		translated into the number of \db3s on the string theory side of the
		correspondence. The stack of \db3s merely accounts for the emergence of
		the \adsfivesfive spacetime, see section \ref{sec:D3andAdS}. Moreover,
		the following arguments strictly only apply to BPS states.
		
		Comparing the remaining symmetry groups of $\N=4$ SYM theory and
		type~IIB supergravity we indeed observe a matching of symmetries. In
		section~\ref{sec:N4theory} we noted the symmetry group of the gauge
		theory to be $\PSU(2,2\mathop|4)$. The bosonic subgroup of this is
		$\SU(2,2)\times\SU(4)_R\cong\SO(2,4)\times\SO(6)$. These are precisely
		the isometry groups of \adsfivesfive, where $\SO(2,4)$ is the isometry
		group of the $\ads_5$ part, while the five-sphere is invariant under
		$\SO(6)$ transformations.  The fermionic symmetries can be shown to coincide as
		well, leading to the overall symmetry group
		$\PSU(2,2\mathop|4)$.
		
		In fact the isometries of \adsfivesfive act as the conformal group on
		the boundary \cite{Susskind:2005js}. Any gauge invariant field theory
		operator $\mathcal O$ does transform under some representation of the
		conformal group. Since the boundary theory is invariant under conformal
		transformations, the supergravity field $\phi$ in the source term
		\eqref{eq:boundaryInterAction} has to transform in the dual (conjugate)
		representation. Conformal invariance of the theory restricts the field
		$\phi$ further. For instance, in the coordinates where the boundary is
		located at $u=0$ the supergravity equations of motion for a scalar
		$\phi$ have two linear independent solutions at asymptotically small
		$u=\epsilon$ near the boundary,
		\begin{equation}
		\label{eq:fieldNearBoundary}
			\phi(\vec x,\epsilon) = \phi_0(\vec x)\,\epsilon^{d-\Delta} +\phi_1(\vec x)\,\epsilon^{\Delta}. 
		\end{equation}
		Here $d$ denotes the number of dimensions of the boundary, which in our
		case is $d=4$. Generically, the value of $\Delta$ for a scalar
		supergravity field $\phi$ depends on the mass $m_\phi$ of the field
		\cite{Witten:1998qj} as
		\begin{equation}
		\label{eq:sugraMasses}
			m_\phi^2 = \Delta(\Delta-d),
		\end{equation}
		with $\Delta>0$. The second term of \eqref{eq:fieldNearBoundary}
		vanishes at the boundary $\epsilon\to0$ while the first term may
		diverge. The existence of a well defined boundary value $\phi_0(\vec x)$
		tells us that this function has scaling dimension $d-\Delta$, i.\,e.\ 
		$\phi_0(\vec x) \mapsto \phi_0(\vec x)/\epsilon^{d-\Delta}$ on
		rescalings. From the interaction term of the conformally invariant
		action \eqref{eq:boundaryInterAction} we thus see that the boundary
		value of $\phi(\vec x,u)$ acts as the source to an operator $\mathcal
		O(\vec x)$ of scaling dimension $\Delta$.
		
		In summary, to identify the supergravity field $\phi$ dual to an
		operator $\mathcal O$ we have to spot all supergravity fields
		transforming in the dual representation to that of the operator under
		consideration. The conformal weight $\Delta$ of the operator determines
		the mass of the supergravity field by \eqref{eq:sugraMasses}. The mass
		spectrum of $\N=2$ supergravity compactified on \adsfivesfive has been
		computed \cite{Kim:1985ez}, and therefore the field can be identified
		uniquely. Examples of computations of dual field--operator pairs can be
		found e.\,g.\  in refs.\
		\cite{Aharony:1999ti,D'Hoker:2002aw,Gubser:1998bc,Witten:1998qj}.
		
		We also identified the boundary of \ads\ space with four-dimensional
		Minkowski spacetime. This spacetime was only defined up to conformal
		transformations, and we will identify it from now on with the domain of
		the conformally invariant $\N=4$ SYM theory. Notice that near the
		boundary all processes occurring in the field theory directions can be
		thought of as being scaled in such a way that all lengths of the \ads\
		theory, even long distance or IR phenomena, are mapped to short scales,
		i.\,e.\  the UV limit on the conformal field theory side. To see this,
		consider the metric \eqref{eq:AdS5S5metricInr} in the near horizon
		limit. Then distances $\ds^2_{\text{CFT}}$ in the field theory, which
		are measured along $\vec x$ appear with a warp factor relative to the
		distance $\ds^2_{\ads}$ in \ads\ space,
		\begin{equation}
			\ds^2_{\text{CFT}} = \frac{R^2}{r^2}\, \ds^2_{\ads}.
		\end{equation}
		At large $r\gg 1$ ($\text{IR}_\text{\ads}$) close to the boundary, short
		scale phenomena of the CFT ($\text{UV}_\text{CFT}$) match the events in
		\ads\ space, while at small $r\ll 1$ ($\text{UV}_\text{\ads}$), far from
		the boundary, long scale phenomena in the CFT ($\text{IR}_\text{CFT}$)
		match the \ads\ distances. The radial coordinate in this way sets the
		renormalization scale of the field theory which is holographically
		described by the supergravity theory. This phenomenon is called the
		\emph{UV/IR duality}.
		
		In fact the behavior of correlation functions under renormalization
		group flows can be computed holographically. The UV divergences known
		from field theory translate into IR divergences on the gravity side.
		The procedure to incorporate scale dependence and renormalize $n$-point
		correlation functions is known as \emph{holographic renormalization}. We
		will not review the procedure in detail here, but rather give an idea of
		the procedure and state some results. A nice overview which also
		addresses some subtleties can be found in ref.~\cite{Skenderis:2002wp}.
		
		The correlation functions \eqref{eq:2pointPrescription} in general
		suffer from IR divergences, i.\,e.\  divergent terms at large values of the
		radial coordinate. Analogous to quantum field theory renormalization,
		they can be cured by analyzing the behavior of the field solutions near
		the boundary and adding appropriate counterterms $S_\text{ct}$ to the
		action $S$ which do not alter the equations of motion but render the
		resulting correlators finite.
		
		To analyze the field behavior near the boundary it is convenient to work
		in coordinates $u$ as in \eqref{eq:metricAdSonly} where the boundary is
		located at $u\to 0$. The solution for the second order equation of
		motion of any field $\mathcal F$ can then be expanded in a series around
		$u=0$. In general there are two independent solutions scaling as $u^m$
		and $u^{m+n}$ near the boundary. The general solution can be written as
		\begin{equation}
			\mathcal F(x,u) = u^m\left( f^{(0)}(x) + u^2\,f^{(2)}(x)+\ldots+u^n\left(f^{(2n)} + \tilde f^{(2n)}\ln u \right) +\ldots\right)
		\end{equation}
		in a well defined manner where the coefficients $f^{(n)}(x)$ carry the
		dependence on the other coordinates. The values of $m$ and $n$ are
		determined by the mass of the supergravity field and related to the
		conformal dimension of the dual operator, as in the example above. The
		coefficient $f^{(0)}$ determines the boundary behavior of the two
		independent solutions for the equation of motion of $\mathcal F$.
		Solving these equations order by order in $u$ determines the relevant
		coefficients $f^{(k)}$ for $k<2n$ as functions of $f^{(0)}$, which
		thereby can be used as the initial value of the first of the two
		linearly independent solutions. The second parameter needed to define
		the full solution of the second order equation of motion to $\mathcal F$
		is the coefficient $f^{(2n)}$, which in turn determines the remaining
		higher order coefficients. It then is possible to extract the divergent
		terms in the regularized action $S_\text{reg}$, which is given by the on
		shell action with respect to the $u$ dependence of the solution,
		evaluated at the cutoff $\epsilon\ll1$,
		\begin{equation}
			S_\text{reg} = \int\!\dd^4x\; a^{(0)}\,u^{-\nu} + a^{(1)}\,u^{-(\nu+1)} + \dots\Big|_{u=\epsilon}.
		\end{equation}
		The coefficients $a^{(n)}$ now are functions of the coefficient
		$f^{(0)}$, and the $\nu>0$ solely depend on the scale dimension of the
		operator in the conformal field theory. Defining the counterterm action
		as
		\begin{equation}
			S_\text{ct} = -\text{divergent terms from~}S_\text{reg}
		\end{equation}
		The renormalized action is given by
		\begin{equation}
			S_\text{ren} = \lim_{\epsilon\to 0}\left( S_\text{reg}+S_\text{ct}\right).
		\end{equation}
		Finding the renormalized action therefore involves a careful analysis of
		the equations of motion. An extremely useful result of holographic
		renormalization is the fact that the solutions to the equations of
		motion of a supergravity field $\mathcal F$ can directly be related to
		the source and the vacuum expectation value of the dual operator in the
		field theory \cite{Skenderis:2002wp,Haro:2000xn}. In particular
		holographic renormalization unveils that the mode $f^{(0)}$ is
		proportional to the source of the dual operator, while the mode
		$f^{(2n)}$ is proportional to the vacuum expectation value of the same
		operator. In general, the mode that is proportional to the source scales
		in a non-normalizable way with $u$, while the mode proportional to the
		vacuum expectation value is normalizable. An example is given in
		\eqref{eq:fieldNearBoundary} for the case of a scalar field. The
		integers $m$ and $n$ are determined in terms of the supergravity field's
		mass, which in turn translates to the conformal dimension of the dual
		operator. In \eqref{eq:fieldNearBoundary} the value of $\phi_0$ is
		proportional to the source and $\phi_1$ is proportional to the vacuum
		expectation value of the dual operator. We will encounter an explicit
		example for the source and vacuum expectation value when we introduce
		the prominent pair of a D-brane embedding function and its field theory
		dual operator in section~\ref{sec:addingFlavor}.
		\begin{table}
			\begin{tabular*}{\linewidth}{@{}p{.5\linewidth}@{}p{.5\linewidth}@{}}
				Supergravity on \adsfivesfive            & \hfill $4$ dim.\; $\N=4$ CFT\\
				\hline
				\rule{0mm}{1.5em}boundary of $\ads_5$ 	 & \hfill field theory domain\\
				isometry of $\ads_5$ 					 & \hfill conformal symmetry\\
				isometry of $\S^5$ 						 & \hfill R-symmetry\\
				weak coupling in $\gs$ 					 & \hfill strong coupling in $\lambda$\\
				variations in radial coordinate 		 & \hfill renormalization group flow\\
				field boundary value $\phi_0$ 			 & \hfill source for operator $\mathcal O$\\
				field mass $m$ 							 & \hfill conformal weight $\Delta$\\
				IR normalizable mode 					 & \hfill $\vev{\mathcal O}$ of dual operator\\
				IR non normalizable mode 				 & \hfill source of dual operator\\
				quantum corrections of $\order\gs$ 		 & \hfill corrections in $1/N$\\
				stringy corrections of $\order{\alpha'}$ & \hfill corrections in $1/\lambda$\\[\smallskipamount]
				\hline
				\end{tabular*}
				\caption[Entries of the \ads/CFT  dictionary]{
					A few examples for entries of the \ads/CFT  dictionary.
					Precise operator--field pairings can be found
					e.\,g.\  in refs.\ 
					\cite{Aharony:1999ti,D'Hoker:2002aw,Gubser:1998bc,Witten:1998qj}.
				}
		\end{table}
	\subsubsection{Tests and evidence}
		
		Although a rigorous mathematical proof of the correspondence is not
		derived so far, there is increasing evidence for the Maldacena
		conjecture to hold. Soon after the discovery of the correspondence
		correlation functions of field theory operators where computed from
		gravity. A direct comparison to results obtained from field theory
		calculations is not straightforward since the results from gravity
		calculations are valid in the strongly coupled regime of the gauge
		theory while the gauge field theory computations are performed in
		the perturbatively accessible regime of weak coupling.
		
		Nevertheless, it was early realized that the answers obtained by gravity
		calculations gave the correct scaling behavior of $n$-point correlation
		functions, which is dictated by conformal invariance
		\cite{Witten:1998qj}. Moreover, certain correlation functions satisfy
		non-renormalization theorems, which state that the results are
		independent of the coupling constant. Examples are the two- and
		three-point functions of $1/2$-BPS operators, which show a
		perfect matching of gauge and gravity results
		\cite{Freedman:1998tz,Lee:1998bxa}. Also the conformal anomaly of $\N=4$
		SYM theory which is present in curved background spacetimes could be
		reproduced exactly from \ads/CFT, which even provided methods to obtain
		the anomaly in six-dimensional field theories for the first time
		\cite{Henningson:1998gx}.
		
		Generalizations of the correspondence, which will be partly
		discussed below, relate different backgrounds to different gauge
		theories. Such modifications allowed for comparison of finite
		temperature $\N=4$ SYM with calculations obtained from the gauge
		gravity theory. These calculations show agreement of correlation
		functions in the hydrodynamic limit of low frequency/long distance
		\cite{Policastro:2002se,Son:2002sd}. The recently most
		enthusiastically discussed hydrodynamic result from gauge/gravity
		calculations is the observation of one universal value for the lower
		bound on the ratio $\eta/s$ of shear viscosity $\eta$ over entropy
		density $s$ for all known gauge theories with gravity duals
		\cite{Kovtun:2003wp,Kovtun:2004de,Son:2006em}. In the
		large $N$ limit the result in SI units is
		\begin{equation}
			\frac{\eta}{s} \geq \frac{\hbar}{4\pi k_{\scriptscriptstyle B}}.
		\end{equation}
		If it should turn out that QCD has a gravity dual and is in the same
		universality class as the known generalizations of the
		correspondence, this result (including large $N$ corrections) could
		be the first prediction from string theory within reach of
		experiment. Possibly the experiments at the RHIC and future collider
		experiments will give answers on the value of $\eta/s$ for QCD. So far
		the data seems to be in agreement with the above bound, which means
		that the QGP appears as the most perfect fluid ever observed.

\subsection{Generalizations and extensions}
	
	Before we discuss the features of QCD and the sQGP which can be described by
	holographic duals, we introduce those generalizations of the correspondence
	which are most relevant for this work. These generalizations are necessary
	to incorporate features which are missing in the $\N=4$ SYM field theory
	described by supergravity on \adsfivesfive. Most important for a description
	of the strongly coupled quark-gluon plasma are the inclusion of fundamental
	degrees of freedom and a way to describe systems at finite temperature.
	
	The introduction of finite temperature allows to model interesting
	qualitative features like the confinement/deconfinement phase transition at
	some critical temperature. In holographic models the transition occurs at
	different temperatures for the gauge fields and the fundamental degrees of
	freedom. The realization of finite temperature is achieved by a modification
	of the background geometry of the gravity theory. Moreover, we will make
	some comments about the subtleties that are related to the computation of
	Green functions at finite temperature. While correlation functions at zero
	temperature can be obtained in Euclidean spacetimes and a subsequent Wick
	rotation, this procedure generally cannot be applied at finite temperature.
	
	The inclusion of fundamental degrees of freedom is a generalization in
	the sense of additional fields we add to the gauge theory. All fields in
	$\N=4$ SYM transform in the adjoint representation of the gauge group
	and therefore rather account for the gauge degrees of freedom (gluons)
	than for the quarks. The added quark degrees of freedom will be
	represented by additional D-branes on the supergravity side. The incorporation
	of quarks into the theory furthermore allows to investigate the spectra
	of their bound states, which the subsequent section is devoted to.

	\subsubsection{Finite temperature and AdS black holes}
	\label{sec:finiteTempBlackHoles}
		
		At finite temperature $T$ any quantum mechanical system can be found 
		in one of the possible states of energy $E$ with the probability
		distribution in equilibrium described by the density matrix
		\begin{equation}
		\label{eq:BoltzmannDensity}
			\hat\rho = \frac{\ee^{-\beta H}}{\tr \ee^{-\beta H}}\,,
		\end{equation}
		where $\beta=1/T$ is the inverse of the temperature and $H$ is the
		Hamiltonian of the system. The \emph{statistical} partition function
		of the ensemble of systems at temperature $T$ can then be defined as
		\begin{equation}
		\label{eq:partFuncThermal}
			Z_\text{stat} = \tr \ee^{-\beta H} = \sum\limits_n\bra{\phi_n}\ee^{-\beta H}\ket{\phi_n},
		\end{equation}
		where the $\ket{\phi_n}$ form a basis of the state space of the system.
		The statistical partition function defines the weight of each state that
		contributes to ensemble averages. Expectation values of some observable
		$A$ in a thermal ensemble are calculated with respect to $Z_\text{stat}$
		by
		\begin{equation}
			\vev{A} = \tr\left(\hat\rho A\right) = \frac{\tr\left( Ae^{-\beta H}\right)}{Z_\text{stat}}.
		\end{equation}
		In the quantum mechanical formalism of path integrals, transition
		amplitudes are given by
		\begin{equation}
		\label{eq:pathIntegralAmplitude}
			\vev{\phi_f(t_f)|\phi_i(t_i)} = \bra{\phi_f(t_f)}e^{-i(t_f-t_i)H}\ket{\phi_i(t_i)} = \int\limits_{\phi=\phi_i}^{\phi=\phi_f}\!\!\!\!\mathrm D\phi\; \ee^{\ii S[\phi]}.
		\end{equation}
		The idea is to sum up all possible paths $\phi(t)$ that evolve from
		the initial configuration $\phi_i$ to the final $\phi_f$. The
		complex phases give the weight for each possible configuration that
		contributes to the evolution. If we would not only consider one
		initial and one final state but sum over an ensemble of many
		possible states, we should therefore recover the sum of all weights,
		the partition sum. The action $S[\phi]$  above is defined as
		\begin{equation}
		\label{eq:action}
			S[\phi] = \int\limits_{t_i}^{t_f}\!\dd t \int\!\dd^{d-1}x \;\L(t,x,\phi).
		\end{equation}
		In quantum mechanics the standard method to obtain expectation values
		is the evaluation of functional derivatives of
		generating functionals, which are commonly also referred to as
		 partition functions. They are defined for some functional
		$S_E[\phi]$ by
		\begin{equation}
			Z_\text{gen} =  \int \mathrm D\phi\; \ee^{- S_E[\phi]},
		\end{equation}
		where we need to specify which functions $\phi$ we have to integrate
		over and what the functional $S_{\scriptscriptstyle E}[\phi]$ is. The
		\emph{imaginary time formalism} gives a prescription which exactly
		reproduces the thermal equilibrium probability weights with the
		Boltzmann factor given in \eqref{eq:BoltzmannDensity}. The prescription
		is to analytically continue the time coordinate into the complex plane,
		such that $t$ in \eqref{eq:action} integrates over complex times.
		Additionally, we introduce a new time coordinate $\tau=\ii t$ as a Wick
		rotation of $t$. If we now restrict the system to such field
		configurations $\phi$ that are periodic (in fact fermionic fields would
		have to satisfy anti-periodicity) along the imaginary axis in complex
		time $t$ with $t_f - t_i=-\ii\beta$ and $\beta\in\mathbbm R$ between
		the initial $t_i$ and the final $t_f$, 
		we can reproduce the Boltzmann weights by setting
		\begin{equation}
				S_{\scriptscriptstyle E}[\phi] = \int\limits_{0}^{\beta}\dd\tau\int\dd^{d-1} x\;\L_{\scriptscriptstyle E}(\tau,x,\phi).
		\end{equation}
		The index $E$ refers to the fact that we use the Euclidean, i.\,e.\  the Wick
		rotated version, of the action. Then the integration from $0$ to $\beta$
		translates into integration over complex times $t_i$ to $t_f$ and
		therefore introduces the factor $t_f - t_i=-\ii\beta$ in
		\eqref{eq:pathIntegralAmplitude}. As we restrict to periodic states on
		the integration intervall, the final states $\bra{\phi_f}$ match the
		initial states $\ket{\phi_i}$. Thus the path integral resembles a trace
		\cite{Das:1997gg},
		\begin{equation}
			Z_\text{gen} = \int\limits_{\text{all $\beta$-periodic}\atop\text{states}}\hspace{-2.7ex} \mathrm D\phi\; e^{- S_E[\phi]}  
			= \sum\limits_{\text{all $\beta$-periodic}\atop\text{states}}\!\!\bra{\phi_\beta}e^{-\beta H}\ket{\phi_\beta}
			= Z_\text{stat}.
		\end{equation}
		Adding source terms to the action, which are set to zero after
		functional derivation, yields the Boltzmann factors as weights from the
		Euclidean generating functional. The imaginary time formalism in this
		way trades time for temperature and imposes boundary conditions. The
		result no longer depends on a real valued time interval but only on the
		purely imaginary time interval $t_f-t_i=-\ii\beta$, which we interpret
		as the temperature $T$ by identifying $T=\frac{1}{\beta
		k_{\scriptscriptstyle B}}$. Abandoning time dependence is in accordance
		with the fact that we investigate a system in equilibrium, where
		expectation values do not change with time. Another consequence of the
		periodic boundary conditions is that any solution admits a discrete
		spectrum in its Fourier transformation. A propagator $G(\tau)$ can be
		decomposed according to
		\begin{equation}
			G(\tau) = \frac{1}{\beta}\sum_n e^{-\ii\omega_n\tau}\,G(\omega_n).
		\end{equation}
		The frequencies $\omega_n$ are called \emph{Matsubara frequencies}.
		
		The link between field theory at finite temperature and gravity is known
		to be related to black hole physics, which has many parallels to
		thermodynamics. From the moment of the discovery of the \ads/CFT
		correspondence it was expected that the finite temperature description
		of the field theory must be given by gravity in the \ads\ black hole
		background \cite{Maldacena:1997re}. We will now introduce the $\ads_5$\
		black hole background and give arguments for the relation between the
		horizon radius and the temperature of the dual field theory along the
		lines of arguments in refs.~\cite{Nastase:2007kj,Witten:1998zw}. The
		generalization of the \adsfivesfive metric \eqref{eq:AdS5S5metricInr} to
		the black hole solution with a horizon at $r=\rh$ is given by,
		\begin{equation}
		\begin{gathered}
			\label{eq:adsBHmetricInr}
			\ds^2 = \frac{r^2}{R^2}\left( - f(r)\,\dd t^2 + \dd \threevec{x}^2\right) + \frac{R^2}{r^2} \frac{1}{f(r)}\,\dd r^2+R^2\dd\Omega_5^2,\\
			f(r) = 1-\frac{\rh^4}{r^4}.\hfill~
		\end{gathered}
		\end{equation}
		The signature of the metric again depends on our choice of working in
		either Lorentzian or Euclidean \ads\ space. To draw the connection to
		finite temperature imaginary time formalism we again work in the Wick
		rotated coordinate $\tau=\ii t$, where the metric has Euclidean
		signature. Euclidean signature however is only given outside the
		horizon, inside we would introduce negative signs from $f(r)$ (in the
		Lorentzian case the signs of $t$ and $r$ would change). On the other
		hand it is known that the spacetime can be continued beyond the horizon
		and therefore the spacetime can be regularized at $\rh$.
		
		The idea is to show that periodicity in the Euclidean time $\tau$
		that leads to thermal probability distributions in field theory
		corresponds to a regularization of the Euclidean spacetime at the
		horizon. The period $\beta=1/T$ is then identified with the inverse
		temperature as in field theory. We concentrate on the $\tau$ and $r$
		coordinates in the above metric and observe how they behave near the
		horizon at $r\approx\rh$,
		\begin{equation}
			\ds^2 = \frac{4\rh}{R^2}\,(r-\rh)\, \dd \tau^2 + \frac{R^2}{4\rh}\, (r-\rh)^{-1} \dd r^2.
		\end{equation}
		We know that the Euclidean spacetime is well defined only on and
		outside the horizon and therefore introduce a new coordinate
		$\varrho^2=r-\rh$, which casts the metric into the form
		\begin{equation}
			\ds^2 = \frac{R^2}{\rh}\left( \dd\varrho^2 + \varrho^2\;\frac{4\rh^2}{R^4}\,\dd \tau^2 \right).
		\end{equation}
		The factor in front is merely a constant. The metric in parentheses
		is the metric of a plane in polar coordinates, $\ds^2=\dd\varrho^2 +
		\varrho^2\dd\theta^2$. The angular variable in our case is $\theta=\tau
		2\rh/R^2$. The space in polar coordinates, however, is regular only
		for an angular variable that is periodic with period $2\pi$, otherwise
		a conical singularity is located at $\varrho=0$, which is the horizon
		of the \ads\ black hole. Periodicity of $\theta$ with period $2\pi$
		then translates into periodicity of $\tau$ with period $\beta$ by
		\begin{equation}
			\frac{2\rh}{R^2}\,\tau +  2\pi \sim \frac{2\rh}{R^2}\,(\tau+\beta).
		\end{equation}
		Where we used the symbol $\sim$ to denote the equivalence relation
		of identified points.  Regarding identified points as equal and
		using $\beta=1/T$ we obtain the relation between the temperature
		of the field theory and the horizon radius on the gravity side,
		\begin{equation}
		\label{eq:blackHoleTemperature}
			\rh = T\pi R^2.
		\end{equation}
		This result exactly reproduces the expression for the Hawking
		temperature of the \ads Schwarzschild black hole described by the
		metric \eqref{eq:adsBHmetricInr}.
		
		The \ads\ black hole background allows to investigate strongly coupled
		gauge theories at finite temperature by performing calculations in the
		gravity theory. Moreover, finite temperature defines an energy scale (or
		length scale $\rh$ on the gravity side) in the theory and therefore
		breaks global scale invariance, which will have consequences for the
		property of (de)confinement of the field theory, which we discuss in
		section~\ref{sec:confinement}. Despite the detour to a background of
		Euclidean signature with time $\tau$, the field theory we want to use as
		a model for real world QCD is defined on a background of Minkowski
		signature with time $t$, which we will use predominantly from now on.
		Some subtleties and consequences regarding Minkowski signature \ads/CFT
		are discussed in the following.

	\subsubsection*{Thermal real time Green functions}
	\label{sec:thermalGreenFunctionsIntro}
		
		The \ads/CFT correspondence was originally formulated and
		successfully applied in backgrounds of Euclidean signature, i.\,e.\  in
		the imaginary time formalism. Results in real time can be derived by
		subsequent Wick rotation. However, many common situations require
		the formulation of a problem and its solution in real time. One
		mathematical reason for the need of a real time formulation arises
		from simplifications which are often introduced by deriving
		solutions only for certain limits of the parameter space. In many
		cases for instance, solutions are obtained in the hydrodynamic limit
		of low frequency/long distance physics. In this case only the low
		Matsubara frequencies are known and therefore analytic continuation
		of results to real time is somewhere between difficult and
		impossible. Physical arguments against the imaginary time formalism
		arise whenever deviations or even far from equilibrium scenarios are
		considered. We discussed that the imaginary time formalism mimics
		thermal equilibrium probability distributions. For systems out of
		equilibrium the restriction of the path integral to periodic paths
		is not justified. Moreover the solutions considered in the imaginary
		time formalism are periodic, it is doubtful whether such solutions
		can model long time evolutions. Summing up, it is desirable to have
		a real time prescription for computations of correlation functions
		in Minkowski space.
		
\begin{figure}
			\centering
			\includegraphics{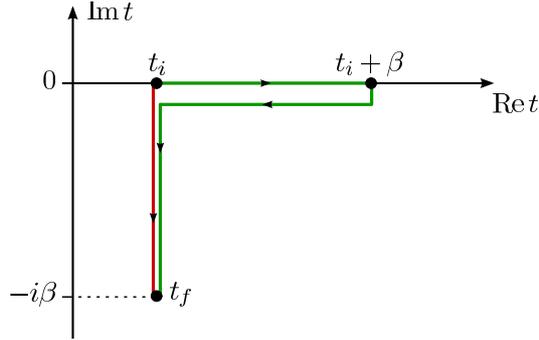}
			\caption[Complex time integration contours in finite
			temp.\ field theory]{Integration contour in the complex time
			plane for the imaginary time (red) and the
			Schwinger-Keldysh (green) formalism for finite
			temperature field theory.} 
			\label{fig:finiteTintegrationContour}
		\end{figure}

		In field theory such a prescription is given by the Schwinger-Keldysh
		formalism \cite{Das:1997gg}. The difference to the imaginary time
		formalism is a modification of the time integration in the complex time
		plane. The time coordinate is still defined on the complex plane.
		However, instead integrating from $t_i$ to $t_i-\ii\beta$ along the
		negative imaginary axis, this time a detour along the real axis is
		taken. The path first proceeds along the real axis from $t_i$ to
		$\beta$, which marks the end of the physical real valued time interval
		of interest. Then, the integration contour enters the negative half
		plane arbitrarily far and leads back below the real axis to $\Re t=\Re
		t_i$ to continue parallel to the imaginary axis and finally end in
		$t_f=t_i-\ii\beta$. Figure~\ref{fig:finiteTintegrationContour} shows the
		integration paths of the imaginary time and Schwinger-Keldysh
		formalisms. We are interested in the analog of the latter prescription
		in the gauge/gravity duality.
		
		A recipe for the derivation of holographic Minkowski space Green
		functions in real time was derived by Son and Starinets together
		with Herzog and Policastro
		\cite{Son:2002sd,Policastro:2002se,Herzog:2002pc}. The resulting
		prescription is concise and amounts in only small changes from the
		prescription given by \eqref{eq:2pointPrescription}. The difference
		is given by the way we compute the on shell action $S_\text{sugra}$.
		The action is commonly obtained by writing the solution of a field
		$\phi$ such that the ``bulk contributions'' in radial direction $r$
		factorize from the ``boundary contributions'' along the field theory
		directions $x$ on the boundary $r_\text b$. Usually we work in 
		momentum space with momentum $k$ instead of position space with
		coordinate $x$ and write
		\begin{equation}
			\label{eq:separateBoundaryBehaviour}
			\phi(r,k) = f(r,k)\, \phi^\text{bdy}(k), \qquad \text{with}\quad \lim_{r\to r_\text b} f(r,k) = 1.
		\end{equation}
		In the coordinates introduced so far the boundary was located at
		$r_b=\infty$. The on shell action can then be written as
		\begin{equation}
			\label{eq:recipeOnShell}
			S_\text{sugra} = \left.\int\!\frac{\dd^4k}{(2\pi)^4}\; {\phi^\text{bdy}}'(-k)\, \mathcal F(r,k)\, \phi^\text{bdy}(k) \right|_{r=\rh}^{r=r_\text b}.
		\end{equation}
		Here, we carried out the integration over radial coordinates from the
		black hole horizon to the boundary. The two $\phi_0$ arise from the
		kinetic term and the function $\mathcal F(r,k)$ collects the remaining
		factors of $f(r,k)$ and $\del_r f(r,k)$ and possibly other factors
		appearing in the action under consideration. A detailed and explicit
		calculation can be found in section~\ref{chap:specFuncs} where we apply
		the recipe, and in refs.~\cite{Son:2002sd,Policastro:2002se}. The two point
		Green function of the operator dual to $\phi$ would then be given by the
		second functional derivative of the action,
		\begin{equation}
			\label{eq:recipeCorrelator}
			G = -\mathcal F(r,k)\Big|_{\rh}^{r_\text b}-\mathcal F(r,-k)\Big|_{\rh}^{r_\text b}. 
		\end{equation}
		So far we followed the method we already introduced for zero
		temperature. The only difference is that now $\rh\neq 0$. The evaluation
		of this expression is mathematically possible, but gives physically
		wrong answers. For example, the resulting Green functions would be real
		functions, opposed to physical solutions, which are in general complex
		valued. This behaviour is due to the boundary conditions that have to be
		imposed on the fields. In the Schwinger-Keldysh formalism they arise
		from the periodicity of the fields and from the orientation of the
		integration contour. This introduces a contour ordering prescription,
		which translates to time ordering in physical processes. Loosely
		speaking, a causal propagator in the \ads black hole background
		describes propagation of a field configuration that has to obey the
		infalling wave boundary condition at the black hole horizon. This
		boundary condition imposes the physically given fact that at the
		horizon, positive energy modes can only travel inwards, while negative
		energy modes only travel outwards. It can be shown that upon imposing
		these boundary conditions, the time ordered retarded part of the
		propagator in momentum space is determined by the boundary behaviour of
		the fields alone \cite{Herzog:2002pc}. The prescription of Son and
		Starinets for fields obeying the infalling wave boundary condition then
		reads
		\begin{equation}
		\label{eq:correlatorFromBoundaryContribution}
			\GR = -2\mathcal F(r,k)\Big|_{r_\text b}.
		\end{equation}
		The contributions from the horizon are neglected. This method was
		used to great extent in subsequent publications. At zero temperature
		it agrees with the analytic continuation of Euclidean results
		\cite{Son:2002sd}.

	\subsubsection{Fundamental matter\,---\,adding flavor}
	\label{sec:addingFlavor}
		
		We so far recognized that closed string excitations in the vicinity
		of the stack of $N$ \db3s gives rise to fields transforming in the
		adjoint representation of $\SU(N)$, which was identified with the
		color group. We could criticize a lack of fundamental degrees of
		freedom, such as quarks in QCD. Karch and Katz introduced a way to
		add fundamental fields to the theory \cite{Karch:2002sh}. The cure
		can be obtained from modes of \emph{open} string excitations with
		\emph{one} end of the string on the stack of $N$ \db3s and the other
		end on a different stack of $N_f$ D$p$-branes. In this work we will
		restrict to stacks of coinciding branes. We cannot distinguish
		between the branes of a stack on which a string ends. We thus
		encounter a $\U(N)$ and a $\U(N_f)$ symmetry which reflects the
		invariance of the theory under the exchanges of the branes
		\cite{Becker:2007zj,Johnson:2003gi}. The modes of these strings
		transform in the fundamental representation of $\SU(N)\subset\U(N)$
		and $\SU(N_f)$, respectively. We interpret these as the color and
		flavor groups. So the fundamental fields (quarks) of our gauge
		theory correspond to strings that have one end on the stack of $N$
		\db3s, and the other on an additional stack of $N_f$ D$p$-branes
		which may be separated from the color branes.
		
		Throughout this work we will consider the so-called probe limit in
		which $N_f\ll N$. This ensures that the backreaction of the
		additional branes on the near horizon geometry of the \db3s can
		consistently be neglected. In this way we do not have to worry about
		how the new D$p$-branes might alter the background geometry but
		stick to \adsfivesfive. The Maldacena limit of infinitely many colors
		$N\to\infty$ is then also  called the \emph{probe limit}, since
		we add some neglectable amount of $N_f$ \emph{probe branes}.
		
		The string modes stretching from the \db3s to the probe D$p$-branes
		also transform under the fundamental representation of the probe
		branes' gauge group $\SU(N_f)$. However in the Maldacena limit with
		$N\to\infty$ the 't~Hooft coupling $\lambda_f=2\pi\gs N_f$ of the
		stack of probe branes can be neglected with respect to the color
		gauge group coupling $\lambda=2\pi\gs N$. The probe brane gauge
		group in this way decouples from the color gauge group. 
		We will identify the probe gauge group as the flavor group and
		interpret strings stretching from the stack of $N_f$ D$p$-branes to
		the stack of $N$ \db3s as fundamental matter which comes in $N$
		varieties of color and $N_f$ flavors. The additional D$p$-branes are
		therefore also called \emph{flavor branes}. For finite $N_f$, the
		large $N$ limit then is the equivalent to the quenched limit of
		lattice QCD, which allows to neglect fermion loops in all amplitudes
		relative to effects of the glue.
		
		The global $\U(N_f)$ flavor symmetry of the field theory translates into
		a gauge symmetry on the supergravity side. The conserved currents of
		the field theory are dual to the gauge fields on the supergravity side.
		We will elaborate on this issue further when we introduce finite
		particle density. For now we only stress that the introduction of $N_f$
		flavor brane accounts for a gauge field on these branes which acquires
		values in a $\U(N_f)$ Lie algebra. We denote the field strength tensor
		of this gauge field by $F$. The components of this tensor are labeled
		by $F^a_{\mu\nu}$, where the $\mu$ and $\nu$ denote spacetime indices
		while $a=1,2,\ldots,N_f^2$ is an index in the vectorspace of the
		$\U(N_f)$ generators.
		
		The remaining issues then are, what dimensions the flavor branes should
		have and how they have to be positioned with respect to the \db3s.
		Generically, D-branes couple to the field strengths of type~IIB
		supergravity, cf.\ \eqref{eq:WZaction}. Karch and Randall showed that
		there are stable probe brane solutions which span topologically trivial
		cycles and are determined by the DBI action alone \cite{Karch:2000gx}.
		There are several such solutions which then give rise to fundamental
		degrees of freedom in the dual field theory \cite{Karch:2002sh}.

		\subsubsection*{Dirac-Born-Infeld action}
			
			The dynamics of D$p$-branes is crucial for the calculations
			performed in the following. Here, we introduce an action which allows
			to derive the equations of motion for D-branes. Later we will deal
			with stacks of D$p$-branes, for now we consider the simpler case of
			a single brane.
			
			The interpretation of a single D-brane as the surface on which the
			endpoints of strings lie implies Dirichlet boundary conditions for
			the positions of these points. It is the Polyakov action that
			describes the dynamics of the strings. In the presence of background
			fields, a generalization of this action is given by a non-linear
			sigma model \cite{Johnson:2003gi}. The extremization of such an
			action respecting the Dirichlet boundary conditions is equivalent to
			the extremization of the \emph{Dirac-Born-Infeld} action
			\cite{Leigh:1989jq}. This action captures the low energy dynamics of
			the string mode corresponding to the open string excitations of the
			D$p$-brane. For a single D$p$-brane with a worldvolume $\mathcal M$
			parametrized by worldsheet coordinates $\xi^i$ with
			$i=0,1,\ldots,p$, the DBI action is given by
			\begin{equation}
				\label{eq:DBIaction}
				S_\text{\tiny DBI} = -T_p \int\limits_{\mathcal M}\!\!\dd^{p+1}\xi \; \ee^{-\Phi}  \sqrt{\left|\det\left( P[g + B] + 2\pi\alpha' F\right)\right|}\:.
			\end{equation}
			Here $g(\xi)$, $B(\xi)$ and $F(\xi)=\dd A(\xi)$ are the background
			metric, the Kalb-Ramond $B$-field and the gauge field strength
			tensor on the brane. The operator $P[\,\cdot\,]$ denotes the
			pullback on the brane worldvolume. The field $\Phi$ is the dilaton.
			The brane tension $T_p$ was given in \eqref{eq:braneTension}. We
			will make extensive use of the DBI action.
			
			The DBI action is a low energy effective action that includes
			stringy corrections in $\alpha'$ up to arbitrary order. An expansion
			of the DBI action in powers of $\alpha'$ reproduces the Maxwell
			action in order $F^2$ and introduces higher powers of $F$ as
			corrections. However, this action does not include any powers of the
			derivative of the field strength and therefore is strictly valid
			only for constant field strengths. For a \D0-brane the DBI action
			resembles the worldline action of a pointlike particle.
			
			In cases where the Ramond-Ramond sector contributes non-vanishing
			$n$-forms $C_n$ the full action for a D$p$-brane is given by
			\begin{equation}
				S = \SDBI + S_\text{\tiny WZ},
			\end{equation}
			where $S_\text{\tiny WZ}$ is the Wess-Zumino action
			\begin{equation}
				\label{eq:WZaction}
				S_\text{\tiny WZ} = T_p \int P\left[\sum_n C_n \,\ee^{B}\right]\ee^{2\pi\alpha'F}.
			\end{equation}
			However, all problems discussed in this work restrict to cases where
			there are no contributions from the Wess-Zumino action,
			$S_\text{\tiny WZ}=0$. For the case of the \D0-brane, the WZ action
			resembles the coupling of a pointlike particle to an electromagnetic
			field. 
		
		\subsubsection*{The \D3/\D7 setup}
		
			Obviously, in the quark-gluon plasma the fundamental degrees of
			freedom move freely throughout the directions which are interpreted
			as Minkowski spacetime. Thus we will restrict our attention to probe
			branes which span at least all the Minkowski directions, i.\,e.\  we
			consider spacetime filling D$p$-branes with $p\geq 3$. There are two
			heavily used models on the market, the \D3/\D7 setup and the
			Sakai-Sugimoto model. Throughout this work we will use the \D3/\D7
			setup, which we introduce here. The Sakai-Sugimoto model will be
			addressed in a novercally short section afterwards.
			
			In the \D3/\D7 configuration, the background is generated by a stack
			of $N$ \db3s in the way introduced above, which is then probed by
			$N_f$ flavor \db7s. The \db3s account for the background geometry
			and in the near horizon geometry give rise to the \emph{closed}
			string excitations of type IIB supergravity, accounting for the dual
			field theory $\N=4$ SYM. The additional degrees of freedom
			introduced by \emph{open} string oscillations of string stretching
			between the \db3s and the $N_f$ probe \db7s give rise to an $\N=2$
			hypermultiplet in the fundamental representation of $\SU(N)$
			\cite{Karch:2002sh}. The fermionic fields in this multiplet,
			which we will denote by $\psi$, are interpreted as the analogon to
			quarks in QCD. The dynamics of quarks and mesons will therefore
			depend on the dynamics of D-branes in the holographic dual.
			An explicit realization of \db7s embedded into \adsfivesfive is
			given below. Applications of such flavor branes to describe quarks
			and mesons, their spectra and stability will be the subject of the
			following sections.

		\subsubsection*{Other \D3/D$q$ setups}
			
			Type IIB string theory exhibits \D3, \D5, \D7 and \D9
			branes. We will investigate the \D3/\D7 model subsequently.
			Constructions with other types of branes can be interesting. 
			However, with the quark-gluon plasma in mind, we do not investigate
			other probes than \db7 for the following reasons.
			
			\db9 can not be separated from the \db3s since the former ones span
			the entire background spacetime and therefore cannot be separated
			from the \db3s in order to generate massive quarks. \db3 and \db5
			branes do not have this caveat. On the other hand they still have to
			span a certain cycles in the dimensions transverse to the \db3s.
			Consider for example the embedding scheme of the \D3/\D5 setup.
			Here, we split the directions of the $\mathbbm{R}^6=\mathbbm
			R\times\S^5$ transverse to the \db3s, given by the radial coordinate
			of \ads space and the directions along the $\S^5$, into a radial
			coordinate $\varrho$ and a cycle $\S^2$ on the \db5, and $\varphi$,
			$\phi$ and $L$ transverse to all branes.
			
			\noindent
			\parbox{\linewidth}{
			\vspace{\bigskipamount}
			\begin{tabular}{rl}
				 & $t$\quad $x_1$\quad $x_2$\quad $x_3$\quad $\varrho$\quad $\S^2$\quad $\varphi$\quad $L$\quad $\phi$\;\\
				\db3 & \raisebox{0.5ex}{\rule{6.75em}{.25em}}\\
				\db5 & \raisebox{0.5ex}{\rule{4.75em}{.25em}}\hspace{2em}\raisebox{0.5ex}{\rule{3.25em}{.25em}}\\
			\end{tabular}
			\vspace{\bigskipamount}}
			
			In this setup the ends of the string cannot move freely in the
			$x_3$ direction of the Minkowski spacetime. Configurations like
			the \D3/\D5 and \D3/\D3 setup therefore describe defect
			theories, in which the fundamental degrees of freedom are
			confined to lower dimensional hyperplanes.

		\subsubsection*{The Sakai-Sugimoto model}
			
			The Sakai-Sugimoto model, introduced in
			refs.~\cite{Sakai:2004cn,Sakai:2005yt}, describes the gravity
			dual to a Yang-Mills field theory in $3+1$ dimensions where the
			gauge fields transform in the adjoint representation of the
			color group $\SU(N)$, supplemented by $N_f$ additional chiral
			fermions and $N_f$ antichiral fermions which transform in the
			fundamental representation of the $\U(N)$ and in the fundamental
			representation of a $\U(N_f)$ flavor group. Supersymmetry is
			completely broken in this theory.
			
			The geometric realization of this setup is given by a
			\D4/\D8/$\overline{\text{\D8}}$ construction. A stack of $N$ \db4s
			in the near horizon limit gives rise to the background geometry of
			the (type IIA) supergravity theory, analogous to the \D3 setup. This
			time however one of the directions along the \db4s has to be
			compactified in order to avoid a conical singularity in the
			resulting background. The matter fields are introduced by a number
			of $N_f\ll N$ probe \db8s and anti-\db8s. These branes introduce the
			chiral symmetry groups $\U(N_f)_R$ and $\U(N_f)_L$ which account for
			fermions of opposite chirality. A caveat of this model is that the
			bare quark masses of these fields are vanishing.

		\subsubsection*{Embedding D-branes}
		\label{sec:embeddingDbranes}
			
\begin{figure}
				\centering
				\includegraphics[width=\linewidth]{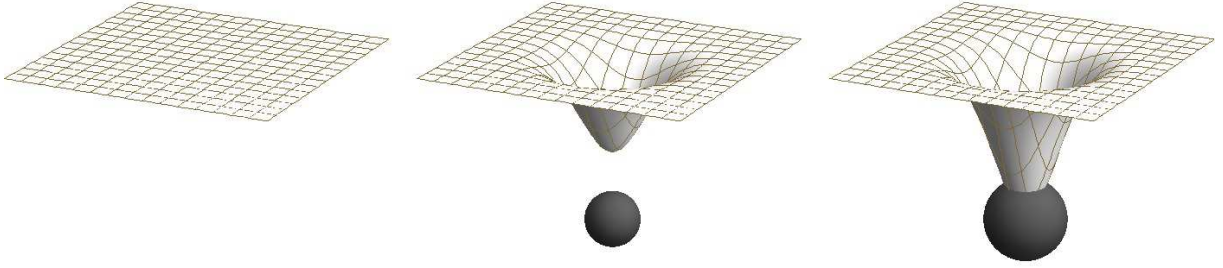}
				\caption[Sketch of brane embeddings for different temperatures]{%
					Sketch of brane embeddings in the directions transverse
					to the \db3 for different values of the temperature
					(relative to quark mass). Left: zero temperature,
					center: small temperature, right: high temperature, here
					the brane crosses the horizon.}
				\label{fig:fancyBranes}
			\end{figure}

			As a concrete realization of the \D3/\D7 setup we now consider the
			embedding of a \db7 into \adsfivesfive and its thermal
			generalization, the \adsfivesfive black hole background. We will
			perform the calculation in the black hole background and can obtain
			pure \adsfivesfive solutions as the limit of vanishing horizon
			radius, $\rh\to0$. Intuitive expectations would lead to embeddings
			which are influenced by the attractive gravitational force of the
			black hole as drawn in the cartoon of figure~\ref{fig:fancyBranes},
			which we will quantify now.
			
			The action of a probe \db7 is given by the DBI action
			\eqref{eq:DBIaction}. For now we consider the case of vanishing
			field strengths $2\pi\alpha'F=B=0$. Therefore we are left with
			\begin{equation}
				\label{eq:D7DBI}
				\SDBI = -T_p \int\!\!\dd^8\xi \;\sqrt{\left|\det G\right|}\:.
			\end{equation}
			The constant prefactor $T_p$ (cf.\ \eqref{eq:braneTension}) is
			not important for the following discussion. The action is
			determined by the induced metric $G(\xi)=P[g(x)]$ on the worldsheet,
			where $\xi$ are coordinates of the worldvolume of the \db7. The
			elements of the induced metric are given by
			\begin{equation}
			\label{eq:inducedMetric}
				G_{\mu\nu}(\xi) = \frac{\partial x^a}{\partial\xi^\mu}\frac{\partial x^b}{\partial\xi^\nu}\,g_{ab}.
			\end{equation}
			Here $g$ is the metric of the \adsfivesfive black hole
			background with coordinates $x^a(\xi)$ into which we embed the
			\db7. Two of these coordinates can be interpreted as functions
			which determine the position of the eight-dimensional
			worldvolume of the probe brane in the two directions transverse
			to the brane. These functions have to be determined in order to
			minimize the action \eqref{eq:D7DBI}.
			
			It is convenient for this purpose not to work in the coordinates
			of \eqref{eq:adsBHmetricInr} but to change to a new radial
			coordinate $\varrho$ given by
			\begin{equation}
				\varrho^2 = r^2 + \sqrt{r^4-\rh^4}.
			\end{equation}
			The metric now is
			\begin{equation}
			\begin{gathered}
			\label{eq:adsBHmetricInw}
				\ds^2=\frac{\varrho^2}{2R^2}\left(-\frac{f^2(\varrho)}{\tilde f(\varrho)}\,\dd t^2 +\tilde f(\varrho)\,\dd\threevec{x}^2\right)+\frac{R^2}{\varrho^2}\left( \dd \varrho^2 + \varrho^2\dd\Omega^2_5\right)\\
				f(\varrho) = 1-\frac{\rh^4}{\varrho^4}, \quad \tilde f(\varrho) = 1+\frac{\rh^4}{\varrho^4}.\hfill~
			\end{gathered}
			\end{equation}
			In this way we can identify the transverse part to the \db3s as
			nothing else than $\mathbbm R^6$ and we write it as
			\begin{equation}
			\label{eq:transverseMetric}
				\dd \varrho^2 + \varrho^2\dd\Omega^2_5=\sum_{i=1}^6\dd \varrho_i^2 = \underbrace{\dd w^2 + w^2\dd\Omega_3^2}_{\mathbbm R^4(\varrho_{1,\ldots,4})} + \underbrace{\dd L^2 + L^2 \dd \phi^2}_{\mathbbm R^2(\varrho_{5,6})}.
			\end{equation}
			In these coordinates we parametrized the domain of the \db3s by
			$t$ and the three spatial coordinates $\threevec x$. The part of
			the spacetime transverse to it is parametrized by the six
			coordinates $\varrho_i$, with radial coordinate $\varrho=(\sum\varrho_i^2)^{1/2}$.
			Equivalently, we wrote the transverse space as a product space
			of a four-dimensional $\mathbbm R^4$ in polar coordinates with
			radial coordinate $w$ and a two-dimensional $\mathbbm R^2$
			with radial coordinate $L$, such that $\varrho^2=w^2+L^2$.
			
			An embedding of the eight-dimensional worldvolume of the \db7
			into \adsfivesfive is then given by two functions which describe
			the positions in the two dimensions transverse to the brane.
			Stability of the \db7 solution demands that the brane spans a
			trivial three-cycle in the transverse direction to the \db3s
			\cite{Karch:2002sh}. We thus embed the brane along the following
			directions.

			\noindent
			\parbox{\linewidth}{
			\vspace{\bigskipamount}
			\begin{tabular}{rl}
				 & $t$\quad $x_1$\quad $x_2$\quad $x_3$\quad $w$\quad $\S^3$\quad $L$\quad $\phi$\;\\
				\db3 & \raisebox{.5ex}{\rule{6.5em}{.25em}}\\
				\db7 & \raisebox{.5ex}{\rule{10.25em}{.25em}}\\
			\end{tabular}
			\vspace{\bigskipamount}}\\
			The worldvolume of the \db7 is then parametrized by coordinates
			$\xi$,
			\begin{equation}
				\xi^0 = t,\quad \xi^{1,2,3}=x^{1,2,3}, \quad \xi^4=w, \quad \xi^{5,6,7}\text{\:along\;}\Omega_3,
			\end{equation}
			they determine the position of the \db7 by the embedding
			functions $L(\xi)$ and $\phi(\xi)$. However, to ensure
			Poincar\'e invariance the embedding functions cannot depend on
			$\xi^{0,\ldots,3}$. Moreover, the rotational $\SO(4)$ symmetry
			along the directions of the internal $\mathbbm R^4$ of the
			worldvolume results in embedding functions which only depend on
			$\xi^4=w$. The induced metric \eqref{eq:inducedMetric} on
			the \db7 then reads
			\begin{equation}
			\label{eq:inducedD7rho}
			\begin{gathered}
				\ds_{\text{\D7}}^2 = \frac{ w^2+L^2}{2R^2}\left(-\frac{f^2}{\tilde f}\,\dd t^2 +\tilde f\,\dd\threevec{x}^2\right)+\frac{R^2}{ w^2+L^2}\left( \dd  w^2 +  w^2\dd\Omega^2_3\right),\\
				f = 1-\frac{\rh^4}{( w^2+L^2)^2}, \quad \tilde f = 1+\frac{\rh^4}{\left( w^2+L^2\right)^2}.\hfill~
			\end{gathered}
			\end{equation}
			This metric is $\ads_5\times\S^3$ at asymptotically large
			$w$. Note that the embedding function $\phi(w)$ does
			not appear. This reflects the rotational symmetry of the setup
			in the space perpendicular to the brane. Further inserting the
			result into \eqref{eq:D7DBI} allows to derive the equation of
			motion for the	embedding $L(w)$,
			\begin{equation}
			\begin{gathered}
			\label{eq:D7embeddingBH}
				0=\del_ w\left[ \frac{\mathcal W( w,L)}{\sqrt{1+(\del_ w L)^2}}\,\del_ w L \right]-\sqrt{1+(\del_ w L)^2}\frac{8\rh^8 w^3}{( w^2+L^2)^5}\,L,\\[\medskipamount]
				\mathcal W( w,L) =  w^3 \left(1-\frac{\rh^8}{( w^2+L^2)^4} \right).\hfill~
			\end{gathered}
			\end{equation}
			From the asymptotic form of the equation of motion we see that
			the solution near the boundary at large $ w$ behaves as
			\begin{equation}
			\label{eq:braneAsymptotics}
				L = m_{\scriptscriptstyle L} + \frac{c}{ w^2} +\dots\:.
			\end{equation}
			The
			embedding profile $L$ asymptotically tends to a constant value
			$m_{\scriptscriptstyle L}=\lim_{ w\to\infty}L$, which we use as a free parameter
			of the setup. Together with the demand for smooth embeddings the
			boundary conditions for the solutions $L( w)$ can be
			written as
			\begin{equation}
			\label{eq:embeddingBoundaryCond}
				\lim_{ w\to\infty}L( w) = m_{\scriptscriptstyle L}, \qquad \del_ w L(0)=0
			\end{equation}
			for embeddings that reach $ w=0$ and
			\begin{equation}
				\lim_{ w\to\infty}L( w) = m_{\scriptscriptstyle L}, \qquad L( w)\Big|_\text{horizon} \perp \text{horizon}
			\end{equation}
			for embeddings which enter the horizon. Any other boundary
			condition than orthogonality to the black hole horizon would
			lead to a transverse component of the gravitational force on the
			brane, which would deform the embedding until orthogonality is
			reached in the final equilibrium state.
			
\begin{figure}%
				\centering
				\includegraphics[width=.5\linewidth]{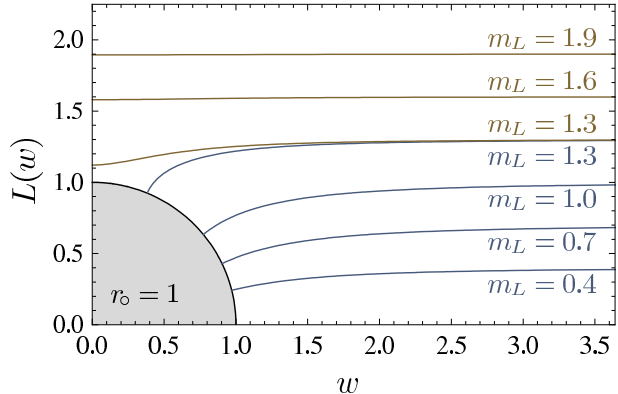}
				\caption[\db7 embeddings in \adsfivesfive black hole background]{
					Solutions to \eqref{eq:D7embeddingBH} yield
					Black hole (blue lines) and
					Minkowski embeddings (brown lines) of \db7s in the
					\adsfivesfive black hole background. The jump between black
					hole and Minkowski embedding at $m_{\scriptscriptstyle
					L}=1.3$ induces a change of the worldvolume topology,
					reflected in a first order phase transition of the dual
					field theory, see refs.~\cite{Babington:2003vm,Kirsch:2004km}.
				\label{fig:embeddingExamples}}
			\end{figure}

			The differential equation \eqref{eq:D7embeddingBH} generally has
			to be solved numerically \cite{Babington:2003vm}. At zero
			temperature, however, where $\rh=0$, as well as in the limit of
			large $\varrho$ the equation of motion is solved analytically by
			a constant embedding function. Some brane profiles are shown in
			figure~\ref{fig:embeddingExamples}. The embeddings which do not
			touch the horizon have a regular worldvolume metric. They are
			called \emph{Minkowski embeddings}. Note that these embeddings
			do not span the whole range of the coordinate $\varrho$ in
			\adsfivesfive, since $\varrho^2=L^2+w^2$ and $\min \varrho=\min
			L(\varrho) > \rh$. From the induced metric
			\eqref{eq:inducedD7rho} we see that these branes ``end'' at
			finite $\varrho$ before reaching the black hole horizon, since
			the $\S^3$ wrapped by the \db7 probe shrinks to zero size as in
			ref.~\cite{Karch:2002sh}. Those embeddings that end on the
			horizon exhibit a black hole on their worldvolume and are
			therefore called \emph{black hole embeddings}.
			
			Note that the supergravity scalar $L$ is part of the radial
			coordinate $r$ of \ads space. At asymptotically large values of
			the radial coordinate, where $L=m_{\scriptscriptstyle L}$, the relation is
			$2r^2=L^2+w^2$. So at fixed $w$ near the boundary
			$L\propto r$. In the inverse radial coordinate $u$ it scales
			like $u^1=u^{d-\Delta}$. According to the \ads/CFT dictionary,
			the dual operator of the $d=4$ dimensional field theory
			therefore is of dimension $\Delta=d-1=3$. This operator is
			the bilinear $\bar\psi\psi$. From holographic renormalization we
			learned that the mode of the solution $L$ scaling like
			$u^{d-\Delta}$ is proportional to the source term of this
			operator. From the field theory Lagrangian $\L=
			m_q\bar\psi\psi+\ldots$ we see that this source is the mass
			$m_q$ of the ``quark field'' $\psi$. An exact calculation
			relates the parameter $m_{\scriptscriptstyle L}$ to the quark mass $m_q$
			by\footnote{Equation \eqref{eq:mqFromEmbedding} does not look
			like the formula for $m_q$ given in the original paper,
			ref.~\cite{Karch:2002sh}, where the concept was introduced. The
			relative factor of $\sqrt 2$ arises from the different
			coordinate systems used here and in ref.~\cite{Karch:2002sh}.
			The transformation between them introduces that factor in the
			embeddings and therefore also in the quark masses, cf.\ 
			appendix~\ref{chap:coordinates}.}
			\begin{equation}
			\label{eq:mqFromEmbedding}
				m_q = \frac{m_{\scriptscriptstyle L}}{2^{\tiny 3/2}\pi\alpha'}\,.
			\end{equation}
			The embedding function in this way determines the mass of the
			quarks in the dual field theory \cite{Karch:2002sh}.
			
			Accordingly, the coefficient $c$ in \eqref{eq:braneAsymptotics}
			that scales like $u^\Delta=u^3$ is proportional to the vacuum
			expectation value $c_{c}=\vev{\bar\psi\psi}$, known as the chiral
			condensate. We will not discuss this quantity in detail in this
			work.

\subsection{Holographic quantum chromodynamics}
\label{sec:adsAndQCD}
	
	In this section we want to point out some of the most important features
	of QCD and whether they can be described by gravity duals or not. We will
	see that the conditions at which the quark-gluon plasma exists, in
	particular finite temperature, allow for an at least qualitative
	description of many aspects via gravity duals. 
	
	The gauge/gravity correspondence is a remarkable tool for the
	investigation of the strongly coupled regime of gauge theories.
	Depending on the choice of parameters, the features of the gauge field
	theory will be more or less close to what we expect from quantum
	chromodynamics.
	
	There is hope that the \ads/CFT correspondence may provide some pre- and
	postdictions even though the exact gravity dual to QCD is not known.
	During the last years it turned out that there exist some quantities,
	like the celebrated ratio $\eta/s=1/(4\pi)$ of shear viscosity
	to entropy density, that are universal in the sense that they do not
	depend on a particular supergravity background. Instead they are valid
	for all theories that have a gravity dual. If QCD is within this
	universality class, the results from other gauge theories than QCD may
	be applied to quantum chromodynamics as well. 
	
	\subsubsection*{Field content and supersymmetry}
	
		On the field theory side of the setup we stated that there is the gauge
		multiplet of $\N=4$ super Yang-Mills theory supplemented by $N_f$ $\N=2$
		hypermultiplets. The gauge degrees of freedom show up in a multiplet
		together with scalar and fermionic superpartner particle fields. In the
		Maldacena limit these multiplet exists in infinitely many colors, i.\,e.\ 
		the rank of the gauge group is infinite. The fundamental fermionic
		flavor degrees of freedom on the other hand exist only in a finite
		number $N_f$ of flavors, mostly we will restrict to $N_f=2$. The models
		studied in the following do not explicitly break supersymmetry further,
		while other setups may allow to break supersymmetry completely
		\cite{Constable:1999ch}.
		
		Compared to QCD we deal with infinitely times more degrees of
		freedom. However, as we saw the large $N$ limit simplifies the
		theory drastically in the way that it allows us to neglect string
		loops. In fact corrections of results obtained from large $N$
		expansions with expansion parameter $1/N$ tend to be small, as e.\,g.\ 
		lattice calculations have shown.
		
		Moreover, at all finite values of temperature $T\neq0$ supersymmetry
		is broken spontaneously. Unlike most other spontaneously broken
		symmetries it is not restored at high temperatures
		\cite{Das:1978rx}. For a system at equilibrium this can be seen from
		the fact that the fermionic degrees of freedom contained in a
		supermultiplet obey antiperiodic boundary conditions along the
		imaginary axis in the complex time plane, while bosons are periodic.
		Therefore the Fourier decompositions and therewith the masses of
		these fields will differ. Fields with different masses however are
		not related to each other by supersymmetry transformations. This
		shows that supersymmetry is broken at finite temperature.
		
		To great extent we will be interested in the behavior of
		fundamental matter and the bound states of the fundamental fields.
		As we have done above when we identified the mass of the quarks with
		the mass of the fundamental fermions in the hypermultiplet, we will
		think of these fields as the cousins of the quarks in
		QCD\,---\,keeping in mind that gauge invariant operators receive
		contributions from the scalar superpartners.

	\subsubsection*{Conformal symmetry}
		
		By definition the \ads/CFT correspondence relates gravity to a
		\emph{conformal}, i.\,e.\  scale invariant, quantum theory. We introduced
		it as $\N=4$ SYM theory. This is in vast contrast to QCD where we
		know several scales which break conformal symmetry.
		
		The masses of the quarks which have to be regarded as fundamental
		parameters of QCD break scale invariance explicitly. The analogon
		to quark mass is realized geometrically in terms of the brane embedding
		profile.
		
		The dynamically determined momentum scale $\Lambda_\text{QCD}$ at
		which the coupling constant diverges arises from quantization
		effects and therefore is a manifestation of scale anomalies. This
		scale has its dual in the background geometry of the supergravity
		theory. Temperature is geometrically realized by introducing a black
		hole into the spacetime of the gravity theory. The radius $\rh$ of
		the black hole introduces the scale dual to the finite temperature
		in field theory. In the way a finite temperature can be interpreted
		as a lower bound on the momentum of particles, the black hole horizon
		radius introduces a cutoff. We interpreted the radial coordinate in
		\ads as the scale of a renormalization group flow. The horizon
		radius then works as a momentum cutoff. Geometrically, it introduces
		a scale and thereby explicitly breaks global conformal invariance.

	\subsubsection*{Bound states of quarks}
	\label{sec:boundStates}
		
		One of the great successes of the gauge/gravity duality is the
		possibility to derive spectra of bound states of fundamental matter from
		first principles. The string tension $\alpha'$ is a fundamental
		parameter of the theory and determines the quark mass, which naturally
		appears as a parameter in the spectra of bound states. Another parameter
		is the 't~Hooft coupling $\lambda$. Since we work in the low energy
		limit where we expect not to resolve the string scale we cannot
		explicitly assign any value to $\alpha'$, and in the Maldacena limit we
		cannot assign a finite value to $\lambda$. We are thus unable to derive
		numerical values for meson masses. Nevertheless, the ratio of meson
		masses we obtain from holographic models can be compared to observations
		from experiment. Experimentally observed ratios of meson masses are
		reproduced with an accuracy of about 10\% \cite{Erdmenger:2007cm}. With
		respect to the various limits in which the according calculations are
		performed, this is an astonishing accuracy.
		
		\paragraph{Baryons}
			
			The possibilities to model baryons is very limited to this day.
			This is partly due to the fact that we are restricted to the
			limit of infinitely many colors. Since baryons are colorless
			composite particles made out of $N_c$ quarks, we would have to
			describe an object made of an infinite number of particles.
			
			There are however Skyrmion like solutions in the Sakai-Sugimoto
			model \cite{Sakai:2005yt}. Recently, baryon like operators where
			considered in Chern-Simons-matter field theory derived in an
			$\ads_4/\text{CFT}_3$ model \cite{Park:2008bk}.
		
		\paragraph{Mesons}
		\label{par:mesons}

			As mesons are composed operators containing one quark and one
			antiquark field, they transform in the adjoint representation of
			the flavor gauge group $\SU(N_f)$, which can also be expressed
			as a bifundamental representation with one index in the
			fundamental representation $\mbold N_{\!f}$
			and the other in the antifundamenal $\mbold{\bar N}_{\!f}$, as
			we did in section~\ref{sec:largeNlimit} for the color gauge
			group. This transformation property on the string theory side is
			given by a string that has both ends on the probe \db7. So the
			mesons of the field theory are dual to the excitations of
			\D7-\D7 strings. The endpoints of these strings determine
			position of the probe \db7. Consequently the excitations of the
			\D7-\D7 strings describe fluctuations of the probe branes. The
			meson masses can then be obtained from the solutions to the
			linearized equations of motion of these fluctuations of the
			probe branes around the embedding.
			
			As a very short sketch of the procedure and for later reference we
			outline the calculation of the spectrum of scalar mesons at zero
			temperature, first published in ref.~\cite{Kruczenski:2003be}.
			Above, the embedding of the flavor \db7s at zero temperature was
			shown to be described by constant functions $L(w)=m_{\scriptscriptstyle L}$ and
			constant $\phi(w)$. We now allow for small deviations from this
			embedding by adding small fluctuations $\tilde\varphi_{L,\phi}(\xi)$ to the
			embedding functions,
			\begin{equation}
				\label{eq:fluctAnsatz}
				L\mapsto L+ \tilde\varphi_L(\xi),\qquad \phi\mapsto\phi+\tilde\varphi_\phi(\xi).
			\end{equation}
			We want to consider small deviations from the brane profile and
			therefore may restrict our attention to the linearized equations of
			motion for the fluctuations $\varphi_{L,\phi}$. In the same way in
			which the embedding is determined by the equations of motion
			obtained from the DBI action, we can derive the linearized
			equations of motion for the fluctuations from the same action
			\eqref{eq:D7DBI}. Analogous to the derivation of the equation of
			motion \eqref{eq:D7embeddingBH} for the embedding functions the
			equations for the fluctuations around $L$ in the zero temperature
			case of $\rh=0$ are obtained by plugging in the ansatz
			\eqref{eq:fluctAnsatz} into the action. The resulting linearized
			equation of motion for the fluctuation $\tilde\varphi$ was calculated in
			ref.~\cite{Kruczenski:2003be} for zero temperature. To make the
			connection to the computation of the derivation of the embedding
			functions, we stick to the coordinates \eqref{eq:adsBHmetricInw} with
			\eqref{eq:transverseMetric} in which the fluctuation equation for
			fluctuations $\tilde\varphi_L$ are derived to be
			\begin{equation}
			\label{eq:fluctEOM}
				0=\frac{2R^4}{(w^2+L^2)^2}\,\del^i\del_i\tilde\varphi + \frac{1}{w^3}\,\del_w\left(w^3\del_w\tilde\varphi\right) + \frac{2}{w^2}\,\nabla^a\nabla_a\tilde\varphi.
			\end{equation}
			Here $i$ is summed over the Minkowski directions, $\nabla_a$ are the
			covariant derivatives along the directions of the $\S^3$ spanned by
			the probe \db7, and the radial coordinate of \ads is given by
			$\varrho^2=w^2+L^2$. Use the ansatz
			\begin{equation}
				\tilde\varphi = \varphi(w)\:\ee^{-\ii\,\fourvec k\fourvec x}\:Y^l(\S^3)
			\end{equation}
			with $Y^l(\S^3)$ as the spherical harmonics along the three sphere,
			such that $\nabla^a\nabla_a\tilde\varphi=-l(l+2)\tilde\varphi$. Here
			$l=0,1,2,\ldots$ is the angular momentum number on the $\S^3$.
			However, in this work we only consider the solutions with $l=0$. The
			plane wave factor is responsible for
			$\del^i\del_i\tilde\varphi=-k^2\tilde\varphi$, with momentum vour
			vector $\vec k$ which determines the meson mass $M$ by $M^2=-k^2$.
			Therefore the above ansatz transforms \eqref{eq:fluctEOM} into an
			ordinary differential equation for the radial part $\varphi(w)$,
			\begin{equation}
			\label{eq:fluctEOMradial}
				0=\frac{2R^4}{(w^2+L^2)^2}\,M^2\varphi(w) + \frac{1}{w^3}\,\del_w\left(w^3\del_w\varphi(w)\right).
			\end{equation}
			It can be solved in terms of hypergeometric functions. However,
			normalizable solutions only exist for
			\begin{equation}
			\label{eq:mesMassDimensionful}
				M_n = m_q\frac{4\pi\alpha'}{R^2}\,\sqrt{(n+1)(n+2)}\:,\qquad n=0,1,2,\ldots,
			\end{equation}
			where the quark mass $m_q$ enters through the embedding $L$ by
			\eqref{eq:embeddingBoundaryCond} and \eqref{eq:mqFromEmbedding},
			with constant $L=m_{\scriptscriptstyle L}$. This is the mass spectrum of mesons at
			zero temperature and vanishing particle density. In fact this is
			the form of the spectrum for scalar, pseudo scalar and vector
			mesons \cite{Kruczenski:2003be}. We will compare later results
			at finite temperature and finite density to this formula.
			
			Various aspects of meson spectroscopy have been under
			investigation, among these are the discrete meson spectra of
			stable quark-anti quark mesons at zero temperature
			\cite{Kruczenski:2003be}, the decreasing stability and melting
			of these states at finite temperature and finite particle
			density \cite{Erdmenger:2007ja,Erdmenger:2008yj,Hoyos:2006gb},
			and the investigation of the spectra of heavy-light mesons
			\cite{Karch:2002xe,Erdmenger:2007vj}.
			
			We will derive meson spectra for various
			purposes. On the one hand side we are interested in the
			dependence of the spectra under variation of temperature and
			particle density in order to understand the behavior of bound
			states of quarks. On the other hand we will observe the
			influence of external fields on the mass spectra to derive the
			polarizability of the mesons, which in turn influence their
			diffusion behavior inside the quark-gluon plasma.
			
			For mesons in the Sakai-Sugimoto model we again refer to
			refs.~\cite{Sakai:2004cn,Sakai:2005yt}.

	\subsubsection*{Confinement/deconfinement}
	\label{sec:confinement}
		
		The probably most prominent feature of QCD is the running coupling
		constant, meaning the change in the value of the coupling constant of
		the strong interaction under variations of the energy scale of the
		interactions. Processes involving high momentum transfer are influenced
		less by the strong interactions that those which occur at low momentum.
		Mathematically, the value of the coupling constant even diverges at a
		momentum scale known as $\Lambda_\text{QCD}$. Such a running of the
		coupling constant with respect to the energy scale is obviously only
		possible in the absence of conformal invariance, which would forbid the
		existence of a characteristic scale. As a result, quarks at low
		energies, e.\,g.\  low temperature, are confined to bound states which appear
		as colorless entities to a far away located observer. At high
		energies/high temperatures, the quarks may escape from these states and
		travel through spacetime independently.
		
		Experiments show us that at energies above approximately 175 MeV quarks
		and gluons start to enter the deconfined regime. The exact value depends
		on various parameters. So far there is no analytic proof for these
		properties of quarks and gluons. To great extent this is due to the fact
		that the confinement/deconfinement transition occurs in the strongly
		coupled regime of the gauge theory. Traditional perturbative methods may
		not be applied here.
		
		In the framework of the gauge/gravity duality, however, one can hope to
		see effects of the confinement/deconfinement transition, since we can
		work in the strongly coupled regime of the gauge theory. The original
		correspondence contained adjoint matter fields, given by the gauge
		multiplet of $\N=4$ SYM theory. At finite temperature, the gauge fields
		undergo a first order phase transition at a temperature
		$T_\text{gauge}$. It coincides with the Hawking-Page temperature and can
		be interpreted as the confinement/deconfinement transition
		\cite{Witten:1998zw,Karch:2006bv}.
		
		The fundamental matter existing in probe brane setups also exhibits a
		phase transition at finite temperature, which though occurs at a
		different temperature $T_\text{fund}$ than the transition of the gauge
		fields. Various different models exist where the fundamental degrees of
		freedom indeed undergo a phase transition from stable bound states to
		dissociating ones
		\cite{Babington:2003vm,Kirsch:2004km,Erdmenger:2007ja,Hoyos:2006gb,Myers:2007we}.
		We will come back to this transition when we discuss the QCD phase
		diagram in section~\ref{chap:phaseDiag}.

		It is interesting to note a difference between holographic and lattice
		models. The deconfinement temperature for fundamental matter associated
		with the destabilization of mesons derived from holographic models is
		proportional to the mass of the constituent quarks of the meson,
		$T_\text{fund}\propto m_q$. Lattice results in the quenched
		approximation, in contrast, suggest a scaling of the transition
		temperature of meson destabilization with the transition temperature for
		the gauge fields $T_\text{fund}\propto T_\text{gauge}$
		\cite{Asakawa:2003re,Datta:2003ww}. This in principle allows for
		interesting comparison of lattice and holographic models with
		experimental data.

	\subsubsection*{Chiral symmetry}
	
		Soon after the introduction of fundamental matter to the gauge/gravity
		duality it was shown that various probe brane setups are capable of
		realizing chiral symmetry breaking at finite temperature holographically
		\cite{Babington:2003vm,Kirsch:2004km,Evans:2004ia,Kruczenski:2003uq,Sakai:2004cn}.
		The order parameter for the transition between the chiral symmetric
		phase and the phase of spontaneously broken chiral symmetry is given by
		the chiral condensate $c_c$ for massless quarks, i.\,e.\  the vacuum
		expectation value of the bifundamental $c_c=\vev{\bar\psi\psi}$.
		  In some models this
		transition coincides with the confinement/deconfinement transition
		\cite{Evans:2008tu}. 
		
		In the Sakai-Sugimoto model, chiral symmetry breaking is
		realized by the merging of the embeddings of the \D8 and
		$\overline{\text{\D8}}$ in the low temperature phase. In this way the
		flavor groups $\U(N_f)_L$ and $\U(N_f)_R$ originating from string
		excitations on the respective brane combine to a single vector subgroup
		$\U(N_f)_V$.

\section{Thermal vector meson spectra at finite particle density}
\label{chap:specFuncs}

In this section we address the first two questions raised in the introduction.
We wondered whether bound states of quarks can be observed in holographic models
of the thermal quark-gluon plasma, and how they are influenced by the medium.
From experiment we know that temperature and particle density influence the
interaction between particles, a prominent example is the transition of quarks
and gluons from the confined to the deconfined phase at increasing interaction
energy, or equivalently at high temperature.

Temperature and quark density are the most important parameters of the model we
make use of in this work. Together with the mass of the fundamental fields they
define the axes of the phase diagram of fundamental matter in the holographic
QGP. In the context of gauge/gravity duality, there has been an intensive study
of the phase diagram of $\N=4$ supersymmetric $\SU(N)$ Yang-Mills theory in the
large $N$ limit, with fundamental degrees of freedom added by considering the
\ads-Schwarzschild black hole background with \db7 probes \cite{Apreda:2005yz,
Kirsch:2004km,Babington:2003vm,Mateos:2006nu,Albash:2006ew,Hoyos:2006gb}.
Another approach was pursued by studying string worldsheet instantons
\cite{Faulkner:2008qk}. Subsequently, particular interest has arisen in the more
involved structure of the phase diagram when the baryon chemical potential is
present, taking finite density effects into account \cite{Kobayashi:2006sb}. We
contemplate the phase diagram and its parameters more detailed in
section~\ref{chap:phaseDiag}. In this section we concentrate on the mesonic
bound states of quarks and their dependence on temperature and density.

The aim of this section is the combination of both, finite temperature and
finite density effects in the description of a thermal holographic plasma. In
$\N=4$ SYM theory with finite baryon density, we relate our work to the phase
diagram shown in figure 2 of ref.~\cite{Mateos:2007vc}, reproduced below in
figure~\ref{fig:phaseDiagram}. We restrict to setups with non-vanishing particle
density. Here, fundamental matter is described solely by probe branes with the
geometry of black hole embeddings \cite{Kobayashi:2006sb,Karch:2007br}.
 The holographic realization of finite particle
density is discussed below, where we introduce the setup.

The concept of mesonic bound states in $\N=4$ SYM theory in the Maldacena limit
at vanishing temperature and particle density, and an outline of how to obtain
their spectra was sketched in section~\ref{sec:boundStates}. Here, we extend
these calculations to incorporate \emph{in-medium effects} of finite temperature
and particle density, giving rise to non-vanishing baryon or isospin chemical
potential. The motivation to do so stems from the possibility to conduct
experiments at non-vanishing isospin density \cite{Tsang:2003td}, as well as the
better accessibility by lattice methods of the finite isospin region in the
phase diagram of QCD compared to finite baryon chemical potential. The work
presented here restricts to the calculation of vector meson bound states with
vanishing spatial momentum. At finite momentum the vector mesons couple to
scalar mesons. Extensions of our work to finite momentum can be found in
refs.~\cite{Myers:2008cj,Mas:2008jz}.

The meson spectra will be represented in terms of \emph{spectral functions}.
These functions of an energy variable will exhibit resonance peaks of finite
width, corresponding to decay rates, at energies corresponding to the meson
masses. The necessary extensions of the setup to finite particle density and the
concept of spectral functions are introduced in the subsequent sections. In
order to determine the spectral function at finite temperature and finite baryon
density, we make use of the methods developed in the context of \ads/CFT applied
to hydrodynamics, c.\,f.\  for instance
refs.~\cite{Son:2002sd,Teaney:2006nc,Kovtun:2006pf}. For vanishing chemical potential,
a similar analysis of mesons has been performed in ref.~\cite{Myers:2007we}. There it
was found that the mass spectrum is discrete for quarks with masses
significantly above the energy scale set by the temperature. At lower quark
mass, a quasiparticle structure is seen which displays the broadening decay
width of the mesons. As the mass decreases or temperature rises, the mesons are
rendered unstable, reflected in broad resonance peaks. These excitations
dissipate their binding energy into the plasma. Note that for this case, there
are also lattice gauge theory results \cite{Aarts:2007wj}.

The achievements of the work presented in this section are the successful
incorporation of either baryonic or isospin chemical potential at finite
temperature. Before the results of this section where published as
refs.~\cite{Erdmenger:2007ja,Erdmenger:2008yj} these aspects where investigated
separately in the literature. As we will see, the simultaneous incorporation of
both temperature and particle density leads to spectra which can be compared to
previous publications consistently in the appropriate limits. In particular, we
find that at low temperature to quark mass ratio, i.\,e.\  close to the Minkowski
phase, where the characteristic energy scale of the system is given by the quark
mass, the spectrum is asymptotically discrete and coincides with the
zero-temperature supersymmetric meson mass formula found in
ref.~\cite{Kruczenski:2003be} and rephrased in equation
\eqref{eq:mesMassDimensionful}. However, away from this regime the dominant
energy scale is either the finite temperature or the chemical potential. Here
the observed spectra differ qualitatively from the above in some respects and
resemble aspects of mesonic excitations in QCD. They also show interesting
similarities to phenomenological models. We elaborate on the physical
characteristics of our results in the summary of this section.

In the case of an \emph{isospin} chemical potential, previous work in the
holographic context has appeared in refs.~\cite{Apreda:2005yz,Parnachev:2007bc}. In
this case, two coincident \db7 probes are considered, which account for
fundamental matter of opposite isospin charge. We find that spectral functions
quantitatively deviate from the baryonic background case. A triplet splitting of
quasi-particle resonances in the spectral function is observed, which depends on
the magnitude of the chemical potential.

\subsection{Spectral functions}
\label{sec:retardedThermalGreen}
	
	The spectral function $\R(\omega,\threevec q)$ of an operator $J$ describes
	the probability density in $(\omega,\threevec q)$-space to detect the
	quantity encoded in the eigenvalues of the operator $J$ at given energy
	$\omega$ and spatial momentum $\threevec q$. In our case, we want to
	describe quarkonium states and are interested in the mass/energy spectrum of
	the stable bound states and their lifetimes. In other words, we want to
	compute the spectral functions $\R(\omega,\threevec q)$ of a quark-antiquark
	operator corresponding to vector mesons. This operator appears in the field
	theory as the flavor current $\fourvec
	J(x)=\bar\psi(x)\fourvec\gamma\psi(x)$ of fundamental fields $\psi(x)$ (and
	their superpartners). For simplicity, let us restrict to the case of
	vanishing momentum, $\threevec q=0$, where the remaining parameter is the
	energy $\omega$ alone. Peaks in the spectral function at an energy $\omega$
	indicate that there is a large probability to find a quark-antiquark state,
	which is denoted as a \emph{quasiparticle} if the width of the peak is small
	compared to the height. The position $\omega$ of the peak gives the energy
	or mass of the quasiparticle while the width of the peak translates into the
	lifetime of this particle in position space. According to Fourier
	transformation, a broad peak, which is a large object in momentum space,
	corresponds to an event of short lifetime in position space, and vice versa
	a narrow peak in the spectral function is a signal for a particle with a
	long lifetime.
	
	We very briefly comment on how to derive the spectral function from two
	point functions, and how to extract the relevant information from them. See
	textbooks like ref.~\cite{Zagoskin} for details. The formulation of spectral
	densities in terms of two point Green functions is convenient because we can
	compute the latter holographically.
	
	We think of $J$ being the operator that describes the free mesonic
	quasiparticle as an excitation of one of the possible QGP many-particle
	states $\ket{n}$. There are infinitely many different of such states in the
	thermal ensemble that represents the QGP. The probability to occupy one of
	them is given by the density matrix $\hat\rho$, described in
	section~\ref{sec:finiteTempBlackHoles}. These states form a basis of the
	Hamiltonian $H$ of the ensemble, such that $\sum_n\ket{n}\bra{n}=\mathbbm1$.
	
	The probability of propagation from an initial spacetime point $x_i$, which
	we define as $x_i=(0,\threevec0)$, to some final point $x_f=(t,\threevec x)$
	is given by the time ordered Green function
	\begin{equation}
	\label{eq:mesonCorr}
		G_n(t,\threevec x) = -\ii\smatrix{n}{\theta(t)\, J(x_f) J^\dagger(0)}{n},
	\end{equation}
	where the step function $\theta(t)$ accounts for time ordering. The index
	$n$ shall remind us that this probability is not an ensemble average. It
	just gives the probability for the event if the QGP is in the state
	$\ket{n}$. By switching to the Schr\"odinger representation of the meson
	operators and denoting the momentum operator by $\hat{\threevec k}$,
	\begin{equation}
		J(t,\threevec x) = \ee^{-\ii(\hat{\threevec k}\threevec x- Ht)} J \ee^{\ii(\hat{\threevec k}\threevec x-Ht)},
	\end{equation}
	and insertion of a full set of eigenstates $\ket{n'}$ we arrive at
	\begin{equation}
		G_n(t,\threevec x) = -\ii\sum_{n'}\theta(t)\, \ee^{\ii(E_{n}-E_{n'})t-\ii\threevec k\cdot\threevec x}\smatrix{n}{J^{\vphantom\dagger}}{n'}\smatrix{n'}{J^\dagger}{n}.
	\end{equation}
	We denote the energy difference of the excited system to the QGP ground
	state as the energy $\omega_{nn'}$ of the mesonic excitation. This energy
	certainly depends on the state of the plasma with the excited mesonic state
	$\ket{n'}$, and the state $\ket{n}$ it was created from,
	$\omega_{nn'}=E_{n}-E_{n'}$. We write
	$\smatrix{n}{J^{\vphantom\dagger}}{n'}\smatrix{n'}{J^\dagger}{n}=\left|\smatrix{n'}{J^\dagger}{n}\right|^2$,
	and perform a Fourier transformation with respect to energy and momentum
	$(\omega,\threevec q)$,
	\begin{equation}
		G_n(\omega,\threevec q) =  \sum_{n'} \frac{\delta(\threevec k-\threevec q)}{\omega_{nn'}-\omega+\ii\varepsilon}\,\left|\smatrix{n}{J^\dagger}{n'}\right|^2,
	\end{equation}
	where the small $\varepsilon\in\mathbbm{R}$ accounts for proper convergence.
	The delta function reflects that the momentum is conserved in the
	multiparticle system. To get the probability for the detection of a meson
	with energy $\omega$ and momentum $\threevec q$ in the QGP, we have to
	perform the ensemble average. This eventually leads to the relation
	\begin{equation}
	\label{eq:corrLongSpecDens}
		G(\omega,\threevec q) = \frac{1}{Z} \sum_{n,n'} \frac{1+\ee^{-\beta \omega_{nn'}}}{\omega_{nn'} - \omega+\ii\varepsilon}\,\delta(\threevec k-\threevec q)\left|\smatrix{n'}{J^\dagger}{n}\right|^2.
	\end{equation}
	It is convenient to write this Green function as
	\begin{equation}
	\label{eq:corrShortSpecDens}
		G(\omega,\threevec q)=\int\dd\omega' \frac{\R(\omega',\threevec q)}{\omega'-\omega+\ii\varepsilon}.
	\end{equation}
	Here, we defined a weight function $\R(\omega',\threevec k)$ for the
	propagation of the meson state, which assigns different probabilities to the
	propagation according to the Green function $G(\omega,\threevec
	q)=1/(\omega'-\omega)$. The probability density $\R$ is called the
	\emph{spectral density} or \emph{spectral function}. From
	\eqref{eq:corrLongSpecDens} and \eqref{eq:corrShortSpecDens} we see that the
	spectral function is given by
	\begin{equation}
		\R(\omega',\threevec k)=\frac{1}{Z}\sum_{n,n'}\delta\left(\omega'-\omega_{nn'}\right)\delta\left(\threevec k-\threevec q\right) \left|\smatrix{n'}{J^\dagger}{n}\right|^2 \left(1+\ee^{-\beta \omega'}\right).
	\end{equation}
	This notation reflects the physical interpretation we gave earlier. The
	probability to observe the multiparticle system in a state $\ket{n'}$ with a
	mesonic excitation, created by acting with $J^\dagger$ on the initial QGP
	state $\ket n$, obeys energy and momentum conservation, and depends on the
	temperature, given by $\beta$.
	
	Most important for our purpose is the retarded Green function $\GR$. We
	where not explicit about the retarded and advanced contributions to the
	Green function in the above discussion. Nevertheless, the
	Sokhatsky-Weierstrass theorem in complex analysis allows to derive the
	following relation between the spectral function and the retarded Green
	function that we will make use of,%
	\begin{equation}
	\label{eq:specFuncFromGreenFunc}
		\R(\omega, \threevec k) = -2\Im \GR(\omega,\threevec k).
	\end{equation}
	The large probability density for the propagation of a quark-antiquark pair
	with the right energy content to form a bound state directly translates into
	an excess of the spectral function at that particular value of $\omega$. In
	the rest frame of the particle, which we are restricting our attention to,
	the energy can directly be translated into the mass $M$ of the meson by
	$\omega=M$. The calculation of the Green functions of flavor currents $J$ in
	this way yields information about the quasiparticle spectrum of a given
	theory\,---\,the meson spectrum.

	From the relation \eqref{eq:specFuncFromGreenFunc}, we see that a convenient
	way to obtain the spectral function is to compute the retarded Green
	function of the mesonic operator. A way to achieve this was sketched in
	section~\ref{sec:retardedThermalGreen}. We see that all information about
	the spectrum is contained in the correlation function $\GR$. The correlation
	function in turn is determined by the residues of its poles in the complex
	plane. From field theory we know that the poles of the correlation function
	in the complex $\omega$-plane can directly be translated to the energies of
	the states. We will consider spectral functions at vanishing spatial
	momentum $\threevec q$, determined by the energy $\omega$ alone.
	
	The standard example in field theory is Klein-Gordon theory which amounts to
	the equation of motion for a scalar field $\phi$ given by
	\begin{equation}
		\left(\Box - m^2\right)\phi = 0.
	\end{equation}
	In terms of the formalism of Green functions the evolution of a delta-shaped
	initial perturbation of $\phi$ is given by the inverse of the differential
	operator $\left(\Box - m^2\right)$. The modes of this solutions are then
	obtained from the Fourier transform (with $\Box\mapsto\omega^2$ in our
	example of a particle at rest),
	\begin{equation}
		G(\omega) \propto \frac{1}{\omega^2-m^2}\,.
	\end{equation}
	The Green function exhibits poles at $\omega=\pm m$, corresponding to modes
	with the energy $\omega$ of the stable particle at rest. These real valued
	poles are less frequently referred to as normal modes. The solution to more
	complicated systems than Klein-Gordon theory, where we have unstable
	excitations which dissipate energy, we encounter \emph{quasinormal modes}
	$\Omega$ which are \emph{complex valued}. Quasinormal modes (QNM) where
	introduced in the context of metric fluctuations in black hole background.
	The black hole geometry accounts for the attenuation since any amount of
	energy that crosses the event horizon irreversibly disappears from the
	system outside. This mechanism also works in the case we will investigate.
	The difference to the simple example of Klein-Gordon theory is that we
	consider correlators of gauge field fluctuations in an \adsfivesfive black
	hole background rather than a scalar field in flat space. The fluctuations
	of the gauge field on the brane can transport energy into the black hole,
	but no energy can escape from the horizon. This introduces dissipation,
	which is then described by the imaginary part of the quasinormal modes.
	
	For demonstration, presume the solution $\phi(t)$ of the equation of motion
	for a mesonic excitation may be decomposed into quasinormal modes
	$K(\Omega)$ for complex frequencies $\Omega$. Suppose we only excite one
	mode for a single complex $\Omega'$, such that
	$K(\Omega)=k\,\delta(\Omega-\Omega')$. This mode describes an attenuated
	oscillation as long as the imaginary value of $\Omega$ is negative,
	\begin{equation}
		\phi(t) = \int\!\dd \Omega\; K(\Omega)\,\ee^{-\ii\,\Omega t} = k\,\ee^{-\ii\Re(\Omega') t} \: \ee^{\Im(\Omega')t}.
	\end{equation}
	For positive imaginary parts of the quasinormal modes, we encounter the
	unphysical case of infinite amplification of any fluctuation of the field.
	Therefore, in a physical setup one may find singularities of the retarded
	Green functions $\GR(\omega, \threevec q)$ only in the lower half of the
	complex $\omega$-plane. Those with the lowest absolute value of the
	imaginary part are referred to as the hydrodynamic poles of the retarded
	real-time Green function since they determine the long time behavior.
	Consider the made up example in which a Green function exhibits a pole at
	$\Omega=\omega_0-\ii\Gamma$,
	\begin{equation}
	\label{eq:quasiNormalModes}
		\GR(\omega) \propto \frac{1}{\omega-\Omega}.
	\end{equation}
	The pole emerges as a peak in the spectral density of real valued energies
	$\omega$,
	\begin{equation}
			\R(\omega) =-2\Im \GR(\omega) \propto \frac{2\, \Gamma}{(\omega-\omega_0)^2+\Gamma^2}\,,
	\end{equation}
	located at $\omega_0$ with a width given by $\Gamma$. These peaks are
	interpreted as quasi-particles if their lifetime $1/\Gamma$ is
	considerably long, i.\,e.\  if $\Gamma \ll \omega_0$ and thus the peaks in the
	spectral function are narrow.
	
	We compute the spectral function for real and complex values of the energy
	$\omega$ in this section. However, we focus on the meson spectra, i.\,e.\  the
	$\R(\omega,0)$ for $\omega\in\mathbbm{R}$, and postpone the discussion of
	the physical consequences and a detailed analysis of the behavior and
	location of the quasi normal modes to section~\ref{chap:phaseDiag}. In-depth
	analytical and numerical investigations of the behavior of quasinormal modes
	in gauge/gravity duality can be found in refs.~\cite{Amado:2007yr,Hoyos:2006gb}.
	
\subsection{Holographic setup}
\label{sec:backAndBranes}
	
	For an explicit calculation of the flavor current Green functions, we have
	to find the solutions to the equations of motion for the supergravity field,
	which is holographically dual to the flavor currents in the field theory.
	This field, which we called $\phi$ in the general example
	\eqref{eq:separateBoundaryBehaviour}, is the gauge field $A$ on the probe
	brane \cite{Kruczenski:2003be}. This means that we have to include the
	contributions of the gauge field strength tensor $F=\dd A$ in the DBI
	action.
	
	Because of the non-linearity of the DBI action, the embedding functions and
	gauge fields couple. We no longer can expect to determine the embedding
	functions in terms of the quark mass alone. Instead, we first introduce a
	suitable coordinate system to describe the background geometry, then derive
	the equations of motion and solutions of the background fields, which are
	the embedding and the gauge field on the \db7s. In the subsequent section we
	investigate the fluctuations of the gauge fields on this background to
	eventually derive the meson spectra from them. To account for finite
	temperature, we consider finite values $\rh\neq0$ in the \ads black hole
	metric.
	
	We compute the functional dependence of the gauge field $A(\rho,\fourvec k)$
	and the embedding functions numerically in the limit of vanishing spatial
	momentum $\threevec q\to 0$ for the fluctuations. In this limit the momentum
	four vector simplifies, $\fourvec k=(\omega,\threevec
	q)=(\omega,\threevec0)$ and there are no couplings between the vector and
	scalar mesons.
	
	\subsubsection{Background geometry and supergravity action}
		
		We work in the \D3/\D7 setup introduced above, i.\,e.\  we consider
		asymptotically $AdS_5\times S^5$ space-time which arises as the near
		horizon limit of a stack of $N$ coincident \db3s. More precisely, our
		background is the $AdS$ black hole geometry discussed in
		section~\ref{sec:finiteTempBlackHoles}, which is the geometry dual to a
		field theory at finite temperature \cite{Policastro:2002se}. In this
		background, we encountered \db7 embeddings of Minkowski type as well as
		black hole embeddings. The phase transition between both classes of
		embeddings is of first order \cite{Babington:2003vm,Kirsch:2004km}. The
		analysis of the meson spectrum shows that it corresponds to a transition
		between a phase of stable bound states of the fundamental degrees of
		freedom and a phase in which these mesons have finite lifetime. In
		physical parameters from the field theory point of view, we are in the
		deconfined phase at high temperatures at which mesons are unstable and
		said to be melting. For a well defined notion of high versus low
		temperatures, we need to compare the temperature to some energy scale.
		In our setup, the only available scale for comparison is the quark mass.
		Whether we are in the stable or in the melting phase, i.\,e.\  whether one or
		the other type of embedding is realized on the gravity side, depends on
		the ratio of quark mass $m_q$ to temperature $T$, c.\,f.\ 
		figure~\ref{fig:embeddingExamples}. This can be seen from the equation
		of motion for the embeddings \eqref{eq:D7embeddingBH}. It is invariant
		under scale transformations by a factor of $a$, resulting in $L\mapsto
		aL$, $w\mapsto aw$ and $\rh\mapsto a\rh$. Scaling $L$ by $a$ amounts to
		scaling the quark mass $m_q$ by the same factor $a$, while scaling of
		$\rh$ is equivalent to scaling the temperature $T$ by this factor.
		Because of the scale invariance of the equation of motion the functional
		behavior of the embedding, and therewith the physics of the \D3/\D7
		setup, is identical for all setups with the same ratio of quark mass and
		temperature. From \eqref{eq:mqFromEmbedding} and
		\eqref{eq:blackHoleTemperature} we infer
		\begin{equation}
			\frac{m_q}{T}=\frac{m_{\scriptscriptstyle L}}{\rh}\frac{\sqrt\lambda}{2}\,.
		\end{equation}
		The free parameters of our setup appear on the right hand side as the
		asymptotic value $m_{\scriptscriptstyle L}$ of the \db7 embedding and
		the black hole horizon $\rh$. The ratio of quark mass and temperature is
		defined by the ratio of these parameters, which we will henceforth
		denote by the dimensionless quantity
		\begin{equation}
		\label{eq:quarkMass}
			m=\frac{m_{\scriptscriptstyle L}}{\rh}=\frac{2m_q}{\sqrt\lambda T}\,.
		\end{equation}
		It was found that the transition to the melting meson phase occurs at a
		value of approximately $m=1.3$ \cite{Babington:2003vm}. At this value
		there is a change in the topology of the probe brane, which changes
		between the black hole type with a singularity and the regular Minkowski
		embedding. We demonstrated this in figure~\ref{fig:embeddingExamples}.
		
		We use $m$ as the parameter which defines whether we are in the regime
		of high or low temperature. However, this will not affect the topology
		of the embedding in our setup. Below, we follow an argumentation which
		reveals that we may restrict to black hole embeddings, since this is the
		thermodynamically favored configuration in setups with finite particle
		density. Black hole embeddings are conveniently described in the
		coordinate system \eqref{eq:adsBHChi}, derived in
		appendix~\ref{chap:coordinates} and also used in ref.~\cite{Kobayashi:2006sb},
		\begin{equation}
		\label{eq:AdSBHmetric}
				\ds^2 =\frac{\varrho^2}{2R^2}\left( -\frac{f^2(\varrho)}{\tilde f(\varrho)}\,\dd t^2 + \tilde f(\varrho)\,\dd \threevec x^2 \right)
						 +R^2\left( \frac{\dd \varrho^2}{\varrho^2} + \left(1-\chi^2\right)\dd\Omega_3^2 + \left(1-\chi^2\right)^{-2}\,\dd\chi^2 + \chi^2\dd\phi^2  \right).
		\end{equation}
		with
		\begin{equation}
		\label{eq:metricDefinitions}
			f(\varrho)=1-\frac{\rh^4}{\varrho^{4}},\qquad\tilde f(\varrho)=1+\frac{\rh^4}{\varrho^{4}}.
		\end{equation}
		In the following, some equations may be written more conveniently in
		terms of the dimensionless radial coordinate $\rho=\varrho/\rh$, which
		covers a range from $\rho_{\scriptscriptstyle H}=1$ at the event horizon
		to $\rho \to \infty$, representing the boundary of $\ads_5$ space.
		
		As in section~\ref{sec:addingFlavor}, we embed $N_f$ \db7s in this
		spacetime, such that they extend in all directions of $\ads_5$ space
		and along the directions of the three-sphere $\S^3$, which is part
		of the $\S^5$. Due to the symmetries of this background, the embeddings
		depend only on the radial coordinate $\varrho$ and are parametrized by
		the functions~$\chi(\varrho)$. Due to our choice of the gauge field
		fluctuations in the next subsection, the remaining three-sphere in this
		metric will not play a prominent role. The induced metric on the \db7 is
		given by
		\begin{equation}
		\label{eq:inducedD7Metric} 
			\dd s_{\text{\D7}}^2 = \frac{\varrho^2}{2R^2} \left(-\frac{f^2}{\tilde f}\,\dd t^2+
				\tilde f\, \dd \threevec x^2 \right) 
				+\frac{R^2}{\varrho^2}  \frac{1-\chi^2+\varrho^2{\chi'}^2}{1-\chi^2}\, \dd \varrho^2 
				+R^2(1-\chi^2)\, \dd \Omega_3^2.  
		\end{equation}
		Here and in what follows we use a prime to denote a derivative with
		respect to $\varrho$ (resp.\ to $\rho$ in dimensionless equations).
		We write $\sqrt{-G}$ to denote the square root of the determinant of
		the induced metric on the \db7, which is given by
		\begin{equation}
		\label{eq:inducedD7MetricDeterminant}
				\sqrt{-G} = \varrho^3 \frac{f \tilde f}{4}\, \left(1-\chi^2\right) \sqrt{1-\chi^2 + \varrho^2 {\chi'}^2}.
		\end{equation}

		Note that in general the branes are not necessarily coincident, and
		thus there will be one embedding function $\chi^a$ per brane, i.\,e.\ 
		$a=1,\ldots,N_f$. We will make use of the DBI action to derive the
		embedding profiles $\chi^a(\varrho)$. However, we postpone this task to
		the following subsection because the action also depends on the gauge
		field on the probe brane. We will see that the embedding function
		couples to this field on the brane.
		
		Each of the branes features a $\U(1)$ ``flavor gauge field'' $A^a$, with
		$a=1,\ldots,N_f$. This gauge field is arises from the fluctuation modes
		of an open string with both ends attached to the probe brane. For branes
		at arbitrary positions we therefore have an overall $\U(1)^{N_f}$
		symmetry which is promoted to an $\U(N_f)$ in the case of coinciding
		branes. This symmetry enhancement comes from the fact that we then can
		no longer distinguish the branes and therefore cannot tell on which
		brane a string ends. Each of the two ends of a string can be assigned a
		label, also refered to as a Chan-Paton factor, which identifies the
		brane on which the string ends. There are $N_f^2$ possible
		configurations, matching the degrees of freedom of the non-Abelian
		symmetry group $\U(N_f)$. The correct action for such a configuration of
		coinciding branes is given by the \emph{non-Abelian DBI action}
		\cite{Myers:1999ps},
		\begin{equation}
			S = -T_p \int\!\!\dd^p\xi\:
			\str\left[\det{}_{\!\!\scriptscriptstyle \perp} Q\: \det\left(P\left[E+E_{\cdot i}(Q^{-1}-\mathbbm1)^{ij}E_{j\cdot}\right] +2\pi\alpha'F \right)\right]^{\frac12}
		\end{equation}
		with $\cdot$ as a placeholder for a spacetime index, and
		\begin{align}
			E_{\mu\nu} &= g_{\mu\nu} + B_{\mu\nu},\\
			\label{eq:nonCommuteEmbeddings}
			{Q^i}_j &= \delta^i_j+\ii2\pi\alpha'\commute{\Phi^i}{\Phi^k}E_{kj}.
		\end{align}
		The Greek indices label the background spacetime coordinates, while
		Latin labels $i,j,k$ denote the directions perpendicular to the brane.
		Note that $Q$ is a matrix with labels denoting solely these directions,
		and the determinant operator $\det{}_{\!\!\scriptscriptstyle \perp}$
		acts with respect to them. The determinant $\det$ acts with respect to
		the labels of the directions along the brane. Additionally to this
		Lorentz structure, all the operators in the action above are elements of
		the $\U(N_f)$ algebra on which the symmetrized trace $\str$ acts. The
		operators $\Phi$ are operator valued analogon to the scalar embedding
		functions.
		
		The non-Abelian nature of the embedding functions $\Phi$ introduces
		non-commutativity of the spacetime coordinates. The physical
		consequences of the non-Abelian DBI action are not entirely
		understood by now. However, non-commutative spacetimes are
		candidates for the cure of UV divergences of quantum field theories
		and are applied in M-theory to describe spherical D-brane
		configurations \cite{Myers:2003bw}. The non-commutative
		contributions are manifest in \eqref{eq:nonCommuteEmbeddings}, and
		hidden in the gauge covariant pullback of $g$, which introduces
		gauge covariant derivatives of the embedding,
		\begin{equation}
		\begin{aligned}
			P[g]_{ab} &= G_{ab} =\nabla_{\!a}\,x^\mu\: \nabla_{\!b}\, x^\nu \:g_{\mu\nu}\\
			&= g_{ab} + g_{ai}\nabla_{\!b}\Phi^i + g_{bi}\nabla_{\!a}\Phi^i + g_{ij}\nabla_{\!a}\Phi^i\,\nabla_{\!b}\Phi^j
		\end{aligned}
		\end{equation}
		with
		\begin{equation}
		\label{eq:nonAbelianPullback}
			\nabla_{\!a}\Phi^i = \del_a\Phi^i + \ii\, 2\pi\alpha' \commute{A_a}{\Phi^i},
		\end{equation}
		where the indices $i,j$ are transverse and $a,b$ are along the
		worldvolume of the probe brane. The non-Abelian DBI action features
		commutator terms $\commute{A}{\Phi}$ and $\commute{\Phi}{\Phi}$ of gauge
		fields and embedding functions. These commutator terms can be thought of
		as corrections to the Abelian DBI action, which is reproduced if the
		commutators in \eqref{eq:nonCommuteEmbeddings} and
		\eqref{eq:nonAbelianPullback} are vanishing.
		
		All setups we consider, feature a symmetry in the directions transverse
		to the \db7s which allows to set one of the two $\Phi^i$ to zero, i.\,e.\ 
		the embedding function in this direction is constantly zero. Thus, the
		commutators in \eqref{eq:nonCommuteEmbeddings} vanish. Moreover, we
		restrict to background configurations, arising from fields which are
		part of the Cartan subalgebra of $\U(N_f)$. As a justification for this
		restriction, we claim our freedom to define a basis in the vector space
		of the $\U(N_f)$ algebra and choose the non trivial embedding to define
		the direction of the generator $\mathbbm1$, which is the generator
		of $\U(1)\subset\U(N_f)$. The embedding matrices $\Phi$ thereby are
		diagonal. A usual interpretation of the Eigenvalues on the diagonal is
		that they give the embedding functions for each of the $N_f$ branes. The
		generators of the $\U(1)$ symmetry affect all flavor branes in an
		identical manner. The $\U(1)$ is therefore interpreted as the symmetry
		associated to \emph{baryon charge}.
		
		Recall that the embeddings determine the quark masses of the dual field
		theory. The construction of coinciding branes in our setup therefore
		implies that the flavor eigenstates coincide with the mass eigenstates
		of the particles in the dual field theory. Generalizations to distinct
		bases in the flavor and mass vector spaces should be possible.
		
		In the following, we are especially interested in two different setups,
		both of which feature a vanishing Kalb-Ramond field $B=0$ but different
		$N_f$. The case of $N_f=1$ allows investigation of the effects of baryon
		charge and the corresponding baryonic chemical potential. The
		symmetrized trace in the action is trivial in this case. The choice of
		$N_f=2$ features the diagonal generator proportional to
		$\mathrm{diag}(1,-1)$ which can be interpreted to charge the two flavors
		oppositely and therefore models \emph{isospin} symmetry. The non-Abelian
		DBI action simplifies to
		\begin{equation}
		\label{eq:simplifiedNonAbelianDBI}
			S = -T_p \int\!\!\dd^p\xi\;\str\sqrt{ \left|\det\left(G +2\pi\alpha'F \right)\right|}\:.
		\end{equation}
		According to the arguments above, the $\U(N_f)$ matrix structure of $G$
		and $F$ for $N_f=2$ is given by
		\begin{align}
			G &= G\,\sigma^0,\\
			F &= F^B\,\sigma^0 + F^I\sigma^3,
		\end{align}
		where we use the Cartan subalgebra of the $\U(2)$, given by
		two Pauli matrices 
		\begin{equation}
			\sigma^0=\left( \begin{array}{cc}1&0\\ 0&1\end{array}\right),\qquad
			\sigma^3=\left( \begin{array}{cc}1&0\\ 0&-1\end{array}\right).
		\end{equation}
		All operators in the action therefore are diagonal. The two probe
		branes are coincident and the diagonal entries of the field strength
		tensor $F$ determine the net charges of the branes. We have
		\begin{equation}
			F = \left( \begin{array}{cc}F^B+F^I&0\\ 0&F^B-F^I\end{array}\right) = \left( \begin{array}{cc}F^{(1)}&0\\ 0&F^{(2)}\end{array}\right).
		\end{equation}
		
		In any case, the restriction to the diagonal Cartan subalgebra of
		$\U(N_f)$ simplifies the non-Abelian DBI action, e.\,g.\  symmetrization
		of the trace is trivial in the sense that all commutators vanish and
		$\str=\tr$. Expansion of the square root and evaluation of the trace
		with subsequent restoration of the square root eventually leads to
		the following action for two \db7s in the \adsfivesfive black hole
		background with vanishing $B$ field,
		\begin{equation}
		\label{eq:multibraneSimpleDBI}
			S=-T_7\sum_{k=1}^{N_f}\int\dd^8\xi\;\sqrt{\det\left(G+2\pi\alpha'F^{(k)}\right)}.
		\end{equation}
		We will concentrate on the cases of $N_f=1$ and $N_f=2$, and
		separately switch on either the baryonic $U(1)$, parametrized by
		field strengths $F^B$, or the isospin subgroup $\SU(2)$ along the
		direction of $\sigma^3$, parametrized by the field strengths $F^I$.
		Thus, the action for each brane is the same as long as we do not
		switch on both fields simultaneously.
		
		To sum up, the background geometry described so far is dual to thermal
		$\N=4$ supersymmetric $\SU(N)$ Yang-Mills theory with $N_f$ additional
		$\N=2$ hypermultiplets. These hypermultiplets arise from the lowest
		excitations of the strings stretching between the \db7s and the
		background-generating \db3s. The particles represented by the
		fundamental fields of the $\N=2$ hypermultiplets model the quarks in our
		system. Their mass $m_q$ is given by the asymptotic value of the
		separation of the \D3- and \db7s. Since the physics of the thermal
		\D3/\D7 setup is determined by the ratio of quark mass to temperature,
		we use the parameter $m$, which is proportional to this ratio.

	\subsubsection{Background gauge fields\,---\,finite particle density}
		
		In addition to the parameter $m$, we aim for a description of the system
		at finite baryon density $n_{\scriptscriptstyle B}$, which in turn
		accounts for a finite chemical potential $\mu$. In the thermal $\SU(N)$
		gauge theory, a baryon is composed of $N$ quarks, such that the baryon
		density~$n_{\scriptscriptstyle B}$ can be directly translated into a
		quark density~$n_q=n_{\scriptscriptstyle B}\, N$. The thermodynamic dual
		quantity of the quark density is the quark chemical potential~$\mu$,
		which is realized by a non-dynamical time component of the gauge field.
		The chemical potential is the source of the charge density operator
		$J_0$, i.\,e.\  the time component of the current $J$, of the particles
		charged with respect to the potential under consideration. The time
		component of the current of fundamental spin~$0$ fields $\phi$ and the
		spin~$1/2$ fields $\psi$ in the $\N=2$ hypermultiplet is given by
		\begin{equation}
			J_0 = \bar\psi\gamma_0\psi + \phi\del_0\phi.
		\end{equation}
		In   the
		dual holographic formulation the source $\mu$ of this charge density
		then corresponds to the non-renormalizable mode of the according
		holographically dual field in the supergravity theory. This field is the
		time component of the supergravity gauge field on the probe brane, which
		we denote by $\bar A_0$. The normalizable mode will yield the
		expectation value of particle density. We consider a constant chemical
		potential in space and time, i.\,e.\  there is no spacetime dependence.
		Instead we work with a gauge field background $\bar A_0(\rho)$ which
		only depends on the radial \ads coordinate.
		
		We do not rederive the dictionary entries here, but rather rephrase what
		is important for the following developments. The holographic
		interpretation of the embedding $\chi$ of the probe \db7s was discussed
		(in a different coordinate system) in section~\ref{sec:addingFlavor}.
		The probe branes account for holographic duals of fundamental quarks with
		mass $m_q$, determined by the non-renormalizable mode of the embedding
		function. The asymptotic form of the fields $\chi(\rho)$ and $\bar
		A_0(\rho)$ can be found from the equations of motion in the boundary
		limit~$\rho\to\infty$. The expansion coefficients in an expansion in
		powers of $\rho$ are given by
		\begin{align}
		\label{eq:boundaryAt}
			\bar A_0 & =\mu_q - \frac{1}{\rho^2}\cdot\frac{\rh}{2\pi\alpha'} \frac{2^{5/2}\,n_q}{N_fN_c\sqrt\lambda T^3} + \cdots, \\
		\label{eq:boundaryChi}
			\chi & =\frac{m}{\rho}+\frac{c_c}{\rho^3}+ \cdots .
		\end{align}
		Here $\mu_q$ is the quark chemical potential, $n_q$ is the quark
		density, $m$ is the dimensionless quark mass parameter given in
		\eqref{eq:quarkMass} and $c_c$ is the quark condensate mentioned earlier
		(but irrelevant in this work). We made use of the dimensionless
		$\rho$-coordinate that runs from the horizon value~$\rho=1$ to the
		boundary at $\rho\to\infty$. The chemical potential and density of
		baryons are simply
		\begin{equation}
			\mu_{\scriptscriptstyle B} = \frac{\mu_q}{N_c},\qquad n_{\scriptscriptstyle B}=\frac{n_q}{N_c}\,.
		\end{equation}
		
		Once we have found the solutions $\bar A_0$ to the equations of
		motion for the gauge field, the value $\mu_q$ of the chemical
		potential in the dual field theory can be extracted as
		\begin{equation}
		\label{eq:chemPotLimit}
			\mu_q = \lim_{\rho\to\infty} \bar A_0(\rho) = \frac{\rh}{2\pi\alpha'} \tilde\mu_q,
		\end{equation}
		where we introduced the dimensionless quantity $\tilde\mu$ for
		convenience. We apply the same normalization to the gauge field and
		distinguish the dimensionful quantity $\bar A$ from the
		dimensionless
		\begin{equation}
			\tilde A_0=\frac{2\pi\alpha'}{\rh}\,\bar A_0 
		\end{equation}
		(we save the symbols without diacritics for later use). Analogously,
		the solutions of the embedding functions carry information about the
		quark mass parameter $m$,
		\begin{equation}
			\label{eq:mFromChi}
			m=\lim_{\rho\to\infty}\:\rho\,\chi(\rho).
		\end{equation}
		
		We mentioned that for non-vanishing baryon density, there are no
		embeddings of Minkowski type, and all embeddings reach the black hole
		horizon. This is due to the fact that a finite baryon density in an
		infinite volume of Minkowski spacetime requires an infinite number of
		strings in the dual supergravity picture. These strings have one end on
		the stack of \db3s and the other on the stack of $N_f$ probe \db7s.
		These strings pull the brane towards the black hole
		\cite{Kobayashi:2006sb}. Such spike configurations are common for
		configurations in which branes of different dimensionality connect
		\cite{Karch:2000gx}.
		
\begin{figure}
			\centering
			\includegraphics[width=.5\linewidth]{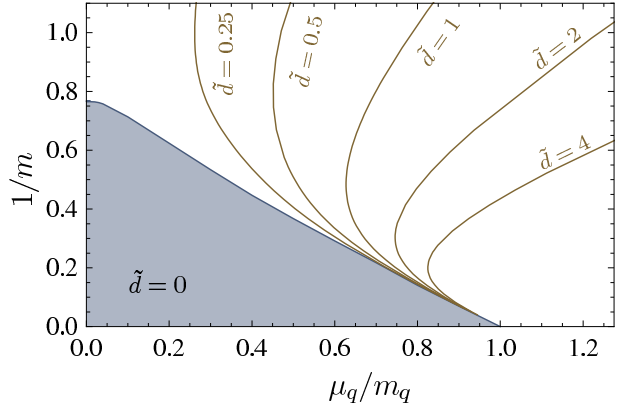}
			\caption[Lines of constant $\tilde d$ in the phase diagram]{%
				The phase diagram of fundamental matter in the \D3/\D7 setup.
				Horizontal axis: Chemical potential normalized to the quark
				mass. Vertical axis $m^{-1}\propto T/m_q$. In this work we
				analyze the white region of finite particle density $\tilde d$,
				for which we show some lines of constant values for $\tilde d$.
				The relation between $\tilde d$ and $n_{\scriptscriptstyle B}$
				is given by \eqref{eq:dtFromnB}
			}
			\label{fig:phaseDiagram}
		\end{figure}

		Very recently, however, it was found that for a vanishing baryon
		number density, there may indeed be Minkowski embeddings if a
		constant vacuum expectation value of $\tilde A_0$ is present, which
		does not depend on the holographic coordinate
		\cite{Karch:2007br,Mateos:2007vc,Nakamura:2006xk,Nakamura:2007nx,Ghoroku:2007re}.
		The phase diagram found there is reproduced in
		figure~\ref{fig:phaseDiagram}.
		In the shaded region, the baryon density
		vanishes~($n_{\scriptscriptstyle B}=0$) but temperature, quark mass and
		chemical potential can be nonzero. This low temperature region only
		supports Minkowski embeddings with the brane ending before reaching the
		horizon. In contrast, the unshaded region supports black hole embeddings
		with the branes ending on the black hole horizon. In this regime the
		baryon density does not vanish~($n_{\scriptscriptstyle B}\neq 0$). At
		the low-$\mu$ end of the line separating~$n_{\scriptscriptstyle B}=0$
		from~$n_{\scriptscriptstyle B}\not=0$ in figure~\ref{fig:phaseDiagram},
		there exists also a small region of multivalued embeddings, which are
		thermodynamically unstable~\cite{Mateos:2007vc}. In the black hole phase
		there is a phase transition between different black hole
		embeddings~\cite{Kobayashi:2006sb}, resembling the meson melting phase
		transition for fundamental matter at vanishing density. This first order
		transition occurs in a region of the phase diagram close to the
		separation line between the two regions with vanishing (shaded) and
		non-vanishing (unshaded) baryon density. This transition disappears
		above a critical value for the baryon density $n_{\scriptscriptstyle B}$
		given by
		\begin{equation}
		\label{eq:dcrit}
			n_{\scriptscriptstyle B} =\frac{N_f\sqrt{\lambda}T^3}{2^{5/2}}\,\tilde d ,\quad\text{with critical~} \tilde d^*= 0.00315\,.
		\end{equation}
		
		In this work we exclusively explore the region in which
		$n_{\scriptscriptstyle B} > 0$, i.\,e.\  we examine thermal systems in the
		canonical ensemble. For a detailed discussion of this aspect see
		refs.~\cite{Karch:2007br,Mateos:2007vc}.
		
		To determine the solutions of the supergravity fields on the probe
		branes we have to extremize the DBI action
		\eqref{eq:multibraneSimpleDBI}, we write shortly as
		\begin{equation}
				\label{eq:dbiAction}
				S_{\text{DBI}} = -T_7\sum_{k=1}^{N_f}\int\!\! \dd^8 \xi\; \sqrt{| \det ( G + \tilde F^{(k)} )|}.
		\end{equation}
		The induced metric $G(\xi)$ on the stack of $N_f$ coincident branes is
		given by \eqref{eq:inducedD7Metric}, $\tilde F$ is the dimensionless
		field strength tensor of the gauge fields on the brane.
		
		For now we consider the simpler case of a baryonic chemical potential
		modeled by the $\U(1)$ subgroup of $U(N_f)$. In this case, the sum
		amounts to an overall factor of $N_f$. In ref.~\cite{Kobayashi:2006sb}
		the dynamics of such a system of branes and gauge fields was analyzed in
		view of describing phase transitions at finite baryon density. Here, we
		use these results as a starting point which gives the background
		configuration of the probe branes' embedding function and the gauge
		field values at finite baryon density. To examine vector meson spectra,
		we will then investigate the dynamics of fluctuations in this gauge
		field background.
		
		In the coordinates introduced above, the action $S_{\text{DBI}}$ for
		the embedding $\chi(\varrho)$ and the field strength $F$ is obtained
		by inserting the induced metric and the field strength tensor into
		\eqref{eq:dbiAction}. From now on we make use of the dimensionless
		coordinates and reproduce the action found in
		ref.~\cite{Kobayashi:2006sb}. To do so, we remember that the only
		non-vanishing component of the background field is the
		$\rho$-dependent time component. Therefore, the only non-vanishing
		components of the field strength tensor are $\tilde
		F^{(k)}_{40}=-\tilde F^{(k)}_{04}$. We evaluate the determinant and
		arrive at
		\begin{equation}
			S_{\text{DBI}} = -T_7 N_f \int\!\! \dd^8 \xi\; \sqrt{-G}\:\sqrt{1 + G^{00}G^{44} \big(\tilde F_{40} \big)^2}
		\end{equation}
		with components $G^{ab}$ of the inverse metric $G^{-1}$. After inserting
		these components we get
		\begin{equation}
		\label{eq:actionEmbeddingsAt} 
			S_{\text{DBI}}=-N_f T_7\,\rh^3 \int\!\! \dd^8 \xi \; \frac{\rho^3}{4} f \tilde{f} (1-\chi^2) 
			 \sqrt{1-\chi^2+\rho^2 {\chi'}^2-2 \frac{\tilde{f}}{f^2}(1-\chi^2) \big(\tilde F_{40}\big)^2} \; ,
		\end{equation}
		where $\tilde F_{40}=\partial_\rho \tilde A_0$ is the field
		strength on the brane. The background fields $\chi$ and $\tilde A_0$
		depend solely on $\rho$. This action only depends on derivatives of
		the gauge field. We therefore can identify the constant of motion
		$\tilde d$ satisfying $\del_\rho \tilde d=0$,
		\begin{equation}
			\tilde d = \frac{\del \SDBI}{\del\,\big(\del_\rho \tilde A_0\big)}.
		\end{equation}
		Evaluation of this formula and insertion of the asymptotic expansion
		of $\tilde A$ reveal that this dimensionless constant is related to
		the parameters of our setup by~\cite{Kobayashi:2006sb}
		\begin{equation}
		\label{eq:dtFromnB}
			\tilde d = \frac{2^{5/2}\:n_{\scriptscriptstyle B}}{N_f\sqrt\lambda\,T^3}.
		\end{equation}
		We can therefore think of the constant $\tilde d$ as parametrizing
		the baryon density $n_{\scriptscriptstyle B}$.
		
\begin{figure}
			\centering
			\includegraphics[width=.8\linewidth]{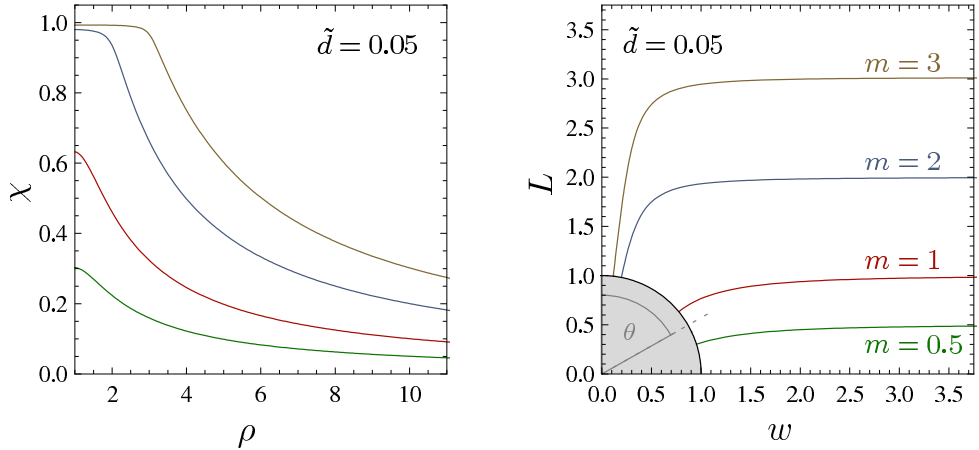}
			\caption[Examples for black hole embeddings at finite baryon density]{%
				Examples for the embedding function $\chi(\rho)$ with
				$\chi=\cos\theta$ and the according profile $L(w)$.
				Matching colors indicate corresponding curves.}
			\label{fig:embeddingsChiAndL}
		\end{figure}

\begin{figure}
			\centering
			\includegraphics[width=.5\linewidth]{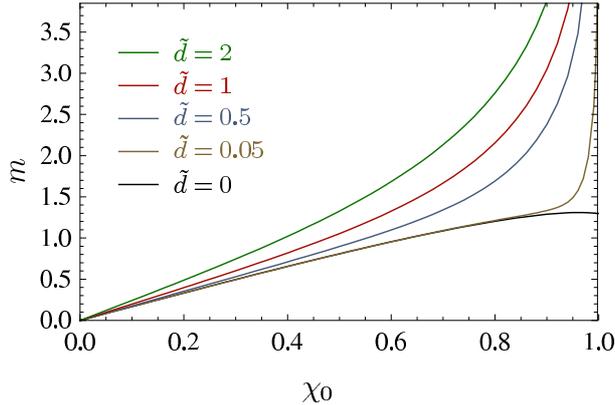}
			\caption{Dependence of the quark mass parameter $m$ on the
				initial value $\chi_0$ of the embedding.
			}
			\label{fig:mFromChi0}
		\end{figure}

		The equations of motion for the background fields are conveniently
		obtained after Legendre transforming the
		action~\eqref{eq:actionEmbeddingsAt} to $\hat S=S-\delta S/\delta
		F_{40}$ in order to eliminate dependence on the gauge field in
		favor of dependence on the constant of motion $\tilde d$
		\cite{Kobayashi:2006sb}. Varying
		this Legendre transformed action with respect to the field~$\chi$
		gives the equation of motion for the embeddings~$\chi(\rho)$,
		\begin{equation}
		\begin{aligned}
		\label{eq:eomChi}
			 & \partial_\rho\left[\frac{\rho^5 f \tilde{f} (1-\chi^2)
			{\chi'}}{\sqrt{1-\chi^2+\rho^2{\chi'}^2}} \sqrt{1 +
			\frac{8 \tilde{d}^2}{\rho^6 \tilde{f}^3 (1-\chi^2)^3}}\right] \\
			= & - \frac{\rho^3 f \tilde{f} \chi }{\sqrt{1-\chi^2+\rho^2{\chi'}^2}}
			\sqrt{1 +\frac{8 \tilde{d}^2}{\rho^6 \tilde{f}^3 (1-\chi^2)^3}}
			  \left[3 (1-\chi^2) +2 \rho^2 {\chi'}^2 -24 \tilde{d}^2
			\frac{1-\chi^2+\rho^2{\chi'}^2}{\rho^6 \tilde{f}^3 (1-\chi^2)^3+8 \tilde{d}^2}
			\right] .
		\end{aligned}
		\end{equation}
		This equation for $\chi(\rho)$ can be solved numerically for given
		$\tilde d$ and initial value~$\chi_0$. We impose boundary conditions
		such that the branes cross the horizon perpendicularly
		\begin{equation}
			\chi(\rho=1)=\chi_0,\qquad  \partial_\rho \chi(\rho) \Big|_{\rho=1}=0.
		\end{equation}
		Figure~\ref{fig:embeddingsChiAndL} shows some examples. The embeddings
		at finite density resemble the large $\rho$ asymptotics of the
		embeddings found at zero density. For small $\rho$ however, at
		finite particle density there always is the spike reaching down to the
		event horizon.
		The initial value of $\chi_0$ determines the position on which the brane
		reaches the horizon and in this way determines the quark mass parameter
		$m$, c.\,f.\  equation~\eqref{eq:mFromChi}. It is zero for $\chi_0=0$ and
		tends to infinity for $\chi_0 \to 1$. Figure~\ref{fig:mFromChi0} shows
		this dependence of $m$ on $\chi_0$ for different values of the baryon
		density $\tilde d$. In general, a small (large)~$\chi_0$ is equivalent
		to a small (large) $m$. For $\chi_0\lesssim 0.5$, we nearly observe
		proportionality. For vanishing~$\tilde d=0$, we only can model quarks
		with $m\le1.3$, heavier quarks at vanishing density are described by
		embeddings of Minkowski type, which we do not discuss here. (The trained
		eye can see that there is a maximum of $m=1.3$ in
		figure~\ref{fig:mFromChi0} before $m$ drops to smaller values towards
		$\chi_0=1$. This is reflecting the existence of a phase transition to
		Minkowski embeddings at $\tilde d=0$).
		
\begin{figure}
			\centering
			\includegraphics[width=.8\linewidth]{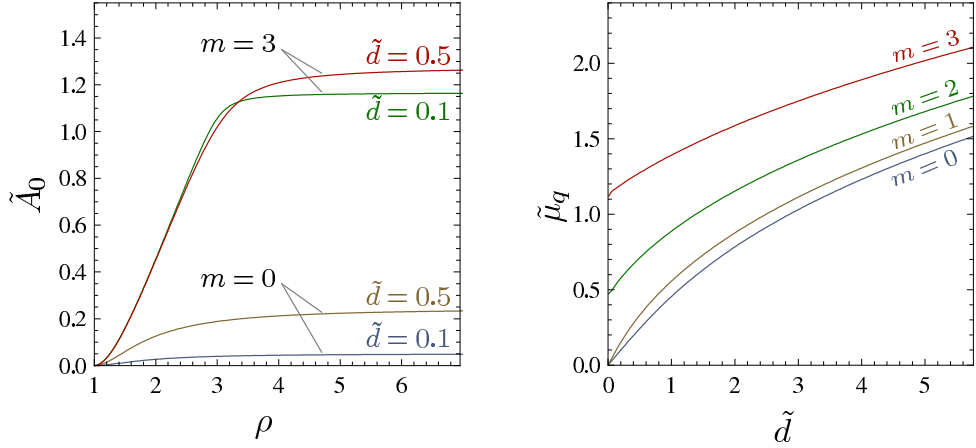}
			\caption[Examples for the background gauge field and the
				resulting chemical potential.]{%
				Examples for the background gauge field time component
				$\tilde A_0$ (left) and the resulting chemical potential
				$\tilde \mu$ (right). Note that the chemical potential is
				not zero at asymptotically small but non-zero $\tilde d$ for
				$m>1.3$, reproducing the phase transition line in
				figure~\ref{fig:phaseDiagram}.
			}
			\label{fig:backgAt}
		\end{figure}

		The equation of motion for the background gauge field $\tilde A$ is 
		given by
		\begin{equation}
		\label{eq:eomD}
			 \partial_\rho\tilde{A}_0 = 2 \tilde{d}\,
			\frac{f \sqrt{1-\chi^2+\rho^2 {\chi'}^2}}
			{\sqrt{\tilde{f}(1-\chi^2) \left[\rho^6 \tilde{f}^3 (1-\chi^2)^3+8 \tilde{d}^2\right]}}. 
		\end{equation}
		Integrating both sides of the equation of motion from
		$\rho_{\text{H}}=1$ to some $\rho$, and respecting the boundary
		condition $\tilde A_0(\rho=1)=0$ \cite{Kobayashi:2006sb}, we obtain
		the full background gauge field
		\begin{equation}
			\label{eq:backgroundAt}
			 \tilde{A}_0(\rho) = 2\tilde{d}
			\int\limits_{1}^\rho\!\! \dd\rho\;
			\frac{f\sqrt{1-\chi^2+\rho^2{\chi'}^2}}{\sqrt{\tilde{f}
			(1-\chi^2)\left[\rho^6 \tilde{f}^3 (1-\chi^2)^3 +8 \tilde{d}^2\right]}}\,.
		\end{equation}
		Examples for the functional behavior of $\tilde A_0(\rho)$ are shown
		in figure~\ref{fig:backgAt}.
		While there is a significant slope of $\tilde A(\rho)$ near the
		horizon at $\rho=1$, the gauge field tends to a constant at large
		$\rho$. From~\eqref{eq:chemPotLimit} we recall that this value is
		the chemical potential of the field theory. We will henceforth
		compute $\tilde \mu$ by evaluating the formula above for large
		$\rho$. Note that at any finite baryon density $\tilde d\propto
		n_{\scriptscriptstyle B}\neq 0$ there exists a minimal chemical potential which is
		reached in the limit of massless quarks.
		
\subsection{Meson spectra at finite baryon density}
\label{sec:baryonMu}
	
	\subsubsection{Equations of motion}
	\label{sec:eomFiniteDensitySpectra}
	
		We now compute the spectral functions of flavor currents at finite
		baryon density~$\tilde d$, and temperature $T$ in the black hole phase.
		Compared to the limit of vanishing density treated
		in~\cite{Myers:2007we}, we discover a qualitatively different behavior
		of the finite temperature excitations corresponding to vector meson
		resonances.
		
		To obtain the mesonic spectral functions, we compute the correlations of
		flavor currents $J$ by means of the holographically dual gauge field
		fluctuations $A_\mu$ about the background given
		by~\eqref{eq:actionEmbeddingsAt}. We denote the full gauge field by
		\begin{equation}
		\label{eq:hatA}  
				\hat A_\mu(\rho, \vec{x}) = \delta^0_\mu \tilde A_0(\rho) + A_\mu(\vec{x},\rho)\,.
		\end{equation}
		According to section~\ref{sec:backAndBranes}, the background field has a
		non-vanishing time component, which depends solely on $\rho$. The
		fluctuations in turn are gauged to have non-vanishing components along
		the Minkowski coordinates $\fourvec x$ only, and only depend on these
		coordinates and on $\rho$. Additionally, the fluctuations are assumed to
		be small, such that it suffices to consider their linearized equations
		of motion. At this point we simply neglect the fluctuation of the scalar
		and pseudoscalar modes and their coupling to the vector fluctuations.
		This procedure is justified by the restriction to fluctuations with
		vanishing spatial momentum, which is imposed later. In this limit the
		vector mesons do not couple to the other mesonic excitations. A
		generalization of this work which includes these coupling and finite
		momentum spectra can be found in ref.~\cite{Mas:2008jz}.
		
		The equations of motion are obtained from the action
		\eqref{eq:dbiAction}, where we introduce small fluctuations $A$ by
		setting
		\begin{align}
			\tilde A&\mapsto \tilde A + A,\\
			\Rightarrow \quad \tilde F&\mapsto\tilde F + F.
		\end{align}
		The background gauge field~$\tilde A$ is given by~\eqref{eq:eomD}.
		The fluctuations $A$ now propagate on a background $\G$ given by
		\begin{equation}
			\G =G + \tilde F,
		\end{equation}
		and their dynamics is determined by the Lagrangian
		\begin{equation}
			\L = \sqrt{\left| \det \left( \G + F \right)\right|},
		\end{equation}
		with the fluctuation field strength $F_{\mu\nu} =
		2\partial_{[\mu}A_{\nu]}$. In contrast background field $\tilde A$, the
		fluctuations $A$ do depend on the Minkowski directions as well as on the
		radial coordinate of $\ads_5$. Since the fluctuations and their
		derivatives are chosen to be small, we consider their equations of
		motion only up to linear order, and derive these equations from the part
		of the Lagrangian $\L$ which is quadratic in the fields and their
		derivatives. Denoting this part by $\L_2$, we get
		\begin{equation}
			\L_2 =  -\frac{1}{4} \sqrt{\left|\det \G\right|}
							 \left( \G^{\mu\alpha}\G^{\beta\gamma} F_{\alpha\beta}F_{\gamma\mu} -\frac{1}{2}\, \G^{\mu\alpha}\G^{\beta\gamma}F_{\mu\alpha}F_{\beta\gamma}\right).
		\end{equation}
		Here and below we use upper indices on $\G$ to denote elements of
		$\G^{-1}$. The equations of motion for the components of $A$ are
		\begin{equation}
		\label{eq:eomFluct}
			0= \partial_\nu\left[  \sqrt{\left|\det \G\right|}
								 \left( \G^{\mu\nu}\G^{\sigma\gamma}-\G^{\mu\sigma}\G^{\nu\gamma} - \G^{[\nu \sigma]} \G^{\gamma\mu} \right) \partial_{[\gamma}A_{\mu ]} \right].
		\end{equation}
		The terms of the corresponding on-shell action at the
		$\rho$-boundaries are (with $\rho$ as an index for the coordinate
		$\rho$, not summed)
		\begin{multline}
		\label{eq:onShellActionForCorr}
			S^{\text{on-shell}} =  \rh \pi^2 R^3 N_f T_7 \int\!\! \dd^4 x \sqrt{\left|\det \G \right|}\\
		 \times \left( \left( \G^{04}\right)^2 A_0 \partial_\rho A_0 - \G^{44} \G^{ik} A_i \partial_\rho A_k  - A_0 \G^{4 0} \mathrm{tr}(\G^{-1}F)\right)\Bigg|^{\rho_B}_{\rho_{\scriptscriptstyle H}=1}.
		\end{multline}
		
		From this form of the action we can derive the correlation functions by
		means of the procedure outlined in
		section~\ref{sec:thermalGreenFunctionsIntro}. First, we Fourier
		transform the fields as
		\begin{equation}
		\label{eq:fourierA}
		A_\mu(\rho,\vec{x}) = \int\!\! \frac{\dd^4 k}{(2\pi)^4}\, e^{i\vec{k}\vec{x}} A_\mu(\rho,\vec{k}) \,.
		\end{equation}
		As above, we are free to choose our coordinate system to give us a
		momentum vector of the fluctuation with non-vanishing spatial
		momentum only in $x$-direction, $\vec{k}=(\omega,q,0,0)$.
		
		To obtain the correlator $\GR_{ik}$ with indices $i,k$ labeling
		Minkowski directions, we have to consider the second term in the
		parentheses of \eqref{eq:onShellActionForCorr}, including all its
		prefactors. Denote the resulting expression by $\mathcal A(\rho,\fourvec k)$. We
		decompose the gauge field fluctuations into a boundary and a bulk
		contribution, $A(\rho,\fourvec k)=f(\rho,\fourvec
		k)A^\text{bdy}(\fourvec k)$, where $\lim_{\rho\to\infty}f(\rho,\fourvec
		k)=1$. The prescription from section~\ref{sec:thermalGreenFunctionsIntro}
		tells us to divide out the boundary terms $A^\text{bdy}$ from $\mathcal
		A(\rho,\fourvec k)$ in order to get what was denoted by $\mathcal F$ in
		\eqref{eq:correlatorFromBoundaryContribution}. Once we found the
		solutions $A(\rho,\fourvec k)$ we can obtain this expression by
		evaluating
		\begin{equation}
			\mathcal F_{ik}(\rho,\fourvec k)\Big|_{\rho_\text{b}} = \lim_{\rho\to\infty} \frac{\mathcal A(\rho,\fourvec k)}{A_i(\rho,\fourvec k)\,A_k(\rho,\fourvec k)}\,,
		\end{equation}
		where the indices $i,k$ correspond to the Minkowski indices on
		$\GR_{ik}$.
		
		To evaluate this expression, we have to insert the solutions to the
		equations of motion for $A(\rho,\fourvec k)$ into this expression. Note
		that on the boundary $\rho_{\scriptscriptstyle B}$ at $\rho\to\infty$,
		the background matrix $\G$ reduces to the induced \db7 metric $G$.
		Therefore, the analytic expression for boundary contributions to the
		on-shell action is identical to the one found in ref.~\cite{Myers:2007we}.
		In our case of finite baryon density, new features arise through the
		modified embedding and gauge field background, which enter the
		equations of motion \eqref{eq:eomFluct} for the field fluctuations.
		
		We adopt the procedure of ref.~\cite{Myers:2007we}, where the coordinates in
		Minkowski directions where chosen such that the fluctuation four vector
		$\fourvec k$ exhibits only one non vanishing spatial component, e.\,g.\  in
		$x$-direction as $\fourvec{k}=(\omega,q,0,0)$. In addition, the action
		was expressed in terms of the gauge invariant field component
		combinations
		\begin{equation}
			E_x=\omega A_x+ q A_0,\qquad E_{y,z}=\omega A_{y,z}\, .
		\end{equation}
		In the case of vanishing spatial momentum $q\to 0$, the Green
		functions for the different components coincide and were computed as
		\cite{Myers:2007we}
		\begin{equation}
		\label{eq:q0GreenFunction}
			\GR = \GR_{xx} =\GR_{yy} = \GR_{zz} = 
			 \frac{N_f N_c T^2}{8}\; \lim_{\rho\to\infty}\left(\rho^3 \frac{\partial_\rho E(\rho)}{E(\rho)} \right)\, ,  
		\end{equation}
		where the $E(\rho)$ in the denominator divides out the boundary value of
		the field in the limit of large $\rho$. Again, the indices on the Green
		function denote the components of the operators in the correlation
		function, all off-diagonal correlations (as~$G_{yz}$, for example)
		vanish.
		
		In the limit of $q\to 0$, the equations of motion for transverse
		fluctuations~$E_{y,z}$ match those for longitudinal
		fluctuations~$E_x$. For a more detailed discussion see
		ref.~\cite{Myers:2007we}. As an example, consider the equation of
		motion obtained from \eqref{eq:eomFluct} with $\sigma = 2$,
		determining $E_y=\omega A_2$,
		\begin{equation}
		\begin{split}
			\label{eq:eomEq0}
			0 =\,& E''+ \partial_\rho\ln\left(\sqrt{|\det \G|}\G^{22}\G^{44}\right)\,
				  E'- \frac{\G^{00}}{\G^{44}}\, \rh^2 \omega^2 E \\  
			 = \,& E'' + \partial_\rho \ln \left({\frac{\rho^3 f \left(1-\chi^2\right)^2}{\sqrt{1 - \chi^2 + \rho^2 {\chi'}^2 - \frac{2f(1-\chi^2)}{\tilde f^2}  (\partial_\rho \tilde A_0 )^2}}}\right) E'
			   +  8 \wn^2 \frac{\tilde f}{f^2} \frac{1-\chi^2 + \rho^2 {\chi'}^2 }{\rho^4 (1-\chi^2)}\,E.
		\end{split}
		\end{equation}
		Here we introduced the dimensionless frequency
		\begin{equation}
			\wn=\frac{\omega}{2\pi T}\,.
		\end{equation}
		
		In order to numerically integrate the equations of motion
		\eqref{eq:eomEq0}, we determine local solutions of that equation near
		the horizon at $\rho_{\scriptscriptstyle H}=1$, which obey the infalling
		wave boundary condition. This condition ensures causality by demanding
		that the excitations can propagate in inward direction, but nothing can
		exit the horizon. The local
		solutions can be used to compute initial values in order to integrate
		\eqref{eq:eomEq0} forward towards the boundary. The equation of motion
		\eqref{eq:eomEq0} has coefficients which are singular at the horizon.
		According to mathematical standard methods, the local solution of this
		equation behaves as~$(\rho-\rho_{\scriptscriptstyle H})^\beta$,
		where~$\beta$ is a so-called `index' of the differential
		equation~\cite{BenderOrszag:1978}. We compute the possible indices to be
		\begin{equation}
		\label{eq:indices}
			\beta=\pm i\,\wn .
		\end{equation}
		Only the negative sign will be retained in the following, since it
		casts the solutions into the physically relevant incoming waves at
		the horizon and therefore satisfies the incoming wave boundary
		condition. The
		solution $E$ can be split into two factors, which are
		$(\rho-1)^{-i\wn}$ and some function $F(\rho)$, which is regular at
		the horizon. The first coefficients of a series expansion of
		$F(\rho)$ can be found recursively as described
		in~\cite{Teaney:2006nc,Kovtun:2006pf}. At the horizon the local
		solution then reads
		\begin{equation}
		\begin{aligned}
		\label{eq:localSolutions}
		E(\rho)  =\; & (\rho-1)^{-i\wn}\, F(\rho)\\
				  =\; & (\rho-1)^{-i\wn} \left [1+\frac{i\wn}{2}(\rho-1)+
		\cdots \right ].
		\end{aligned}
		\end{equation}
		So, $F(\rho)$ asymptotically assumes values
		\begin{equation}
		\label{eq:startingValues}
		F(\rho=1)=1,\qquad \partial_\rho F(\rho)\Big|_{\rho=1}=\frac{i\wn}{2}\, .
		\end{equation}
		
		To calculate numeric values for $E(\rho)$, we have to specify the baryon
		density~$\tilde d$ and the initial value $\chi_0$, which determines the
		mass parameter $m$. These parameters determine the embeddings $\chi$
		appearing in \eqref{eq:eomEq0}. We can then obtain a solution $E$ for a
		given frequency $\wn$ by numerical integration of the equation of motion
		\eqref{eq:eomEq0}, using the initial values \eqref{eq:localSolutions}
		and \eqref{eq:startingValues}.
		
		Spectral functions are finally obtained by combining
		\eqref{eq:q0GreenFunction} and \eqref{eq:specFuncFromGreenFunc},
		\begin{equation}
				\R(\omega,0) = - \frac{N_f N_c T^2}{4}\; \Im\lim_{\rho\to\infty}\left(\rho^3 \frac{\partial_\rho E(\rho)}{E(\rho)} \right).
		\end{equation}

	\subsubsection{Spectra}
		
		We now discuss the resulting spectral functions at finite baryon
		density, and observe crucial qualitative differences compared to the
		case of vanishing baryon density.
		\begin{figure}
			\centering
			\includegraphics[width=.5\linewidth]{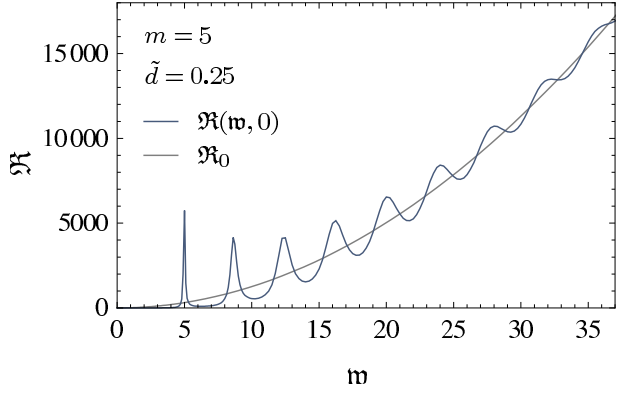}
			\caption[Example of a spectral function]{An example for a spectral function at finite baryon
			density, compared to the zero temperature result.}
			\label{fig:specFuncExample}
		\end{figure}
		In figure~\ref{fig:specFuncExample} an example for the spectral function
		at fixed baryon density $n_{\scriptscriptstyle B}\propto \tilde d$ is
		shown. In the limit of large $\wn$, corresponding to asymptotically
		small temperatures, the spectral function can be derived analytically.
		  This zero temperature result is given by
		\begin{equation}
		\label{eq:zeroTempR}
			\mathfrak{R}_0 = N_f N_c T^2\, \pi\wn^2.
		\end{equation}
		Figure~\ref{fig:specFuncExample} shows this function as well.
		
		All graphs shown here are obtained for a value of $\tilde d$ above
		$\tilde d^*$, given by \eqref{eq:dcrit}, such that we investigate the
		regime in which there is no fundamental phase transition of first
		order. Recall that the parameters of our theory are given by $\tilde
		d\propto n_{\scriptscriptstyle B}/T^3$ and $m\propto m_q/T$. Therefore variations in the
		quark density at fixed temperature and quark mass are introduced by
		tuning $\tilde d$ only. The effects of different quark masses can be
		seen by tuning $m$ alone. The effect of changes in temperature involves
		changes in both $m$ and $\tilde d$.
		
		It is interesting to compare the spectra we obtain at finite
		temperature and density to the vector meson spectrum obtained at
		zero temperature and vanishing quark density. It is given by the
		same relation as the mass spectrum \eqref{eq:mesMassDimensionful}
		which we encountered in the example of scalar mesons
		\cite{Kruczenski:2003be}. In our case, where the mesons do not carry
		spatial momentum, we can translate the mass $M_n$ of the
		$n^\text{th}$ excitation into an energy $\omega_n=M_n$. At this
		energy we would see a resonance in a supersymmetric setup. In terms
		of the dimensionless quantities we use here, these resonance energies
		are given by
		\begin{equation}
			\label{eq:mesMassDimless}
			\wn_n = \frac{M_n}{2\pi T}= m \sqrt{\frac{(n+1)(n+2)}{2}}\, ,\qquad n=0,1,2,\ldots\,,
		\end{equation}
		where $n$ labels the Kaluza-Klein modes arising from the \db7 wrapping
		the $\S^3$.

		\subsubsection*{Finite temperature effects}
		
\begin{figure}
				\centering
				\includegraphics[width=.5\linewidth]{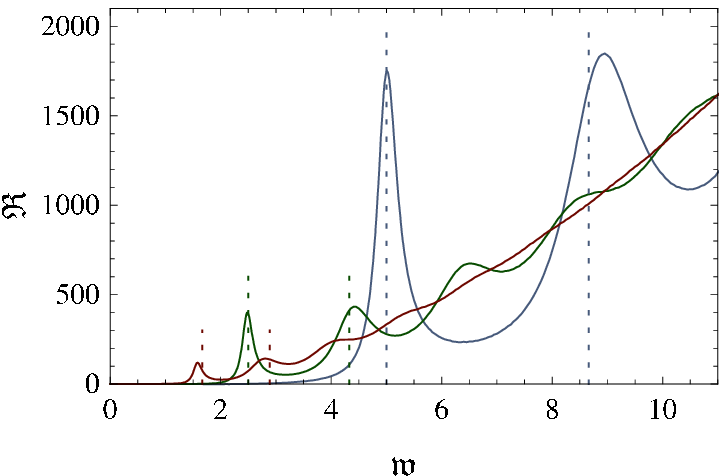}
				\caption[Effect of temperature variations on the meson spectrum]{The effect of variations
				in temperature on the meson spectrum.
				\parbox{\linewidth}{%
				\begin{tabbing}
					low T\quad \= $m=5$,\qquad\= $\tilde d=1$\qquad\qquad\=blue line\\[\smallskipamount]
					med.~T \> $m=5/2$,\> $\tilde d=1/8$\qquad\>green line\\[\smallskipamount]
					high~T \> $m=5/3$,\> $\tilde d=1/27$\qquad\>red line
				\end{tabbing}
				}
				}
				\label{fig:specTempDependence}
			\end{figure}

			We analyze finite temperature effects by choosing two values of
			$m$ and $\tilde d$, which correspond to a given values of quark
			mass, quark density and a temperature $T$. A change in
			temperature amounts to
			\begin{equation}
				T\mapsto\alpha T
			\end{equation}
			and thereby leads to
			\begin{equation}
				\tilde d \mapsto \frac{\tilde d}{\alpha^3}\,,
				\qquad
				m \mapsto \frac{m}{\alpha}\,.
			\end{equation}
			An example is shown in figure~\ref{fig:specTempDependence}.
			There we plot spectra for three different temperatures, which we
			call low ($m=5$, $\tilde d=1$), medium ($m=5/2$,
			$\tilde d=1/8$) and high ($m=5/3$,$\tilde d=1/27$) temperature. We can see
			that at high temperature there is hardly any structure visible
			in the spectral function. However, we have chosen a temperature
			at which already a slight excitation is visible at low energies
			$\wn$. Decreasing temperature leads to more and more pronounced
			peaks in the spectral function. Moreover, at decreasing
			temperature these peaks move closer to the resonance energies
			\eqref{eq:mesMassDimless}, corresponding to zero temperature and
			density (drawn as the corresponding dashed lines in the figure).
			
			The formation of sharp resonances at low temperature indicates the
			intuitively expected behavior of long living mesons in a cold
			medium, which melt, i.\,e.\  decay faster, at high temperatures. However,
			we did not perform an analysis of the quality factor of the resonance
			peaks, i.\,e.\  we did not calculate the lifetimes of the vector mesons.
			From figure~\ref{fig:specTempDependence} we can see that the
			height-to-width ratio of the peaks seems not to improve to a great
			extent at low temperatures.
			
		\subsubsection*{Finite density effects}
		
\begin{figure}
				\centering
				\includegraphics[width=.5\linewidth]{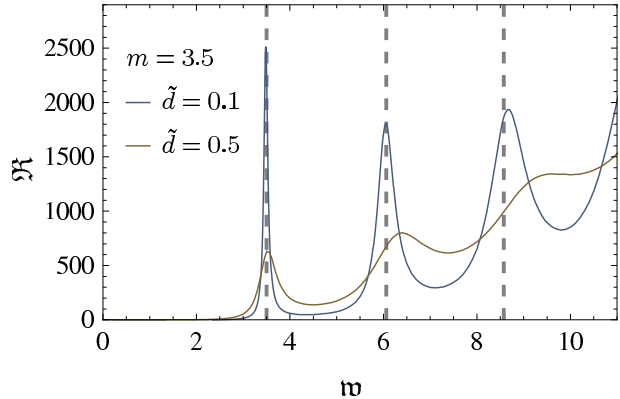}
				\caption[Dependence of the spectra on baryon density]{%
				The dependence of the spectra on baryon density. The
				dashed lines again mark the supersymmetric spectrum.}
				\label{fig:specDensityDependence}
			\end{figure}

			To investigate the effects of finite baryon density
			$n_{\scriptscriptstyle B}$, we tune $\tilde d$ while keeping $m$
			constant. This amounts to varying the quark density at constant
			temperature and quark mass. The effect is shown in
			figure~\ref{fig:specDensityDependence}. We observe that the peak
			width is considerably influenced by baryon density. At low baryon
			density the resonances are close to line-like excitations, while
			they are broadened with increasing particle density. Additionally,
			increasing the particle density also causes a slight shift of the
			resonances to \emph{higher} energies.
			
			These observations are interesting from a phenomenologically
			inclined point of view. The in-medium effects on mesonic bound
			states are important to interpret the results of heavy ion collision
			experiments. Estimations from the early 1990{\small s} based on
			effective models predicted decreasing vector meson masses at
			increasing densities \cite{Brown:1991kk}, known as Brown-Rho
			scaling. Experimental data from experiments at the SPS facility at
			CERN, however, is better described by models like the one found in
			refs.~\cite{Rapp:1999fm,Rapp:1999us}. There the in-medium effects
			also are reflected in peak broadening and shifts to higher energies.
			
			For information on the spectral functions at vanishing particle
			density we refer the reader to ref.~\cite{Myers:2007we}. Where the
			low temperature regime for $\tilde d=0$ was investigated.

		\subsubsection*{Dependence on quark mass}
			
\begin{figure}
				\centering
				\includegraphics[width=.8\linewidth]{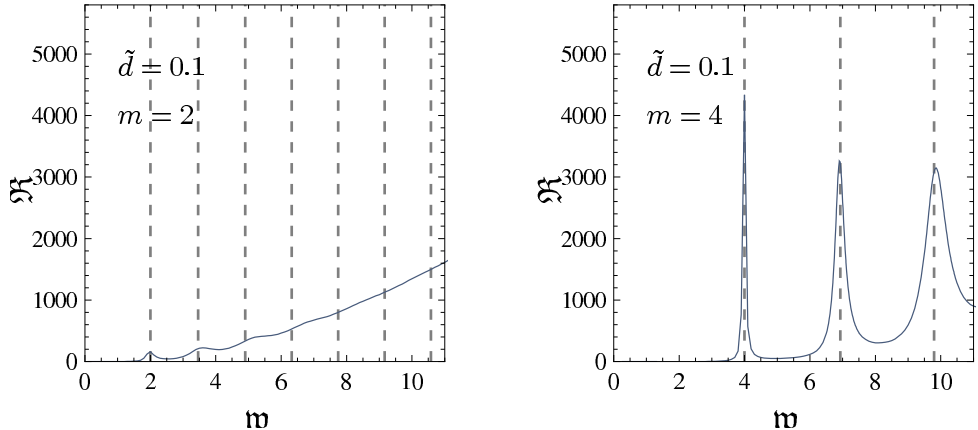}
				\caption[Dependence of the spectra on quark mass]{%
					The dependence of the spectra on quark mass. The
					dashed lines again mark the supersymmetric spectrum.}
				\label{fig:specMassDependence}
			\end{figure}

			To observe the dependence on the mass of the quarks, we plotted
			spectra for different $m$ at constant values of $\tilde d$ in
			figure~\ref{fig:specMassDependence}. We observe more and more
			pronounced resonances as we increase the meson mass. These mesons
			eventually nearly resemble the line spectrum
			\eqref{eq:mesMassDimless} known from the supersymmetric case of zero
			temperature and vanishing quark density. This observation reflects
			the decreasing effect of finite temperature and chemical potential
			with increasing quark mass. In a regime where the scale of the quark
			mass outweighs both additional scales $T$ and $\tilde d$ their
			effects seem to be negligible. This is the case when we observe a
			configuration which is located close to the Minkowski phase in the
			phase diagram, cf.\ figure~\ref{fig:phaseDiagram}.
			
			In ref.~\cite{Erdmenger:2007ja} we elaborate on the spectral
			functions behavior at low quark masses. There we observed that the
			position of the vector meson excitations in the regime of very low
			quark masses \emph{decreased} with increasing quark mass. Further
			increasing the quark mass lead to increasing quark masses as
			described in this section. We omit this discussion here, but resume
			on the topic when we discuss the pole structure of the spectral
			functions. The reason is that the peaks referred to in
			ref.~\cite{Erdmenger:2007ja} are only visible after subtraction of
			the zero temperature part $\R_0$ from the spectral function. To
			interpret the spectral function as a probability density for the
			detection of a quasiparticle, we cannot subtract $\R_0$, as we
			would otherwise produce negative probability densities, which are
			not well defined.
			
			Our setup is a modification of the one used in ref.~\cite{Myers:2007we}.
			There, the authors considered vector meson spectra at vanishing
			baryon density. These spectra only show peaks moving to smaller
			frequency as the quark mass is increased. There is no contradiction
			to the results presented this work. Note that the authors of
			ref.~\cite{Myers:2007we} by construction are restricted to the regime of
			high temperature/small quark mass. Nevertheless, they continue to
			consider black hole embeddings below the temperature of the
			fundamental phase transition where these embeddings are only
			metastable, the \emph{Minkowski embeddings} being thermodynamically
			favored. At small baryon density and small $m$ our spectra are
			virtually coincident with those of~\cite{Myers:2007we}. However, in
			our case, at finite baryon density, black hole embeddings are
			favored for all values of the mass over temperature ratio.
			
	\subsubsection{Pole structure}
		
\begin{figure}
			\begin{minipage}[t]{.45\linewidth}
				\includegraphics[width=\linewidth]{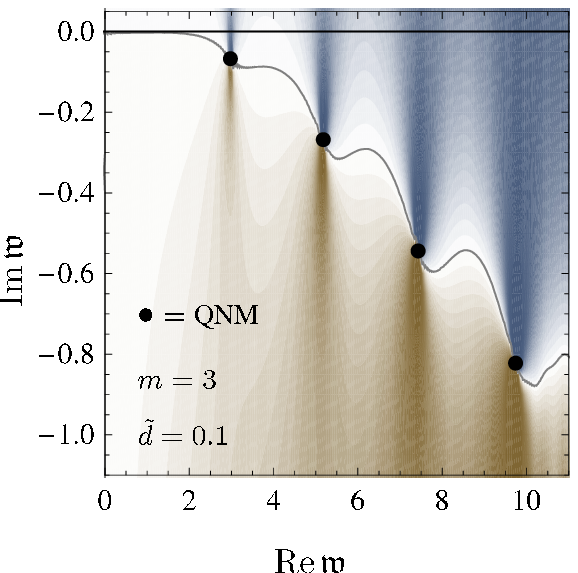}
				\caption[Spectral function in the complex $\wn$-plane]{A contour
					plot of the spectral function in the complex
					$\wn$-plane. Blue shading indicates
					$\R>0$, brown shading indicates
					$\R<0$. The gray contour traces
					$\R=0$. The values along the line $\Im\wn=0$ represent
					the physical spectrum we plotted for several parameters
					above.}
				\label{fig:specFuncComplexPlane}
			\end{minipage}
			\hfill
			\begin{minipage}[t]{.45\linewidth}
				\includegraphics[width=\linewidth]{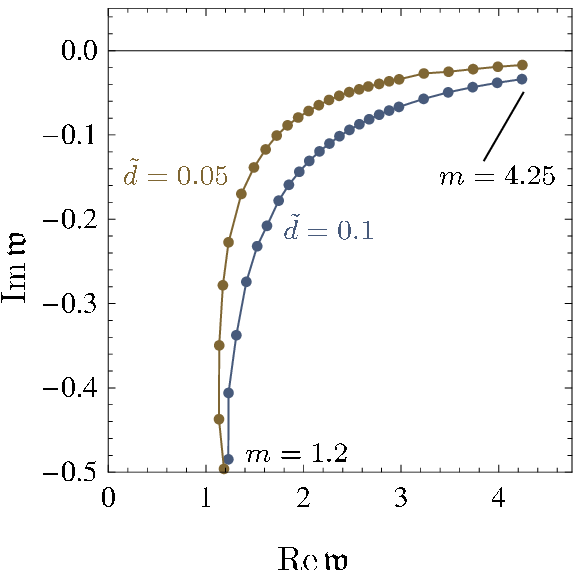}
				\caption[Trajectory of QNMs in the complex $\wn$-plane]{The
					trajectory of the quasi normal mode with lowest $|\wn|$
					in the complex $\wn$-plane parametrized by the quark
					mass parameter $m$. for $\Re\wn<3$ successive points
					have $\Delta m=0.1$, for $\Re\wn>3$ we chose $\Delta
					m=0.25$.}
				\label{fig:poleMovement}
			\end{minipage}
		\end{figure}

		In this section we comment on the quasi normal modes (QNM) of the system
		under investigation. As discussed above, these are the poles in the
		complex frequency plane, i.\,e.\  the spectral function diverges at these
		locations. An impression of the continuation of the spectral function
		into the complex frequency plane is given in
		figure~\ref{fig:specFuncComplexPlane}. The data to plot the graph was
		obtained in exactly the same way as for the spectral functions shown in
		the preceding sections, except the fact that we upgraded the numerics
		to process $\wn\in\mathbbm C$. Therefore the spectral functions shown so
		far are given by the values along the real axis.
		
		Our numerics turn out to be reliable for $|\Im\wn|\lesssim1$ and
		therefore cannot determine poles in the plane of  $\wn\in\mathbbm C$
		which lie beyond this limit. We trust the values within the regions
		shown in the figures of this work, although there possibly is room for
		improvement in accuracy. We checked our code for stability against the
		initial conditions and parameters, and are mainly interested in the
		qualitative behavior of the results.
		
		The spectra presented in preceding sections show that the first
		resonance peaks, i.\,e.\  those for small $n$, are very narrow, while the
		following peaks show a broadening accompanied by decreasing amplitude.
		The physical consequence would be a longer lifetime of the lower~$n$
		excitations. This is reflected in a smaller imaginary part of the
		corresponding quasi normal mode in
		figure~\ref{fig:specFuncComplexPlane}. It is a known fact that the
		quasinormal modes develop larger real and \emph{imaginary} parts at
		higher $n$. So the sharp resonances at low~$\wn$, which correspond to
		quasiparticles of long lifetime, originate from poles with small
		imaginary part. For higher excitations in $n$ at larger~$\wn$, the
		resonances broaden and get damped due to larger imaginary parts of the
		corresponding quasi normal modes.
		
		Above we observed variations in the positions and widths of the peaks in
		the spectral function, depending on the changes in temperature, particle
		density and meson mass. This behavior can be translated into a movement
		of the quasinormal modes in the complex plane.
		Figure~\ref{fig:poleMovement} shows the trajectory of the quasi normal
		mode corresponding to the first peak in the spectral function,
		parametrized by $m$. At small densities, we can see the turning
		behavior of the mode which starts to move in the direction of decreasing
		real part at small $m$, and then turns to asymptotically large real part
		while converging to the real axis for further increasing $m$.
		
		It would be interesting to compare our results to a direct calculation
		of the quasinormal modes of vector fluctuations in analogy to
		ref.~\cite{Hoyos:2006gb}. There, the quasinormal modes are considered for
		scalar fluctuations exclusively, at vanishing baryon density. The
		authors observe that starting from the massless case, the real part of
		the quasinormal frequencies increases with the quark mass first, and
		then turns around to decrease. This behavior agrees with the peak
		movement for scalar spectral functions observed in ref.~\cite{Myers:2007we}
		(above the fundamental phase transition at $\chi_0\le 0.94$) where the
		scalar meson resonances move to higher frequency first, turn around and
		move to smaller frequency increasing the mass further. These results do
		not contradict the present work since we consider vector modes
		exclusively.
		
\subsection{Meson spectra at finite isospin density}
\label{sec:isoSpectra}

	\subsubsection{Equations of motion}
	
		In order to examine the case of two flavors, $\N_f=2$, with opposite
		isospin chemical potential in the strongly coupled plasma, we extend
		our previous analysis of vector meson spectral functions to a
		chemical potential with $\SU(2)$-flavor, i.\,e.\  isospin, structure.
		Starting from the general action \eqref{eq:simplifiedNonAbelianDBI}
		we now consider the non-Abelian field strength tensors
		\begin{equation}
		\label{eq:isoF}
		\hat F_{\mu\nu}=\sigma^a\,\left (2\partial^{\vphantom a}_{[\mu}\hat A^a_{\nu]}
		  +\frac{\rh^2}{2\pi\alpha'}f^{abc}\hat A_\mu^b \hat A_\nu^c\right),
		\end{equation}
		with the Pauli matrices~$\sigma^a$ and $\hat A$ given by
		equation~\eqref{eq:hatA}. The upper index on the gauge field labels
		the component in the vector space of the $\SU(2)$ generators.
		The factor~$\rh^2/(2\pi\alpha')$ is due to
		the introduction of dimensionless fields as described
		below~\eqref{eq:chemPotLimit}. The totally antisymmetric
		$f^{abc}=\varepsilon^{abc}$ with $\varepsilon^{123}=1$ arise from
		the structure constants of $\SU(2)$.
		
		In the non-Abelian field strength tensor, the term quadratic in the
		gauge field describes a self interaction of the gauge field. The
		coupling constant for this interaction may be determined by a
		redefinition of the gauge field, such that the kinetic term of the
		effective four-dimensional theory has the canonical form. In
		appendix~\ref{sec:mesoncoupling} (taken from ref.~\cite{Erdmenger:2008yj}) we
		show that the redefinition is given by
		\begin{equation}
			\hat A\mapsto \frac{c_{\scriptscriptstyle A}}{\sqrt{\lambda}}\hat A\,,
		\end{equation}
		where the dimensionless constant $c_{\scriptscriptstyle A}$ depends on
		the geometry of the \D7 worldvolume directions along $\rho$ and the
		$\S^3$, which are transverse to the directions of the \db3. In
		particular, $c_{\scriptscriptstyle A}$ is independent of the 't~Hooft
		coupling $\lambda$. Determining the exact value of
		$c_{\scriptscriptstyle A}$ is left to further work in terms of the ideas
		presented in appendix~\ref{sec:mesoncoupling}. In the following we chose
		a convenient $c_{\scriptscriptstyle A}=\frac{4\pi}{\sqrt 2}$. The field
		strength tensor in the redefined fields is given by
		\begin{equation}
		  F^a_{\mu\nu}=2\del^{\vphantom{a}}_{[\mu}\hat A^a_{\nu]}+\frac{c_{\scriptscriptstyle A}}{\sqrt{\lambda}}f^{abc}\hat A^b_\mu\hat A^c_\nu
		\end{equation}
		
		In order to obtain a finite isospin-charge density~$n_{\scriptscriptstyle I}$ and its
		conjugate chemical potential~$\mu_{\scriptscriptstyle I}$, we introduce a
		$\rho$-dependent time component of the $\SU(2)$ valued background
		gauge field~$\tilde A$~\cite{Erdmenger:2007ap}. This background
		field is defining a direction in the vector space of the $\SU(2)$
		generators. We choose coordinates such that the direction of the
		background field aligns with the $\sigma^3$ direction while the
		other $\SU(2)$ components are vanishing,
		\begin{equation}
		\label{eq:isoBackgrd}
		\tilde A_0 = \tilde A^3_0\sigma^3= \tilde A^3_0(\rho) 
		\left (\begin{array}{c c}
		1 & 0\\ 
		0& -1
		\end{array}\right ), \qquad \tilde A^1_0=\tilde A^2_0=0.
		\end{equation}
		This specific choice of the 3-direction in flavor space as well as
		spacetime dependence simplifies the isospin background field
		strength, such that we get two copies of the baryonic
		background~$\tilde F_{\rho 0}$ on the diagonal of the flavor matrix,
		\begin{equation}
			\tilde F_{40}=
			\left (
				\begin{array}{c c}
					\partial_ \rho \tilde A_0 &            0\\ 
								0             & -\partial_ \rho \tilde A_0
				\end{array}
			\right ).  
		\end{equation}
		The derivation of the background field configuration leads to the
		same explicit form of the action as \eqref{eq:actionEmbeddingsAt}.
		We can therefore make use of the background field solutions
		$\chi(\rho)$ and $\tilde A_0(\rho)$ found in the baryonic case. As
		before, we collect the induced metric $G$ and the background field
		strength~$\tilde F$ in the background tensor $\G=G+\tilde F$.
		
		For the fluctuations, however, we encounter an additional structure.
		The $\SU(2)$ valued fluctuations in general have components along
		all the directions of this vector space. We make use of
		$\tr(\sigma^i\sigma^j)=2\delta^{ij}$ and apply the background field
		method in analogy to the baryonic case examined in
		section~\ref{sec:baryonMu}. As before, we obtain the quadratic
		action for the fluctuations $A^a_\mu$ by expanding the determinant
		and square root in powers of $A^a_\mu$. The term linear in
		fluctuations again vanishes by the equation of motion for the
		background field. This leaves the quadratic action
		\begin{equation}
		\label{eq:quadIsoAction}
		\begin{gathered}
			S^{(2)}_{\text{iso}} = 2\pi^2R^3\,\rh\,T_\text{R} T_7 
			\int\limits_1^\infty\!\! \dd\rho\,\dd^4x\; \sqrt{\left|\det \G\right|}\hspace{10em}\quad\\
		\begin{aligned}
			 \times\Bigg[ & \G^{\mu\alpha} \G^{\nu\beta} \bigg(
				   \partial^{\hphantom a}_{[\mu}A^a_{\nu]}\,\partial^{\hphantom a}_{[\alpha}A^a_{\beta]} 
				   +\frac{\rh^4}{(2\pi\alpha')^2}\,(\tilde A_0^3)^2 f^{ab3} f^{ac3}\, A_{[\mu}^b\delta^{\hphantom a}_{\nu]0}\,
					A_{[\alpha}^{c}\delta^{\hphantom a}_{\beta]0} \bigg)\\  
			&	   +\left(\G^{\mu\alpha} \G^{\nu\beta}-\G^{\alpha\mu} \G^{\beta\nu}\right)
				   \frac{\rh^2}{2\pi\alpha'}\,\tilde A_0^3 f^{ab3} \,\partial^{\hphantom a}_{[\alpha}A^a_{\beta]}\, A_{[\mu}^{b}\delta^{\hphantom a}_{\nu]0} \Bigg] .
		\end{aligned}
		\end{gathered}
		\end{equation}
		The factor $T_\text{R}$ arises from the trace over the generators of
		$\SU(2)$. If we use the Pauli matrices as generators we get
		$T_\text{R}=2$. Another common choice for the generators is
		$\sigma^i/2$, which amounts to $T_\text{R}=1/2$. We leave the
		explicit choice open, since it merely introduces an unimportant
		finite proportionality constant to the action. Note that besides the
		familiar Maxwell term, two other terms appear
		due to the non-Abelian structure. One of the new terms depends
		linearly, the other quadratically on the background gauge
		field~$\tilde A$ and both contribute non-trivially to the dynamics.
		The equation of motion for gauge field fluctuations on the \db7 is
		\begin{align}
		\label{eq:eomIsoFluct}  
		0=\;&\partial_\kappa \left[ \sqrt{\left|\det \G\right|}
		 \left( \G^{\nu\kappa} \G^{\sigma\mu} - \G^{\nu\sigma} \G^{\kappa\mu} \right)
		 \check F_{\mu\nu}^a \right] 
		 - \sqrt{\left|\det \G\right|}
		  \,\frac{\rh^2}{2\pi\alpha'}\, \tilde A_0^3 f^{ab3} \left( \G^{\nu 0} \G^{\sigma\mu} 
		  - \G^{\nu\sigma} \G^{0\mu} \right)\check F_{\mu\nu}^b
		   \, ,  
		\end{align}
		with the modified field strength linear in fluctuations 
		$\check{F}^a_{\mu\nu}=2\partial^{\hphantom a}_{[\mu}A^a_{\nu]}+c_{\scriptscriptstyle A}/\sqrt\lambda f^{ab3}
		   \tilde A_0^3\left( \delta_{0\mu} A_\nu^b+ \delta_{0\nu} A_\mu^b\right)\rh^2/{(2\pi\alpha')}$.
		
		Integration by parts of~\eqref{eq:quadIsoAction} and application
		of~\eqref{eq:eomIsoFluct} yields the on-shell action
		\begin{equation}
		\label{eq:onShellAction}
			S^{\text{on-shell}}_{\text{iso}}=\; 
			 \rh T_{\text R} T_7 \, \pi^2 R^3 \int\!\! \dd^4x\, \sqrt{\left|\det \G\right|}
			  \times\left. \left (  
			 G^{\nu 4} G^{\beta\mu}- G^{\nu\beta} G^{4\mu}
			  \right ) A_{\beta}^a \check F^a_{\mu\nu}\right |_{\rho_{\scriptscriptstyle H}}^{\rho_B}\, . 
		\end{equation}
		The three field equations of motion (flavor index $a=1,2,3$)
		for fluctuations in transverse Lorentz-directions $\mu=2,3$ can
		again be written in terms of the combination $E^a_T=q A^a_0+\omega
		A^a_\alpha$. At vanishing spatial momentum~$q=0$ we get
		\begin{align}
		\label{eq:eomAalphaSplitFlav1}
			0 =\;& {E_T^1}''+\del_\rho\ln\left(\sqrt{\left|\det G\right|} G^{44} G^{22}\right) {E_T^1}' - \frac{G^{00} (\rh\omega)^2}{G^{44}}E^1_T \\
				 & \hphantom{{E_T^1}''}-\frac{G^{00}}{G^{44}}\left[\left({\frac{\rh^2}{2\pi\alpha'}}\tilde A^3_0\right)^2E^1_T 
				 +2 \ii \rh \omega  {\frac{\rh^2}{2\pi\alpha'}}\tilde A^3_0 E^2_T\right],\nonumber \\[\medskipamount]
		\label{eq:eomAalphaSplitFlav2}
			0 =\;& {E_T^2}'' +\del_\rho\ln\left(\sqrt{\left|\det G\right|} G^{44} G^{22}\right) {E_T^2}' - \frac{G^{00} (\rh\omega)^2}{G^{44}}E^2_T \\
				 & \hphantom{{E_T^2}''}-\frac{G^{00}}{G^{44}}\left[\left({\frac{\rh^2}{2\pi\alpha'}}\tilde A^3_0\right)^2E^2_T 
				  - 2 \ii \rh \omega  {\frac{\rh^2}{2\pi\alpha'}}\tilde A^3_0 E^1_T \right] ,\nonumber \\[\medskipamount]
		\label{eq:eomAalphaSplitFlav3}
			0 =\;&  {E_T^3}'' +\del_\rho\ln\left(\sqrt{\left|\det G\right|} G^{44} G^{22}\right) {E_T^3}' - \frac{G^{00} (\rh\omega)^2}{G^{44}}E^3_T .
		\end{align}
		Note that we use the dimensionless background gauge field $\tilde
		A_0^3=\bar A_0^3 (2\pi \alpha')/\rh$ with $\rh=T\pi R^2$. Despite the
		presence of the new non-Abelian terms, at vanishing spatial momentum the
		equations of motion for longitudinal fluctuations $E_L^a$ acquire the
		same form as the transverse equations~\eqref{eq:eomAalphaSplitFlav1}, to
		\eqref{eq:eomAalphaSplitFlav3}.
		
		Two of the above ordinary second order differential equations are
		coupled through their flavor structure. Decoupling can be achieved
		as\footnote{At this point there is an essential difference which
		distinguishes this setup from the approach with a constant
		potential~$\bar A_0^3$ at vanishing mass followed e.\,g.\  in
		ref.~\cite{Erdmenger:2007ap}. While the metric coefficients for
		massless quarks are identical in both cases, there is a
		$\rho$-dependence of the background gauge field in the present
		setup.} in ref.~\cite{Erdmenger:2007ap} by transformation to the
		flavor combinations
		\begin{equation}
		\label{eq:flavorTrafo}
			X =E^1+ \ii E ^2,\qquad	Y =E^1- \ii E ^2\, .
		\end{equation}
		The equations of motion for these fields are given by
		\begin{alignat}{3}
		\label{eq:eomX}
			0=\; && X ''&+\del_\rho\ln\left(\sqrt{\left|\det \G\right|} \G^{44} \G^{22}\right) X'
			 &&- 4 \frac{\rh^4}{R^4}\frac{\G^{00}}{\G^{44}}\left(\wn- \mathfrak{m}\right)^2\kern-.42ex X, \\[\medskipamount]    
		\label{eq:eomY}
			0=\; && Y'' &+\del_\rho\ln\left(\sqrt{\left|\det \G\right|} \G^{44} \G^{22}\right) {Y}'
			 &&- 4 \frac{\rh^4}{R^4}\frac{\G^{00}}{\G^{44}}\left(\wn+ \mathfrak m\right)^2 Y, \\[\medskipamount]
		\label{eq:eomE3}
			0=\; && {E^3}''&+\del_\rho\ln\left(\sqrt{\left|\det \G\right|} \G^{44} \G^{22}\right) {E^3}' 
			 &&- 4 \frac{\rh^4}{R^4}\frac{\G^{00}}{\G^{44}} \wn^2E^3\, , 
		\end{alignat}
		with $\wn=\omega/(2\pi T)$ and dimensionless but $\rho$-dependent
		$\mathfrak m= \bar A^3_0 /(2\pi T)$. Proceeding as described in
		section~\ref{sec:baryonMu}, we determine the local solution of these
		equations at the horizon. The indices turn out to be
		\begin{equation}
		\label{eq:isospinIndices}
		\beta = \pm \ii \left(\wn \mp\frac{\bar A^3_0(\rho=1)}{2\pi T}\right)\, .
		\end{equation}
		Since $\bar A^3_0(\rho=1)=0$ we are left with the same index as in
		\eqref{eq:indices} for the baryon case, i.\,e.\  the chemical
		potential does not influence the singular behavior of the
		fluctuations at the horizon. The local solution coincides to linear
		order with the baryonic solution given in \eqref{eq:localSolutions}.
		
		For the special case of zero temperature the background geometry is
		\adsfivesfive. For finite chemical potential in the zero
		temperature case we can obtain the gauge field correlators in
		analogy to ref.~\cite{Freedman:1998tz}. The zero temperature result
		$\R_{0,\text{iso}}$ analog to \eqref{eq:zeroTempR} is given by
		\begin{equation}
			\label{eq:zeroTempRiso}
			\mathfrak{R}_{0,\text{iso}}=T_\text{R} N T^2\,\pi\left(\wn\pm \mn_\infty\right)^2 ,  
		\end{equation}
		with the dimensionless chemical potential
		$\mn_\infty=\lim_{\rho\to\infty}\mn$.
		
	\subsubsection{Spectra}
		
		Application of the recipe analog to the case of baryonic chemical
		potential yields the spectral functions of flavor current correlators in
		a medium with finite isospin density. Note that after transforming to
		flavor combinations $X$ and $Y$, given in \eqref{eq:flavorTrafo}, the
		diagonal elements of the propagation submatrix in flavor-transverse $X,
		Y$ directions vanish, $G_{XX}=G_{YY}=0$. Now the off-diagonal elements
		give non vanishing contributions. However, the component $E^3$,
		longitudinal in flavor space, is not influenced by the isospin chemical
		potential, such that~$G_{E^3E^3}$ is nonzero, while other combinations
		with $E^3$ vanish~\cite{Erdmenger:2007ap}.
		
\begin{figure}
			\centering
			\includegraphics[width=.5\linewidth]{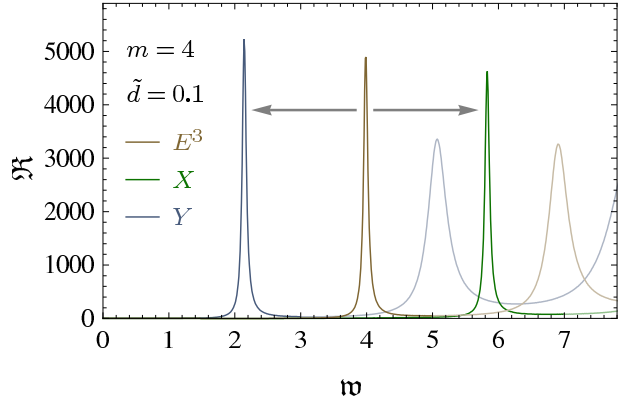}
			\caption[Spectrum at finite isospin density]{%
				The vector meson spectral functions of the three isospin
				components. For a concise image we emphasize the first
				peak of each component by stronger color saturation.
			}
			\label{fig:isoSpectrum}
		\end{figure}

		In figure~\ref{fig:isoSpectrum} we compare spectral functions for
		the isospin case, where we emphasize the first peak of each of the
		three components. Note that the $E^3$ spectrum coincides with the
		baryonic case, as the equation of motion \eqref{eq:eomE3} coincides
		with \eqref{eq:eomEq0}.
		
		While the qualitative behavior of the isospin spectral functions agrees
		with the one of the baryonic spectral functions, there nevertheless is a
		quantitative difference for the flavor-transverse components $X, Y$. We
		find that the propagator for flavor combinations~$G_{YX}$ exhibits a
		spectral function for which the peaks are shifted to higher frequencies,
		compared to the Abelian case curve. For the spectral function computed
		from~$G_{XY}$, the opposite is true, its peaks appear at lower
		frequencies. The quasiparticle resonance peak in the spectral
		function~$\R_{YX}$ appears at higher frequencies than expected from the
		vector meson mass formula~$\eqref{eq:mesMassDimless}$. The other
		flavor-transverse spectral function~$\R_{XY}$ displays a resonance at
		lower frequency than observed in the baryonic case.
		
		This may be interpreted as a splitting of the resonance peak into three
		distinct peaks. This is due to the fact that we explicitly break the
		symmetry in flavor space by our choice of the background field~$\tilde
		A^3_0$. Decreasing the chemical potential reduces the distance of the
		two outer resonance peaks from the one in the middle and therefore the
		splitting is reduced.
		
		The described behavior resembles the mass splitting of mesons in
		presence of a isospin chemical potential expected to occur in QCD
		\cite{He:2005nk,Chang:2007sr}. A linear dependence of the separation of
		the peaks on the chemical potential is expected. Our observations
		confirm this behavior. Since the vector mesons are isospin triplets and
		we break isospin symmetry explicitly, we see that in this respect our
		model is in qualitative agreement with effective QCD models. Note also
		the complementary discussion of this point in~ref.~\cite{Aharony:2007uu}.

\subsection{Summary}
	
	Two distinct setups were examined at non-zero charge density in the black
	hole phase. First, switching on a \emph{baryon} chemical potential through
	non-zero baryon density, we find that nearly stable vector mesons exist
	close to the transition line to the Minkowski phase. Far from this regime,
	at small quark masses or high density, the spectral functions do not show
	distinct resonance peaks.
	
	Moreover, at small quark masses and particle densities we observe that the
	quasi normal modes move to positions with smaller real part in the complex
	$\omega$ plane, in accordance with the observations in the case of vanishing
	chemical potential~\cite{Myers:2007we}. Increasing the quark mass over
	temperature ratio beyond a distinct value, the plasma adopts the behavior
	known from the case of zero temperature. In the spectral functions we
	computed, this zero-temperature-like behavior is found in form of line-like
	resonances, at low particle densities exactly reproducing the
	zero-temperature supersymmetric vector meson mass spectrum.
		
	Besides finite temperature effects it is especially interesting to observe
	the in-medium effects caused by finite particle density. We observe a
	broadening width of the resonance peaks in the spectrum as a function of
	increasing particle density. At the same time, a slight shift of the
	resonance position to higher energies occurs. This result contradicts the
	expectations from the effective QCD models investigated by Brown and Rho
	\cite{Brown:1991kk}. However, experimental data from collision experiments
	at SPS do not support Brown-Rho scaling either. Instead more recent
	effective models, which are in good agreement with experimental data, also
	show a broadening of the $\rho$-meson resonance peaks accompanied by a small
	positive mass shift \cite{Rapp:1999fm,Rapp:1999us}. It would be interesting
	to investigate the mechanisms, that lead to the qualitative agreement of
	effective QCD models and the \D3/\D7 setup that we observed here. Other
	in-medium effects will be studied in section~\ref{sec:quarkoniumDiffusion}.
	There we determine the mass shift of mesons due to polarization of mesons by
	the presence of the gluonic background field in the plasma.
	
	Second, we switched on a nonzero \emph{isospin} density, and equivalently an
	isospin chemical potential arises. The spectral functions in this case show
	a qualitatively similar behavior as those for baryonic potential. However,
	we additionally observe a splitting of the single resonance peak at
	vanishing isospin potential into three distinct resonances. This suggests
	that by explicitly breaking the flavor symmetry by a chemical potential, the
	isospin triplet states, vector mesons in our case, show a mass splitting
	similar to that observed for QCD~\cite{He:2005nk}. It is an interesting task
	to explore the features of this isospin theory in greater detail in order to
	compare with available lattice data and effective QCD
	models~\cite{Kogut:2002zg,Kogut:2002tm,Kogut:2004zg,Splittorff:2000mm,Loewe:2002tw,
	Barducci:2004tt,Sannino:2002wp,Ebert:2005cs,Ebert:2005wr}. In most of these
	approaches, baryon and isospin chemical potential are considered at the same
	time, which suggests another promising extension of this work. Moreover, in
	the context of gravity duals, it will be interesting to compare our results
	for the isospin chemical potential to the work presented in
	ref.~\cite{Aharony:2007uu}.
	
	Alternatively, instead of giving the gauge field time component a
	non-vanishing vacuum expectation value, one may also switch on $B$-field
	components and combine the framework developed
	in~\cite{Erdmenger:2007bn,Albash:2007bk,Albash:2007bq} with the calculation
	of spectral functions for the dual gauge theory.
	
	Our spectra also show that for given quark mass and temperature, lower $n$
	meson excitations can be nearly stable in the plasma, while higher $n$
	excitations remain unstable. At vanishing baryon density, the formation of
	resonance peaks for higher excitations has also been observed
	in~\cite{Mateos:2007yp}.

\section{Diffusion in the holographic plasma}
\label{chap:transport}

Based on the observation that the many particle system observed at RHIC is well
described by hydrodynamics with very low viscosity, the quark-gluon plasma is
widely regarded as an almost perfect liquid
\cite{Huovinen:2001cy,Teaney:2000cw,Kolb:2000fha}. In principle, the theory of
hydrodynamics should be capable of modeling the collective dynamics of the
plasma. This requires knowledge of initial conditions and the equations of state
of the system. The hydrodynamic description then yields the dynamics of the
system in terms of collective quantities such as currents, densities and
entropy. For a comprehensive understanding of the quark-gluon plasma and related
systems it would be pleasing to be able to derive the thermodynamic and kinetic
properties of the system from first principles. In this section we present work
which was conducted with the motivation to advance towards this goal. Adopting
different points of view, we contemplate one particular attribute characteristic
for fluids: diffusion.

The diffusion coefficient is a transport coefficient which parametrizes the
ability of a fluid to reach an equilibrium state by transport of some initially
unevenly distributed quantity through currents. In the quark-gluon plasma, these
currents are the color and flavor currents, which account for the transport of
quarks and gluons through the plasma. A particle of high diffusivity (high
mobility) looses only a small part of its energy while traversing a given
distance the medium and will transport its associated charge much faster than a
particle which looses much momentum by interaction with the medium.

Close to thermodynamic equilibrium, transport coefficients such as that for
diffusion can be derived from first principles by so called
\emph{Kubo-formulae}, which describe the coefficients in terms of correlation
functions of the current which accounts for equilibration of the system. This
approach has been used successfully in the past to derive transport coefficients
and conductivities from holographic models
\cite{Chesler:2007sv,CasalderreySolana:2006rq,Bigazzi:2009tc,
Ejaz:2007hg,Gubser:2006bz,Herzog:2006gh,Hosoya:1983id,Karch:2007pd,Kovtun:2003wp,Landsteiner:2007bd,Mas:2006dy,Mateos:2006yd,Myers:2007we}.

In this section, once more the adjoint and fundamental matter described by the
gauge multiplet of $\N=4$ and the $\N=2$ hypermultiplet of thermal SYM theory in
the limit of a large number of gauge degrees of freedom serves as a model for
the quark-gluon plasma.

The energy loss of heavy quarks and mesons in media has been a subject of
intense experimental interest
\cite{Adare:2006nq,Bielcik:2005wu,Adare:2008sh,Adare:2006ns,Adler:2005ph,Arnaldi:2006ee,Alessandro:2004ap}.
The suppression of charm and bottom quarks observed at RHIC motivated several
groups to utilize the gauge-gravity duality
\cite{Maldacena:1997re,Witten:1998qj,Gubser:1998bc,Aharony:1999ti} to compute
the drag of fundamental heavy quarks in $\N=4$ super Yang-Mills theory at strong
coupling \cite{Herzog:2006gh,CasalderreySolana:2006rq,Gubser:2006bz}. In this
approach, the heavy quark is given by a classical string attached to the \db7
probe. First studies of flavors in thermal \ads/CFT beyond the quenched
approximation, i.\,e.\  with non-zero $N_f/N$, were performed
in~\cite{Bertoldi:2007sf,Bigazzi:2009bk}.

We pick up the previous efforts and generalize them by including the effects of
finite density, respectively chemical potential, on diffusion properties. The
first part of this section very shortly addresses consequences of finite
baryon density on a certain holographic method to derive the diffusion
coefficient of baryons in a holographic plasma. Subsequently, in
section~\ref{sec:isospinDiffusion}, we study the gauge/gravity dual of a finite
temperature field theory at finite \emph{isospin} chemical potential.
The isospin chemical potential is obtained by giving a
finite vacuum expectation value to the time component of the non-Abelian gauge
field on the brane, as in the previous sections. In order to obtain analytical
results, we restrict our attention to the limit of massless quarks.

The consideration of an isospin chemical potential is an interesting field to
study since it is still easier accessible by lattice calculations than setups at
baryonic chemical potential. Hopefully holographic models one day allow for
comparison to e.\,g.\  large $N$ lattice calculations. Moreover, isospin diffusion
has been measured in heavy ion reactions~\cite{Liu:2006xs,Tsang:2003td}.

Eventually in section~\ref{sec:quarkoniumDiffusion} we adopt a different point
of view on diffusion in the holographic quark-gluon plasma. From the technical
point of view, we do not pursue the approach of Kubo to obtain the diffusion
coefficient directly from current correlation functions. Instead, we make use of
a stochastic Langevin model, which determines the momentum broadening of
particles due to random kicks from interactions with the medium.

Moreover, conceptually we extend the area of research on diffusion processes in
the holographic QGP to \emph{mesons} traversing the plasma. This effort bears
two interesting aspects. One is the aim for a description of the kinetics of
heavy mesons in the QGP, since observation show that heavy mesons like the
$J/\psi$ survive the deconfinement transition. The other is the estimation of
the effects of strong coupling on the plasma. The particular effective model we
use does not rely on any weak or strong coupling limit for the interaction of
the mesons with the medium. This allows for comparison of perturbative results
for momentum broadening at weak coupling with holographic results for the strong
coupling regime. In this way we present a method that may allow to estimate the
effect of strong coupling on dynamic effects in the QGP. 

\subsection{Baryon diffusion}
\label{sec:baryonDiffusion}
	
	In this section, we calculate the baryon diffusion coefficient $D$ and its
	dependence on the baryon density in the thermal holographic plasma. The
	coordinates and parameters we use are the same as in
	section~\ref{chap:specFuncs}, they are discussed in detail in
	section~\ref{sec:backAndBranes}.
	
	In the context of holography, the idea is to describe the conserved current
	$J$ of the gauge field theory in terms of the dual gauge field in the
	supergravity theory, as we did in section~\ref{chap:specFuncs}. This current
	in our case is the current $J$ which transports baryon charge and therefore
	is non-zero when baryon diffusion occurs. The dual supergravity field was
	identified as the gauge field fluctuation $A$ on the probe brane. Any gauge
	field configuration of $A$ that satisfies the equations of motion also
	generates a conserved current in terms of the field strength $F=\dd A$, as
	$\dd F=0$. As in electrodynamics one can identify the columns of the field
	strength tensor with vector currents. Fick's law of diffusion $\threevec
	J=D\threevec\nabla J^0$ can be shown to be satisfied in the long distance
	limit for the on-shell field strength tensor of a supergravity gauge field.
	The constant $D$ is then identified with the diffusion constant
	\cite{Kovtun:2003wp}. This constant describes how strong the currents
	$\threevec J$ are which drive a hydrodynamic system into equilibrium, as a
	reaction on gradients in the charge distribution $J^0$.
	
	To solve for $D$, one therefore has to find the solutions of the gauge field
	fluctuations which are holographically dual to the relevant current. As soon
	as a solution is found and gauge/gravity duality is invoked to compute the
	current $\threevec J$, one may solve Fick's law for $D$. At vanishing
	particle density, the gauge field solution can determined from the DBI
	action in terms of metric coefficients alone \cite{Kovtun:2003wp}. Because
	we use the same coordinate system as the authors of ref.~\cite{Myers:2007we}, we
	arrive at the same explicit form of the induced metric
	\eqref{eq:inducedD7Metric} on the probe \db7 in our setup. We therefore
	reuse the result for the diffusion coefficient derived there.
	
	A very concise formula for the diffusion coefficient $D$ of R-charges was
	derived in reference to the \emph{membrane paradigm} in ref.~\cite{Kovtun:2003wp}.
	It was later directly translated to the diffusion of flavor currents in the
	\D3/\D7 setup in ref.~\cite{Myers:2007we}. The name ``membrane paradigm'' does not
	refer to D-branes but instead alludes to the fact that the analogies between
	black hole physics and thermodynamics very often can be expressed in terms
	of events taking place at the event horizon (or slightly outside the
	horizon, then referred to as the ``stretched horizon''), which has no
	materialistic manifestation but still appears as a significant surface, a
	membrane, to an observer or in the relevant formulae.
	
	We extend previous efforts by introducing finite baryon density. This quantity explicitly
	enters the solution to the gauge field \eqref{eq:backgroundAt} which in turn
	explicitly enters the equation of motion \eqref{eq:eomEq0} for the
	fluctuations. Therefore, one should expect a modified result for the
	diffusion coefficient including the explicit occurrence of the baryon density
	$\tilde d$. We rely on the fact that these terms vanish for $\tilde d\to0$,
	restrict to the small density regime and stick to the expression found in
	ref.~\cite{Myers:2007we}.
	Nevertheless we still cover finite density effects in this way, since the
	probe brane embeddings $\chi$ are different for vanishing and finite
	density, as we see from their equation of motion
	\eqref{eq:eomChi}. This difference
	should therefore translate into a dependence of the diffusion constant on
	the parameter $\tilde d$, which is proportional to the baryon density by
	\eqref{eq:dtFromnB}.
	We know that in the case of finite baryon density, black hole embeddings
	describe the entire parameter range of temperature and quark mass
	\cite{Karch:2007br,Kobayashi:2006sb}. 
	
	Finally, the formula for the diffusion coefficient $D$ found in
	ref.~\cite{Myers:2007we} is given by
	\begin{equation}
		D = \left .\frac{\sqrt{-G}}{G_{11}\sqrt{-G_{00}\, G_{44}}}\right |_{\rho=1}
		\int\limits_{1}^{\infty}\!\! \dd\rho\;\frac{-G_{00}\, G_{44}}{\sqrt{-G}},  
	\end{equation}
	where the metric coefficients $G_{\mu\nu}$ can be obtained from
	\eqref{eq:inducedD7Metric} with the square root of the absolute value of the
	metric determinant $\sqrt{-G}$ given by
	\eqref{eq:inducedD7MetricDeterminant}. Insertion of these coefficients and
	$\rh=T\pi R^2$ yields
	\begin{equation}
	\label{eq:diffusionConstant} 
		D = \frac{2(1-\chi_0^2)^{3/2}}{T \pi}\int\limits_1^\infty\!\dd\rho\; \frac{\rho(\rho^4-1)\sqrt{1-\chi^2(\rho)+\rho^2{\chi'}^2(\rho)}}{(\rho^4+1)^2\left(1-\chi^2(\rho)\right)^2}\,.
	\end{equation}
	The embeddings $\chi$ are determined as in section~\ref{chap:specFuncs} by
	solving equation \eqref{eq:eomChi} in terms of the parameters $\tilde d$ for
	baryon density and initial value $\chi(1)=\chi_0$, which determines the
	quark mass normalized to temperature, c.\,f.\  figure~\ref{fig:mFromChi0}.
	
\begin{figure}
		\centering
		\includegraphics[width=.5\linewidth]{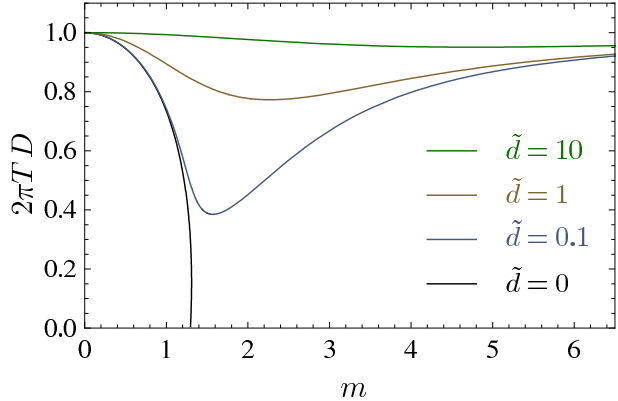}
		\caption[Baryon diffusion coefficient]{Approximate baryon diffusion
			coefficient $D$ as a function of the quark mass to temperature ratio
			$m$. The curve for $\tilde d=0$ reproduces the result from
			ref.~\cite{Myers:2007we}. The exact results derived in ref.~\cite{Mas:2008qs}
			show deviations most notably in the limit of large $m$.
		}
		\label{fig:diffusionCoeffBaryon}
	\end{figure}

	The results for $D$ are shown in figure~\ref{fig:diffusionCoeffBaryon},
	where we compare to the result at vanishing baryon density found in in
	ref.~\cite{Myers:2007we}. There is a phase transition, at approximately $m=1.3$
	which we briefly address in section~\ref{sec:baryonDiffusionPhaseTrans}.
	
	The diffusion coefficient never vanishes in the medium with non-zero
	density. Both in the limit of $T/m_q\to 0$ and $T/m_q\to\infty$, $D$
	converges to $1/(2\pi T)$ for all densities, i.\,e.\  to the same value as for
	vanishing baryon density, as given for instance in \cite{Kovtun:2003wp} for
	R-charge diffusion. In the regime of moderate to low temperatures the
	diffusion constant develops a nonzero minimum.
	
	In order to give a physical explanation for this behavior, we focus on the
	case without baryons first. We see that the diffusion coefficient vanishes
	at the temperature of the fundamental deconfinement transition. This is
	simply due to the fact that at and below this temperature, all charge
	carriers are bound into mesons not carrying any baryon number.
	
	For non-zero baryon density however, there is a fixed number of charge
	carriers (free quarks) present at any temperature. This implies that the
	diffusion coefficient never vanishes. Switching on a very small baryon
	density, even below the temperature of the phase transition, where most of
	the quarks are bound into mesons, by demanding $\tilde d\neq0$ there will
	still be a finite amount of free quarks. By increasing the baryon density,
	we increase the amount of free quarks, which at some point outnumber the
	quarks bound in mesons. In the large density limit the diffusion
	coefficient approaches $D_0 = 1/(2\pi T)$ for all values of $T/m_q$, because
	only a negligible fraction of the quarks is still bound in this limit.
	
	As a final comment, we point out that after the publication of these results
	in refs.~\cite{Erdmenger:2007ja,Erdmenger:2008yj} a more careful analysis
	of the calculation of the diffusion coefficient was performed in
	ref.~\cite{Mas:2008qs}. Here the diffusion coefficient was identified with
	the proportionality coefficient $D$ in the dispersion relation for the
	hydrodynamic quasi normal mode $\omega$ (the so called diffusion pole or
	hydrodynamic pole, cf.\ section~\ref{sec:retardedThermalGreen}) of the gauge
	field fluctuations at finite spatial momentum $\threevec k$, given by
	\begin{equation}
		\omega = -\ii D \threevec k^2 + \order{\threevec k^3}.
	\end{equation}
	The inclusion of finite spatial momentum introduces several new aspects we
	circumvented in the limit of $k\to0$. In accordance with the results
	obtained from this analysis, the results for $D$ in the low quark mass/high
	temperature regime agree well with our na\"ive approach \cite{Mas:2008qs}.
	
\subsection{Isospin diffusion}
\label{sec:isospinDiffusion}

	We study the diffusion coefficient of particles charged under isospin
	chemical potential. The results where originally published as
	ref.~\cite{Erdmenger:2007ap}. Physically, the isospin chemical potential
	corresponds to the energy necessary to invert the isospin of a given
	particle. Within nuclear physics, such a chemical potential is of relevance
	e.\,g.\  for the description of neutron stars. In two-flavor QCD, effects of a
	finite isospin chemical potential have been discussed for instance in
	refs.~\cite{Son:2000xc,Splittorff:2000mm,Toublan:2003tt}.
	
	In the following paragraphs we outline the procedure and comment on the
	restrictions we imposed. Recent work revealed that some of these can
	actually be considered as shortcomings. In particular the consideration of a
	constant background gauge field on the brane can be justified only for very
	small chemical potentials. Due to the assumed smallness of the chemical
	potential we neglected second order terms in the background gauge field on
	the brane. Both limitations however can be cured, based on the insights
	published in refs.~\cite{Karch:2007br,Kobayashi:2006sb}. A reviewed version of our
	results can be found in ref.~\cite{Kaminski:2008ai}.
	
	As in section~\ref{chap:specFuncs}, we introduce an isospin chemical
	potential $\mu$ by defining the vacuum expectation value of the $\SU(2)$
	gauge field on the coinciding probe \db7s to be
	\begin{equation}
	\label{eq:backg}
		A_0 = \left(
		\begin{array}{cc}
			\mu & 0 \\ 0 & -\mu
		\end{array}\right),
	\end{equation}
	For simplicity, we work with a constant background field configuration. The
	above $A_0$ is a solution to the \db7 equations of motion and is present
	even for the \db7 embedding corresponding to massless quarks. However, We
	consider $\mu$ to be small, such that the Bose-Einstein instability observed
	in ref.~\cite{Apreda:2005yz}, which is of order $\order{\mu^2}$, does not affect
	our discussion here.
	
	Even though, this constant gauge field given by \eqref{eq:backg} is a
	solution to the equation of motion, it does not represent the
	thermodynamically preferred configuration \cite{Karch:2007br}. Instead we
	should rather make use of the solution presented in \eqref{eq:backgroundAt}
	with constant embedding function $\chi=0$, for which the authors of
	ref.~\cite{Karch:2007br} present an analytical solution. Nevertheless, for
	simplicity we stick with \eqref{eq:backg} in this section, which may be
	justified in the case of very small densities, where the derivative of
	$A_0(u)$ is very small, c.\,f.\  figure~\ref{fig:backgAt}.
	
	Again, we work in the \D3/\D7 setup. For simplicity, we consider only the
	\D7 probe embedding for vanishing mass $m=0$. This embedding is constant and
	terminates at the horizon. This simplification allows us to derive our
	results purely analytically. We establish the $\SU(2)$ non-Abelian action
	for a probe of two coincident \db7s and obtain the equations of motion for
	fluctuations about the background \eqref{eq:backg}. These fluctuations are
	dual to the $\SU(2)$ flavor current $J^{\mu a}$. We find an ansatz for
	decoupling the equations of motion for the different Lorentz and flavor
	components, and solve them by adapting the method developed in
	refs.~\cite{Policastro:2002se,Son:2002sd}. This involves Fourier transforming to
	momentum space, and using a power expansion ansatz for the equations of
	motion. We discuss the approximation necessary for an analytical solution,
	which amounts to considering frequencies with $\omega<\mu<T$.  With
	this approach we obtain the complete current-current correlator. The key
	point is that the constant chemical potential effectively replaces a time
	derivative in the action and in the equations of motion. In the Fourier
	transformed picture, this leads to a dependence of physical observables on
	the square root $\sqrt{\omega}$ of the frequency. This non-linear behavior
	goes beyond linear response theory. We discuss the physical properties of
	the Green functions contributing to the current-current correlator. In
	particular, for small frequencies we find a frequency-dependent diffusion
	coefficient $D(\omega) \propto \frac{1}{T} \sqrt{\omega/ \mu}$. Whereas
	frequency-dependent diffusion has\,---\,to our knowledge\,---\,not yet been
	discussed in the context of the quark-gluon plasma, it is well-known in the
	theory of quantum liquids and therefore may possibly also apply to the quark
	gluon plasma. For instance, for small frequencies the square-root behavior
	we find agrees qualitatively with the results of refs.~\cite{PhysRevLett,Rabani}
	for liquid para-hydrogen. Generally, frequency-dependent diffusion leads to
	a non-exponential decay of time-dependent fluctuations
	\cite{Bhattacharjee:1980tf}.
	
	The approach used in this section is related but different from the
	procedure we implemented to obtain the spectral functions at finite isospin
	density in section~\ref{chap:specFuncs}. The non-analytical behavior we
	derive is due to the limits we consider. Coming from equation
	\eqref{eq:isospinIndices}, the difference between a constant non-vanishing
	background gauge field and the varying one becomes clear. Here, the field is
	chosen to be small and constant in~$\rho$, such that terms quadratic in the
	background gauge field~$\tilde A_0^3\ll 1$ can be neglected. This implies
	that the square~$(\wn\mp\mn)^2$ in~\eqref{eq:eomX} and~\eqref{eq:eomY} is
	replaced by~$\wn^2\mp 2\wn\mn$, such that we obtain the indices~$\beta=\pm
	\wn \sqrt{1\mp\frac{\bar A^3_0(\rho=1)}{(2\pi T) \wn}}$ instead
	of~\eqref{eq:isospinIndices}. If we additionally assume $\wn\ll \tilde
	A^3_0$, then the $1$ under the square root can be neglected. In this case
	the spectral function develops a non-analytic structure coming from the
	$\sqrt \omega$ factor in the index.
	
	This section is organized as follows. We start with a comment on
	frequency-dependent diffusion within hydrodynamics and the method we use to
	compute the diffusion constant holographically. Thereafter, we establish the
	\D7 probe action in presence of the isospin chemical potential, derive the
	corresponding equations of motion and solve them. Finally, we obtain the
	associated Green functions in the hydrodynamical approximation. From their
	pole structure we can read off the frequency-dependent diffusion
	coefficient. We comment on our results briefly where it is appropriate in
	this section and leave a summarizing discussion for
	section~\ref{sec:summaryDiffusion}, including an interpretation of our
	results.

	\subsubsection{Diffusion coefficients from Green functions}
		
		Thermal Green functions have proven to be a useful tool not only to
		derive spectra, as above, but also for analyzing the structure of
		hydrodynamic theories and for calculating hydrodynamic quantities such
		as transport coefficients. In this section we once more use the
		gauge/gravity dual prescription of ref.~\cite{Son:2002sd} for
		calculating Green functions in Minkowski spacetime. These correlators
		can be thought of as being determined by their pole structure, in the
		way discussed in connection with the spectral functions, c.\,f.\ 
		section~\ref{sec:retardedThermalGreen}. From these poles at complex
		frequencies we derive the diffusion coefficient $D$ of isospin charge
		with charge density $J^0$ and conserved current four vector $\fourvec
		J=(J^0,\threevec J)$.
		
		Considering systems governed by hydrodynamics, such as the quark-gluon
		plasma, we are eager for solutions to the hydrodynamic equations of
		motion. Regarding diffusion, we pay special interest to the Green
		function for the diffusion equation
		\begin{equation}
		\label{eq:diffusionEquation}
			\partial_0 J^0 (t, \threevec x) = D\,\nabla^2 J^0(t, \threevec x) ,
		\end{equation}
		with $J^0$ the density, given by the time component of a diffusive
		current four vector $\fourvec J$, and $D$ is the diffusion constant we
		are interested in. In Fourier space this equation reads
		\begin{equation}
		\label{eq:constantD}
			\ii \omega J^0(\omega, \threevec k) = D \threevec  k^2 J^0(\omega, \threevec k ).
		\end{equation}
		This determines the dispersion relation $\omega=-\ii D \threevec k^2$ of
		the mode with energy $\omega$. In the language of Green functions we
		will observe solutions of the form
		\begin{equation}
			G(\omega, \threevec k) \propto \frac{1}{\ii \omega - D \threevec k^2} \, .
		\end{equation}
		Here, the dispersion relation determines the poles of the Green function.
		Finding the correct dependence of the poles of the correlator on
		$\omega$ and $\threevec k$ therefore allows to determine the diffusion
		constant $D$.
		
		The diffusive quantity we are interested in the isospin charge with
		density $J^0$ as part of the four vector $\fourvec J=(J^0,\threevec J)$.
		These currents are holographically dual to the gauge field on the brane.
		We therefore have to solve the equations of motion \emph{for the gauge
		fields} to obtain the relevant field theory Green function $G$ of
		isospin diffusion of $J^0$ by following the recipe for retarded
		correlation functions \cite{Son:2002sd}. We are interested in the
		hydrodynamic properties and therefore restrict to the hydrodynamic long
		wavelength/low energy limit such that we restrict to the lowest order in
		$\threevec k$.
		
		For the non-Abelian case with an isospin chemical potential, in section
		\ref{sec:greenResults} we will obtain retarded Green functions of the
		form
		\begin{equation}
		\label{eq:omegadiff}
			G(\omega, \threevec k) \propto \frac{1}{\ii \omega - D(\omega)  \threevec k^2 + \order{\threevec k^3}}\,.
		\end{equation}
		Retarded Green functions of this type have been discussed for instance
		in ref.~\cite{Bhattacharjee:1980tf}. Equation \eqref{eq:omegadiff}
		describes frequency-dependent diffusion with coefficient $D(\omega)$,
		such that \eqref{eq:constantD} becomes
		\begin{equation} \label{eq:omegaD}
			\ii \omega J^0(\omega, \threevec k) = D (\omega)  \threevec k^2 J^0(\omega, \threevec k) \, .
		\end{equation}
		In our case, $J^0$ is the isospin density at a given point in the
		liquid.
		
		This is a non-linear behavior which goes beyond linear response theory.
		In particular, when Fourier-transforming back to position space, we have
		to use the convolution for the product $D \cdot J_0$ and obtain
		\begin{equation}
		\label{eq:convolution}
			\del_0 J^0 (t,\threevec x) + \nabla^2 \int\limits_{-\infty}^{t} \dd s \, J^0(s,\threevec x) D(t-s) = 0
		\end{equation}
		for the retarded Green function. This implies together with the
		continuity equation $\del_0 J^0+ \threevec\nabla\cdot \threevec J=0$,
		with $\threevec J$ the three-vector current associated to $J^0$, that
		\begin{equation}
			\threevec J = - \threevec\nabla (D*J^0),
		\end{equation}
		where $*$ denotes the convolution. This replaces the linear response
		theory constitutive equation $\threevec J = - D \threevec\nabla J^0$.
		Note that for $D(t-s)= D \delta(t-s)$ with $D$ constant,
		\eqref{eq:convolution} reduces to \eqref{eq:diffusionEquation}.
		
	\subsubsection{Holographic setup}
	\label{sec:isoSpinDiffusionSetup}
		
		In this subsection we use the \ads black hole coordinates given in
		appendix~\ref{chap:coordinates}, equation \eqref{eq:adsBHu},
		\begin{equation}
			\label{eq:AdSBHmetricIso}
			\begin{gathered}
					\ds^2 = \frac{\rh^2}{R^2u}\left( -f(u)\dd t^2 + \dd \threevec x^2 \right) +\frac{R^2}{4u^2\,f(u)} \dd u^2 +R^2 \dd\Omega_5^2,\\[\smallskipamount]
						f=1-u^2,\qquad\rh=T\pi R^2,\\[\smallskipamount]
				  0 \leq u \leq 1,\qquad x_i\in\mathbb{R}
			\end{gathered}
		\end{equation}
		with the metric $\dd\Omega_5^2$ of the unit $5$-sphere. This geometry is
		asymptotically \adsfivesfive with the boundary of the \ads part at
		$u=0$, the black hole horizon is located at $u=1$.
		
		Into this ten-dimensional spacetime we embed $N_f=2$ coinciding \db7s,
		hosting the flavor gauge fields $A$. We choose the same embedding as
		in the previous calculations, which extends the \db7s in all directions
		of $\ads_5$ space and wraps an $\S^3$ on the $\S^5$. Here, we restrict
		ourselves to the most straightforward case, that is the trivial constant
		embedding of the branes through the origin along the \ads radial
		coordinate $u$. This corresponds to massless quarks in the dual field
		theory. On the brane, the metric simply reduces to
		\begin{equation}
			\label{eq:finiteTMetric} 
			\begin{gathered}
				\ds^2 = \frac{\rh^2}{R^2u} \left(-f(u)\,\dd t^2+
				  \dd x_1^2+\dd x_2^2+\dd x_3^2 \right) +  
				  \frac{R^2}{4 u^2\,f(u)}\, \dd u^2 + R^2 \dd\Omega_3^2.
			\end{gathered}
		\end{equation}
		Due to the choice of our gauge field in the next subsection, the remaining
		three-sphere in this metric will not play a prominent role. We use
		labels $\mu,\nu,\ldots$ to denote any direction, $i,j,\ldots$ to refer
		to Minkowski directions, $u$ is used as a label for the radial
		coordinate, and $\alpha$ will be used to refer to the $x^{1,2}$
		directions. 
		
		\subsubsection*{Introducing a non-Abelian chemical potential}
			
			A gravity dual description of a chemical potential amounts to a
			non-dynamical time component $A_0$ of the gauge field in the action
			for the \db7 probe embedded into the background given above. There
			are essentially two different ways to realize a non-vanishing
			contribution from a chemical potential to the field strength tensor
			$F=2\partial_{[\mu}A_{\nu]}+f^{abc}A^b_\mu A^c_\nu$. The first is to
			consider a $u$-dependent baryon chemical potential, as we did in the
			preceding section. We work with a constant chemical potential of the
			form
			\begin{equation}
			\label{eq:nonAbelianMu}
				A_0^{\vphantom{a}} = A^a_0 \, T^a ,
			\end{equation}
			where we sum over indices which occur twice in a term and denote the
			gauge group generators by $T^a$. The brane configuration described
			above leads to an $\SU(N_f)$ gauge group with $N_f=2$ on the brane,
			which corresponds to a global $\SU(N_f)$ in the dual field theory.
			For $N_f=2$, the generators of the gauge group on the brane are
			given by $T^a=\frac{\sigma^a}{2}$, with Pauli matrices $\sigma^a$.
			We will see that \eqref{eq:nonAbelianMu} indeed produces non-trivial
			new contributions to the action.
			
			Using the standard background field method of quantum field theory,
			we consider the chemical potential as a fixed background and study
			gauge field fluctuations around it. We single out a particular
			direction in flavor space by taking $A^3_0=\mu$ as the only
			non-vanishing component of the background field. From now on we use
			the symbol $A^a_\nu$ to refer to gauge field fluctuations around the
			fixed background,
			\begin{equation}
			\label{eq:backgroundRule}
				A^a_\nu \mapsto \mu \delta_{\nu 0}\delta^{a3} + A^a_\nu .
			\end{equation}
			
			We gauge the component along the radial coordinate to $A_u=0$ and
			assume that $A_\mu = 0$ for $\mu=5,6,7$. Due to the symmetries of
			the background, we effectively examine gauge field fluctuations
			$A_\mu$ depending on the five-dimensional subspace on the brane
			spanned by the coordinates $x^{0,1,2,3}$ and by the radial \ads
			coordinate $u$. The magnitude of all components of $A$ and the
			background chemical potential $\mu$ are considered to be small. This
			allows us to simplify certain expressions by dropping terms of
			higher order in $A$ and in the chemical potential $\mu$.

	\subsubsection{Equations of motion and their solutions}
		
		The action describing the dynamics of the flavor gauge fields in the
		\D3/\D7 setup is the Dirac-Born-Infeld action. Since we work with
		vanishing gauge field components in all of the directions perpendicular
		to the \db3s, there are no contributions from the Chern-Simons action.
		As mentioned, we consider the constant \D7 probe embedding corresponding
		to vanishing quark mass, $m=0$. The metric on the brane is then given by
		\eqref{eq:finiteTMetric}. Since we are interested in two-point
		correlators only, it is sufficient to consider the DBI action to second
		order in $\alpha'$,
		\begin{equation}
		\label{eq:dbiActionIso}
			S_{\text{D7}} = -T_{7}T_\text{R}\, (2 \pi^2 \alpha')^2 R^3 
			 \int\limits_{u=0}^{u=1} \dd u\,
			  \dd^4 x\; \sqrt{-G}\,G^{\mu\sigma}\,G^{\nu\beta}\, F_{\mu\nu}^a
			  \, F_{\sigma\beta}^a \, ,
		\end{equation}
		where we use the \db7 tension $T_7$ as in \eqref{eq:braneTension},
		performed the integration over the $5,6,7$-directions, which are the
		directions along the $S^3$, and the factor $T_\text{R}$ arising from the
		trace over the representation matrices $T^a$,
		\begin{equation}
		\label{eq:actionDefinitions}
			\tr (T^a\, T^b) = T_\text{R}\, \delta^{ab}.
		\end{equation}
		In our case we have $T_\text{R}=1/2$. 
		
		Evaluating the DBI action given in~\eqref{eq:dbiActionIso} with the
		substitution rule~\eqref{eq:backgroundRule}, we arrive at
		\begin{equation}
		\begin{aligned}
		\label{eq:dbiActionWithMu}
			S_{{\mathrm D7}} =\;& -T_{7}T_\text{R}\,(2 \pi^2 \alpha')^2 R^3 \int\limits_{u=0}^{u=1} \dd u\,
			  \dd^4 x\; \sqrt{-G}G^{\mu\sigma}G^{\nu\beta}\\
					& \times\left(
			  4 \del_{[\mu}^{\vphantom{a}} A^a_{\nu]}\,
			  \partial_{[\sigma}^{\vphantom{a}} A^a_{\beta]}  -
			  8 \delta_{0\nu} \delta_{0\beta} f^{abc} \partial_{[0}^{\vphantom{a}} A^a_{\mu]}\,
			  A^b_{\sigma}\, \mu^c  \right),
		\end{aligned}
		\end{equation}
		where we use the short-hand notation $\mu^c=\mu \delta^{3c}$ and neglect
		terms of higher than linear order in $\mu$, and higher than quadratic
		order in $A$ since both are small in our approach.
		
		Up to the sum over flavor indices $a$, the first term in the bracket in
		\eqref{eq:dbiActionWithMu} is reminiscent of the Abelian super-Maxwell
		action in five dimensions, considered already for the R-charge current
		correlators in ref.~\cite{Policastro:2002se}. The new second term in
		our action arises from the non-Abelian nature of the gauge group, giving
		terms proportional to the gauge group's structure constants $f^{abc}$ in
		the field strength tensor
		$F^a_{\mu\nu}=2\partial_{[\mu}^{\vphantom{a}}A^a_{\nu]}+f^{abc}
		A^b_\mu\, A^c_{\nu}$.
		
		We proceed by calculating the retarded Green functions for the action
		\eqref{eq:dbiActionWithMu}, following the prescription of
		ref.~\cite{Son:2002sd} as outlined in section
		\ref{sec:thermalGreenFunctionsIntro}. According to this prescription, as
		a first step we consider the equations of motion obtained from the
		action~\eqref{eq:dbiActionWithMu}, which are given by
		\begin{equation}
		\begin{aligned}
		\label{eq:eom}
			0 =\; & 2 \partial_{\mu}\left(
			  \sqrt{-G}\, G^{\mu\sigma} G^{\nu\beta}\: \partial^{\vphantom{a}}_{[\sigma} A^a_{\beta]}
			  \right)  \\
			  &+ f^{abc}\Big[ \sqrt{-G} G^{00} G^{\nu\beta}\:\mu^c
			   \left(\partial^{\vphantom{b}}_{\beta} A_{0}^b-2\partial^{\vphantom{b}}_0 A_{\beta}^b\right)
			  +\delta^{0\,\nu} \partial^{\vphantom{b}}_\mu\left(
			  \sqrt{-G}\, G^{00} G^{\mu\sigma} A_{\sigma}^b \mu^c \right)\Big].
		\end{aligned}
		\end{equation}
		It is useful to work in momentum space from now on. We therefore expand
		the bulk gauge fields in Fourier modes in the $x^i$ directions,
		\begin{equation}
		\label{eq:fourierTrafo}
		A_\mu (u,\vec{x})=\int \!\!\frac{\dd^4 k}{(2\pi)^4}\;
		  \ee^{-\ii \omega x_0 + \ii \threevec{k}\cdot \threevec{x}} A_\mu(u,\fourvec{k}).
		\end{equation}
		As we work in the gauge where $A_u=0$, we only have to take care of the
		components $A_i$ with $i=0,1,2,3$.
		
		For the sake of simplicity, we choose the momentum of the fluctuations
		to be along the $x^3$ direction, so their momentum four-vector is
		$\fourvec{k}=(\omega,0,0,q)$. With this choice we have specified to
		gauge fields which only depend on the radial coordinate $u$, the time
		coordinate $x^0$ and the spatial $x^3$ direction.

		\subsubsection*{Equations for $A_1^a$- and $A_2^a$-components}
		
			Choosing the free Lorentz index in the equations of
			motion~\eqref{eq:eom} to be $\nu=\alpha=1,2$ gives two identical
			differential equations for $A_1$ and $A_2$,
			\begin{equation}
			\label{eq:eomA1A2}
				0= {A^a_\alpha}''+\frac{f'}{f}{A^a_\alpha}'+
				   \frac{\wn^2-f \qn^2 }{u f^2}A^a_\alpha+
				   2\ii\frac{\wn}{u f^2} f^{abc} \frac{\mu^b}{2\pi T} A^c_\alpha,
			\end{equation}
			where we indicated the derivative with respect to $u$ with a prime and have
			introduced the dimensionless quantities
			\begin{equation}
			\label{eq:definitionWQM}
				\wn=\frac{\omega}{2 \pi T} \, ,
				\qquad \qn=\frac{q}{2 \pi T} \, ,\qquad \mn=\frac{\mu}{2 \pi T} \, .
			\end{equation}
			We now make use of the structure constants of $\SU(2)$, which are
			$f^{abc}=\varepsilon^{abc}$, where $\varepsilon^{abc}$ is the
			totally antisymmetric epsilon symbol with $\varepsilon^{123}=1$.
			Writing out \eqref{eq:eomA1A2} for the three different choices of
			$a=1,2,3$ results in
			\begin{align}
			\label{eq:eomA1A2Flavor1}
				0&= {A_\alpha^1}''+\frac{f'}{f} {A_\alpha^1}'
				  +\frac{\wn^2 - f \qn^2}{u f^2} A_\alpha^1-
				  2\ii\frac{\mn\wn}{u f^2} A_\alpha^2 \, ,    \\
			\label{eq:eomA1A2Flavor2}
				0&= {A_\alpha^2}''+\frac{f'}{f} {A_\alpha^2}'
				  +\frac{\wn^2 - f \qn^2}{u f^2} A_\alpha^2+
				  2\ii\frac{\mn\wn}{u f^2} A_\alpha^1 \, ,    \\
			\label{eq:eomA1A2Flavor3}
				0&= {A_\alpha^3}''+\frac{f'}{f} {A_\alpha^3}'
				  +\frac{\wn^2 - f \qn^2}{u f^2}  A_\alpha^3 \, .
			\end{align}
			The first two of these equations for the gauge field directions
			transverse to the background field are coupled, the third
			longitudinal one is the same equation that was solved in the Abelian
			Super-Maxwell case~\cite{Policastro:2002se}. Note that these
			equations are influenced by one of the addressed oversimplifications
			in a way that will turn out to be crucial at the end. We neglect
			terms quadratic in $\mn$. If we would allow for these quadratic
			terms we could complete the square and avoid the non-analytical
			behavior we observe later \cite{Kaminski:2008ai}.
		
		\subsubsection*{Equations for $A_0^a$- and $A_3^a$-components}
		
			The remaining choices for the free Lorentz index $\nu=0,3,u$
			in~\eqref{eq:eom} result in three equations which are also not
			independent. The choices $\nu=0$ and $\nu=u$ give
			\begin{align}
			\label{eq:eomA0A3nuIs0}
				0&= {A_0^a}''-\frac{\qn^2}{u f}A_0^a-\frac{\qn\wn}{u f}A_3^a
				  - \ii \frac{\qn}{u f} f^{abc} \frac{\mu^b}{2 \pi T}A_3^c \, ,  \\
			\label{eq:eomA0A3nuIsu}
				0&= \wn {A_0^a}'+\qn f {A_3^a}' + \ii f^{abc} \frac{\mu^b}{2\pi T}
				{A_0^c}' \, .
			\end{align}
			Solving~\eqref{eq:eomA0A3nuIsu} for ${A_0^a}'$, differentiating it
			once with respect to $u$ and using \eqref{eq:eomA0A3nuIs0} results
			in equation \eqref{eq:eom} for $\nu=3$,
			\begin{align}
			\label{eq:eomA0A3nuIs3}
				0 = \;&{A_3^a}''+\frac{f'}{f}{A_3^a}'+\frac{\wn^2}{u f^2}\, A_3^a
				+
				  \frac{\qn\wn}{u f^2}A_0^a
				  +  \ii \frac{\qn}{u f^2} f^{abc} \frac{\mu^b}{2\pi T}\, A_0^c
				  + 2\ii \frac{\wn}{u f^2} f^{abc} \frac{\mu^b}{2\pi T}\, A_3^c \, .
			\end{align}
			We will make use of the equations \eqref{eq:eomA0A3nuIs0} and
			\eqref{eq:eomA0A3nuIsu} which look more concise. These equations of
			motion for $A_0^a$ and $A_3^a$ are coupled in Lorentz and flavor
			indices. To decouple them with respect to the Lorentz structure, we
			solve \eqref{eq:eomA0A3nuIsu} for ${A_3^a}'$ and insert the result
			into the differentiated version of \eqref{eq:eomA0A3nuIs0}. This
			gives
			\begin{equation}
			\label{eq:eomA0}
					0 = {A^a_0}''' + \frac{(uf)'}{uf}\,{A^a_0}'' + \frac{\wn^2-f\qn^2}{uf^2}\,{A^a_0}'
						 + 2\ii\frac{\wn}{uf^2}\,f^{abc} \frac{\mu^b}{2 \pi T}\, {A^c_0}'.
			\end{equation}
			
			The equations for $a=1,2$ are still coupled with respect to their
			gauge structure. The case $a=3$ was solved in
			ref.~\cite{Policastro:2002se}. We will solve \eqref{eq:eomA0} for
			${A^a_0}'$ and can obtain ${A^a_3}'$ from \eqref{eq:eomA0A3nuIsu}.
			Note that it is sufficient for our purpose to obtain solutions for
			the \emph{derivatives} of the fields. These contribute to equations
			\eqref{eq:separateBoundaryBehaviour} to
			\eqref{eq:correlatorFromBoundaryContribution} that give the retarded
			thermal Green functions, while the functions $A=f(u,\vec{k})
			A^{\text{bdy}}(\vec{k})$ themselves simply contribute a factor of
			$f(u,-\vec{k})$ which merely gives a factor of unity at the
			boundary.

		\subsubsection*{Solutions}
		\label{sec:solutions}
			
			Generally, we follow the methods developed in
			ref.~\cite{Policastro:2002se}, and our differential equations are very
			similar to the ones considered there. Additionally, we need to
			respect the flavor structure of the gauge fields, only the equations
			for flavor index $a=3$ resemble those analyzed in
			ref.~\cite{Policastro:2002se}. Those for $a=1,2$ involve extra terms,
			which couple the equations. Coupling occurs not only via their
			Lorentz indices, but also with respect to the flavor indices. We
			already decoupled the Lorentz structure in the previous section. As
			typical for an explicitly broken $\SU(2)$, the equations of motion
			which involve different gauge components will decouple if we
			transform to the variables
			\begin{equation}
			\label{eq:flavorTrafoIso}
			\begin{aligned}
				X_i            & = A_i^1 + \ii A_i^2,\\
				\widetilde X_i & = A_i^1 - \ii A_i^2.
			\end{aligned}
			\end{equation}
			Here, the $A_i^1$, $A_i^2$ are the generally complex gauge field
			components in momentum space. Note that up to $\SU(2)$
			transformations, the combinations \eqref{eq:flavorTrafo} are the
			only ones which decouple the equations of motion for $a=1,2$. These
			combinations are reminiscent of the non-Abelian $\SU(2)$ gauge field
			in position space,
			\begin{equation}
			\label{eq:gaugeFieldMatrix}
				A_i= A^a_i \frac{\sigma^a}{2} = \frac{1}{2}
				\begin{pmatrix}
					A^3_i & A^1_i - \ii A^2_i \\
					A^1_i + \ii A^2_i  & - A^3_i \\
				\end{pmatrix}.
			\end{equation}
			The equations of motion for the flavor index $a=3$ were solved in
			ref.~\cite{Policastro:2002se}. To solve the equations of motion for the
			fields $A^a_i$ with $a=1,2$, we rewrite them in terms of $X_i$ and
			$\widetilde X_i$. Applying the transformation~\eqref{eq:flavorTrafo}
			to the equations of motion \eqref{eq:eomA1A2Flavor1} and
			\eqref{eq:eomA1A2Flavor2} and the $a=1,2$ versions of
			\eqref{eq:eomA0} and \eqref{eq:eomA0A3nuIsu} leads to
			\begin{align}
			\label{eq:eomXalpha}
				0 & =  X_\alpha'' +\frac{f'}{f} X_\alpha'
						+\frac{\wn^2-f \qn^2 \mp 2 \mn\wn}{u f^2} X_\alpha,\\
			\label{eq:eomX0}
				0 & =  X_0''' + \frac{(uf)'}{uf}\,X_0'' 
							+ \frac{\wn^2-f\qn^2\mp 2\mn\wn}{uf^2} X_0',\\
			\label{eq:eomX3}
				0 & =  \left(\wn \mp \mn\right) X_0' + \qn f X_3',
			\end{align}
			where again $\alpha=1,2,$ and the upper signs correspond to $X$ and
			the lower ones to $\widetilde X$.
			
			As in the section on spectral functions, we observe that some
			coefficients of these functions are divergent at the horizon $u=1$.
			We hark back to the ansatz
			\begin{equation}
			\label{eq:ansatzA1A2}
					X_i=(1-u)^\beta\, F(u), \qquad 
					\widetilde X_i=(1-u)^{\widetilde \beta}\, \widetilde F(u),
			\end{equation}
			with regular functions $F(u)$ and $\widetilde F(u)$. To cancel the
			singular behavior of the coefficients, we have to find the adequate
			$\beta$ and $\widetilde \beta$, the so-called indices, given by
			equations known as the indicial equations for $\beta$ and
			$\widetilde \beta$. We eventually get for all $X_i$ and $\widetilde
			X_i$
			\begin{equation}
			\label{eq:indicesIso}
				\beta=\pm\frac{\ii\wn}{2}\sqrt{1-\frac{2\mn}{\wn}}, \qquad
				\widetilde \beta=\pm\frac{\ii\wn}{2}\sqrt{1+\frac{2\mn}{\wn}}.
			\end{equation}
			
			Note that these exponents differ from those of the Abelian
			Super-Maxwell theory~\cite{Policastro:2002se} by a dependence on
			$\sqrt{\wn}$ in the limit of small frequencies~($\wn<\mn$). In the
			limit of vanishing chemical potential~$\mn\to 0$, the indices given
			in~ref.~\cite{Policastro:2002se} are reproduced from~\eqref{eq:indices}.
			Again, however, if we include the quadratic order in $\mn$ it is
			possible to complete the square to get
			$\beta=\mp\ii/(2\wn\mp\mn)$
			\cite{Kaminski:2008ai}.
			
			In order to solve \eqref{eq:eomXalpha},~\eqref{eq:eomX0}
			and~\eqref{eq:eomX3} analytically, we introduce a series expansion
			ansatz for the function $F$ in the momentum variables~$\wn$
			and~$\qn$. In section~\ref{sec:eomFiniteDensitySpectra} we solved
			the resulting equations up to first order in the radial coordinate
			in order to get initial conditions for the subsequent numerical
			integration. Here we are not interested in the dependence along $u$,
			but in the dependence on the lowest order in $\wn$ and $\qn$ in
			order to extract the dispersion relation that determines the poles
			in the according correlators. In fact, the physical motivation
			behind this expansion is that we aim for thermodynamical quantities
			which are known from statistical mechanics in the hydrodynamic limit
			of small four-momentum $\fourvec{k}$. So the standard choice would
			be
			\begin{equation}
			\label{eq:standardHydroLimit}
				F(u)= F_0 + \wn F_1 + \qn^2 G_1 + \ldots \, .
			\end{equation}
			On the other hand, we realize that our indices will appear linearly
			(and quadratically) in the differential equations' coefficients
			after inserting~\eqref{eq:ansatzA1A2}
			into~\eqref{eq:eomXalpha},~\eqref{eq:eomX0} and~\eqref{eq:eomX3}.
			The square root in~$\beta$ and~$\widetilde \beta$ mixes different
			orders of~$\wn$. In order to sort coefficients in our series ansatz,
			we assume $\wn<\mn$ and keep only the leading $\wn$ contributions to
			$\beta$ and $\widetilde \beta$, such that
			\begin{equation}
			\label{eq:approximateIndices}
				\beta\approx \mp \sqrt{\frac{\wn\mn}{2}},\qquad
				\widetilde \beta\approx \pm i \sqrt{\frac{\wn\mn}{2}}.
			\end{equation}
			This introduces an additional order $\order{\wn^{1/2}}$, which we
			include in our ansatz \eqref{eq:standardHydroLimit}, giving
			\begin{equation}
			\label{eq:hydroLimit}
				F(u)= F_0 + \wn^{1/2} F_{1/2} + \wn F_1 + \qn^2 G_1 + \ldots\, ,
			\end{equation}
			and analogously for the tilded quantities. If we had not included
			$\order{\wn^{1/2}}$ the resulting system would be overdetermined. On
			the other hand this procedure of including non-analytical square
			root terms would be superseded by including the second order terms
			in $\mn$ from the beginning. The results we obtain by using the
			approximations \eqref{eq:approximateIndices} and
			\eqref{eq:hydroLimit} have been checked against the numerical
			solution for exact $\beta$ with exact~$F(u)$. These approximations
			are useful for fluctuations with $\qn,\wn < 1$.
			 
			Note that by dropping the $1$
			in~\eqref{eq:indices} we also drop the Abelian limit.
			
			Consider the indices~\eqref{eq:indicesIso} for positive frequency
			first. In order to meet the incoming wave boundary condition, we
			restrict the solution $\widetilde \beta$ to the negative sign only.
			For the approximate~$\widetilde \beta$
			in~\eqref{eq:approximateIndices} we therefore choose the lower
			(negative) sign. This exponent describes a mode that travels into
			the horizon of the black hole. In case of $\beta$, we demand the mode
			to decay towards the horizon, choosing the lower (positive) sign
			in~\eqref{eq:approximateIndices} consistently. Note that for
			negative frequencies~$\omega <0$ the indices $\beta$ and~$\widetilde
			\beta$ exchange their roles.
			
			Using \eqref{eq:approximateIndices} in \eqref{eq:ansatzA1A2} and
			inserting the ansatz into the equations of motion, we find equations
			for each order in $\qn^2$ and $\wn$ separately. After solving the
			equations of motion for the coefficient functions $F_0$, $F_{1/2}$,
			$F_1$ and $G_1$, we eventually can assemble the solutions to the
			equations of motion for $X$ as defined in \eqref{eq:flavorTrafoIso},
			\begin{equation}
			\begin{aligned}
			\label{eq:fullX}
					X(u) &= (1-u)^\beta \, F(u)\\
						&= (1-u)^\beta \, 
			  \left( F_0 + \sqrt{\wn} F_{1/2} + \wn F_1 + \qn^2 G_1 + \ldots \right).
			  \end{aligned}
			\end{equation}
			and a corresponding formula for $\widetilde X(u)$ from the ansatz
			\eqref{eq:ansatzA1A2}.
			
			\bigskip
			
			Illustrating the method, we now write down the equations of motion
			order by order for the function $X_\alpha$. To do so, we use
			\eqref{eq:fullX} with \eqref{eq:approximateIndices} in
			\eqref{eq:eomXalpha} with the upper sign for $X_\alpha$. Then we
			examine the result order by order in $\wn$ and $\qn^2$,
			\begin{align}
			\label{eq:eomFSortedByOrders1}
				\order{\text{const}}:\qquad
				0&= F_0''+\frac{f'}{f} F_0' \, ,\\
			\order{\sqrt{\wn}}:\qquad
			\label{eq:eomFSortedByOrders2}
				0&= F_{1/2}''+\frac{f'}{f}F_{1/2}' -\frac{\sqrt{2\mn}}{1-u} F_0'
				  -\sqrt{\frac{\mn}{2}}\frac{1}{f} F_0\, ,\\
				\order{\wn}:\qquad
			\label{eq:eomFSortedByOrders3}
				0&= F_1''+\frac{f'}{f} F_1'-\frac{\sqrt{2\mn}}{1-u}F_{1/2}'
				  - \sqrt{\frac{\mn}{2}} \frac{1}{f} F_{1/2}
				   \hphantom{\;=} - \mn\frac{4-u(1+u)^2}{2 u f^2} F_0 \, ,\\
			\label{eq:eomFSortedByOrders4}
				\order{\qn^2}:\qquad
				0&= G_1''+\frac{f'}{f} G_1'- \frac{1}{u f} F_0 \, .
			\end{align}
			
			At this point we observe that the differential equations we have to
			solve for each order are shifted with respect to the solutions found
			in ref.~\cite{Policastro:2002se}. The contributions of order
			$\wn^n$ in ref.~\cite{Policastro:2002se} now show up in order
			$\wn^{n/2}$. Their solutions will exhibit factors of order
			$\mu^{n/2}$. Again, we emphasize that inclusion of $\order{\mn^2}$
			terms would just result in a shift of $\wn\mapsto\wn\pm\mn$.
			
			Solving the system \eqref{eq:eomFSortedByOrders1} to
			\eqref{eq:eomFSortedByOrders4} of coupled differential equations is
			straightforward in the way that they can be reduced to several
			uncoupled first order ordinary differential equations in the
			following way. Note that there obviously is a constant solution
			$F_0=C$ for the first equation. Inserting it into
			\eqref{eq:eomFSortedByOrders2} and \eqref{eq:eomFSortedByOrders4}
			leaves us with ordinary differential equations for $F_{1/2}'$ and
			$G_1'$ respectively. Using the solutions of $F_0$ and $F_{1/2}$ in
			\eqref{eq:eomFSortedByOrders3} gives one more such equation for
			$F_1'$.
			
			To fix the boundary values of the solutions just mentioned, we
			demand the value of $F(u=1)$ to be given by the constant $F_0$ and
			therefore choose the other component functions' solutions such that
			$\lim_{u\to 1}F_{1/2}=0$, and the same for $F_1$ and $G_1$. The
			remaining integration constant $C$ is determined by taking the
			boundary limit $u\to0$ of the explicit solution \eqref{eq:fullX},
			making use of the second boundary condition
			\begin{equation}
				\lim_{u\to 0} X(u) = X^{\text{bdy}},
			\end{equation}
			see appendix~\ref{sec:solutionsEOM}. Eventually, we end up with all
			the ingredients needed to construct the gauge field's fluctuations
			$X(u)$ as in \eqref{eq:fullX}.
			
			We solve the equations \eqref{eq:eomXalpha} with lower sign for
			$\widetilde X_\alpha$ and \eqref{eq:eomX0} for $X_0'$ and its tilded
			partner in exactly the same way as just outlined, only some
			coefficients of these differential equations differ. The solution
			for $X_3'$ is then obtained from \eqref{eq:eomX3}.
			
			All solutions are given explicitly in
			Appendix~\ref{sec:solutionsEOM} together with all other information
			needed to construct the functions $X_\alpha$, $\widetilde X_\alpha$,
			$X_0'$, $\widetilde X_0'$, $X_3'$ and $\widetilde X_3'$.

	\subsubsection{Current correlators}
	\label{sec:correlators}
		
		In this section we obtain the momentum space correlation functions for
		the isospin currents by means of the holographically dual gauge field
		component combinations $X$ and $\widetilde X$ defined in
		equation~\eqref{eq:flavorTrafoIso}. Recall that the imaginary part of the
		retarded correlators essentially gives the thermal spectral functions
		(c.\,f.\  section~\ref{chap:specFuncs}). The following discussion of the
		correlators' properties can therefore be related to the discussion of
		the corresponding spectral functions.
		
		First note that the on-shell action gets contributions from the
		non-Abelian structure,
		\begin{equation}
		\begin{aligned}
		\label{eq:onShellActionOfA}
			S_\text{D7} =\;& -T_7 T_\text{R}\, (2 \pi^2 \alpha')^2 R^3 \\
					& \times 2 \int\!\!
			  \frac{\dd^4 q}{(2 \pi)^4}\: \Bigg[\left.\sqrt{-G}G^{uu}G^{jk}\:
				{A^a_{j}}'(\vec q) \, 
			   A^a_{k}(-\vec q)   
			  \right|_{u=0}^{u=1}
			   -
			  4 \ii q \, f^{abc} \mu^c \int\limits_0^1 \dd u \:
			  \sqrt{-G}G^{00}G^{33}A_{[3}^{a} A^b_{0]}\,
			  \Bigg],
		\end{aligned}
		\end{equation}
		where $j,k=0,1,2,3$, and the index $u$ denotes the radial \ads-direction.
		Up to the sum over flavor indices, the first term in
		the bracket is similar to the Abelian Super-Maxwell action of
		ref.~\cite{Policastro:2002se}. The second term is a new contribution depending
		on the isospin chemical potential. It is a contact term which we will
		neglect. The correlation functions however get a structure that is
		different from the Abelian case. This is due to the appearance of the
		chemical potential in the equations of motion and their solutions.
		Writing~\eqref{eq:onShellActionOfA} as a function of $X$ and $\widetilde
		X$ results in
		\begin{equation}
		\begin{aligned}
		\label{eq:onShellActionOfX}
			S_{{\mathrm D7}} =\;& -T_7T_\text{R}\, (2 \pi^2 \alpha')^2 R^3
			 2 \int \frac{\dd^4 q}{(2 \pi)^4}\\
			&\times \Bigg[ 
			\left. \sqrt{-G}\,G^{uu}G^{jk}\left[
			 \frac{1}{2}\left({X_j}'\widetilde{X}_{k}
			 +{\widetilde{X}_j}'{X_{k}}\right)  +
			   {A^3_j}' A^3_{k}
			\right]\right|_{u=0}^{u=1}  
			-4  q \mu \int\limits_0^1
			 \dd u \sqrt{-G}G^{00}G^{33}
			\:2 X_{[0} \widetilde{X}_{3]}
			\Bigg]  .
		\end{aligned}
		\end{equation}
		
		In order to find the current correlators, we apply the method outlined
		in section~\ref{sec:thermalGreenFunctionsIntro}
		to~\eqref{eq:onShellActionOfX}, with the solutions for the fields given
		in appendix~\ref{sec:solutionsEOM}. As an example, we derive the
		correlators $G_{0\widetilde 0}=\vev{J_0(\fourvec q)
		\widetilde{J}_0(-\fourvec q)}$ and $G_{\widetilde 0 0}= \vev{
		\widetilde{J}_0(\fourvec q) \,{J_0}(-\fourvec q)}$ of the flavor current
		time components $J_0$ and $\widetilde{J}_0$, coupling to the bulk fields
		$X_0$ and $\widetilde{X}_0$, respectively. Correlation functions of all
		other components are derived analogously.

		\subsubsection*{Green functions: Calculation}
			
			First, we extract the prefactor of $(\del A_0)^2$ from the
			action  \eqref{eq:dbiActionIso} and call it $B(u)$,
			\begin{equation}
			\label{eq:bOfU}
			B(u)=-T_7 T_\text{R}\,(2\pi^2\alpha')^2 R^3
				\sqrt{-G}\,G^{uu}\,G^{00}.
			\end{equation}
			We need this factor below to calculate the Green function,
			\begin{equation}
				\label{eq:GreenFunctionRecipeIso}
				\GR=\lim_{u\to 0} B(u) f(u,-\fourvec k) \del_u f(u,\fourvec k).
			\end{equation}
			
			The second step, finding the solutions to the mode equations of
			motion, has already been performed in section~\ref{sec:solutions}.
			In the example at hand we need the solutions~$X_0$ and~$\widetilde
			X_0$. From~\eqref{eq:fullX} and from appendix~\ref{chap:solutionsEOMISO}
			we obtain
			\begin{align}
			\label{eq:derivativesX0}
				X_0{}'=&-(1-u)^{\sqrt{\frac{\wn\mn}{2}}}\,
				 \frac{\qn^2 \widetilde X_0^{\text{bdy}}+\wn\qn \widetilde X_3^{\text{bdy}}}
				  {\sqrt{2\mn\wn}+\wn\mn \ln 2+\qn^2}\\
				  &\times \Bigg[
				   1-\wn^{1/2}\,\sqrt{\frac{\mn}{2}}
				   \ln\left(\frac{2u^2}{u+1}\right)
				   -\wn\frac{\mn}{12}\Bigg(
				   \pi^2+3\ln^2 2+3\ln^2(1+u)+6\ln 2 \ln\left(
					\frac{u^2}{1+u}\right)   \nonumber\\ 
				   & \qquad +12 \mathrm{Li}_2(1-u)+
					12\mathrm{Li}_2(-u)-12\mathrm{Li}_2\left(\frac{1-u}{2}\right)
				   \Bigg)
				   +\qn^2\ln\left(\frac{u+1}{2u}\right)
				  \Bigg] \nonumber , \\  
			\label{eq:derivativesX0t}
			\widetilde X_0{}'=&\,\hphantom{-}(1-u)^{-\ii\sqrt{\frac{\wn\mn}{2}}}\,
			  \frac{\qn^2 X_0^{\text{bdy}}+\wn\qn X_3^{\text{bdy}}}
			  {i\sqrt{2\mn\wn}+\wn\mn \ln 2-\qn^2}\\
			  &\times\Bigg[
			   1+\wn^{1/2}\,i\sqrt{\frac{\mn}{2}}
			   \ln\left(\frac{2u^2}{u+1}\right)
			   +\wn\frac{\mn}{12}\Bigg(
			   \pi^2+3\ln^2 2+3\ln^2(1+u)+6\ln 2 \ln\left(
				\frac{u^2}{1+u}\right)  \nonumber \\   
			   & \qquad +12 \mathrm{Li}_2(1-u)+
				12\mathrm{Li}_2(-u)-12\mathrm{Li}_2\left(\frac{1-u}{2}\right)
			   \Bigg)
			   +\qn^2\ln\left(\frac{u+1}{2u}\right)\Bigg] .\nonumber  
			\end{align}
			Note that we need the derivatives to apply~\eqref{eq:recipeOnShell}
			and \eqref{eq:recipeCorrelator}.
			
			Now we perform the third step and insert \eqref{eq:bOfU},
			\eqref{eq:derivativesX0} and \eqref{eq:derivativesX0t} into
			\eqref{eq:GreenFunctionRecipeIso}. Our solutions $X_0$ and $\widetilde
			X_0$ replace the solution $f(u,\vec k)$ and $f(u,-\vec k)$
			in~\eqref{eq:recipeCorrelator}. The resulting expression is
			evaluated at $u_b=0$, which comes from the lower limit of the
			$u$-integral in the on-shell action \eqref{eq:onShellActionOfX}. At
			small $u=\epsilon\ll1$, \eqref{eq:derivativesX0} and
			\eqref{eq:derivativesX0t} give
			\begin{align}
			\label{eq:smallUderivativesX0}
			\lim\limits_{u\to 0} X_0{}'=& -
			\frac{\qn^2 \widetilde X_0^{\text{bdy}}+\wn\qn \widetilde X_3^{\text{bdy}}}
			  {\sqrt{2\mn\wn}+\wn\mn \ln 2+\qn^2}
			  -\lim\limits_{\epsilon\to 0} \left(
			 \qn^2 \widetilde X_0^{\text{bdy}}+\wn\qn \widetilde X_3^{\text{bdy}}
			\right)\ln \epsilon \, ,  \\
			\label{eq:smallUderivativesX0t}
			\lim\limits_{u\to 0} \widetilde X_0{}'=  
			&\hphantom{-}
			\frac{\qn^2 X_0^{\text{bdy}}+\wn\qn X_3^{\text{bdy}}}
			  {i\sqrt{2\mn\wn}+\wn\mn \ln 2-\qn^2}
			  +\lim\limits_{\epsilon\to 0} \left(
			 \qn^2 X_0^{\text{bdy}}+\wn\qn X_3^{\text{bdy}}\right)\, \ln \epsilon \, .
			\end{align}
			In the next to leading order of~\eqref{eq:smallUderivativesX0}
			and~\eqref{eq:smallUderivativesX0t} there appear singularities, just
			like in the Abelian Super-Maxwell
			calculation~\cite[equation~(5.15)]{Policastro:2002se}. However, in
			the hydrodynamic limit, we consider only the finite leading order.
	
		\subsubsection*{Green functions: Results}
		\label{sec:greenResults}
		
			Putting everything together, for the two Green functions for the
			field components $X_0$, $\widetilde X_0$ given in
			\eqref{eq:flavorTrafoIso} by
			\begin{equation*}
			X_0 = A_0^1 + \ii A_0^2,\qquad
			\widetilde X_0 =  A_0^1 - \ii A_0^2,
			\end{equation*}
			we obtain
			\begin{align}
			\label{eq:GX0X0Traw}
			G_{\widetilde 0 0}=&\,
			  \frac{N_c T }
			  {8\pi} \: \frac{2\pi T\, \qn^2}{
			  i\sqrt{2\mn\wn}-\qn^2 +
			  \wn\mn\, \ln 2 }\, ,\\
			\label{eq:GX0TX0raw}
			G_{0\widetilde 0}=&\,
			  \frac{N_c T }
			  {8\pi}\:\frac{2\pi T\,\qn^2}{
			  -\sqrt{2\mn\wn}-\qn^2 -
			  \wn\mn\,\ln 2}\, .
			\end{align}
			These are the Green functions for the time components in Minkowski
			space, perpendicular to the chemical potential in flavor space. All
			Green functions are obtained considering hydrodynamic approximations
			in $\order{\wn^{1/2},\wn,\qn^2}$, neglecting mixed and higher orders
			$\order{\wn^{3/2},\wn^{1/2}\qn^2,\qn^4}$.
			
			The prefactor in \eqref{eq:GX0X0Traw}, \eqref{eq:GX0TX0raw} is
			obtained using $T_7$ as in \eqref{eq:braneTension}, $T_\text{R}$
			from \eqref{eq:actionDefinitions}, and carefully inserting all
			metric factors, together with the standard \ads/CFT relation
			$R^4=4\pi g_s N{\alpha'}^2$. As in other settings with flavor
			\cite{Mateos:2006yd}, we concordantly get an overall factor of $N$,
			and not $N^2$, for all correlators. Contrary to those approaches, we
			do not get a factor of $N_f$ when summing over the different
			flavors. This is due to the fact that in our setup, the individual
			flavors yield distinct contributions. Most striking is the
			non-trivial dependence on the (dimensionless) chemical potential
			$\mn$ in both correlators. Note also the distinct structures in the
			denominators. The first one, \eqref{eq:GX0X0Traw}, has an explicit
			relative factor of $\ii$ between the terms in the denominator. In
			the second correlator, \eqref{eq:GX0TX0raw}, there is no explicit
			factor of $\ii$. The correlator \eqref{eq:GX0X0Traw} has a complex
			pole structure for $\omega>0$, but is entirely real for $\omega<0$.
			On the other hand, \eqref{eq:GX0TX0raw} is real for $\omega>0$ but
			develops a diffusion structure for $\omega<0$. So the correlators
			$G_{0 \widetilde 0}$ and $G_{\widetilde 0 0}$ essentially exchange
			their roles as $\omega$ changes sign.
			We find a similar
			behavior for all correlators $G_{j \widetilde l}$ and $G_{\widetilde
			j l}$ with $j,l=0,1,2,3$. Once more, this behavior is a consequence
			of the insertion of $\order{\wn^{1/2}}$ and neglecting of terms of
			order $\order{\mn^2}$ in the hydrodynamic expansion
			\eqref{eq:hydroLimit}.
			
			We assume $\mn$ to be small enough in order to neglect the
			denominator term of order
			$\order{\wn\mn}\ll\order{\sqrt{\wn\mn},\qn^2}$. Moreover, using the
			definitions of $\wn,\qn$ and $\mn$ from \eqref{eq:definitionWQM} we
			may write \eqref{eq:GX0X0Traw} and \eqref{eq:GX0TX0raw} as
			\begin{align} 
			\label{eq:GX0X0T}
				G_{0\widetilde 0}=&  
				  \,-\frac{N T }
				  {8\pi\sqrt{2\mu}}\:\frac{q^2 \sqrt{\omega}}{
				  \omega+q^2 D(\omega)
				  }\, ,\\   
			\label{eq:GX0TX0} 
				G_{\widetilde 0 0}=&   
				  \,\hphantom{-}
				  \frac{N T }
				  {8\pi\sqrt{2\mu}}\:\frac{q^2 \sqrt{\omega}}{
				  i\omega-q^2 D(\omega)
				  }\, ,  
			\end{align}
			where the frequency-dependent diffusion coefficient $D(\omega)$ is
			given by
			\begin{equation} 
			\label{eq:diffusionCoeff}
				D(\omega)=\sqrt{\frac{\omega}{2\mu}}\:\frac{1}{2\pi T}\, .
			\end{equation}
			We observe that this coefficient also depends on the
			inverse square root of the chemical potential $\mu$. Its physical
			interpretation is discussed below in section~\ref{sec:diffusionIso}.
			
			In the same way we derive the other correlation functions
			\begin{align}
			\label{eq:GX3X3}
				G_{3\widetilde 3}=&
				  -\frac{N T}
				  {8\pi\sqrt{2\mu}}
				  \:\frac{\omega^{3/2}\,(\omega-\mu)}{\widetilde{Q}(\omega,q)} \:,&
				G_{\widetilde 3 3}=&
				  \frac{N T}
				  {8\pi\sqrt{2\mu}}
				  \:\frac{\omega^{3/2}\, (\omega+\mu)}{Q(\omega,q)}\:,& \\
			\label{eq:GX0X3}
				G_{0\widetilde 3}=&
				  -\frac{N T}
				  {8\pi\sqrt{2\mu}}
				  \:\frac{\sqrt{\omega}\,q(\omega  -\mu )}{\widetilde{Q}(\omega,q)}\:,&
				G_{\widetilde 0 3}=&
				  \frac{N T}
				  {8\pi\sqrt{2\mu}}
				  \:\frac{\sqrt{\omega}\,q(\omega +\mu )}{Q(\omega,q)}\:, & \\
			\label{eq:GX3X0}
				G_{3\widetilde 0}=&
				  -\frac{N T}
				  {8\pi\sqrt{2\mu}}
				  \:\frac{\omega^{3/2}\,q}{\widetilde{Q}(\omega,q)}\: ,  &
				G_{\widetilde 3 0}=&
				  \frac{N T}
				  {8\pi\sqrt{2\mu}}
				  \:\frac{\omega^{3/2}\,q}{Q(\omega,q)} \:.&
			\end{align}
			with the short-hand notation 
			\begin{equation}
			\label{eq:denominatorQs}
				Q(\omega,q)= \ii\omega-q^2 D(\omega), \qquad
			\widetilde Q(\omega,q) = \omega+q^2 D(\omega) \, .
			\end{equation}
			Note that most of these functions are proportional to powers of $q$
			and therefore vanish in the limit of vanishing spatial momentum
			$q\to 0$. Only the \mbox{$33$-combinations} from \eqref{eq:GX3X3}
			survive this limit. In contrast to the Abelian Super-Maxwell
			correlators from ref.~\cite{Policastro:2002se} given in appendix
			\ref{sec:abelianCorrelators}, it stands out that our results
			\eqref{eq:GX0X0T}, \eqref{eq:GX0TX0} and \eqref{eq:GX3X3} and
			\eqref{eq:GX3X0} have a new zero at $\omega =\pm\mu$. Nevertheless,
			bear in mind that we took the limit $\omega < \mu$ in order to
			obtain our solutions. Therefore the apparent zeros at $\pm \mu$ lie
			outside of the range considered. Compared to the Abelian case there
			is an additional factor of $\sqrt{\omega}$. The dependence on
			temperature remains linear.
			
			In the remaining $X$-correlators we do not find any pole structure
			to order $\sqrt{\omega}$, subtracting an $\order{q^2}$
			contribution as in ref.~\cite{Policastro:2002se},
			\begin{align}
			\label{eq:GXalphaXalphaT}
				G_{1 \widetilde 1}=& G_{2 \widetilde 2}=
				\frac{\sqrt 2 N_c T}{8\pi}\sqrt{\mu\omega}\, , \\
			\label{eq:GXalphaTXalpha}
				G_{\widetilde 1 1}=& G_{\widetilde 2 2}=
				-\frac{\ii\sqrt 2 N_c T}{8\pi}\sqrt{\mu\omega}\, .
			\end{align}
			We can see that the $G_{\alpha \widetilde \alpha}$ (with
			$\alpha=1,2$) are purely imaginary for negative $\omega$ and real
			for positive $\omega$. The opposite is true for $G_{\widetilde
			\alpha \alpha}$, as is obvious from the relative factor of $\ii$.
			  
			The correlators of components, pointing along the isospin potential
			in flavor space ($a=3$), are found to be
			\begin{equation}
			\label{eq:GA03A03}
			\begin{aligned} 
				G_{A_0^3A_0^3} &=  
				  \frac{N_c T}{4 \pi}\:\frac{q^2}{\ii\omega-D_0 q^2},\\
				G_{A_0^3A_3^3} &=G_{A_3^3A_0^3} =
				  \frac{N_c T}{4 \pi}\:\frac{\omega q}{\ii\omega-D_0 q^2}, 
			\end{aligned}
			\end{equation}
			\begin{equation}
			\label{eq:GAalpha3Aalpha3}
			\begin{aligned}
				G_{A_1^3A_1^3} &=G_{A_2^3A_2^3}=-\frac{N_c T\,\ii\omega}{4 \pi},\\
				G_{A_3^3A_3^3} &=\frac{N_c T}{4 \pi}\:\frac{\omega^2}{i\omega-D_0 q^2},
			\end{aligned}
			\end{equation}
			with the diffusion constant $D_0=1/(2\pi T)$. Note that these
			correlators have the same structure but differ by a factor $4/N$
			from those found in the Abelian super-Maxwell
			case~\cite{Policastro:2002se}~(see also
			\eqref{eq:abelianCorrelators1} and \eqref{eq:abelianCorrelators2}).
			In particular the correlators in equation \eqref{eq:GA03A03} do not
			depend on the chemical potential.
			
			To analyze the novel structures appearing in the other correlators,
			we explore their real and imaginary parts as well as the
			interrelations among them,
			\begin{alignat}{3}
			  \label{eq:reGX0X0t>=}
			 \Re G_{0 \widetilde 0}(\omega\ge 0) & = 
			 \hphantom{-}\Re G_{\widetilde 0 0}(\omega < 0)  && = 
				  -\frac{N_c T}{8\pi}
				  \:\frac{q^2}{
				  \sqrt{2\mu\left|\omega\right|}+q^2/(2\pi T) }\, ,\\[\medskipamount]
			\label{eq:reGX0X0t<}
			  \Re G_{0 \widetilde 0}(\omega < 0) & = 
			  \hphantom{-}\Re G_{\widetilde 0 0}(\omega \ge 0) && = 
				  -\frac{N_c T}{16\pi^2} \frac{q^4}{
				  2\mu\left|\omega\right|+q^4/(2\pi T)^2 }\,, \\[\medskipamount]
			\label{eq:imGX0X0t>=}
			   \Im G_{0 \widetilde 0}(\omega< 0) & = 
			  -\Im G_{\widetilde 0 0}(\omega \ge 0) && = 
				  \frac{N_c T}{8\pi}\:\frac{q^2 \sqrt{2\mu\left|\omega\right|}}
				   {
				   2\mu\left|\omega\right|+q^4/(2\pi T)^2 }\,,  \\[\medskipamount]
			\label{eq:imGX0X0t<}
			  \Im G_{0 \widetilde 0}(\omega \ge 0) &= 
			  \hphantom{-}\Im G_{\widetilde 0 0}(\omega  <  0) &&= 0.
			\end{alignat}
			Now we see why, as discussed below \eqref{eq:GX0TX0}, $G_{0
			\widetilde 0}$ and $G_{\widetilde 0 0}$ exchange their roles when
			crossing the origin at $\omega=0$. This is due to the fact that the
			real parts of all $G_{j\widetilde l}$ and $G_{\widetilde j l}$ are
			mirror images of each other by reflection about the vertical axis at
			$\omega=0$. In contrast, the imaginary parts are inverted into each
			other at the origin.
			The real part shows a deformed resonance behavior. The imaginary
			part has a deformed interference shape with vanishing value for
			negative frequencies. All curves are continuous and finite at
			$\omega = 0$. However, due to the square root dependence, they are
			not differentiable at the origin. Parts of the correlator which are
			real for positive $\omega$ are shifted into the imaginary part by
			the change of sign when crossing $\omega=0$, and vice versa.
			
			To obtain physically meaningful correlators, we follow a procedure
			which generalizes the Abelian approach of ref.~\cite{Kovtun:2005ev}. In
			the Abelian case, gauge-invariant components of the field strength
			tensor, such as $E_\alpha = \omega A_\alpha$, are considered as
			physical variables. This procedure cannot be transferred directly to
			the non-Abelian case. Instead, we consider the non-local part of the
			gauge invariant $\tr F^2$ which contributes to the on-shell action
			\eqref{eq:onShellActionOfA}. In this action, the contribution
			involving the non-Abelian structure constant\,---\,as well as
			$\mu$\,---\,is a local contact term. The non-local contribution
			however generates the Green function combination
			\begin{equation}
			\label{eq:sumOfGA0A0}
				G_{A_i^1 A_j^1}+G_{A_i^2 A_j^2}+G_{A^3_i A^3_j}\, .
			\end{equation}
			We take this sum as our physical Green function. This choice is
			supported further by the fact that it may be written in terms of the
			linear combinations \eqref{eq:flavorTrafo} which decouple the
			equations of motion. For example, for the time component, written in
			the variables $X_0,\,\widetilde X_0$ given by
			\eqref{eq:flavorTrafo}, the combination \eqref{eq:sumOfGA0A0} reads
			(compare to \eqref{eq:onShellActionOfX})
			\begin{equation}
			\label{eq:sumOfGX0X0} 
				G_{0\widetilde 0}+G_{\widetilde 0 0}+G_{A^3_0 A^3_0}\,  .
			\end{equation}   
			The contribution from $G_{A^3_0 A^3_0}$ is of order $\order{\mu^0}$,
			while the combination for the first two flavor directions,
			$G_{0\widetilde 0}+G_{\widetilde 0 0}$, is of order $\order{\mu}$.
			
			We proceed by discussing the physical behavior of the Green function
			combinations introduced above.
			Their frequency dependence is of the same form as in the Abelian
			correlator obtained in ref.~\cite{Policastro:2002se}, as can be seen from
			\eqref{eq:abelianCorrelators1}. Since we are interested in effects
			of order $\order{\mu}$, we drop the third flavor direction $a=3$
			from the sum \eqref{eq:sumOfGX0X0} in the following. It is
			reassuring to observe that the flavor directions $a=1,\,2$, which
			are orthogonal to the chemical potential, combine to give a
			correlator spectrum qualitatively similar to the one found in
			ref.~\cite{Policastro:2002se} for the Abelian Super-Maxwell action.
			However, we discover intriguing new effects such as the highly
			increased steepness of the curves near the origin due to the square
			root dependence and a kink at the origin\,---\,which have to be seen
			with skepticism because they vanish upon reinstating terms of order
			$\order{\mn^2}$.
			
			We observe a narrowing of the inverse resonance peak compared to the
			form found for the Abelian Super-Maxwell action (and also compared
			to the form of our $G_{A_0^3A_0^3}$.
			At the origin, the real and imaginary part are
			finite and continuous, but they are not continuously differentiable.
			However, the imaginary part of $G_{A_0^3A_0^3}$ has finite slope
			at the origin. The real part though has vanishing derivative at
			$\omega=0$.
			
			The correlators $G_{3 \widetilde 3}$, $G_{\widetilde 3 3}$, $G_{0
			\widetilde 3}$ and $G_{\widetilde 0 3}$ have the same interrelations
			between their respective real and imaginary parts as $G_{0
			\widetilde 0}$ and $G_{\widetilde 0 0}$. Nevertheless, their
			dependence on the frequency and momentum is different, as can be
			seen from \eqref{eq:GX3X3} to \eqref{eq:GX3X0}. A list of the
			$33$-direction Green functions split into real and imaginary parts
			can be found in appendix~\ref{sec:correlationFunctions}.
			
	\subsubsection{Isospin diffusion coefficient}
	\label{sec:diffusionIso}
		
		The attenuated poles in hydrodynamic correlation functions have specific
		meanings (for exemplary discussions of this in the context of \ads/CFT
		see e.\,g.\  refs.~\cite{Kovtun:2006pf,Policastro:2002tn}). In our case we observe
		an attenuated pole in the sum $G_{0\widetilde 0}+G_{\widetilde 0 0}$ at
		$\omega = 0$.
		The pole lies at
		$\Re\omega=0$. This structure appears in hydrodynamics as the signature
		of a diffusion pole located at purely imaginary $\omega$. Its location
		on the imaginary $\omega$-axis is given by the zeros of the denominators
		of our correlators as (neglecting $\order{\omega,q^4}$)
		\begin{equation} 
		\label{eq:poleLocation12}
			\sqrt{\omega}=-\ii\, \frac{ q^2}{2\pi T \sqrt{2\mu}}.     
		\end{equation}   
		Squaring both sides of \eqref{eq:poleLocation12} we see that this effect
		is of order $\order{q^4}$. On the other hand, looking for poles in the
		correlator involving the third flavor direction $G_{A_0^3 A_0^3}$, we
		obtain dominant contributions of order $\order{q^2}$ and $\order{\mu^0}$
		(neglecting
		$\order{\omega^2,q^2}$)
		\begin{equation}
		\label{eq:poleLocation3}
			\omega=-\ii \frac{ q^2}{2\pi T} \,  .
		\end{equation}
		This diffusion pole is reminiscent of the result of the Abelian result
		of ref.~\cite{Policastro:2002se} given in
		appendix~\ref{sec:abelianCorrelators}. As discussed in
		section~\ref{sec:correlators}, we consider gauge invariant combinations
		$G_{0\widetilde 0}+G_{\widetilde 0 0}+G_{A_0^3 A_0^3}$. In order to
		inspect the non-Abelian effects of order $\order{\mu}$ showing up in the
		first two correlators in this sum, we again drop the third flavor
		direction which is of order $\order{\mu^0}$.
		
		Motivated by the diffusion pole behavior of our correlators in
		flavor-directions $a=1,2$ corresponding to the combinations $X,
		\widetilde X$ (see \eqref{eq:poleLocation12}), we wish to regain the
		structure of the diffusion equation given in \eqref{eq:omegaD}, which in
		our coordinates~($k=(\omega,0,0,q)$) reads
		\begin{equation}
		\label{eq:diffusionEquationFT}
			\ii \omega\, J_0=D(\omega)\, q^2 J_0 .
		\end{equation}
		Our goal is to rewrite \eqref{eq:poleLocation12} such that a term of
		$\order{\omega}$ and one term of order $\order{q^2}$ appears.
		Furthermore there should be a relative factor of $-\ii$ between these
		two terms. The obvious manipulation to meet these requirements is to
		multiply \eqref{eq:poleLocation12} by $\sqrt{\omega}$ in order to get
		\begin{equation}
		\label{eq:poleLocation12timesSqrtW}
			\omega=-\ii q^2  \frac{\sqrt{\omega}}{2\pi T \sqrt{2\mu}} \,  .
		\end{equation}
		
		Comparing the gravity result \eqref{eq:poleLocation12timesSqrtW} with
		the hydrodynamic equation \eqref{eq:diffusionEquationFT}, we obtain the
		frequency-dependent diffusion coefficient
		\begin{equation}
		\label{eq:diffusionCoefficient2}
			D(\omega)= \sqrt{\frac{\omega}{2\mu}} \frac{1}{2\pi T}\,.
		\end{equation}
		Our argument is thus summarized as follows: Given the isospin chemical
		potential as in \eqref{eq:backg}, \eqref{eq:backgroundRule}, $J_0$ from
		\eqref{eq:omegaD} is the isospin charge density
		in \eqref{eq:diffusionEquationFT}. According to
		\eqref{eq:diffusionEquationFT}, the
		coefficient \eqref{eq:diffusionCoefficient2} describes the diffusive
		response of the quark-gluon plasma to a gradient in the isospin charge
		distribution. For this reason we interpret $D(\omega)$ as the isospin
		diffusion coefficient.
		
		Near the pole, the strongly coupled plasma behaves analogously to a
		diffractive medium with anomalous dispersion in optics. In the presence
		of the isospin chemical potential, the propagation of non-Abelian gauge
		fields in the black hole background depends on the square root of the
		frequency. In the dual gauge theory, this corresponds to a
		non-exponential decay of isospin fluctuations with time.
		
		The square root dependence of our diffusion coefficient is valid for
		small frequencies. As long as $\omega/T<1/4$, the square root is larger
		than its argument and at $\omega/T=1/4$, the difference to a linear
		dependence on frequency is maximal. Therefore in the regime of small
		frequencies $\omega/T<1/4$, which is accessible to our approximation,
		diffusion of modes close to $1/4$ is enhanced compared to modes with
		frequencies close to zero.
		
\subsection{Meson diffusion at strong and weak coupling}
\label{sec:quarkoniumDiffusion}
	
	In this section, we consider heavy mesons moving slowly through high
	temperature non-Abelian plasmas. In the context of transport properties of
	the holographic quark-gluon plasma we are mainly interested in the diffusion
	behavior of mesons. The central quantity we discuss here will however not be
	the diffusion coefficient, but its inverse, the momentum broadening
	coefficient $\kappa$, which determines the square of the momentum transfer
	per unit time, as we will see below.
	
	The motivation for considering meson diffusion is twofold. First, future
	experiments at RHIC promise to measure the elliptic flow of $J/\psi$ mesons,
	and it is important to support this experimental program with theoretical
	work. To this end, various groups have studied the thermal properties of
	heavy mesons within the context of the \ads/CFT correspondence
	\cite{Liu:2006nn,Peeters:2006iu,Mateos:2007vn,Ejaz:2007hg}. However, in
	spite of this progress, the transport properties of these mesonic
	excitations are not well understood. Although the kinetics derived in this
	work are not directly applicable to the heavy ion experiments, we believe
	that the results do hold some important information for phenomenology.
	
	The second motivation for this work is theoretical. After the quark drag was
	computed using the correspondence, it was realized that the drag of
	quarkonium is zero in a large $N_c$ limit
	\cite{Liu:2006nn,Peeters:2006iu,Mateos:2007vn}. Since within a thermal
	environment the drag and diffusion of these mesonic states is certainly not
	zero, it remained as a theoretical challenge to compute the kinetics of
	these states using the \ads/CFT setup.
	
	As a central result, we compare the diffusion of mesons at \emph{weak and
	strong coupling}. Using a simple dipole effective Lagrangian which does not
	rely on the value of the coupling, we calculate the in-medium mass shift and
	the drag coefficient of the meson in $\N=4$ Super Yang Mills theory. At weak
	coupling we use perturbative methods, at strong coupling holographic models
	are employed. In the large $N$ limit the mass shift is finite while the drag
	is suppressed by $1/N^2$. We reach the conclusion that relative to
	weak coupling expectations the effect of strong coupling is to reduce the
	momentum diffusion rate and thereby increase the relaxation time, which
	measures the time until the mesons in the plasma equilibrate their momentum
	spectrum to that of the thermal medium.
	
	We also briefly pick up the discussion of in-medium effects on meson
	spectra, subject of section~\ref{chap:specFuncs}. There, the width of mesons
	in hot dense media was holographically determined by extending the analysis
	of meson melting to finite baryon density. In general the meson lifetime
	determined in this way is suppressed by the density of heavy quarks.
	However, we do not address the effects of finite density in this section. We
	are concerned with the thermal effects which capture the rescattering
	between the meson and the surrounding $\N=4$ medium.
	
	We focus on heavy mesons where the binding energy is much
	greater than the temperature. In this tight binding regime, mesons survive
	well above the critical temperature $T_c$ for deconfinement and the meson
	width is sufficiently narrow to speak sensibly about drag and momentum
	diffusion. This behavior was observed for holographic models in
	section~\ref{chap:specFuncs}.
	
	For real charmonium, the binding energy can be estimated from the mass
	splitting $\Delta M_{2s-1s}^{J/\psi} \approx 589 \text{MeV}$ between the $2s$ and
	$1s$ states, and for bottomonium from the $3s$ and $1s$ states with $\Delta
	M_{3s-1s}^{\Upsilon} \approx 895 \text{MeV}$ respectively \cite{Yao:2006px}.
	Therefore it is not really clear that real quarkonia above $T_c \approx
	170$--$190 \text{MeV}$ \cite{Aoki:2006br,Cheng:2006qk} can be modeled as a simple
	dipole which lives long enough to be considered a quasi-particle. Indeed
	weak coupling hot QCD calculations of the spectral function show that over
	the temperature range $\gym^2M$--$\gym M$, the concept of a meson
	quasi-particle slowly transforms from being well defined to being
	increasingly vague
	\cite{Laine:2006ns,Laine:2007gj,Laine:2007qy,Burnier:2007qm,Brambilla:2008cx}.
	There is lattice evidence based on the maximal entropy method (which is not
	without uncertainty) that $J/\psi$ and $\Upsilon$ survive to $1.6\,T_c$ and
	approximately $3\,T_c$ respectively
	\cite{Umeda:2002vr,Asakawa:2003re,Datta:2003ww,Iida:2006mv,Jakovac:2006sf,Aarts:2007pk}.
	However, model potential calculations which fit all the Euclidean lattice
	correlators indicate that the $J/\psi$ and $\Upsilon$ survive only up to
	at most $1.2\,T_c$ and $2.0\,T_c$ respectively
	\cite{Mocsy:2007yj,Mocsy:2007jz}. Clearly, the word ``survive'' in this
	context is qualitative and means that there is a discernible peak in the
	spectral function. Given these facts, our assessment is that the dipole
	approximation might be reasonable for $\Upsilon_{\!1s}$ but poor for
	charmonium states and other bottomonium states.
	
	\smallskip
	
	An overview of this section is as follows. First, in
	section~\ref{sec:perturbative} we review the computation of drag and
	diffusion of heavy quark-antiquark bound states within the setup of
	perturbative QCD. This will outline a two step procedure to determine the
	drag coefficient at strong coupling.
	
	The first step is to determine the in-medium mass shift $\delta M$ (it is
	finite at large $N$ in the quantum field theory), which determines the
	polarizabilities of the meson. As expected from the dipole effective theory,
	the mass shift scales as $T^4/\Lambda_B^3$, with $T$ the temperature and
	$\Lambda_B$ the inverse size of the meson. In the perturbative calculation,
	$\Lambda_B$ is the inverse Bohr radius, while in the \ads/CFT computation the
	meson mass plays this role. In the $\N=4$ field theory the dipole effective
	Lagrangian couples the heavy meson to the stress tensor and the square of
	the field strength, which we denote by the operator $\mathcal O_{F^2}$. In
	\ads/CFT we obtain the mass shifts from the linear response of the meson mass
	by switching on the dual operators. This amounts to consider a black hole
	background or a non-trivial dilaton flow, respectively. For the dilaton flow
	we consider the \D3-\D{(-1)} gravity background of Liu and Tseytlin
	\cite{Liu:1999fc}. This background and the \ads-Schwarzschild background
	allow for an analytic calculation of the meson polarizabilities.
	
	The second step is to compute the force-force correlator on the meson using
	the previously computed polarizabilities. This determines the drag
	coefficient $\eta_{\scriptscriptstyle D}$ and the momentum broadening
	$\kappa$ as reviewed in section~\ref{sec:adsCftCorrelators}. This step
	requires the calculation of two-point functions involving gradients of the
	stress tensor and the field strength squared. Within gauge/gravity duality,
	these are obtained by considering graviton and dilaton propagation through
	the AdS-Schwarzschild black hole background.
	
	Finally, we compare our results to perturbation theory and reach some
	conclusions for the heavy ion collision experiments in
	section~\ref{sec:summaryDiffusion}.
	
	A few passages of this section are adopted from ref.~\cite{Dusling:2008tg} as they
	stand. The phenomenological input and perturbative calculations as well as
	the numerical calculation of the Green functions by holographic methods
	where performed by D. Teaney, K. Dusling and C. Young during our
	collaboration on ref.~\cite{Dusling:2008tg}. The main contribution of the author
	of this work was the \ads/CFT calculation in
	section~\ref{sec:holoGraphicHeavyMesons}, which is described in detail.

	\subsubsection{Effective model for heavy meson diffusion}
	
		We make use of a model for heavy mesons and their interaction with the
		quark-gluon plasma, which was introduced in ref.~\cite{Luke:1992tm}. This
		effective model describes the interaction with the medium by a dipole
		approximation. It relies on the large mass of the meson relative to the
		external momenta of the gauge fields, i.\,e.\  the momentum scale given by
		the temperature of the medium, but does not rely on the smallness of the
		coupling constant. It was used previously to make a good estimate for
		the binding of $J/\psi$ to nuclei \cite{Luke:1992tm}.
		
		Because the model does not rely on the weakness of the coupling
		constant, we can make use of it in both the strong and weak coupling
		regime. At weak coupling we will refer to results from perturbation
		theory, while the results at strong coupling can be calculated from
		holographic duals. Since the exact dual to QCD is not known, we once
		more have to be satisfied with results for $\N=4$ SYM theory. Therefore,
		we have to rephrase the model in terms of supersymmetric fields.

		\subsubsection*{Diffusion in large \boldmath$N$\unboldmath\ QCD } 
		\label{sec:perturbative}

			The heavy meson field $\phi$ describes a scalar meson which has
			a fixed four-velocity $u^{\mu} = (\gamma,\gamma {\threevec v})$. 
			Then the effective Lagrangian for this meson field interacting with the
			gauge fields is \cite{Luke:1992tm}
			\begin{equation}  
			\label{eq:leff}
				 \L_\text{eff} = - \phi^{\dagger} \ii u \cdot\partial \phi  +
				\frac{c_{E}}{N^2} \phi^\dagger \mathcal O_E \phi  + 
				\frac{c_{B}}{N^2} \phi^\dagger \mathcal O_B \phi ,
			\end{equation}  
			where we refer to the last two terms as the
			interaction Lagrangian $\L_\text{int}$, and
			\begin{align}
				\mathcal O_E &= -\frac{1}{2} F^{\mu\sigma a}{F_\sigma}^{\nu a}\: u_\mu u_\nu\,,\\ 
				\mathcal O_B &= -\frac{1}{2} F^{\mu\sigma a}{F_\sigma}^{\nu a}\: u_\mu u_\nu + \frac{1}{4}  F^{\sigma\beta a} {F_{\sigma\beta}}^a\,.   
			\end{align}
			Here, $F$ is the non-Abelian field strength of QCD, with Greek
			letters $\mu,\nu,\ldots$ as Lorentz indices and gauge index $a$, The
			$c_{\scriptscriptstyle E}$ and $c_{\scriptscriptstyle B}$ are matching coefficients (polarizabilities) to be
			determined from the QCD dynamics of the heavy quark-antiquark pair.
			In inserting a factor of $1/N^2$ into the effective
			Lagrangian we have anticipated that the couplings of the heavy meson
			to the field strengths are suppressed by $N^2$ in the large $N$
			limit.
			
			In the rest frame of a heavy quark bound state with $u=(1,\threevec
			0)$ the operators $\mathcal O_E$ and $\mathcal O_B$ simplify to
			\begin{align}
				\mathcal O_E &= \frac{1}{2} \,{\threevec E}^a \cdot {\threevec E}^a\,,\\ 
				\mathcal O_B &= \frac{1}{2} \, {\threevec B}^a \cdot {\threevec B}^a\,,
			\end{align}
			where ${\threevec E}^a$ and ${\threevec B}^a$ are the color electric
			and magnetic fields. If the constituents of the dipole are
			non-relativistic it is expected that the magnetic polarizability
			$c_{B}$ is of order $\mathcal O(\threevec v^2)$ relative to the
			electric polarizability. For heavy quarks, where $c_{\scriptscriptstyle B}$ is neglected,
			and large $N$ these matching coefficients were computed by Peskin
			\cite{Peskin:1979va,Bhanot:1979vb},
			\begin{equation}    
			\label{eqn:peskin}
			  c_{E} = \frac{28\pi}{3\Lambda_B^3}\,, 
			\qquad c_{\scriptscriptstyle B}=0\,.
			\end{equation}
			Here $\Lambda_B\defeq 1/a_0 = (m_q/2) C_F \alpha_s$ is the inverse
			Bohr radius of the mesonic bound state. It is finite at large $N$
			since with $C_F\simeq N/2$ and finite $\lambda$ we have
			$\Lambda_{B}=m_q \lambda /(16\pi)$.
			
			The effective Lagrangian can be used to calculate the in-medium
			mass shift. We will do so in the subsequent by simply consulting
			first order perturbation theory which says that
			\begin{equation}
				\delta M = \vev{H_\text{int}} = -\vev{\L_\text{int}}.
			\end{equation}

		\subsubsection*{Translating the model to $\N=4$ Super Yang-Mills theory}
			
			Our aim is to calculate the heavy meson diffusion coefficient from
			gauge/gravity duality. Subsequent to this subsection we explain the
			Langevin dynamics we use to describe this process, it requires the
			calculation of the two-point correlators as well as of the
			associated polarizabilities $c_{\scriptscriptstyle E}$ and $c_{\scriptscriptstyle B}$. Because we do not now
			the gravity dual to QCD we translate the effective meson model to
			$\N=4$ Super Yang-Mills theory, our standard toy model.
			
			The formalism in $\N=4$ $\SU(N)$ Super Yang-Mills theory is not
			different from the one we introduced in the preceding section. In
			general all operators in $\N=4$ SYM which are scalars under under
			Lorentz transformations and $\SU(4)$ R-charge rotations will couple
			to the meson at some order. The contribution of higher dimensional
			operators is suppressed by powers of the temperature to the inverse
			size of the meson. The lowest dimension operator which could couple
			to the heavy meson field is $\mathcal O_{X^2} = \tr X^i X^i$, where
			$X^i$ denotes the scalar fields of the theory. However, the
			anomalous dimension of this operator is not protected, and the
			prediction of the supergravity description of $\N=4$ SYM is that
			these operators decouple in a strong coupling limit
			\cite{Aharony:1999ti}. The lowest dimension gauge
			invariant local operators which are singlets under $\SU(4)$ and
			which have protected anomalous dimension are the stress tensor
			$\T_{\mu\nu}$ which couples to the graviton, and minus the
			Lagrangian $\mathcal O_{F^2} =-\L_{\N=4}$, which couples to the
			dilaton. (Since we can add a total derivative to the Lagrangian, the
			operator $-\L$ is ambiguous. The precise form of the operator
			coupling to the dilaton is given in ref.~\cite{Klebanov:1997kc}. We
			neglect this ambiguity here.) There also is the operator $\mathcal
			O_{F\hodge F} = \tr F^{\mu\nu}\hodge F_{\mu\nu} + \ldots$, which
			couples to the axion. An interaction involving $\mathcal O_{F\hodge
			F}$ breaks $CP$-symmetry, which is a symmetry of the Lagrangian of
			the $\N=2$ hypermultiplet of the $\N=4$ SYM gauge theory. Thus
			interactions involving $\O_{F\hodge F}$ can be neglected.
			
			Summarizing the preceding discussion, we find that the effective
			Lagrangian describing the interactions of a heavy meson coupling to
			the operators in the field theory is
			\begin{equation}
			\label{eq:LeffN4}
			\L_\text{eff} =\; -\phi^\dagger(t,\threevec x) \ii u\cdot\partial \phi(t,\threevec x)			 
			+\frac{c_{\scriptscriptstyle T}}{N^2}\, \phi^{\dagger} (t,\threevec x) \,\OT\,  \phi(t,\threevec x)  +  \frac{c_{F}}{N^2}\,\phi^\dagger(t,\threevec x) \,\OF\, \phi(t,\threevec x)%
			\,,
			\end{equation}
			which is a linear perturbation of $\N=4$ Super Yang-Mills theory.
			The two composite operators $\OT$ and $\OF$ in the interaction
			Lagrangian $\L_\text{int}$ are
			\begin{align}
				\OT &= \T^{\mu\nu}\:u_{\mu}u_{\nu} \stackrel{\rule{0mm}{5mm}\mbox{\text{\footnotesize$\stackrel{\threevec v=0}{\downarrow}$}}}{=} \T^{00},\\
				\OF &= F^{\mu\nu}F_{\nu\mu}.
			\end{align}
			They account for the interaction of the mesons with the background.
			In gauge/gravity duality the modification of the Lagrangian
			described by $\OT$ is achieved by considering the
			\ads-Schwarzschild black hole background where $\vev\OF=0$. On the
			other hand, a finite $\vev\OF\neq0$ is dual to a non-trivial dilaton
			flow described by Liu and Tseytlin in ref.~\cite{Liu:1999fc}.
			Details follow below.
			
			The polarization coefficients $c_{\scriptscriptstyle T}$ and
			$c_{\scriptscriptstyle F}$ will be determined below from meson mass
			shifts in gauge/gravity duality. This requires breaking some of the
			supersymmetry. We work in the linearized limit of small
			contributions from $\OT$ and $\OF$. This allows to investigate the
			effects of finite temperature and background gauge fields
			separately. Additionally, this justifies the use of first order
			perturbation theory to compute the meson mass shifts in the medium
			as above by setting $\delta M=-\vev{\L_\text{int}}$. For the
			contribution of the energy-momentum tensor, this is achieved by
			switching on the temperature. Then, the mass shift of the meson is
			given by expectation value of the stress tensor. Again we consider
			the rest frame of the mesons,			\begin{equation}  
			\label{eq:mass_finiteT}
				 \delta M =  -\frac{c_{\scriptscriptstyle T}}{N^2} \vev{\T^{00}},
			\end{equation}
			In contrast, for the meson response to $\vev\OF$ the mass shift of a
			heavy meson is given by
			\begin{equation}  
			\label{eq:mass_finiteF}
				 \delta M =  -\frac{ c_{F}}{N^2} \vev\OF.
			\end{equation}
			
		\subsubsection*{Langevin dynamics}       

			We now turn to the kinetics of the slow moving heavy meson with mass
			$M$ in the medium.
			The kinetic energy $E_\text{kin}=pv/2$ of the meson can be assumed 
			to be of order of the temperature $T$ of the medium, such that
			$pv\approx T$. With $p=Mv$ we can estimate the velocity and momentum
			to be
			\begin{equation}
				p\approx\sqrt{MT},\qquad v\approx\sqrt{\frac{T}{M}}\,.
			\end{equation}
			
			For time scales which are long compared to medium correlations, we
			expect that the kinetics of the meson can be modeled as
			Brownian motion and can be described by Langevin equations. These
			are valid for times which are long compared to the inverse
			temperature but short compared to the lifetime of the quasi-particle
			state. We model viscous force and random kicks in spatial directions
			$x_i$ by
			\begin{equation}
			\label{eq:newton_langevin}
				\frac{\dd p_i}{\dd t} = \xi_i(t) - \eta_{\scriptscriptstyle D} p_i\, , 
				\qquad
				\vev{\xi_i(t)\:\xi_j(t')} = \kappa\, \delta_{ij}\,\delta(t-t')\,.
			\end{equation}
			Here, $\xi_i$ is a component of the random force $\threevec \xi$
			with second moment $\kappa$ and $\eta_{\scriptscriptstyle D}$ is the
			drag coefficient. The solution for $p_i(t)$ is given by
			\begin{equation}
				p_i(t)=\int\limits_{-\infty}^t\!\!\dd t'\; \ee^{\eta_{\scriptscriptstyle D}(t-t')}\,\xi_i(t'),
			\end{equation}
			supposed that $\eta_{\scriptscriptstyle D}t\gg1$
			\cite{Moore:2004tg}. This allows to relate the drag and fluctuation
			by
			\begin{equation}
				3MT=\vev{p^2}=\int\limits_{-\infty}^0\!\!\dd t_1\dd t_2\; \ee^{\eta_{\scriptscriptstyle D}(t_1+t_2)}\vev{\xi_i(t_2)\,\xi_i(t_2)}=\frac{3\kappa}{2\eta_{\scriptscriptstyle D}}.
			\end{equation}
			This leads to the Einstein relation
			\begin{equation}
			\label{eq:etad}
				\eta_{\scriptscriptstyle D} = \frac{\kappa}{2 MT}\,.
			\end{equation} 
			
			One of the aims of this section is the calculation of the diffusion
			coefficients $\eta_{\scriptscriptstyle D}$ or $\kappa$,
			equivalently. From \eqref{eq:newton_langevin} we can obtain these
			coefficients once we know the microscopical phenomenological force
			\begin{equation}
			\label{eq:microscopicForce}
			   \mathcal{F}_i(t) = \frac{\dd p_i}{\dd t}
			\end{equation}
			acting on the quasiparticle state. We can then compare the response
			of the Langevin process \eqref{eq:newton_langevin} to the
			microscopic theory \eqref{eq:microscopicForce}. Over a time interval
			$\Delta t$ which is long compared to medium correlations but short
			compared to the time scale of equilibration we can neglect the drag,
			which is small for the heavy meson with $\eta{\scriptscriptstyle
			D}\propto1/M$. Since the considered time interval is
			long compared to medium correlations we can however equate the
			stochastic process, the random kicks $\threevec \xi$, to the
			microscopic theory. We average
			\eqref{eq:newton_langevin}
			\begin{equation}
			   \int\limits_{\Delta t} \! \dd t \int \! \dd t' \, \vev{\xi_i(t)\:\xi_j(t')} = \Delta t \:  \kappa \, \delta_{ij}
			   =\int\limits_{\Delta t} \! \dd t  \int \! \dd t' \,\vev{\mathcal F_i(t)\, \mathcal F_j(t')}.
			\end{equation}
			In a rotationally invariant medium we have for $i=j$
			\begin{equation}
			\label{ff}
				\kappa =  \frac{1}{3} \int \!\dd t \, \vev{\mathcal{F}_j(t) \mathcal{F}_j(0)} .
			\end{equation}
			We now identify the force with the negative of the gradient of the
			potential $V$ that we read off from the Lagrangian or our theory,
			i.\,e.\  the interaction Lagrangian $V=-\L_\text{int}$. For the case of
			QCD with only $\mathcal O_E$ switched on we get
			\begin{equation}
				 \threevec{\mathcal F}(t) = \int\! \dd^3 x\; 
				 \phi^\dagger(t,\threevec x) \: \frac{c_{\scriptscriptstyle E}}{N^2} \threevec\nabla \mathcal{O}_E(t,\threevec x)\: \phi(t,\threevec x),
			\end{equation}
			which is the usual form of a dipole force averaged over the wave
			function of the meson.
			
			In our case $\kappa$ is a constant in space and time, i.\,e.\  we
			consider situations with constant diffusion parameters in a
			homogeneous medium, for instance slight deviations from equilibrium.
			From the point of view of a more general description in Fourier
			space with $\kappa(\omega)$ we therefore are only interested in the
			hydrodynamic limit of $\omega\to0$. The fluctuation dissipation
			theorem relates the spectrum of \eqref{ff} (with the specified time
			order of operators) to the imaginary part of the retarded
			force-force correlation function
			$\GR\propto\vev{\mathcal{F}_j(t)\,\mathcal{F}_j(0)}$ on the right
			hand side. In the hydrodynamic limit we get
			\begin{equation} 
			\label{ff2}
				\kappa =  -\frac{1}{3}\; \lim_{\omega\to 0}  \frac{2T}{\omega} \Im \GR(\omega),
			\end{equation} 
			where the full form of the retarded correlator is
			\begin{equation}  
			  \GR = -\ii\int\! \dd t \; e^{+\ii\omega t} \,  \theta(t) \vev{\commute{\mathcal{F}_j(t)}{\mathcal{F}_j(0)}}.
			\end{equation}
			Integrating out the heavy meson field as discussed in detail in
			ref.~\cite{CasalderreySolana:2006rq}, which treated the heavy quark case,
			we obtain a formula for the momentum diffusion coefficient
			\begin{equation} 
			\label{e2e2formula}
				  \kappa = \frac{1}{3} \frac{c_{\scriptscriptstyle E}^2}{N^4}
				\int \! \frac{\dd^3q}{(2\pi)^3}\,\threevec q^2\left(-\frac{2 T}{\omega}  \Im G^{R} (\omega,\threevec q) \right)\,,
			\end{equation}  
			with the retarded $\mathcal O_E \mathcal O_E$ correlator given by
			\begin{equation}  
			 G^{R}(\omega,\threevec q) = -\ii\int \! \dd^4x \; e^{+i\omega t - i\threevec q\cdot\threevec x}\, 
			\theta(t)\vev{\commute{\mathcal O_E(t,\threevec 0)}{\mathcal O_E(0,\threevec 0)}} 
			\; .
			\end{equation}  
			
			We can understand this result with simple kinetic theory. Examining
			the Langevin dynamics we see that $3\kappa$ is the mean squared
			momentum transfer to the meson per unit time. The factor of three
			arises from the number of spatial dimensions. In perturbation theory
			this momentum transfer is easily computed by weighting the square of
			the transferred momentum of each scattering with the transition rate
			for any gluon in the bath to scatter with the heavy
			quark,
			\begin{equation} 
			\label{eq:kappaqcd}
			 3\kappa =  \int \! \frac{\dd^3p}{(2\pi)^32E_p}\frac{\dd^3p'}{(2\pi)^3 2E_{p'}} \, \left|\mathcal{M}\right|^2\, n_p (1 + n_{p'}) \, \threevec q^2\, (2\pi)^3 \delta^{3}(\threevec q - \threevec p + \threevec p') 
			\; .
			\end{equation}
			Here, $\threevec p$ is the spatial momentum of the incoming gluon,
			$\threevec p'$ is the momentum of the outgoing gluon and $\threevec
			q$ is the momentum transfer $\threevec q=\threevec p-\threevec p'$,
			and $\left|\mathcal{M}\right|^2$ is the gluon meson scattering
			amplitude computed with the effective Lagrangian in \eqref{eq:leff}
			and weighted by the appropriate momentum distributions $n$ of the
			incoming and outgoing gluons,
			\begin{equation}  
			  \left|\mathcal{M} \right|^2 =  \frac{c_{E}^2}{N^2}  \omega^4 \,\left(1 + \cos^2(\theta_{pp'})\right).
			\end{equation}
			Alternatively (as detailed in appendix~A of ref.~\cite{Dusling:2008tg}),
			we can simply evaluate the imaginary part of the retarded amplitude
			written in \eqref{e2e2formula} to obtain the same result.
			
			In $\N=4$ theory the generalized force is given by
			\begin{equation}
				 \threevec{\mathcal F}(t) = -\int\! \dd^3 x\; 
				 \phi^\dagger(t,\threevec x) \threevec\nabla \left( 
				  \frac{c_{\scriptscriptstyle T}}{N^2}\,\mathcal{O}_T(t,\threevec x) +
				  \frac{c_{\scriptscriptstyle F}}{N^2}\,\OF(t,\threevec x) 
				 \right) \phi(t,\threevec x)
			\end{equation}
			which results in a momentum broadening
			\begin{equation} 
			\label{eq:adsCftCorrelator}
				\kappa =  -\frac{1}{3}\; \lim_{\omega\to 0}\int\!\frac{\dd^3q}{(2\pi)^3}\:\threevec q^2 \frac{2T}{\omega}\left( \frac{c_{\scriptscriptstyle T}^2}{N^4} \Im \GR_T(\omega,\threevec q) + \frac{c_{\scriptscriptstyle F}^2}{N^4} \Im \GR_F(\omega,\threevec q)\right),
			\end{equation} 
			where the retarded correlators at vanishing velocity are 
			\begin{align}
			\label{eq:adsCorrelatorT}
				G^{R}_{\T\T} &= -\ii\int\! \dd^4x\; \ee^{+\ii\omega t - \ii \threevec q\cdot \threevec x}\, \theta(t)\vev{\commute{\T^{00}(t,\threevec x)}{\T^{00}(0,\threevec 0)}}, \\
			\label{eq:adsCorrelatorF}
				G^{R}_{FF} &= -\ii\int\! \dd^4x\; \ee^{+\ii\omega t - \ii \threevec q\cdot \threevec x}\, \theta(t)\vev{\commute{\OF(t,\threevec x)}{\OF(0,\threevec 0)}}.
			\end{align} 
			In writing \eqref{eq:adsCftCorrelator} we have implicitly assumed
			that there is no cross term between $\OF$ and $\OT$. In the
			gauge/gravity duality this is reflected in the fact that at tree
			level in supergravity $\frac{\delta^2S_\text{sugra}}{\delta
			g^{00}(x)\,\delta \Phi (y)}=0$.
			
	\subsubsection{Weak coupling\,---\,perturbative results}
			
			We begin with the results for perturbative QCD (pQCD). The mass
			shift is obtained from first order perturbation theory as $\delta M
			= \vev{H_\text{int}} =-\vev{\L_\text{int}}$, yielding
			\begin{equation}
			\label{eq:qcd_massshift}
			\begin{aligned}
				\delta M_\text{pQCD} &= -\frac{c_{E}}{N^2} \left< \mathcal O_E \right>_T \\
						 &= -T \left(\frac{\pi T}{\Lambda_B}\right)^3\, \frac{14}{45}\, .
			\end{aligned}
			\end{equation}
			In the second line we have calculated the thermal expectation value
			$\vev{\mathcal O_{E}}_{T} = \frac{\pi^2}{30} N^2 T^4$ in a free
			gluon gas and used \eqref{eqn:peskin}.
			
			The importance of this result is that it is finite at large $N$ and
			that it is in general suppressed by $(T/\Lambda_B)^3$, i.\,e.\ 
			by powers of the hadron scale to the temperature. If
			higher dimension operators were added to the effective Lagrangian
			their contributions would be suppressed by additional powers of
			$T/\Lambda_B$.
	
			For QCD the integrals written in \eqref{eq:kappaqcd} are
			straightforward and yield the following result for the rate of
			momentum broadening
			\begin{equation} 
			\label{eq:kappaqcd_final}
			\begin{aligned}
				  \kappa_\text{pQCD} &= \frac{1}{N^2}\, c_{\scriptscriptstyle E}^2\, \frac{64 \pi^5}{135} T^9  \\
						 &= \frac{T^3}{N^2} \left( \frac{\pi T}{\Lambda_B} \right)^6\frac{50176\pi}{1215}.
			\end{aligned}
			\end{equation}
			The high power of temperature $T^9$ arises since the dipole cross
			section rises as $\omega^4$. The matching coefficient $c_{\scriptscriptstyle E}$ is
			directly related to the mass shift of the dipole and the inverse
			Bohr radius by \eqref{eqn:peskin} and \eqref{eq:qcd_massshift}. It
			encodes the coupling of the long distance gluon fields to the
			dipole. By taking the ratio between the momentum broadening and the
			mass shift squared, we find a physical quantity which is independent
			of this coupling
			\begin{equation}  
			  \left. \frac{\kappa}{(\delta M)^2}\right|_\text{pQCD}=  \frac{\pi T}{N^2} \, \frac{1280}{3}.
			\end{equation}  
			The large numerical factor $1280/3$ originates from the cross section 
			which grows  as $\omega^4$.  A similarly large factor  appears in
			$\N=4$ SYM below.
			
			For comparison with the \ads/CFT result we list the results for
			$\N=4$ super Yang-Mills theory in the limit of small 't~Hooft
			coupling $\lambda$, again computed in ref.~\cite{Dusling:2008tg},
			\begin{align}
			\label{eq:N4perturbativeResults}
				\delta M_{\lambda\to0} &= c_{\scriptscriptstyle T} \frac{\pi^2T^4}{2}\,,\\[\smallskipamount]
				\kappa_{\lambda\to0} &= c_{\scriptscriptstyle T}^2 \frac{6232\pi^5}{675}\,\frac{T^9}{N^2}\,,\\[\medskipamount]
				\left. \frac{\kappa}{(\delta M)^2}\right|_{\lambda\to0} &= \frac{\pi T}{N^2}\,36.9\: .
			\end{align}

	\subsubsection{Strong coupling\,---\,holographic calculation}
	\label{sec:holoGraphicHeavyMesons}

		We first determine the polarizabilities $c_{F}$ and $c_{T}$ from the
		mass shifts of the meson in two different backgrounds using
		\eqref{eq:mass_finiteT} and \eqref{eq:mass_finiteF}. To accomplish this,
		we will switch on the perturbations of the $\N=4$ Lagrangian which
		correspond to finite $\OT$ and finite $\OF$. Again we consider the linear
		limit such that we can investigate the effects of finite temperature and
		finite background field strengths separately.
		
		Subsequently we will compute the correlators in
		\eqref{eq:adsCorrelatorT} and \eqref{eq:adsCorrelatorF} for strongly
		coupled $\N=4$ theory at finite temperature. The results we obtain in
		this section will be put together in the next subsection using
		\eqref{eq:adsCftCorrelator} to deduce the rate of momentum broadening
		and compare it to the weak coupling result.

		\subsubsection*{Backgrounds dual to finite temperature and field strength}
		
			The gravity background dual to $\N=4$ SYM theory at finite
			temperature is given by the AdS-Schwarzschild black hole with
			Lorentzian signature. This background is needed below both for
			calculating the necessary two-point correlators $\vev{\OT\OT}$ and
			$\vev{\OF\OF}$, as well as for obtaining the meson polarizability
			$c_{T}$, which accounts for meson mass shifts due to finite
			temperature.
			
			We make use of the coordinates derived in
			appendix~\ref{chap:coordinates} as \eqref{eq:adsBHv} to write the
			\ads-Schwarzschild background in Lorentzian signature as
			\begin{equation}
			\label{eq:adsBHMetricQuarkonium}
			\begin{gathered}
				\ds^2=\frac{v^2}{R^2}\left(-\frac{f^2(v)}{\tilde f(v)}\,\dd t^2 +\tilde f(v)\,\dd\threevec{x}^2\right)+\frac{R^2}{v^2}\left( \dd v^2 + v^2\dd\Omega^2_5\right)\\
				f(v) = 1-\frac{\rh^4}{4v^4}, \quad \tilde f(v) = 1+\frac{\rh^4}{4v^4}.\hfill~
			\end{gathered}
			\end{equation}
			In this way we can identify the transverse part to Minkowski space
			as nothing else than $\mathbbm R^6$ and we write it as
			\begin{equation}
			\label{eq:transverseMetricQuarkonium}
				\dd v^2 + v^2\dd\Omega^2_5=\sum_{i=1}^6\dd v_i^2 = \underbrace{\dd y^2 + y^2\dd\Omega_3^2}_{\mathbbm R^4(v_{1,\ldots,4})} + \underbrace{\dd v_5^2 + \dd v_6^2}_{\mathbbm R^2(v_{5,6})}.
			\end{equation}
			with the metric $\dd \Omega_3^2$ of the unit $3$-sphere, and
			$v^2=y^2+v_5^2+v_6^2$. The boundary is reached at asymptotically
			large $y$ while the horizon is located at $\rh/\sqrt2$. Notice that
			the black hole radius $\rh$ is related to the expectation value
			$\vev{\T^{00}}$ by \cite{Gubser:1996de}
			\begin{equation}
			\label{eq:tempFromT}
				\vev{\T^{00}} = \frac{3}{8}\pi^2 N^2 T^4\,, \qquad \rh = T\pi R^2.
			\end{equation}

			\bigskip
			
			The field configuration dual to $\vev\OF \neq 0$ and
			$\vev{T^{\mu\nu}} = 0$ is a non-trivial dilaton background with has
			been given by Liu and Tseytlin and consists of a configuration of
			\db3s with homogeneously distributed \D{(-1)} instantons
			\cite{Liu:1999fc}.  The type~IIB action in the Einstein frame for the dilaton
			$\Phi$, the axion $C$, and the self-dual gauge field strength
			$F_5=\hodge F_5$ reads
			\begin{equation}
			\label{eq:IIBAction}
				S_\text{IIB}= \frac{1}{2\kappa_{10}^2}\int\! \dd^{10}\xi 
				\times \sqrt{-g}\; \left[
				  \mathcal{R}-\frac{1}{2}(\partial \Phi)^2 -\frac{1}{2}\ee^{2\Phi}  
				   (\partial C)^2 -  \frac{1}{4\cdot 5!} (F_5)^2 +\dots
				\right].
			\end{equation}
			The ten-dimensional Newton constant is given by 
			\begin{equation}  
					\label{eq:gravityConst10} 
					\frac{1}{2\kappa_{10}^2}=\frac{2\pi}{(2\pi \ell_s)^8 \gs}= 
					\frac{N^2}{4\pi^5 R^8}\,.
			\end{equation}
			As can be seen from the supersymmetry transformations of the $\N=4$
			fermions, such a background breaks the supersymmetry to $\N=2$.
			Solving the equations of motion derived from~\eqref{eq:IIBAction},
			Liu and Tseytlin obtain the metric \cite{Liu:1999fc}
			\begin{equation}
			\label{eq:liuTseytlinBackground}
			\ds^2_{\text{string}} = \ee^{\Phi/2} \,\ds^2_{\text{Einstein}}   
				= \ee^{\Phi/2}   
				 \left[ 
				  \left(\frac{r}{R}\right)^2 \eta_{\mu\nu}\dd x^\mu \dd x^\nu + 
				  \left(\frac{R}{r}\right)^2 
				  \left(\dd r^2 + r^2 \dd\Omega_5^2\right ) 
				 \right].
			\end{equation}
			and axion-dilaton solution\footnote{Regarding conventions, note that
			in our notation~$q=\frac{R^8}{\lambda}q_{\scriptscriptstyle
			\text{LT}}$, where $q_{\scriptscriptstyle \text{LT}}$ is used in the
			paper of Liu and Tseytlin \cite{Liu:1999fc}.}
			\begin{equation}
					\label{eq:dilatonSolution}  
					e^\Phi= 1 + \frac{q}{r^4} \: ,  \qquad  C=-\ii\left(\ee^{-\Phi} - 1\right)\: .   
			\end{equation} 
			The expectation value $\vev{\OF}$ is given by 
			\begin{equation} 
			\label{eq:qandF2}
			\vev{\OF} = \lim_{r\to\infty} 
			\frac{\delta S_{\text{\tiny IIB}}}{\delta \Phi(r,\fourvec{x})}=  
			  \frac{N^2}{2\pi^2 R^8}\,q.   
			\end{equation}  
			
		\subsubsection*{Computing the polarization coefficients from meson mass shifts}

			We again identify mesons with fluctuations $\tilde \varphi$ of a
			\db7 embedded into the background dual to the field theory under
			consideration. Stable embeddings are obtained if the \db7 spans all
			Minkowski directions as well as the radial AdS coordinate and a
			$3$-sphere in the remaining angular directions. Consider the metric
			\eqref{eq:adsBHMetricQuarkonium} with
			\eqref{eq:transverseMetricQuarkonium} as an example. Here, the \db7
			shall be embedded such that it spans all directions except $v_5$ and
			$v_6$. The meson mass $M$ is then obtained by solving the equation
			of motion for the fluctuations $\tilde \varphi$
			\cite{Kruczenski:2003be}, as outlined in
			section~\ref{sec:boundStates}. Read as an eigenvalue equation, the
			equation of motion for the fluctuation gives the meson mass as the
			eigenvalues $M$ to the corresponding eigenfunctions $\tilde
			\varphi$. The discrete values of $M$ describe the Kaluza-Klein mass
			spectrum of mesons for any given quark mass.
			
			To see this explicitly and generalize to the backgrounds of
			interest below, we rephrase this procedure for the vacuum case
			$\vev{\T^{00}}=\vev{\OF}=0$ in a notation suitable for the
			subsequent generalization. Subsequently we will introduce a non-zero
			$\vev{\OF}$ and $\vev{\T^{00}}$, respectively.
			
			In the case of a \db7 embedded in a ten-dimensional background, the
			brane embedding is described by the locations $v_5$ and $v_6$ in the
			two directions transverse to the brane. This setup was introduced in
			section~\ref{sec:embeddingDbranes}. In general these locations
			depend on all eight coordinates $\xi^i$ of the eight-dimensional
			\db7 worldvolume and are determined by extremizing the DBI-action
			\eqref{eq:D7DBI}. We rephrase it here including the dilaton which we
			where free to set to unity in \eqref{eq:D7DBI},
			\begin{equation}
			\label{eq:DBIactionQuarkonium}
				S_{\text{DBI}} = -T_7 \int\!\dd^8\xi \;\ee^{-\Phi} \sqrt{-\det G}\; ,
				\qquad
				G_{ab} = \frac{\partial X^{\mu}}{\partial \xi^a} \frac{\partial X^{\nu}}{ \partial \xi^{b} }\,g^{\text{st}}_{\mu\nu},
			\end{equation}
			with $T_7$ the \db7 tension and $g^{\text{st}}$ is the string frame
			metric of the ten-dimensional background with coordinates $X^\mu$.
			It is related to the Einstein metric as in
			\eqref{eq:liuTseytlinBackground}. The distinction between the
			Einstein and string frame is ultimately important below to account
			for the effects of the non-trivial dilaton flow. The pullback $G$
			contains the functions $v_5(\xi)$ and $v_6(\xi)$, which are
			determined by solving their equations of motion, derived from
			$S_{\text{DBI}}$.
			
			The background \adsfivesfive dual to $\vev{\T^{00}} =\vev\OF =0$ is
			obtained e.\,g.\  from \eqref{eq:adsBHMetricQuarkonium} with $\rh=0$. It
			is well known that for this background a probe brane embedding is
			given by the functions
			\begin{align}
					v_5 &= 0,\\
					v_6 &= m_v = \text{const},
			\end{align}
			and the constant $m_v$ determines the quark mass
			$m_q=m_v/(2\pi\alpha')$. In terms of the example in
			section~\ref{sec:embeddingDbranes} this is instantly derived from
			the equation of motion \eqref{eq:D7embeddingBH} for the embedding
			$L$ in the zero temperature limit $\rh\to0$. Note however that there
			is a factor $\sqrt2$ differing in the definition of the quark mass
			due to the choice of coordinates.
			
			We are interested in the meson spectrum, which can be obtained from
			the brane fluctuations as in the previous sections, c.\,f.\  for
			instance page~\pageref{par:mesons} in section~\ref{sec:boundStates}.
			We thus allow for small fluctuations $\tilde \varphi$ around this
			solution. Here we consider the fluctuation of the radial part $v_6$,
			dual to the scalar meson excitations,
			\begin{equation}
					v_6 \mapsto v_6 + 2\pi\alpha'\, \tilde\varphi(\fourvec x,w).
			\end{equation}
			By the symmetries of the setup, the fluctuations only depend on the
			Minkowski directions $\fourvec x$ and on the coordinate $y$,
			denoting the radial coordinate on the part of the \db7 which is
			transverse to the Minkowski directions. The resulting equation of
			motion is analog to formula \eqref{eq:fluctEOM} (modulo the
			mentioned factors of $\sqrt2$). The solutions were found by plugging
			in the ansatz
			\begin{equation}
				\tilde\varphi = \varphi(y)\,\ee^{\ii\,\fourvec k \fourvec x}\, Y^l(\S^3)\;,
			\end{equation}
			where $Y^l(\S^3)$ are the scalar spherical harmonics on the $\S^3$
			wrapped by the probe \db7 and $\fourvec k$ denotes a four vector.
			The resulting equation of motion for the function $\varphi(y)$ for
			$l=0$ reads
			\begin{equation}
			\label{eq:eomQ0}
				-\partial_\rho \rho^3 \partial_\rho \varphi(\rho) = \frac{\rho^3}{(\rho^2+1)^2}\,\bar M^2\varphi(\rho) .
			\end{equation}
			Here we introduced the following dimensionless quantities
			\begin{equation}
					\rho = \frac{y}{m_v}\,, \qquad \bar M=\frac{R^2}{m_v} M,
			\end{equation}
			where the dimensionless $\rho$ is not to be mistaken with the same
			symbol of different meaning defined in
			section~\ref{chap:specFuncs}. We moreover identified the meson
			mass squared $M^2$ with the square of the momentum four-vector
			$\fourvec k$ of the fluctuations,
			\begin{equation}
					M^2=-\fourvec k^2.
			\end{equation}
			
			The eigenfunctions $\varphi_n$ solving the Sturm-Liouville equation
			\eqref{eq:eomQ0} are given in terms of the standard hypergeometric
			function ${}_2F_1$,
			\begin{equation}
				\varphi_n(\rho) =  \frac{c_n}{(\rho^2 + 1)^{n+1}}\:{{}_2F_1\left(-(n+1); -n; 2; -\rho^2\right)},
			\end{equation}
			where $n=0,1,2,\ldots$ and $c_n$ is a normalization constant such that
			\begin{equation}
			\label{eq:phiOrthoNorm}
				\int\limits_{0}^{\infty}\! \dd \rho\; \frac{\rho^3}{(\rho^2 + 1)^2}\,\varphi_n(\rho)\,\varphi_m(\rho) = \delta_{nm}.
			\end{equation}
			The lowest mode $\varphi_0$ is given by
			\begin{equation}
			\label{eq:phi0Solution}
				\varphi_0(\rho)=\frac{\sqrt{12}}{\rho^2+1}\,.
			\end{equation}
			The corresponding eigenvalues $M_n$ to the functions $\varphi_n$ are
			given by
			\begin{equation}
			\label{eq:massSpecQ0}
					\bar M_n = 2 \sqrt{(n+1)(n+2)}.
			\end{equation}
			We note that the mass of the lowest state with $n=0$ is
			\begin{equation}  
			\label{eq:massSpecQ0b}
			   M_0  = \frac{m_v}{R^2}\,2\sqrt{2} =m_q\frac{4\pi\alpha'}{R^2}\,\sqrt2 = \frac{4\pi m_q}{\sqrt{\lambda}}\, ,
			\end{equation}
			which will appear frequently below. For a more detailed derivation
			of these results the reader is referred to ref.~\cite{Kruczenski:2003be}.

			\paragraph{Mass shift in the dilaton background}
			Let us now calculate the polarizability $c_{\scriptscriptstyle F}$ which determines the
			change $\delta M$ of the meson mass at a given value of the gauge
			condensate $\vev\OF$ with respect to the meson mass at
			$\vev\OF=0$,
			\begin{equation}
			\label{eq:massShiftF}
					\delta M = - \frac{c_{\scriptscriptstyle F}}{N^2} \vev\OF.
			\end{equation}
			To find $c_{\scriptscriptstyle F}$ we will determine the mass shift $\delta M$ and
			identify $c_{\scriptscriptstyle F}$ with the proportionality constant in front of
			$\vev\OF$.
			
			We are interested in the eigenvalues of fluctuations in the case of
			$q \propto \vev\OF\neq 0$. The ten-dimensional background geometry
			dual to this scenario is given in \eqref{eq:liuTseytlinBackground}
			and the equation of motion for \db7 fluctuations analog to
			\eqref{eq:eomQ0} was derived in ref.~\cite{Brevik:2005fs} to be
			\begin{equation}
			\label{eq:eomQneq0}
				-\partial_\rho \rho^3 \partial_\rho \varphi(\rho) = \bar
				M^2\frac{\rho^3}{(\rho^2+1)^2}\,\varphi(\rho) - 4\bar q
				\frac{\rho^4}{(\rho^2 +1)(\bar q + (\rho^2 +
				1)^2)}\,\partial_\rho\varphi(\rho),
			\end{equation}
			with  the dimensionless
			\begin{equation}
			\label{eq:dimensionlessMandQ}
				\bar q=\frac{q}{L^4}\,.
			\end{equation}
			To obtain analytical results, we consider the case of small $\bar q$
			and linearize in this parameter. Therefore the equation of motion to
			solve is
			\begin{equation}
					-\partial_\rho \rho^3 \partial_\rho \varphi(\rho)
					= \bar M^2\frac{\rho^3}{(\rho^2+1)^2}\,\varphi(\rho) + \Delta(\rho) \varphi(\rho),
			\end{equation}
			where the operator $\Delta(\rho)$ is given by
			\begin{equation}
					\Delta(\rho) = - 4\bar q \frac{\rho^4}{(\rho^2 +1)^3}\,\partial_\rho.
			\end{equation}
			It is this term that describes the difference between the equation
			of motion at non vanishing background perturbation to
			\eqref{eq:eomQ0}, which is reproduced for $q=0$.
			
			To find the solution $\varphi_0(\rho)$ corresponding to the lightest meson 
			with $n=0$ we set up a perturbative expansion. Any deviation
			$\delta\varphi_0$ from the solution $\varphi_0$ of the case $q=0$ may
			be written as a linear combination of the functions $\varphi_n$, which are a
			basis of the function space of all solutions,
			\begin{align}
			\label{eq:phiPerturbed}
					\phi(\rho) &=  \phi_0(\rho) + \sum_{n=0}^\infty a_n \phi_n(\rho) , & a_n &\ll 1 ,\\
					\bar M^2   &= \bar M_0^2 + \delta \bar M_0^2, &  \delta \bar M_0^2 &\ll 1 .
			\end{align}
			Plug this ansatz into the equation of motion \eqref{eq:eomQneq0}, make use of
			\eqref{eq:eomQ0} and keep terms up to linear order in the small parameters
			$a_n$, $\bar q$ and $\delta M_0^2$ to get
			\begin{equation}
					\frac{\rho^3}{(\rho^2+1)^2}\,\sum_{n=0}^\infty a_n \bar M^2_n \varphi_n(\rho)
					= 
					\delta \bar M_0^2 \frac{\rho^3}{(\rho^2+1)^2}\,\varphi_0(\rho) + \bar M_0^2 \frac{\rho^3}{(\rho^2+1)^2} \sum_{n=0}^\infty a_n\,\varphi_n(\rho)
					 + \Delta(\rho) \varphi_0(\rho).
			\end{equation}
			We now multiply this equation by $\varphi_0(\rho)$, integrate over
			$\rho \in [0,\infty]$ and make use of \eqref{eq:phiOrthoNorm} and
			\eqref{eq:phi0Solution} to see that
			\begin{equation}
					\delta \bar M_0^2 = - \int\limits_0^\infty \! \dd\rho \; \varphi_0(\rho)\, \Delta(\rho) \varphi_0(\rho) = -\frac{8}{5}\,\bar q.
			\end{equation}
			From $\delta\bar M^2_0 = 2 \bar M_0 \delta\bar M_0$ 
			we  obtain 
			\begin{equation}
					\delta M_0 = \frac{L}{2R^2} \frac{\delta \bar M_0^2}{\bar M_0} 
					=  -\frac{8}{5\pi} \left( \frac{2\pi}{M_0} \right)^3 \,\frac{1}{N^2} \vev\OF,
			\end{equation}
			where in the last step we used \eqref{eq:massSpecQ0b} for the mass
			and \eqref{eq:qandF2} and \eqref{eq:dimensionlessMandQ} to relate $\bar{q}$ and $\OF$.
			By comparison with \eqref{eq:massShiftF} we  identify
			the polarizability
			\begin{equation}
			\label{eq:alphaF}  
					c_{\scriptscriptstyle F} =  \frac{8}{5\pi} \left( \frac{2\pi}{M_0} \right)^3.
			\end{equation}

			\paragraph{Mass shift in the finite temperature background}
			The calculation of the polarizability $c_{\scriptscriptstyle T}$ is completely analogous.
			We are now looking for the proportionality constant of meson mass
			shifts with respect to deviations from zero temperature, which we 
			noticed to be given by
			\begin{equation}
					\label{eq:deltaMalphaT}
					\delta M = - \frac{c_{\scriptscriptstyle T}}{N^2} \vev{\T^{00}}. 
			\end{equation}
			The background dual to the finite temperature field theory is the
			AdS black hole background given in \eqref{eq:adsBHMetricQuarkonium}
			with \eqref{eq:transverseMetricQuarkonium}.
			
			Again we calculate the meson mass spectrum to identify the
			polarizability by comparison with \eqref{eq:deltaMalphaT}. The
			embedding functions $v_5$ and $v_6$ in this
			background are given by
			\begin{align}
					v_5 &= 0,\\
					v_6 &= v_6(y),
			\end{align}
			where the quark mass is determined by $m_q=\lim_{y\to\infty}
			v_6/(2\pi\alpha')$. The function $v_6(y)$ has to be computed
			numerically \cite{Babington:2003vm}. Some examples of such
			embeddings are shown in figure~\ref{fig:embeddingExamples}.

			We introduce small fluctuations
			$\varphi(\rho)\ee^{\ii\fourvec{k}\fourvec{x}}$ in the $v_5$
			direction,
			\begin{equation}
					v_5 \mapsto v_5(y,\vec{x}) + \varphi(y)\ee^{\ii\fourvec{k}\fourvec{x}}\; .
			\end{equation}
			The linearized equation of motion for the fluctuations
			$\varphi(y)$ in the limit of vanishing spatial momentum and
			$M^2=-\vec k^2 $ can be derived from the DBI action
			\eqref{eq:DBIactionQuarkonium} to be
			\begin{equation}
			\label{eq:eomfluctLorentz}
					\begin{aligned}
							0 = & \; \hphantom{+} \partial_y
							\left[ \mathcal{G} \sqrt{ \frac{1}{1 + \left( \partial_y v_6  \right)^2}}\, 
							\partial_y \varphi(y) \right]
							 -  \sqrt{1 + \left( \partial_y v_6\right)^2} \frac{y^3}{2( y^2 + v_6^2)^5}\, \rh^8 \,\varphi(y)\\
							 & \; +
							\mathcal{G}  \sqrt{ 1 + \left( \partial_y v_6 \right)^2 }\,
							\frac{4 \left( y^2 +v_6^2\right)^2 + \rh^4}{ \left( (y^2 +v_6^2)^2 - \rh^4\right)^2}
							\,4R^4 M^2 \varphi(y)\,,
					\end{aligned}
			\end{equation}
			where we abbreviated
			\begin{equation}
					\mathcal{G} = \varrho^3\left( 1-\frac{\rh^8}{16\left(y^2+v_6^2\right)^4} \right).
			\end{equation}
			
			In the regime of small temperatures, we may linearize in $\rh^4$
			which is the leading order in $\rh$. Furthermore, as may be seen
			from figure~\ref{fig:embeddingExamples}, in the regime of a small temperature $T$
			compared to the quark mass $m_q$, or respectively small ratios of
			$\rh / \lim_{y\to\infty} v_6(y)$, the embeddings become more
			and more constant. So for constant embeddings $v_6 = m_v$ and up to
			order $T^4\propto\rh^4$ the equation of motion simplifies to
			\begin{equation}
			\label{eq:eomfluctLorentzLinear}
				-\partial_\rho \rho^3 \partial_\rho \varphi(\rho) = \bar M^2 \frac{\rho^3}{(\rho+1)^2}\varphi(\rho) + \Delta(\rho) \varphi(\rho),
			\end{equation}
			where we made use of the dimensionless quantities
			\eqref{eq:dimensionlessMandQ} and identify
			\begin{equation}
					 \Delta(\rho) = \frac{3}{4}\frac{\rh^4}{m_v^4}\frac{\rho^3}{(\rho^2+1)^4} \bar M^2.
			\end{equation}
			For the lightest meson, the ansatz \eqref{eq:phiPerturbed} 
			this time leads to
			\begin{equation}
			\begin{aligned}
					\delta \bar M_0^2 &= -\int\limits_0^\infty \! \dd\rho \, \varphi_0(\rho)\, \Delta(\rho) \varphi_0(\rho) \\
							&= -\frac{9}{40} \frac{\rh^4 \bar M_0^2}{m_v^4}\,.
			\end{aligned}
			\end{equation}
			Reinstating units and solving for $\delta M_0$ leads to
			\begin{equation}
					\delta M_0 =  -\frac{12}{5\pi} \left(\frac{2\pi}{M_0}\right)^3 \,\frac{1}{N^2} \left< \T^{00} \right>.
			\end{equation}
			From this we can read off the polarizability $c_{\scriptscriptstyle T}$ as
			\begin{equation}
			\label{eq:alphaT}  
			c_{T} = \frac{12}{5\pi}  \left(\frac{2\pi}{M_0}\right)^3 .
			\end{equation}
			
		\subsubsection*{Computing finite temperature correlators} 
		\label{sec:adsCftCorrelators}
		
			According to \eqref{eq:adsCftCorrelator} we need to compute the
			correlators \eqref{eq:adsCorrelatorT} and \eqref{eq:adsCorrelatorF}
			at finite temperature. We do so by once more performing the
			calculation along the lines of
			refs.~\cite{Policastro:2002se,Son:2002sd}, sketched in
			section~\ref{sec:thermalGreenFunctionsIntro}. The dual supergravity
			field to the energy momentum tensor $\T$ is the graviton $h$, and
			the corresponding field to the operator $\OF$ is the dilaton $\Phi$.
			Therefore, correlators of $\T^{00}$ are associated with graviton
			propagation and obtained from the supergravity field solution to
			$h$. Respectively, correlators of $\OF$ are associated with dilaton
			propagation and obtained from the supergravity solution to the
			dilaton $\Phi$. The calculational procedure for these two
			correlators is standard and has been discussed in
			refs.~\cite{Son:2002sd,Kovtun:2005ev}, and first applied in
			ref.~\cite{Policastro:2002se}, in order to find two-point Minkowski
			correlators as discussed in
			section~\ref{sec:thermalGreenFunctionsIntro}.
			
			On the gravity side, both field correlators are computed in the
			black hole background to account for finite temperature correlation
			functions of the dual gauge theory operators. In this subsection it
			is convenient to work in the coordinates derived as
			\eqref{eq:adsBHu}, with radial \ads coordinate $u$. For an explicit
			calculation we are more specific here as in
			\eqref{eq:correlatorFromBoundaryContribution} by writing
			\begin{equation}
			\label{eq:retardedThermalGreen}
			\begin{aligned}
				\GR(\omega, \threevec q) &\stackrel{\hphantom{\text{\tiny\ads/CFT}}}{=} -\ii\int\dd^4x\;\ee^{-\ii \fourvec k\cdot\fourvec x}\theta(t)\vev{\commute{\OF}{\OF}}\\
					&\stackrel{\text{\tiny\ads/CFT}}{=} \lim_{u\to0}  \mathcal A(u)\, f(u,-\fourvec k)\, \del_uf(u,\fourvec k).    
			\end{aligned}
			\end{equation}
			Here once more the function~$f(u,\vec k)$ relates the boundary and
			bulk values of a gravity field to each other. For example the
			dilaton field~$\Phi$ is related to its value at the
			boundary~$\phi^{\text{bdy}}$ by
			\begin{equation}
			\label{eq:rhoIs0Boundary}
				\Phi(u,\vec k) = f(u,\vec k)\, \phi^{\text{bdy}}(\vec k),
			\end{equation}
			and $f$ is normalized to one at the boundary, i.\,e.\  $f(0,\vec k)=1$.
			For fluctuations $h^{00}$ of the metric component $g^{00}$,
			$\Phi(u,\vec k)$ is replaced by $h^{00}(u,\fourvec k)$ and the field
			theory operators $\OF$ in \eqref{eq:retardedThermalGreen} are
			replaced by $\T^{00}$. The factor $\mathcal A(u)$ can be read off
			from the classical supergravity action
			\begin{equation}
			\label{eq:classicalAction}
			S_\text{cl}= \,\frac{1}{2} \int\! \dd u \, \dd^4x \, 
			  \mathcal A(u)\,(\partial_u \Phi)^2+ \dots\; .
			\end{equation}
			The classical five-dimensional gravity action for the graviton and 
			dilaton is obtained from~\eqref{eq:IIBAction} as   
			\begin{equation}
			\label{eq:dilatonGravitonAction}
			S_\text{cl}=\frac{1}{2\kappa_{5}^2}\int\! \dd u\, \dd^{4}x\sqrt{-g_5} \left(
			  (\mathcal{R}- 2\Lambda)- \frac{1}{2}(\partial \Phi)^2 
				+\dots\right), 
			\end{equation}
			where
			\begin{equation} 
				\frac{1}{\kappa_5^2} = \frac{R^5\Omega_5}{\kappa_{10}^2} 
				  = \frac{N^2}{4\pi^2 R^3}\,.  
			\end{equation}
			So comparing to~\eqref{eq:classicalAction} we get  
			\begin{equation}  
				\mathcal A= -\frac{\sqrt{-g_5}}{2\kappa_5^2}\, g^{uu}.    
			\end{equation}   
			The equation of motion derived from~\eqref{eq:dilatonGravitonAction} 
			in momentum space reads
			\begin{equation} 
				\label{eq:dilatonEom}  
				\Phi ''-\frac{1+u^2}{u(1-u^2)}\Phi' + 
				 \frac{\wn^2-\qn^2 (1-u^2)}{u (1-u^2)^2}\Phi = 0,   
			\end{equation}
			with dimensionless frequency~$\wn=\omega/(2\pi T)$ and spatial
			momentum component~$\qn=q/(2\pi T)$. The prime denotes the
			derivative with respect to the radial coordinate $u$. Note that in
			momentum space the function $\Phi(u)$ is akin to the radial part we
			denoted by $f(u)$ above. The equation of
			motion~\eqref{eq:dilatonEom} has to be solved numerically with
			incoming wave boundary condition at the black hole horizon.
			Computing the indices and expansion coefficients near the
			\emph{boundary} $u=0$, as done in
			refs.~\cite{Teaney:2006nc,Kovtun:2006pf}, we obtain the asymptotic
			behavior as a linear combination of two solutions $\Phi_{1,2}$ with
			asymptotic behavior for small $u$ as
			\begin{align}
				\Phi_1 &= (1 + \dots)\,,\\
				\Phi_2 &= (u^2 + \dots)\,.
			\end{align}
			The general solution to the equation of motion therefore is given by
			the linear combination
			\begin{equation}
			\label{eq:dilatonBoundary}    
				\Phi(u) = \Phi_1 + \mathcal B \Phi_2\,, 
			\end{equation}
			where we normalized the functions $\Phi$ such that the coefficient
			for $\Phi_1$ is $1$, which we are free to do since
			\eqref{eq:dilatonEom} is a homogeneous equation. In this way we
			achieve that the correct normalization for the radial part
			$\lim_{u\to0}\Phi(u)=1$ is implemented.
			
			At the \emph{horizon} the asymptotic solution satisfying the
			incoming wave boundary condition is
			\begin{equation}
			\label{eq:dilatonHorizon}    
			\Phi (u) = (1-u)^{-\frac{\ii\wn}{2}} (1+\dots). 
			\end{equation}
			As discussed in refs.~\cite{Teaney:2006nc,Kovtun:2006pf} we find
			the coefficient~$\mathcal B$ by integrating the two boundary
			solutions from~\eqref{eq:dilatonBoundary} forward towards the
			horizon and by matching the linear combination of the numerical
			solutions $\Phi (u) = \Phi_1^{\text{num}} + \mathcal B
			\Phi_2^{\text{num}}$ to the solution~\eqref{eq:dilatonHorizon} at
			the horizon. We recognized that the solution of the radial part
			$\Phi(u)$ found in this way is equivalent to the radial part of the
			total solution $\Phi$ and can therefore be plugged in for $f$ in
			\eqref{eq:retardedThermalGreen}.
			
			The imaginary part of the retarded correlator then is given by
			\begin{equation}  
			\frac{-2 T}{\omega}\Im  \GR_{FF} = 
			   \frac{N^2 (\pi T)^4}{4\pi^2} 
			\frac{2}{\pi} \frac{\Im  \mathcal{B}}{\wn}\, .     
			\end{equation}    
			Solving~\eqref{eq:dilatonEom} and matching the asymptotic solutions
			as described above, thus enables us to obtain
			\begin{equation}  
			\label{eq:forceCorrelatorF} 
			\lim_{\omega\to 0}\int\!\frac{\dd^3 q}{(2\pi)^3}\,
			  \frac{\threevec q^2}{3}\left(\frac{-2 T}{\omega} \Im  \GR_{FF}(\omega,q)\right)         
			  = N^2 T^9 \,67.258\,.   
			\end{equation}
			The corresponding result for the energy-momentum tensor correlator
			is obtained in an analogous way but the analysis is significantly
			more complicated. Fortunately it has been extensively and carefully
			analyzed in ref.~\cite{Kovtun:2005ev}. The final result is
			\begin{equation}
			\label{eq:forceCorrelatorT} 
				\lim_{\omega\to 0}\int\!\frac{\dd^3 q}{(2\pi)^3}
				  \frac{\threevec q^2}{3}\left(\frac{-2 T}{\omega} \Im  \GR_{\T\T}(\omega,q)\right)         
				  = N^2 T^9\, 355.169\, . 
			\end{equation}    

	\subsubsection{Comparing weak and strong coupling}
	\label{sec:quarkoniumSummary}
		
		We now have the results to compare momentum broadening of the heavy
		meson in a hot medium at weak and strong coupling. Over the duration of
		the lifetime of the heavy meson state it will loose momentum on average
		and simultaneously receive random kicks as codified by the Langevin
		equations of motion \eqref{eq:newton_langevin}.
		The drag and momentum broadening rates are related by the Einstein
		relation
		\begin{equation}  
		\label{einstein_final}
			\eta_{\scriptscriptstyle D} = \frac{\kappa}{2 T M_0}\,,
		\end{equation}
		with $M_0$ the meson mass. 
		
		For strongly coupled $\N=4$ SYM theory we obtain our principle result by
		collecting the results for polarizabilities \eqref{eq:alphaF},
		\eqref{eq:alphaT} and force correlators \eqref{eq:forceCorrelatorF},
		\eqref{eq:forceCorrelatorT}, and use \eqref{eq:adsCftCorrelator},
		\begin{equation}  
		\label{eq:finalKappaN4strong}
		\begin{aligned}
			  \kappa_{\lambda\to\infty} &= \frac{T^3}{N^2} \left(\frac{2\pi T}{M_0} \right)^6 \left( \left(\frac{8}{5\pi}\right)^2 67.258 + \left(\frac{12}{5\pi}\right)^2 355.169 \right) \\
			   &= \frac{T^3}{N^2} \left(\frac{2\pi T}{M_0} \right)^6 224.726\,.
		\end{aligned}
		\end{equation}
		The mass shift in strongly coupled $\N=4$ SYM is given by the sum of the
		mass shift due to dilaton and graviton contributions, respectively.
		However, the exact value of the dilatonic mass shift
		\eqref{eq:massShiftF} is determined by \eqref{eq:qandF2} in terms of
		$q$, which we do not want to speculate about here. Nevertheless, from the
		definition of $q$ in ref.~\cite{Liu:1999fc} we know that it is positive
		for positive instanton numbers.
		In this case the dilatonic mass shift contributes with the same sign as
		the graviton mass shift \eqref{eq:deltaMalphaT} and we write
		\begin{equation}
		\label{eq:finalDeltaMN4strong}
		\begin{aligned}
			\delta M_0\Big|_{\lambda\to\infty} &\leq -\frac{c_{\scriptscriptstyle T}}{N^2} \vev{\T^{00}} \\
				  &= - T \, \left(\frac{2\pi T}{M_0}\right)^3\frac{9\pi}{10}\,.
		\end{aligned}
		\end{equation}
		Here we made use of relations \eqref{eq:alphaT} and
		\eqref{eq:tempFromT}. Comparing these formulas with the analogous
		formulas in weak coupling large $N$ QCD given in
		equations~\eqref{eq:kappaqcd_final} and \eqref{eq:qcd_massshift},
		\begin{equation}  
		   \kappa_{\mbox{\tiny pQCD}} =\, \frac{T^3}{N^2} \left(\frac{\pi T}{\Lambda_B}\right)^6
		   \frac{50176}{1215} \pi \,,
		\end{equation}  
		and 
		\begin{equation}  
		   \delta M_{\mbox{\tiny pQCD}} =-  T\left(\frac{\pi T}{\Lambda_B} \right)^3\, \frac{14}{45}\,,
		\end{equation}
		we see that the meson mass $M_0$ plays the role of the inverse Bohr
		radius $\Lambda_B=(m_{q}/2) \alpha_s C_F$ in the strong coupling dipole
		effective Lagrangian. This is as expected for relativistic bound states.
		Since these prefactors are different we do not compare the numerical
		values. Below we will compare the values of the ratio $\kappa / (\delta
		M^2)$ at strong and weak coupling. We moreover observe that the drag
		coefficient for heavy mesons is suppressed by $N^2$ in the large $N$
		limit.
		
		The phenomenological model we used modeled the interaction of the mesons
		with the medium as dipole interaction terms, which capture short
		distance phenomena. From the field theory side one may wonder if these
		dipole interactions indeed are the dominant interaction mechanism of the
		medium and the mesons. Considering \ads/CFT, we admit that a dipole
		picture of a meson is a short distance description an we can not
		guarantee to describe such UV effects of the field theory well by
		holographic models. On the other hand, even if we cannot describe the
		interaction from first principle, the diffusion of the mesons and the
		scattering of gluons which then propagate with modified momentum into
		the medium is a long distance effect, which we can hope do describe by
		our approach. We therefore consider a quantity that is independent of
		short scale dipole interpretation of the medium, which is parametrized
		by the connection coefficients $c$. To construct such a quantity, we
		remember
		from \eqref{eq:adsCftCorrelator} that (for the simple case of $c_{\scriptscriptstyle F}=0$)
		\begin{equation}
			\kappa = \left(\frac{c_{\scriptscriptstyle T}}{N^2}\right)^2 \lim_{\omega\to 0}\int\!\frac{\dd^3 q}{(2\pi)^3}
				  \frac{\threevec q^2}{3}\left(\frac{-2 T}{\omega} \Im  \GR_{\T\T}(\omega,\threevec q)\right).
		\end{equation}
		Together with \eqref{eq:mass_finiteT} we can then construct a quantity
		that does not depend on the connection coefficient $c_{\scriptscriptstyle T}$,
		\begin{equation}
			\frac{\kappa}{(\delta M)^2} = \frac{1}{\vev{\T^{00}}^2} \lim_{\omega\to 0}\int\!\frac{\dd^3 q}{(2\pi)^3}
				  \frac{\threevec q^2}{3}\left(\frac{-2 T}{\omega} \Im  \GR_{\T\T}(\omega,\threevec q)\right).
		\end{equation}
		It is independent from the connection coefficient $c_{\scriptscriptstyle T}$. All quantities
		in this expression have been calculated previously for weak coupling as
		well as for strong coupling. The strong coupling result is obtained from
		inserting the relations \eqref{eq:tempFromT} and
		\eqref{eq:forceCorrelatorT} and yields
		\begin{equation}
			\left.\frac{\kappa}{(\delta M)^2} \right|_{\T,\lambda \rightarrow \infty} = \frac{\pi T}{N^2}\,8.3\,.
		\end{equation}
		To include the dilaton contribution, we make use of the above results
		\eqref{eq:finalKappaN4strong} and \eqref{eq:finalDeltaMN4strong} to
		get
		\begin{equation}
		\label{eq:strong}
			\left.\frac{\kappa}{(\delta M)^2} \right|_{\lambda \rightarrow \infty} \leq \frac{\pi T}{N^2}\,8.9\,. 
		\end{equation}
		It is then reasonable to use \ads/CFT to estimate to what degree strong
		coupling physics modifies this ratio in QCD. In the free finite
		temperature $\N=4$ theory, the result was computed in appendix~A of
		ref.~\cite{Dusling:2008tg} and cited in
		\eqref{eq:N4perturbativeResults} to be
		\begin{equation}
		\label{eq:weak2}
			\left. \frac{\kappa}{(\delta M)^2} \right|_{
			\lambda\rightarrow 0} \approx  \frac{\pi T}{N^2} \, 36.9\,.
		\end{equation}
		Thus, comparing the strong coupling result \eqref{eq:strong} with the
		weak-coupling result \eqref{eq:weak2}, we conclude that strong coupling
		effects actually \emph{reduce} the momentum transfer rate relative to
		the mass shift.
			
\subsection{Summary}
\label{sec:summaryDiffusion}

	We studied transport properties of baryon charge, isospin charge and heavy
	mesons with three different methods. For the charge diffusion we derived the
	diffusion coefficient in dependence on temperature and particle density (or
	chemical potential, equivalently). In the case of meson diffusion in the
	quark-gluon plasma we observed how strong coupling effects the
	equilibration.
	
	\paragraph{Baryon diffusion} We made use of the most na\"ive formulation of
	the membrane paradigm and derived a dependence of the diffusion coefficient
	for baryons that was qualitatively confirmed by recent and more
	comprehensive studies \cite{Mas:2008qs}. The results, shown in
	figure~\ref{fig:diffusionCoeffBaryon}, indicate that for very large values
	of the baryon density, the diffusion constant asymptotes to its maximal
	value $D=1/(2 \pi T)$. This reflects the fact that in this case, the free
	quarks outnumber the quarks bound in mesons. For finite densities the
	diffusion coefficient is reduced and we observe a minimum at values of the
	quark mass to temperature ratio $m$, which lie above the fundamental phase
	transition at zero density. We will draw a connection between the phase
	transition and the behavior of the diffusion in
	section~\ref{chap:phaseDiag}.

	\paragraph{Isospin diffusion} To obtain the isospin diffusion constant we
	analyzed the dispersion relation of the lowest lying quasi normal mode of
	the isospin current. We have considered a relatively simple gauge/gravity
	dual model for a finite temperature field theory, consisting of a constant
	isospin chemical potential $\mu$ obtained from a time component vacuum
	expectation value for the $\SU(2)$ gauge field on two coincident brane
	probes. We have considered the constant \db7 embedding corresponding to
	vanishing quark mass. 
	
	Within the strong restrictions of our analytic derivation, the main result 
	is that this model, despite its simplicity, leads to a hydrodynamical
	behavior of the dual field theory which goes beyond linear response theory.
	We find in particular a frequency-dependent diffusion coefficient with a
	non-analytical behavior. Frequency-dependent diffusion is a well-known
	phenomenon in condensed matter physics. Here it originates simply from the
	fact that due to the non-Abelian structure of the gauge field on the brane
	probe, the chemical potential replaces a time derivative in the action and
	in the equations of motion from which the Green functions are obtained.
	For a more comprehensive study one should include terms of quadratic order
	in the chemical potential $\mn$, which then would cancel the non-analytic
	behavior.

	\paragraph{Quarkonium diffusion} In the last part of this section we
	studied the diffusion of heavy mesons in the holographic plasma, by setting
	up a Langevin model. The forces on the meson where deduced from an effective
	dipole model for mesons that allowed to carry out our computations at
	\emph{strong and weak} coupling. We therefore where able to estimate the
	effects of strong coupling on meson diffusion.
	
	On the field theory side, we deduced the meson couplings to the stress tensor
	$\T$ and the operator $\OF$ at strong coupling, which are the only relevant
	operators coupling to the heavy dipole.	
	The couplings where deduced from mass shifts, which can be computed on the
	gravity side of the correspondence as a change in the normal
	vibrational modes of the \db7 in the presence of an external gravitational,
	or respectively dilatonic, field.
	
	Because the gravitational and dilatonic fields shift the spectrum of the
	\db7 excitations, gradients such as those from fluctuations of these fields
	give rise to a net force on the mesonic modes. We used the fluctuation
	dissipation theorem to relate the spectrum of these fluctuations in the
	plasma to the momentum broadening of the meson.
	
	The result for $\kappa/(\delta M)^2$, the momentum broadening relative to
	the square of the in medium mass shift, for strongly coupled $\N=4$ SYM
	theory is roughly four to five times smaller than the result in the weakly
	coupled limit. We therefore conclude that in this model the effect of strong
	coupling is to \emph{reduce} the momentum broadening and increase diffusion
	of mesons relative to weak coupling.
	
	From a phenomenological perspective the current calculation was limited to
	very heavy mesons, which survive above $T_c$, where dipole interactions
	between the meson and the medium are dominant. It is certainly unclear if
	this is the relevant interaction mechanism above $T_c$ even for bottomonium.
	Furthermore, the dipole coupling between a heavy meson and the medium is
	dominated by short distance physics which is not well modeled by \ads/CFT.
	
	From the supergravity perspective, a better understanding of how
	gravitational and dilatonic fields fluctuate in bulk would give a
	straightforward procedure to calculate the drag of a finite mass meson.
	Specifically, fluctuations in the bulk would force motion of meson wave
	functions which extend into the holographic fifth dimension.

\section{Exploring the phase diagram}
\label{chap:phaseDiag}

While the theory of quantum chromodynamics is concise in its mathematical
formulation, it exhibits a rich phenomenological structure, including several 
phase transitions. For instance, we
frequently referred to the confinement/deconfinement transition quark matter is
supposed to undergo when it is heated up and/or exposed to high chemical
potential.

Since the development of QCD it was discovered that quark matter can exhibit
numerous qualitatively different behaviors. The most prominent example in QCD
is the change in the coupling constant $\gym$ of the strong interaction with
respect to the momentum scale at which the theory is probed
\cite{Gross:1973id,Politzer:1973fx}. It eventually accounts for the transition
from the zero temperature regime (ground state with minimum momentum), where
quarks are confined to the hadronic color singlets that make up the nuclei of
atoms, to a deconfined state of matter at asymptotically high temperatures (and
high thermal momenta) where quarks and gluons roam freely throughout spacetime.
The value of the momentum interchange between quarks and gluons determines their
interaction potential, which may change qualitatively, e.\,g.\  by increasing
temperature, from a confining shape of infinite depth to a potential with finite
binding energy.

Another example is the discovery of the so called color-flavor-locked phase of
QCD with three color and three flavor degrees of freedom at high chemical
potential, equivalent to high particle densities. It was observed that the
thermodynamically favored ground state of QCD changes with the chemical
potential. While we observe color and flavor symmetry separately for QCD at low
chemical potential, there is some critical value of the chemical potential at
which symmetry breaking occurs and a relation between color and flavor degrees
of bound states of quarks is established \cite{Alford:1997zt,Alford:1998mk}.

The motivations to explore the properties of QCD at high temperature and large
chemical potentials include the struggle for a deeper understanding of the
features of QCD itself, the formation of hadronic matter during the evolution of
the universe and the description of matter inside dense astrophysical objects
such as neutron stars.

We may wonder which parameters determine if we observe confinement and/or color
flavor locking, which other symmetries and observables undergo qualitative
changes and whether these changes occur abruptly or smoothly in the parameter
space. This raises the question about the structure of the phase diagram of the
theory. It describes the regions in the parameter space of the state variables
in which the thermodynamic potentials and their derivatives behave analytically.
The change from one phase to another is often accompanied by symmetry breaking
and can be described by an order parameter which assumes finite values in one
phase and vanishes in the other. For QCD, the quark condensate
$\vev{\bar\psi\psi}$ is frequently used as an order parameter. Non vanishing
values break chiral symmetry. We can distinguish the hadronic phase, where
$\vev{\bar\psi\psi}\neq0$, from the quark-gluon plasma with a higher amount of
symmetry, here $\vev{\bar\psi\psi}=0$. However, strictly speaking the chiral
condensate is non-zero in both phases for finite quark masses and should only be
used as an order parameter in models with vanishing masses for the light quarks.

\begin{figure}
	\centering
	\includegraphics[width=.5\linewidth]{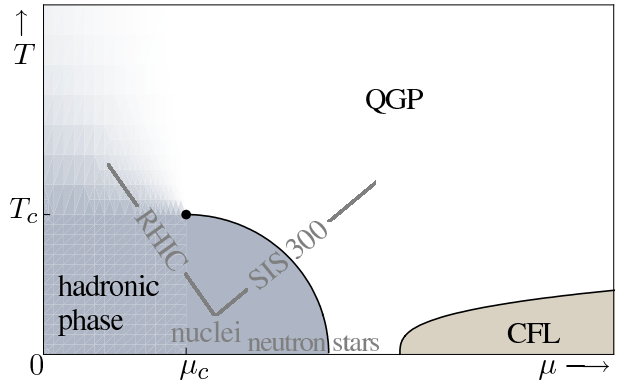}
	\caption[The conjectured QCD phase diagram in $(\mu,T)$]{
		The conjectured QCD phase diagram in the plane of temperature $T$ and
		baryon chemical potential $\mu$. The black lines indicate first order
		phase transitions. The first order confinement/deconfinement transition
		ends in a critical point at unknown $(\mu_c,T_c)$. Here the transition
		becomes second order. Lattice simulations suggest $170 \text{MeV} \lesssim
		T_c\lesssim190\text{MeV}$, nuclear matter has $\mu\approx1 \text{GeV}$
		\cite{Stephanov:2004wx}. CFL is the color-flavor-locked phase.
	}
	\label{fig:phaseDiagQCD}
\end{figure}

The exact overall structure of the phase diagram is not known yet. Especially at
low temperature and chemical potential where the coupling is strong, theoretical
treatments rely on lattice gauge theory, which on the other hand has its
problems with modeling the QCD dynamics at finite temperature and finite baryon
density. Figure~\ref{fig:phaseDiagQCD} shows a sketch of the most basic features
of the theoretically conjectured phase diagram of QCD based on combinations of
analytical and numerical predictions \cite{Schafer:2005ff,Stephanov:2004wx}.

In huge volumes that exist for long periods of time, such as the cubic kilometer
volumes of quark matter in neutron stars, one can expect the thermodynamic limit
to be an appropriate approximation to describe matter, i.\,e.\  thermodynamics is
applicable. As hydrodynamic models have successfully been applied to describe
the collective motion of the fireball produced in heavy ion collisions, it is
reasonable to assume that even the quark-gluon plasma observed in experiments
reaches thermal equilibrium. The equilibration time is estimated to
approximately $1\text{fm}/c$, the plasma state then exists for about another $4\text{fm}/c$
\cite{Muller:2006ee}. Though in equilibrium, the multiparticle system in a
collision experiment certainly does evolve along some trajectory in the phase
diagram, which leads from some point in the QGP phase to a system of hadrons,
which can eventually be detected.

For a thermodynamic description, the state variables can be chosen to be for
instance the temperature $T$, the pressure $P$, and the chemical potentials
$\mu_j$ for the different species of particles, labeled by the index $j$. The
chemical potential $\mu_j$ is the conjugate variable to the particle number
$N_j$. It is therefore only well defined if the number of particles of the
species $j$ is well defined. 

There are only a few charges that are conserved by all standard model processes.
They allow for the definition of particle numbers and thereby determine the
parameters of the phase diagram. These are quark number (which can be translated
into baryon number), lepton number, electric charge, and color charge. Each of
them has a chemical potential associated to it. In the processes we discuss in
heavy ion collisions, there are no leptons (and in neutron stars they are
radiated off by neutrino emission), i.\,e.\  the lepton chemical potential can be set
to zero. The system furthermore is color-neutral, i.\,e.\  the chemical potentials
associated to color charge can also be set to zero. The
Gell-Mann-Nishijima-relation allows to rephrase the remaining parameters, quark
number and electric charge, in terms of isospin and hypercharge. Hypercharge in
turn is determined by the number of particles of the various quark species.

One may argue that in an equilibrium state weak interactions could account for
flavor changing processes, and therefore there would be no well defined particle
number associated to each quark flavor. As a result, the notion of a chemical
potential would not be well defined. In heavy ion collisions however, the system
in equilibrium state has not enough time to undergo weak interactions and
therefore the flavor numbers are conserved, and the different quark flavors have
to be assigned individual chemical potentials.

Thus, the remaining degrees of freedom in the phase diagram are temperature and
the chemical potentials of the interacting quark species. In two-flavor setups,
which we elaborate on in this work, we are free to express these potentials in
terms of the baryon and isospin chemical potential. This is why we used and
continue to use these charges as the parameters throughout this work. In
principle we therefore consider a three-dimensional phase diagram in
$(T,\mu^{\scriptscriptstyle B},\mu^{\scriptscriptstyle I})$. For sake of
simplicity, however, we restrict to cases of either non-vanishing baryonic or
isospin chemical potential.

In the holographic context, several publications where dedicated to the
investigation of the structure of the phase diagram of theories with gravity
duals. For instance, the meson melting transition at finite temperature for the
fundamental matter introduced by \db7s in a background generated by \db3s, was
considered at zero as well at finite particle density. Studies of the behavior
of \db7 probes in the \ads Schwarzschild black hole background revealed a phase
transition, which occurs when the \db7 reaches the black hole horizon. This
transition was shown to be of first order in ref.~\cite{Kirsch:2004km}, see
ref.~\cite{Kruczenski:2003uq} for a similar transition in the \D4/\D6 system, further
details can be found in refs.~\cite{Mateos:2006nu,Albash:2006ew}. Related phase
transitions appear in refs.~\cite{Karch:2006bv,Kajantie:2006hv,Schnitzer:2006xz}.
Subsequently this phase transition was investigated at finite particle density
\cite{Babington:2003vm,Kobayashi:2006sb,Mateos:2007vc}. The transition is
characterized by a stable quasiparticle spectrum in one of the phases and
melting mesons at finite density in the deconfined phase.
Figure~\ref{fig:phaseDiagram} illustrates the two phases. A way to derive the
values of the transition line was mentioned in the same section, see the caption
of figure~\ref{fig:backgAt} and ref.~\cite{Mateos:2007vc} for details. For the phase
diagram in the model given by \D8-$\overline{\text{\D8}}$ probes in a near
horizon limit of \db4s with non vanishing chemical and isospin potential see
ref.~\cite{Parnachev:2007bc}.

In this section we contribute to these studies by different means. In the
subsequent section we observe a phase transition in the baryon diffusion
coefficient and determine the critical baryon density at which this transition
vanishes. The observations we make can be related to the results from
refs.~\cite{Myers:2007we,Mateos:2007vc} and confirm the observations described
therein.

In section~\ref{sec:phaseTransIso} we describe the occurrence of a new phase
transition at finite isospin density, which we published in
ref.~\cite{Erdmenger:2008yj}.

\subsection{Phase transition of the baryon diffusion coefficient}
\label{sec:baryonDiffusionPhaseTrans}

	In this section we pick up the discussion from
	section~\ref{sec:baryonDiffusion}, where we introduced a simple yet
	incomplete derivation of the baryon diffusion coefficient. As we mentioned
	there, the qualitative behavior of the result captures the physics correctly
	and the quantitative evaluation bears only little error in the regime of
	masses up to the temperature scale \cite{Mas:2008qs}. On the other hand, the
	evaluation of the simple formula \eqref{eq:diffusionConstant} is a
	considerable simplification, compared to the exact treatment, described in
	ref.~\cite{Mas:2008qs}. We will therefore stick with our approximation.
	
	We focus on the temperature regime near the phase transition of fundamental
	matter. In the zero density limit, a phase transition of fundamental matter
	was observed to take place simultaneously to the geometric transition from
	Minkowski embeddings at small temperature to black hole embeddings at high
	temperatures. In the low temperature phase the mesonic spectral functions
	exhibit discrete delta peaks, and thereby describe stable mesons, while the
	spectrum becomes continuous with finite width peaks in the high temperature
	phase. The value of the quark mass to temperature ratio at which the phase
	transition can be observed was found to be near $m=1.3$, c.\,f.\ 
	figure~\ref{fig:embeddingExamples}. The transition can be seen as a first
	order phase transition in the quark condensate \cite{Babington:2003vm}. In
	the zero density limit the authors of ref.~\cite{Myers:2007we} observed another
	manifestation of the fundamental phase transition at $m=1.3$ as a phase
	transition in the baryon diffusion coefficient $D$. We are interested in the
	finite density effect on this transition.
	
	At finite density, we know that black hole embeddings capture the physics at
	all temperatures, i.\,e.\  the entire parameter regime of $m$. The fundamental
	phase transition in this case is a transition between two different black
	hole embeddings \cite{Kobayashi:2006sb}. As discussed in
	ref.~\cite{Mateos:2007vc}, the baryon density affects the location and the
	presence of the fundamental phase transition. The transition is of first
	order only very close to the separation line between the regions of zero and
	non-zero baryon density shown in figure~\ref{fig:phaseDiagram}. Note that as
	discussed in refs.~\cite{Karch:2007br,Kobayashi:2006sb,Mateos:2007vc} there exists
	a region in the $(\tilde d,T)$ phase diagram at small $\tilde d$ and $T$
	where the embeddings are unstable. This instability disappears for large
	$\tilde d$.
	
	\begin{figure}
		\centering
		\includegraphics[width=.5\linewidth]{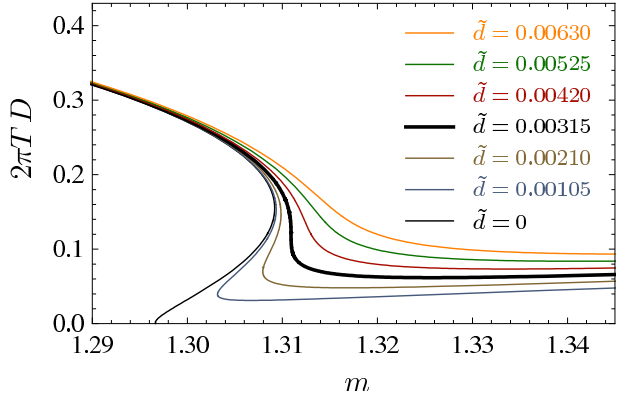}
		\caption[Phase transition of the baryon diffusion coefficient]{
			The normalized baryon diffusion coefficient as a function of
			normalized inverse temperature. At densities below $\tilde
			d=0.00315$ we observe a multivalued dependence on $m$, signaling a
			phase transition. For the behavior of the coefficient in a larger
			range of $m$ see figure~\ref{fig:diffusionCoeffBaryon}.
		}
		\label{fig:phaseTransBaryonDiffusion}
	\end{figure}
	
	We study the baryon diffusion coefficient at different baryon densities.
	Figure~\ref{fig:phaseTransBaryonDiffusion} shows the Diffusion coefficient
	$D$ as a function of the ratio of quark mass to temperature $m$. By fixing
	the quark mass we may think of $m$ as the inverse of the temperature.

	We find that the phase transition is slightly shifted towards smaller
	temperatures when we increase the density. At a critical density of $\tilde
	d^*=0.00315$ the phase transition temperature is given by
	$m=1.31$. Beyond the critical density the transition
	vanishes, in agreement with the critical density $\tilde d^*$ for the phase
	transition in the quark condensate, discussed in ref.~\cite{Kobayashi:2006sb}.

\subsection{A new phase transition at finite isospin potential}
\label{sec:phaseTransIso}
	
	In this section we have a different look at the mesonic spectral functions,
	introduced in section~\ref{chap:specFuncs}. There we where interested in the
	behavior of the quasiparticle resonances and the modification of the
	particle spectrum when we leave the limit of zero temperature and
	vanishing particle density. Here, we focus on a new phenomenon occurring at
	high densities.
	
	We recall that the three solutions $X$, $Y$ and $E^3$ to the fluctuation
	equations of motion \eqref{eq:eomX} to \eqref{eq:eomE3} constitute the
	isospin triplet of mesons which may be constructed out of the isospin
	$1/2$ quarks of the field theory. This is analog to the $\rho$-meson
	in QCD. We discovered that the mode $E^3$ coincides with the solution in
	case of a pure baryonic chemical potential, while the other two solutions
	have peaks in the spectral function at lower and higher values of $\wn$,
	c.\,f.\  figure~\ref{fig:isoSpectrum}. The magnitude of this splitting of the
	spectral lines is determined by the chemical potential and the undetermined
	coupling $c_{\scriptscriptstyle A}$.
		
	In the limit of zero frequency $\wn\to 0$, equations \eqref{eq:eomX} and
	\eqref{eq:eomY} coincide and will result in identical solutions $X$ and $Y$.
	In this limit the solution $E^3$, though, differs from $X$ and $Y$, by means
	of the last term. So for small frequencies $\wn$, we expect differences
	between the solutions $E^3$ and $X$, $Y$. All three equations of motion
	depend on the particle density $\tilde d$ parametrically, since the
	density has influence on the background fields.

	\subsubsection*{Spectral functions and quasi normal modes at high densities}
		
\begin{figure}
			\centering
			\includegraphics[width=.8\linewidth]{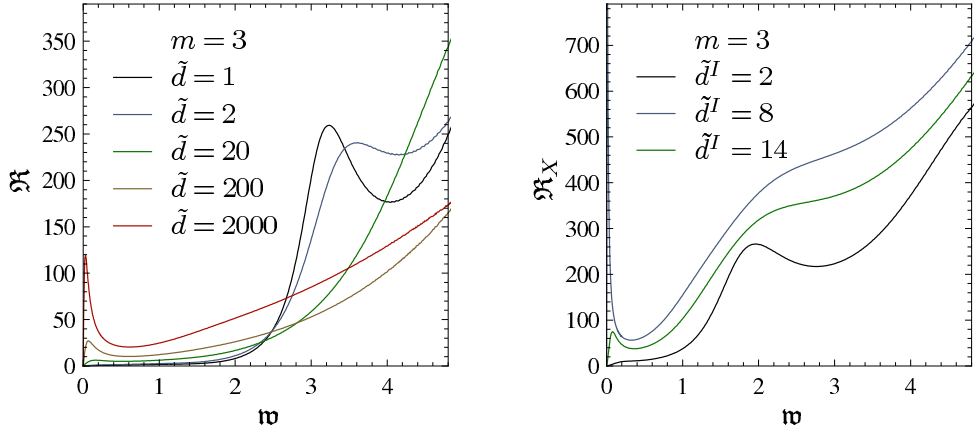}
			\caption[New Structures appearing in the spectra at high densities]{%
				Spectral functions for various baryon densities (left) and
				isospin densities (right), again normalized to $N_fN_cT^2/4$. At
				increasing densities $\tilde d$ the peaks are smeared out, as we
				saw in the discussion of the spectral functions in
				section~\ref{chap:specFuncs}. At very high densities a new
				structure forms at small $\wn$.}
			\label{fig:newStructure}
		\end{figure}

		We work in the canonical ensemble and will now investigate the effects
		of variations in $\tilde d$. Spectral functions for various finite
		baryonic and isospin densities $\tilde d$ are shown in
		figure~\ref{fig:newStructure}. As in section~\ref{sec:baryonMu}, the
		peaks in these spectral functions indicate that quarks form bound
		states. At low baryon densities we recognized the positions of the peaks
		to agree with the supersymmetric result \eqref{eq:mesMassDimless}.
		Increasing the quark density leads to a broadening of the peaks, which
		indicates decreasing stability of mesons at increasing baryon density
		\cite{Aharony:2007uu,Paredes:2008nf}. At the same time the positions of
		the peaks change, which indicates a dependence of the meson mass on the
		baryon density. Now, further increasing the quark density leads to the
		formation of a new structure at $\wn<1$. We will discuss this structure
		below together with the results at finite isospin density.
		
		We now turn to the effects of finite isospin density on the spectrum.
		The peaks in the spectral functions again correspond to mesons. An
		interesting feature at finite isospin chemical potential is the
		formation of a new peak in the spectral function in the regime of small
		$\wn$ at high density/high chemical potential, see
		figure~\ref{fig:newStructure}. Notice that compared
		to the baryonic case, the density at which the new peak forms is about
		two orders of magnitude smaller. As in the baryonic case, the
		excitations related to the supersymmetric spectrum broaden, the
		corresponding mesons become unstable.
		
		We pointed out that the structure of the spectral function is determined
		by the pole structure of the retarded correlator, see
		section~\ref{sec:retardedThermalGreen}. The poles of this function are
		located in the complex $\omega$-plane at positions $\Omega_n \in
		\mathbb{C}$. The spectral functions show the imaginary part of the
		correlator at real valued $\omega$. Any pole in the vicinity of the real
		axis will therefore introduce narrow peaks in the spectral function,
		while poles far from the real axis have less influence and merely
		introduce small and broad structures.
		
\begin{figure}
			\begin{minipage}[t]{.45\linewidth}
				\includegraphics[width=\linewidth]{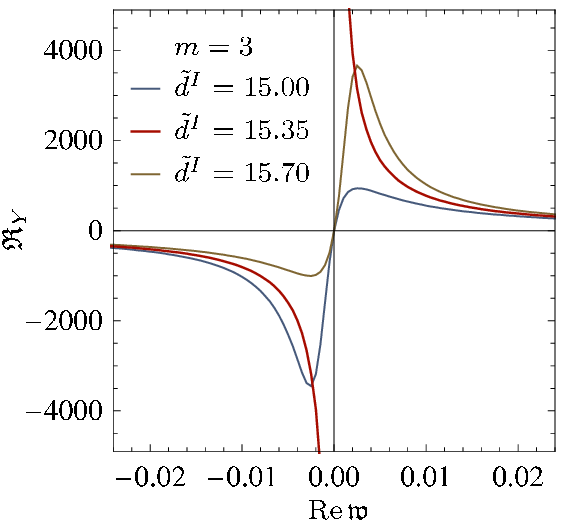}
				\caption[Pole in the spectral function at critical density]{%
					Plot of the spectral function for the mode $Y$ around
					$\wn=0$. At a value of $\tilde d^I = 15.35$ a pole appears at
					the origin. This behavior is due to the movement of poles in
					the complex $\wn$-plane, illustrated in
					figure~\ref{fig:threeExamples}.
				}
				\label{fig:poleExample}
			\end{minipage}
			\hfill
			\begin{minipage}[t]{.45\linewidth}
				\includegraphics[width=\linewidth,height=.925\linewidth]{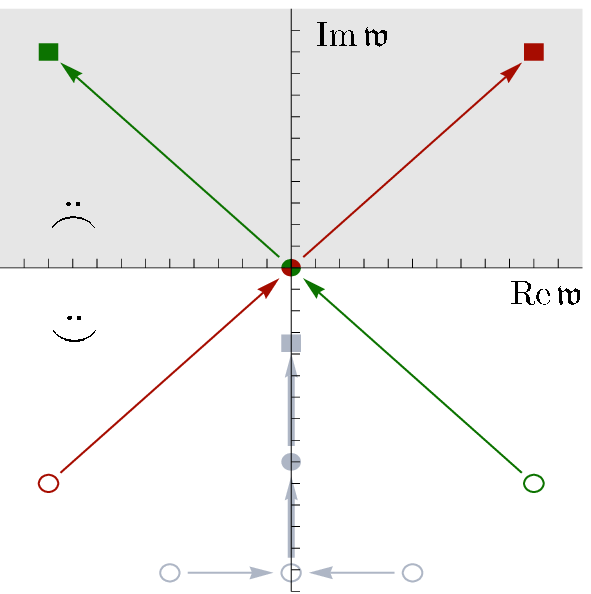}
				\caption[Sketch of pole movements under variations of isospin
					density]{A sketch of the positions and movements of the
					quasinormal frequencies under changes of $\tilde d^I$. Color
					indicates the function: $\text{red}=Y$,
					$\text{green}=X$,
					$\text{blue}=E^3$. The symbols
					indicate the range of $\tilde d^I$: $\circ<\tilde
					d_{\text{crit}}$, $\bullet=\tilde d_{\text{crit}}$,
					$\text{\tiny$\blacksquare$}>\tilde d_{\text{crit}}$. Poles
					in the gray region introduce instabilities.
				}
				\label{fig:qnmSketch}
			\end{minipage}
		\end{figure}

		In section~\ref{sec:retardedThermalGreen} we outlined how the imaginary
		part of the quasinormal modes describes damping, as long as $\Im
		\Omega_n < 0$. The short note on the pole structure demonstrated the
		dependence of the position of the quasinormal modes on the chemical
		potential/particle density. From figure~\ref{fig:newStructure} we deduce
		that at higher densities than studied so far, a quasinormal mode
		approaches the origin of the complex $\omega$ plane as the particle
		density is increased. We observe a pole at $\wn=0$ for a certain
		particle density $\tilde d_{\text{crit}}$, the value depends on $m$. An
		impression of the variation in the spectral function is given in
		figure~\ref{fig:poleExample}.
		
\begin{figure}
			\begin{minipage}[t]{.1\linewidth}
				\raisebox{3ex}{\includegraphics[height=2.02\linewidth]{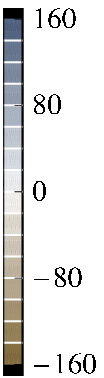}}
			\end{minipage}\hfill
			\begin{minipage}[t]{.275\linewidth}
				\includegraphics[width=\linewidth]{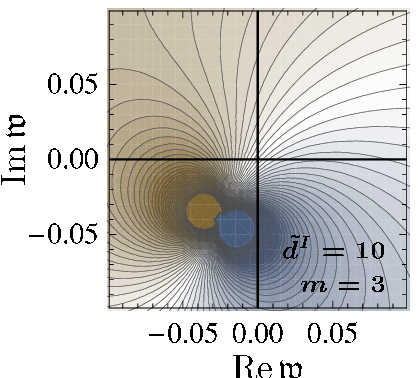}
			\end{minipage}\hfill
			\begin{minipage}[t]{.275\linewidth}
				\includegraphics[width=\linewidth]{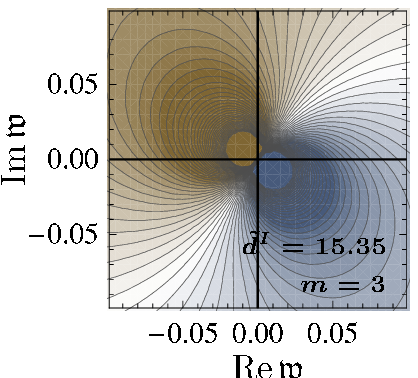}
			\end{minipage}\hfill
			\begin{minipage}[t]{.275\linewidth}
				\includegraphics[width=\linewidth]{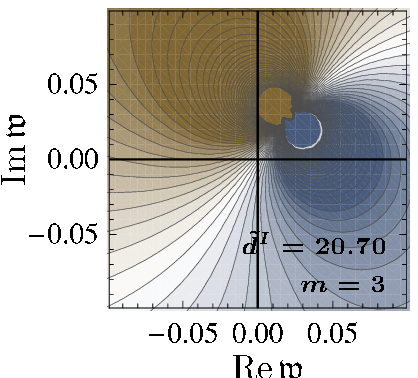}
			\end{minipage}\hfill
			\caption[Example of a QNM entering the upper half plane]{%
				Contour plots of the spectral function for the mode
				$Y$ around $\wn=0$ in the complex $\wn$-plane. The density
				increases from the left plot at sub-critical density to the right
				one at super-critical density. Here, the pole in the upper half
				plane introduces an instability.
			}
			\label{fig:threeExamples}
		\end{figure}

		In figure~\ref{fig:qnmSketch} we qualitatively sketch the result from
		the investigation of the behavior of the quasinormal modes closest to
		the origin of the complex $\wn$-plane. These modes do \emph{not} produce
		the peaks corresponding to the spectrum \eqref{eq:mesMassDimless}.		%
		At low densities all quasinormal modes are located in the lower half
		plane. When increasing the isospin density, the lowest frequency modes
		of the solutions $X$ and $Y$ to \eqref{eq:eomX} and \eqref{eq:eomY} move
		towards the origin of the frequency plane. At the same time two
		quasinormal modes of $E^3$ move towards each other and merge on the
		negative imaginary axis, then travel along the axis towards the origin
		as one single pole. At the critical value of $\tilde d=\tilde
		d_\text{crit}$ the modes from $X$ and $Y$ meet at the origin, the
		quasinormal modes from $E^3$ still reside in the lower half plane. This
		observation matches the discussion at the beginning of this section,
		where we expected $X$ and $Y$ to behave similarly at small $\wn$, while
		$E^3$ should differ from this behavior. Upon further increasing the
		isospin density, the modes $\Omega$ from $X$ and $Y$ enter the upper
		half plane, maintaining their distinct directions. The sign change in
		$\Im\Omega$ from $\Im\Omega<0$ to $\Im\Omega>0$ indicates that a damped
		resonance changes into a self-enhancing one, and thus introduces an
		instability to the system. Figure~\ref{fig:threeExamples} illustrates
		the transition of a quasinormal mode of $Y$ from the lower half plane to
		the upper half plane. The $E^3$-mode does not enter the upper half plane
		at any value of $\tilde d$ we considered. Compare this to the values of
		$\tilde d$ in figure~\ref{fig:newStructure} at which the pole induces
		visible structures at small $\wn$. A comparable movement of poles in a
		different but related setup was found in ref.~\cite{Gubser:2008wv}. There the
		quasinormal modes of correlation functions of electromagnetic currents
		were investigated as a function of temperature.
		
		\medskip
		
		In the following we interpret the observation of decaying mesons and
		the emergence of a new peak in the spectral function in terms of
		field theory quantities. In particular we speculate on a new phase
		in the phase diagram for fundamental matter in the \D3/\D7 setup.
		
		In the far UV, the field theory dual to our setup is supersymmetric,
		thus containing scalars as well as fermions, both of which contribute to
		the bound states we identified with mesons, even when supersymmetry is
		eventually broken. The meson decay at non-vanishing particle densities
		may be explained by the change of the shape of the potential for the
		scalars in the field theory upon the introduction of a non-vanishing
		density. As outlined in appendix~\ref{sec:scalarPotential}, a chemical
		potential may lead to an instability of the theory, since it induces a
		runaway potential for the scalar fields at small field values
		\cite{Harnik:2003ke}. Nevertheless, interactions of $\phi^4$-type lead
		to a Mexican hat style potential for larger field values. In this way
		the theory is stabilized at finite density $\tilde d$ while the scalar
		fields condensate. This squark condensate presumably contributes to the
		vev of the scalar flavor current,
		\begin{equation}
			\tilde d \propto \vev{J^0} \propto \vev{\bar \psi\,\gamma^0\,\psi} + \vev{ \phi \;\partial^0\phi}. 
		\end{equation}
		In the \ads/CFT context, the presence of an upside-down potential for
		the squark vev has been shown in ref.~\cite{Apreda:2005yz} using an
		instanton configuration in the dual supergravity background.
		
		The occurrence of a pole in the upper half plane of complex frequencies
		at finite $\tilde d_{\text{crit}}$ indicates an instability of the
		theory. A comparable observation was made in ref.~\cite{Aharony:2007uu},
		where in fact the vector meson becomes unstable by means of negative
		values for its mass beyond some critical chemical potential. The
		difference between this work and ref.~\cite{Aharony:2007uu} is that our model
		includes scalar modes in addition to the fundamental fermions.
		Nevertheless, in both models an instability occurs at a critical value
		of the chemical potential. The theory may still be stabilized
		dynamically by vector condensation \cite{Buchel:2006aa}. In this case
		the system would enter a new phase of condensed vectors at densities
		larger than $\tilde d_{\text{crit}}$, in accordance with the expectation
		from QCD calculations
		\cite{Lenaghan:2001sd,Sannino:2002wp,Voskresensky:1997ub}.
		
\begin{figure}
			\centering
			\includegraphics[width=.5\linewidth]{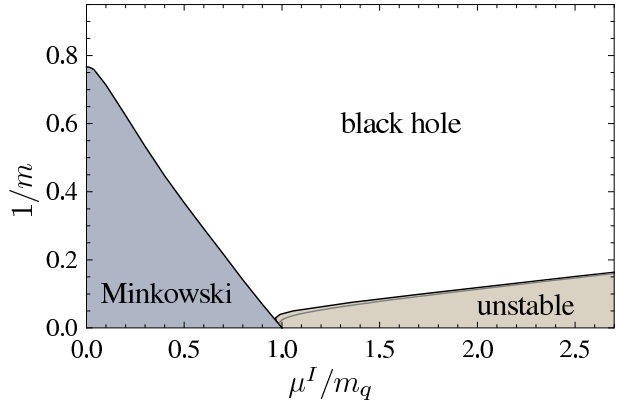}
			\caption[Phase diagram of fundamental matter in $(\mu^I,T)$-plane]{
				the $(\mu^I,T)$-plane. In the blue shaded region \db7s have the
				topology of Minkowski embeddings, the white and brown regions
				are modeled by black hole embeddings. These become unstable in
				the brown region. The boundary of the unstable region
				asymptotically seems to agree with the thin gray line of
				constant density $\tilde d^I=20.5$.
			}
			\label{fig:phaseDiagramIsoTrans}
		\end{figure}

		We perform the analysis of the pole structure at $\wn=0$ for various
		$m$, and interpret the phenomenon of the transition of poles into the
		upper half plane at finite critical particle density as a sign of the
		transition to an unstable phase. We relate the critical particle density
		$\tilde d_{\text{crit}}$ to the according chemical potential $\tilde
		\mu^I$ by $\mu^I=\lim_{\rho\to\infty} A^3_0(\rho)$ and use the pairs of
		$m$ and critical dimensionful $\mu^I$ to trace the line of the phase
		transition in the phase diagram of fundamental matter in the \D3/\D7
		setup. The result is drawn in figure~\ref{fig:phaseDiagramIsoTrans}. The
		picture shows the $(\mu^I,T)$-plane of the phase diagram and contains
		three regions, drawn as blue shaded, white, and brown shaded, as well as
		solid lines, separating the different regions.
		
		The blue shaded region marks the range of parameters, in which
		fundamental matter is described by \db7s with Minkowski embeddings. The
		line, delimiting the blue region, marks the line of phase transitions to
		the black hole phase, where fundamental matter is described by \db7s
		which have black hole embeddings. Using the symmetry of the DBI
		action, this phase transition line can be mapped to the line of phase
		transitions between Minkowski and black hole embeddings, present at
		finite baryon chemical potential
		\cite{Mateos:2007vc,Karch:2007br,Erdmenger:2007cm}.
		
		The brown shaded region in the phase diagram in
		figure~\ref{fig:phaseDiagramIsoTrans} marks the observation made in this
		section. The line delimiting the brown region marks the values of
		$\tilde d_{\text{crit}}$ at which the pole in the spectral function
		appears at $\wn=0$. Beyond this line we enter the brown shaded unstable
		region.
		
		We observe that the separation line of the unstable phase asymptotes to
		a straight line at high temperatures. Within the values computed by us,
		this line agrees with the asymptotic behavior of the contour of particle
		density with $\tilde d\approx 20.5$, drawn as a thin gray line in the phase diagram. We
		thus speculate on a finite critical particle density beyond which the
		black hole phase is unstable. This interpretation is supported by
		analogous studies of the phase diagram of $\N=4$ super-Yang-Mills theory
		with R-symmetry chemical potentials, where a similar line in the phase
		diagram was discovered \cite{Yamada:2006rx,Cvetic:1999ne}. The remaining
		question is whether the brown shaded phase in
		figure~\ref{fig:phaseDiagramIsoTrans} indeed is unstable in the sense
		that it inaccessible for any physical setup, or if there is a way to
		stabilize the system in the parameter range of question. Recent
		publications revealed that the introduction of a further vev for a
		different gauge field component on the stack of probe branes leads to a
		stabilization of the system \cite{Ammon:2008fc,Ammon:2009fe}. The
		resulting setup exhibits a second order phase transition to the new
		phase, which bears analogies to the theories of superfluidity and
		superconductivity
		\cite{Ammon:2008fc,Basu:2008bh,Herzog:2009ci,Ammon:2009fe}.

		Note that the location of the transition line to the unstable phase in
		figure~\ref{fig:phaseDiagramIsoTrans} as well as the results shown in
		figure~\ref{fig:threeExamples} and figure~\ref{fig:qnmSketch} are
		obtained from the analysis of poles in the spectral functions. These
		functions in turn are obtained as solutions to equations~\eqref{eq:eomX}
		to~\eqref{eq:eomE3}, which do depend on the so far unknown factor
		$c_{\scriptscriptstyle A}$ in determining the self coupling of the gauge
		field on the brane. The computation of this factor is left to future
		work. It will determine the exact position of the boundary of the brown
		shaded region in figure~\ref{fig:phaseDiagram}. This will answer the
		question whether there is a triple point in the phase diagram and if the
		color shaded regions meet at a common border. Moreover, other poles than
		the ones investigated here may have influence on the stability of this
		system.
		
\subsection{Summary}
	
	We made two observations concerning the thermodynamic behavior of
	fundamental matter in the \D3/\D7 setup.
	
	First, in our simple approximation of the baryon diffusion coefficient, we
	observe the fundamental phase transition and its dependence on the
	baryon density. We find that increasing the density from zero, where the
	transition temperature is given by $m=1.3$, the transition temperature
	is lowered slightly until the transition vanishes at a critical value of
	$\tilde d^*=0.00315$, where the transition occurs at $m=1.31$. This
	confirms the results of ref.~\cite{Kobayashi:2006sb}, where the transition was
	observed in the quark condensate.
	
	Second, we observe a new phase transition which renders the \D3/\D7 setup
	unstable at values above a critical isospin density. This becomes manifest
	by quasinormal modes of the fluctuations which develop positive imaginary
	parts in this region of the phase diagram. The exact position of the phase
	transition line cannot be determined yet. However, we speculate that the
	instability is due to a modification of the potential for the scalar fields
	in the field theory. This instability can be cured by vector meson
	condensation.
	
	It is tempting to compare figures~\ref{fig:phaseDiagramIsoTrans}
	and~\ref{fig:phaseDiagQCD}. However, we point out that we cannot interpret
	the brown shaded region in fig.~\ref{fig:phaseDiagramIsoTrans} as a direct
	analogon of the color-flavor-locked phase (CFL) in
	fig.~\ref{fig:phaseDiagQCD}, since the parameter range scanned by us only
	allows to observe the phase transition to the new phase only at finite
	isospin chemical potential and not at finite baryon potential. Also the
	critical point at finite $(T^c,\mu^c)$ is not reproduced so far from the
	thermodynamics of the \D3/\D7 model. Nevertheless, the sheer appearance of the
	phase diagram in figure~\ref{fig:phaseDiagramIsoTrans} may serve as a
	motivation for further efforts in exploring the phase diagrams of
	holographic models.

\section{Conclusion}
\label{chap:conclusion}

We considered different generalizations of the \ads/CFT correspondence in order
to shed light on in-medium effects on the fundamental matter in holographic
models for the quark-gluon plasma. The influence of the medium was parametrized
by the values of the temperature and particle density.

\medskip

The first aspect we considered in ~\ref{chap:specFuncs} was the influence
of the medium on bound states of quarks, in particular vector mesons. The
description of mesons from first principles is interesting in its own right,
because the strong coupling parameter forbids to apply well established
perturbative methods in QCD. In the holographic setups, mesonic excitations
arise more or less naturally as vibrational modes of open strings on \db{{}}s. In
the low energy limit, they account for fluctuations of supergravity fields. We
presented the capabilities of a certain realization of a \D3/\D7 brane
configuration by deriving the spectral functions for vector mesons from it. In
the limit of vanishing temperature and density the derived spectra agree with
the previously known results. The main achievement of our efforts, however, was
the extension of the spectral description of vector mesons into the finite
density and temperature regime for all values of quark masses and temperature.
We observe the melting of mesons at high temperature and at the same time
studied the effects of finite particle density. Technically, we related the
characteristics of the spectra to the behavior of the quasi normal modes of the
excitations that holographically account for mesonic bound states of quarks.

The main contribution to a better understanding of in-medium effects from this
project is the derivation of in-medium effects on the spectra. We observe a
destabilization of mesonic bound states with increasing particle density in the
quark-gluon plasma, which is simultaneously accompanied by a slight shift of the
meson masses to higher energies. Due to the fact that the holographic models are
too complex to be solved by analytical methods alone, the precise mechanisms
that account for this behavior are difficult to reveal. A probable physical
explanation for the destabilization certainly can be seen in the fact that in
the strongly coupled medium the surrounding free quarks alter the binding
interquark potential of the mesonic bound state. The closer a quark of the
medium comes to a constituent quark of the meson under consideration, and the
higher the amount of such perturbing spectator quarks is, the more influence can
be expected from the medium on the mesons. Therefore increasing the baryon
density (which can be seen as a measure for the amount of free quarks in the
medium which are not bound into mesons) accounts for accumulating perturbation
of the binding quark-antiquark potential, eventually leading to a dissociation
of the meson. The shift in the meson mass may also be a consequence of the
modification of the interquark potential, which in turn shifts the binding
energies and therewith the energy content of a meson.

Without speculating further on the mechanisms that lead to the observation we
made, we note that our results are in qualitative agreement with
phenomenological models and observation from experiment. The fact that our
result is a non-trivial consequence derived from the \D3/\D7 setup can be seen
as an affirmative answer to the question whether string theory motivated  models can
capture phenomenologically relevant physics.

Another such example was also derived in the context of meson spectra. Namely,
we have shown that the introduction of finite isospin chemical potential indeed
leads to a mass splitting of the different components of the isospin triplet,
constituted by the three possible isospin one combinations of quark-antiquark
pairs. While this is a success on the one hand side, the quantitative evaluation
of the mass difference remains as a task for future investigation, as the
magnitude of the mass splitting heavily relies on meson coupling constants which
are not determined yet and where chosen arbitrarily in our setup. The
qualitative observation of the mass splitting, however, can be explained
entirely analytically. We notice that the degenerate spectrum at vanishing
isospin density stems from the fact that we have an $\SU(2)$ isospin symmetry in
our system. By introducing a finite vacuum expectation value for one single
generator, we break this symmetry and thereby suspend the degeneracy. From the
equations of motion we can read off that the vev does not affect the
longitudinal component in flavor space but shifts the energy eigenvalues of the
transverse modes by identical absolute amount with opposite sign.

\medskip 

The second observable we studied was the diffusion coefficient of both baryon
and isospin charge as well as the diffusion coefficient of mesons. The
motivation to consider baryon and isospin diffusion apparently is to understand
the transport processes of quarks and antiquarks in the QGP and quark matter as
e.\,g.\  expected to exist in neutron stars. The interest in mesons stems from the
fact that there is experimental evidence for mesons to survive the deconfinement
transition to the QGP. We capture this effect in our setup, as we have shown by
observing discernible peaks in the mesonic spectral functions at finite
temperature, discussed above.

The results for the baryon diffusion coefficient where derived in an extremely
simplified manner by plugging in the results for the embedding functions of the
\db7 in our background into the formula for the diffusion coefficient derived
from the membrane paradigm. The main purpose of this task was to show that the
\D3/\D7 setup at finite density is able to yield baryonic diffusion parameters
for the plasma for ratios of the quark mass to temperature in both regimes,
below and above the phase transition for fundamental matter. At vanishing
density, the diffusivity of quarks normalized to the inverse temperature was
known to be almost independent from the mass in the regime of light mesons
(compared to the deconfinement temperature) and to be reduced monotonically with
increasing quark mass. In our simple extension, we have shown that the effect of
finite density on the normalized diffusion of baryon charge is leading to a
dependence on the diffusion coefficient that exhibits a minimum for quarks with
masses $m_q$ slightly heavier than the scale determined by the critical melting
temperature $T_c$ as $m_q^\text{crit}=\frac{1.3\sqrt\lambda}{2}T_c$. We identify
the origin of this behavior as the fact that neither asymptotically heavy nor
massless quarks will be influenced from the thermal momentum scale. A particle,
however, with intermediate mass is certainly sensitive to momentum transfer by
e.\,g.\  collisions with particles in the medium.

Moreover, we observe that with increasing density, the mass dependence of the
diffusivity becomes smaller. We address this to the fact that an extremely high
density, accompanied by a likewise high chemical potential, outweighs the
energy scale set by the finite temperature and in this way suppresses the
intermediate mass scale dependence.

Although we knew that our simple ansatz could not capture all effects of finite
density in this way, our results where proven to be qualitatively correct in
later publications. These, by the way, support the above comment on the rivaling
scales of quark mass and chemical potential. The results of ref.~\cite{Mas:2008qs}
show that at large quark masses (which then outweigh the energy scale of the
chemical potential) the diffusion constant indeed depends on the quark
mass but is almost independent of the particle density.

The situation at non-vanishing isospin chemical potential was analyzed by means
of the dispersion relation for particles carrying isospin charge. Here, we
restricted to massless quarks and small chemical potential. Within the tight
restriction of our setup we where able to derive a frequency dependence of the
diffusion coefficient, which can be interpreted as a dependence of the diffusion
coefficient on the energy of the diffusion massless particles.

The most extensively investigated transport coefficient however is the diffusion
coefficient for heavy scalar mesons in the quark-gluon plasma. We set up a
kinetic model that allowed for a derivation of the (inverse of the) diffusion
coefficient at both strong and weak coupling. This enabled us to compare the
perturbatively obtained weak coupling result to the holographic strong coupling
result. Moreover, we where able to derive the polarizability of mesons from
holographic duals. The latter results for the polarizabilities have to be read
with care, as they rely heavily on the short-distance dipole approximation of
the underlying effective model. As the short distance dipole interaction with
the medium most likely rely heavily on large momentum transfer and thereby on
weak coupling contributions, the validity of the \ads/CFT contributions may be
vulnerable to serious criticism. Nevertheless, the long range effects on the
momentum distribution of the scattered medium particles should be captured by
our gauge/gravity model. We therefore divide out the effects due to
polarizability, and parametrized by $(\delta M)^2$ and compare the quotient of
momentum broadening (inverse diffusion coefficient) $\kappa$ and $(\delta M)^2$.
As a result, within the limits of the validity of our assumptions, we observe a
reduction of the momentum transfer from the mesons to medium particles at strong
coupling compared to weak coupling. This has consequences on the equilibration
of the meson momentum distributions into thermal equilibrium, which we expect to
be slowed down at strong coupling. Hopefully, measurements of the heavy meson
momentum distributions in heavy ion collisions at RHIC and LHC will allow for a
comparison of experimental data with our theoretical expectations.

\medskip

Finally, we devoted one  to the analysis of the phase structure of
fundamental matter in the holographic description of quark matter.

One result was the observation of a phase transition in the baryon diffusion
constant, which shows parallels to a previously observed phase transition in the
the quark condensate, which vanishes at a critical baryon density $d^*$. We
observe a dependence of the value of the quark mass to temperature ration at
which this phase transition occurs on the density. And we identify a critical
density at which the phase transition vanishes. This density matches the value
$d^*$ mentioned above, which makes us believe that we observe the same physical
transition in just an other parameter.

More important is the observation of a new phase transition at finite isospin
density. Above we described the observation that the mass eigenvalues of two of
the three components of the isospin triplet vector mesons experience shifts due
to finite chemical potential. Increasing the isospin chemical potential
$\mu^{\scriptscriptstyle I}$, we observe an instability of our setup at a
critical value $\mu^{\scriptscriptstyle I}(T)$ of the chemical potential. This
value depends on the temperature of the medium. By numerical evaluation of
$\mu^{\scriptscriptstyle I}(T)$ we are able to trace out the boundary of the
stable phase in the $(\mu^{\scriptscriptstyle I},T)$-plane of the phase diagram.
Recent publications indicate that the theory can be stabilized even beyond this
line if additional gauge field components on the flavor branes acquire finite
vacuum expectation values. This indicates that the boundary we traced out in the
pase diagram marks the border between two different phases. The exact position,
however, is subject to the same open questions we addressed when we discussed
the splitting of the vector meson spectrum.

\medskip

As a general conclusion, we ascertain that the \D3/\D7 setup for holographic
duals to strongly coupled gauge theories, provides the capability to describe a
rich amount of phenomenology of the dual field theory. In this article we
highlighted a small part of it. Alluding to the general motivation behind
applications of \ads/CFT to bridge the gap between string theory and
phenomenologically relevant field theories, we finally end with a satisfactory
statement. The analyses and observations described in this work show that the
gauge/gravity duality is not a one-way street. In one direction we where able to
confirm many observations in holographic models by known results and
expectations from established field theories such as QCD, and even experiment.
In this way our confidence in the applicability of the correspondence to real
world phenomena was strengthened. In the other direction, using the example of
meson diffusion, we discovered ways to derive results in regimes of field
theories, which so far where inaccessible, and hopefully bear at least
qualitative truth when compared to field theory results or experiments in
future.

\section*{Acknowledgments}

The author is deeply grateful to all the people he worked with during his time
as a Ph.\;D. student at Johanna Erdmenger's group at the Max-Planck-Institut
f\"ur Physik in Munich. These are in particular Johanna Erdmenger, Johannes
Gro{\ss}e, Riccardo Apreda, Matthias Kaminski, Jeong-Hyuck Park, Stephan Hoehne,
Ren\'e Meyer, Martin Ammon, Patrick Kerner, Hai Ngo, Viviane Grass, Constantin
Greubel and Andrew O'Bannon. Thanks also to Derek Teaney for the enjoyable
collaboration in the interesting project \cite{Dusling:2008tg}.

\medskip
Part of this work was funded by the \emph{Cluster of Excellence for Fundamental
Physics\,---\,Origin and Structure of the Universe}.

\appendix

\section{Coordinates for the AdS black hole background}
\label{chap:coordinates}

Numerous local coordinate systems are used to parameterize the \adsfivesfive
Schwarzschild black hole background. The problem at hand determines which of
them is most useful. Here we list some of the common coordinates and the
transformations between them. For all of the following coordinate systems we use
the same symbols
\begin{equation*}
	R^4=4\pi\gs N\alpha'^2,\qquad\rh=T\pi R^2.
\end{equation*}

\newcounter{coordSystem}
\setcounter{coordSystem}{0}
\newcommand{\newCoords}{\addtocounter{coordSystem}{1}\subsection*{Coordinate system \arabic{coordSystem}}}

\newCoords

\begin{equation}
\label{eq:adsBH1}
	\ds^2 = \frac{r^2}{R^2}\left(-f(r)\,\dd t^2 + \dd \threevec x^2\right) + \frac{R^2}{r^2}\frac{1}{f(r)}\,\dd r^2 + R^2 \dd\Omega_5^2
\end{equation}
with
\[
	f(r)=1-\frac{\rh^4}{r^4}
\]
and
\[
	t,x^1,x^2,x^3\in\mathbbm{R},\qquad r\geq0,\qquad\dd\Omega_5^2=\text{metric of the unit $5$-sphere}
\]
\begin{tabbing}
horizon at\hspace{3em} \= $r=\rh$\\
boundary at \> $r\to\infty$
\end{tabbing}

\newCoords

Introduction of a new radial coordinate
\[
	\varrho^2 = r^2 + \sqrt{r^2-\rh^2} 
\]
transforms \eqref{eq:adsBH1} into
\begin{equation}
\label{eq:adsBH2}
	\ds^2 = \frac{\varrho^2}{2R^2}\left( -\frac{f^2(\varrho)}{\tilde f(\varrho)}\,\dd t^2 + \tilde f(\varrho)\,\dd \threevec x^2 \right) +\frac{R^2}{\varrho^2} \left( \dd \varrho^2 + \varrho^2 \dd\Omega_5^2\right) 
\end{equation}
with
\[
	f(\varrho)=1-\frac{\rh^4}{\varrho^4},\qquad\quad \tilde f(\varrho)=1+\frac{\rh^4}{\varrho^4}
\]
\begin{tabbing}
horizon at\hspace{3em} \= $\varrho=\rh$\\
boundary at \> $\varrho\to\infty$
\end{tabbing}

\subsubsection*{Parametrization of the radial part}

We can identify the part in the last pair of parentheses of \eqref{eq:adsBH2} as
nothing else than $\mathbbm R^6$ and we write it as
\[
	\dd \varrho^2 + \varrho^2\dd\Omega^2_5=\sum_{i=1}^6\dd \varrho_i^2 = \underbrace{\dd w^2 + w^2\dd\Omega_3^2}_{\mathbbm R^4(\varrho_{1,\ldots,4})} \:+\: \underbrace{\dd L^2 + L^2 \dd \phi^2}_{\mathbbm R^2(\varrho_{5,6})}
\]
with radial coordinate $\varrho=\left(\sum_i\varrho_i^2\right)^{\scriptscriptstyle 1/2}$. We write this
space as a product space of a four-dimensional $\mathbbm R^4$ in polar
coordinates with radial coordinate $w\geq 0$ and a two-dimensional $\mathbbm
R^2$ with radial coordinate $L\geq 0$, such that $\varrho^2=w^2+L^2$. The
subspace parametrized by $(w,L)$ is the first quadrant of a Cartesian
coordinate system and can also be parametrized by its radial part $\varrho$ and
an angle $0\leq\theta\leq\pi/2$,
\[
\begin{aligned}
	L &= \varrho\cos\theta,\\
	w &= \varrho\sin\theta
\end{aligned}
\]
such that
\[
	\dd L^2 + \dd w^2 = \dd\varrho^2 + \varrho^2 \dd\theta^2.
\]
Finally, we introduce $\chi=\cos\theta$ and thus can write \eqref{eq:adsBH2} as
\[
\tag{\ref{eq:adsBH2}a}
\label{eq:adsBHChi}
	\ds^2 = \frac{\varrho^2}{2R^2}\left( -\frac{f^2(\varrho)}{\tilde f(\varrho)}\,\dd t^2 + \tilde f(\varrho)\,\dd \threevec x^2 \right)
	        +R^2\left( \frac{\dd \varrho^2}{\varrho^2} + \left(1-\chi^2\right)\dd\Omega_3^2 + \left(1-\chi^2\right)^{-2}\,\dd\chi^2 + \chi^2\dd\phi^2  \right).
\]

\newCoords

Introduction of a new radial coordinate
\[
	v^2 = \frac{1}{2} \left(r^2 + \sqrt{r^2-\rh^2} \right)
\]
transforms \eqref{eq:adsBH1} into
\begin{equation}
\label{eq:adsBHv}
	\ds^2 = \frac{v^2}{R^2}\left( -\frac{f^2(v)}{\tilde f(v)}\,\dd t^2 + \tilde f(v)\,\dd \threevec x^2 \right) +\frac{R^2}{v^2} \left( \dd v^2 + v^2 \dd\Omega_5^2\right) 
\end{equation}
with
\[
	f(v)=1-\frac{\rh^4}{4v^4},\qquad\quad \tilde f(v)=1+\frac{\rh^4}{4v^4}
\]
\begin{tabbing}
horizon at\hspace{3em} \= $v=\frac{\rh}{\sqrt 2}$\\
boundary at \> $v\to\infty$
\end{tabbing}

\newCoords

Introduction of a new \emph{dimensionless} radial coordinate
\[
	u = \frac{\rh^2}{r^2}
\]
transforms \eqref{eq:adsBH1} into
\begin{equation}
\label{eq:adsBHu}
	\ds^2 = \frac{\rh^2}{R^2u}\left( -f(u)\,\dd t^2 + \dd \threevec x^2 \right) +\frac{R^2}{4f(u)\, u^2} \dd u^2 +R^2 \dd\Omega_5^2,
\end{equation}
where
\begin{equation*}
	f(u)=1-u^2,
\end{equation*}
\begin{tabbing}
horizon at\hspace{3em} \= $u=1$,\\
boundary at \> $u=0$.
\end{tabbing}

\newCoords

Introduction of a new radial coordinate
\[
	z = \frac{R^2}{r}
\]
transforms \eqref{eq:adsBH1} into
\begin{equation}
	\ds^2 = \frac{R^2}{z^2}\left( -f(z)\,\dd t^2 + \dd \threevec x^2 + \frac{1}{f(z)}\,\dd z^2 \right) + R^2 \dd\Omega_5^2
\end{equation}
with
\[
	f(z)=1-\frac{z^4}{z_{\!\circ}^4}
\]
\begin{tabbing}
horizon at\hspace{3em} \= $z=z_{\!\circ}=\frac{R^2}{\rh}$\\
boundary at \> $z=0$
\end{tabbing}

\section{Isospin diffusion related equations}
\label{chap:solutionsEOMISO}

{\allowdisplaybreaks

\subsection{Solutions to equations of motion}
\label{sec:solutionsEOM}

	Here we explicitly write down the component functions used to construct the
	solutions to the equations of motion for the gauge field fluctuations up to
	order $\wn$ and $\qn^2$. The functions themselves are then composed as in
	\eqref{eq:fullX}.
	
	The solutions for the components with flavor index $a=3$ where obtained in
	ref.~\cite{Policastro:2002se}.

	\subsubsection{Solutions for $X_\alpha$, $\widetilde X_\alpha$ and $A^3_\alpha$}
		\label{sec:solXalpha}
		
		The function $X_\alpha(u)$ solves \eqref{eq:eomXalpha} with the upper
		sign and is constructed as in \eqref{eq:fullX} from the following
		component functions,
		\begin{align}
				\label{eq:XalphaBeta}
				\beta   =\;& \sqrt{\frac{\wn\,\mn}{2}}+\order{\omega},\\[\bigskipamount]
				F_0      =\;& C,\\
				F_{1/2}  =\;& - C \sqrt{\frac{\mn}{2}}\;\ln\frac{1+u}{2},\\
				F_1      =\;&               - C \frac{\mn}{12}\, \Bigg[ \pi^2-9 \ln^22  + 3 \ln(1-u)\left( \ln 16 - 4 \ln(1+u)\right)\nonumber\\
							 & + 3 \ln(1+u) \left( \ln(4(1+u)) -4\ln u \right)\\
							 & - 12\left( \mathrm{Li}_2(1-u)+\mathrm{Li}_2(-u) + \mathrm{Li}_2\left(\frac{1+u}{2}\right)\right) \Bigg]\nonumber,\\
				G_1      =\;&\frac{C}{2}\,\left[ \frac{\pi^2}{12}+ \ln u\ln(1+u) + \mathrm{Li}_2(1-u)+\mathrm{Li}_2(-u) \right],\\
		\end{align}
		where the constant $C$ can be expressed it in terms of the field's
		boundary value $X^{\text{bdy}} = \lim_{u\to 0} X(u,k)$,
		\begin{multline}
		\label{eq:XalphaC}
			C = X^{\text{bdy}}\\ \times \bigg(1 + \sqrt{\frac{\mn\,\wn}{2}} \ln 2 + \mn\,\wn\left(\frac{\pi^2}{6} + \frac{\ln^2 2}{4}\right) + \frac{\pi^2}{8} \qn^2 + \order{\wn^{3/2},\qn^4} \bigg)^{-1}.
		\end{multline}
		
		The solutions of the equations of motion \eqref{eq:eomXalpha} with lower
		sign for the functions $\tilde X_\alpha(u)$ are given by
		\begin{align}
				\tilde \beta   =\;&-i\sqrt{\frac{\wn\,\mn}{2}}+\order{\omega},\\[\bigskipamount]
				\tilde F_0     =\;& \tilde C,\\
				\tilde F_{1/2} =\;& i \tilde C \sqrt{\frac{\mn}{2}}\;\ln\frac{1+u}{2},\\
				\tilde F_1     =\;&               \tilde C \frac{\mn}{12}\, \Bigg[ \pi^2-9 \ln^22  + 3 \ln(1-u)\left( \ln 16 - 4 \ln(1+u)\right)\nonumber\\
							   & + 3 \ln(1+u) \left( \ln(4(1+u)) -4\ln u \right)\\
							   & - 12\left( \mathrm{Li}_2(1-u)+\mathrm{Li}_2(-u) + \mathrm{Li}_2\left(\frac{1+u}{2}\right)\right) \Bigg],\nonumber\\
				\tilde G_1     =\;&\frac{\tilde C}{2}\,\left[ \frac{\pi^2}{12}+ \ln u\ln(1+u) + \mathrm{Li}_2(1-u)+\mathrm{Li}_2(-u) \right],\\
		\end{align}
		with $\tilde C$ given by
		\begin{multline}
				\tilde C = \tilde X^{\text{bdy}}\\ \times \bigg(1-i\sqrt{\frac{\mn\,\wn}{2}} \ln 2 - \mn\,\wn\left(\frac{\pi^2}{6} + \frac{\ln^2 2}{4}\right) + \frac{\pi^2}{8} \qn^2 + \order{\wn^{3/2},\qn^4} \bigg)^{-1},
		\end{multline}
		so that $\lim_{u\to 0} \tilde X(u,k)=\tilde X^{\text{bdy}}$.
		
		The solution for $A^3_\alpha$ solves \eqref{eq:eomA1A2Flavor3} up to
		order $\wn$ and $\qn^2$ with boundary value
		$\left(A^3_\alpha\right)^{\text{bdy}}$. It is
		\begin{equation}
		\begin{aligned}
				A^3_\alpha =\;& \frac{8\;\left(A^3_\alpha\right)^{\text{bdy}} (1-u)^{-\frac{i\wn}{2}}}{8-4i\wn\ln 2 + \pi^2\qn^2}\\
				& \times\Bigg[   1 + i\frac{\wn}{2} \ln\frac{1+u}{2}\\
				& \hphantom{\times\Bigg[\;}
						   + \frac{\qn^2}{2} \left( \frac{\pi^2}{12}+ \ln u\ln(1+u) + \mathrm{Li}_2(1-u)+\mathrm{Li}_2(-u) \right)\Bigg].
		\end{aligned}
		\end{equation}
		
	\subsubsection{Solutions for $X_0'$, $\widetilde{X}_0'$ and ${A^3_0}'$}
		
		Here we state the solutions to \eqref{eq:eomA0}. This formula describes
		three equations, differing in the choice of $a=1,2,3$. The cases $a=1,2$
		give coupled equations which are decoupled by transformation from
		$A^{1,2}_0$ to $X_0$ and $\widetilde{X}_0$. The choice $a=3$ gives a
		single equation.
		
		The function $X_0'$ is solution to \eqref{eq:eomX0} with upper sign. We
		specify the component functions as
		\begin{align}
				\beta    =\;& \sqrt{\frac{\wn\,\mn}{2}}+\order{\omega},\\[\bigskipamount]
				F_0      =\;& C,\\
				F_{1/2}  =\;& - C \sqrt{\frac{\mn}{2}}\;\ln\frac{2u^2}{1+u},\\
				F_1      =\;& - C \frac{\mn}{12}\, \Bigg[ \pi^2 + 3 \ln^22  + 3 \ln^2(1+u) +6 \ln 2 \ln \frac{u^2}{1+u}\\
												  & + 12 \left( \mathrm{Li}_2(1-u) +\mathrm{Li}_2(-u) - \mathrm{Li}_2\left(\frac{1-u}{2}\right) \right)\Bigg],\\
				G_1      =\;& C \ln\frac{1+u}{2u},\\
		\end{align}
		where the constant $C$ can be expressed in terms of the field's boundary
		value $X^{\text{bdy}} = \lim_{u\to 0} X(u,k)$,
		\begin{equation}
				C        = - \frac{\qn^2  X_0^{\text{bdy}} + \wn\qn X_3^{\text{bdy}}}{\sqrt{2\mn\wn} + \mn\wn \ln 2 + \qn^2}.
		\end{equation}
		To get the function $\widetilde{X}_0'$, we solve \eqref{eq:eomX0} with
		the lower sign and obtain
		\begin{align}
				\widetilde{\beta}    &= -i\sqrt{\frac{\wn\,\mn}{2}}+\order{\omega},\\[\bigskipamount]
				\widetilde{F}_0      &= \widetilde C,\\
				\widetilde{F}_{1/2}  &= i \widetilde{C} \sqrt{\frac{\mn}{2}}\;\ln\frac{2u^2}{1+u},\\
				\widetilde{F}_1      &          = \widetilde{C} \frac{\mn}{12}\, \Bigg[ \pi^2 + 3 \ln^22  + 3 \ln^2(1+u) +6 \ln 2 \ln \frac{u^2}{1+u}\\
												  &\hphantom{= \widetilde{C} \frac{\mn}{12}\, \Bigg[\, } + 12 \left( \mathrm{Li}_2(1-u) +\mathrm{Li}_2(-u) - \mathrm{Li}_2\left(\frac{1-u}{2}\right) \right)\Bigg],\\
				\widetilde{G}_1      &=\widetilde{C}\ln\frac{1+u}{2u},\\
		\end{align}
			where the constant $\widetilde{C}$ can be expressed it in terms of
			the field's boundary value $\widetilde{X}^{\text{bdy}} = \lim_{u\to
			0} \widetilde{X}(u,k)$,
		\begin{equation}
				\widetilde{C} = \frac{\qn^2  \widetilde{X}_0^{\text{bdy}} + \wn\qn \widetilde{X}_3^{\text{bdy}}}{i \sqrt{2\mn\wn} + \mn\wn \ln 2-\qn^2}.
		\end{equation}
		The solution for \eqref{eq:eomA0} with $a=3$ is the function ${A^3_0}'$,
		given by
		\begin{equation}
				{A^3_0}' = (1-u)^{-\frac{i\wn}{2}}\, \frac{\qn^2 A_0^{\text{bdy}}+\wn\qn A_3^{\text{bdy}}}{i\wn-\qn^2} \left(1 + \frac{i\wn}{2}\ln \frac{2u^2}{1+u} + \qn^2 \ln \frac{1+u}{2u}\right).
		\end{equation}
		
	\subsubsection{Solutions for $X_3'$, $\widetilde{X}_3'$ and ${A^3_3}'$}
		
		We give the derivatives of $X_3$ and $\widetilde X_3$ as
		\begin{align}
							 X_3' & = -\frac{\wn - \mn}{\qn f}\, X_0'\\
			 \widetilde{X}_3' & = -\frac{\wn + \mn}{\qn f}\, \widetilde{X}_0'.
		\end{align}
		The solution for ${A^3_3}'$ is
		\begin{equation}
				{A^3_3}' = -\frac{\wn}{\qn} \, {A^3_0}'.
		\end{equation}

\subsection{Abelian Correlators}
\label{sec:abelianCorrelators}

	For reference we quote here the correlation functions of the Abelian
	super-Maxwell theory found in ref.~\cite{Policastro:2002se}. The authors start
	from a $5$-dimensional supergravity action and not from a Dirac-Born-Infeld
	action as we do. Therefore there is generally a difference by a factor
	$N/4$. Note also that here all $N_f$ flavors contribute equally. In our
	notation
	\begin{align}  
	\label{eq:abelianCorrelators1}  
		G_{11}^{a b} &= G_{22}^{a b}  =- \frac{i N^2 T\omega\, \delta^{a b}}{16\pi} + \ldots\,,\\ 
		G_{00}^{a b} &= \frac{ N^2 T q^2\; \delta^{a b}}{ 16 \pi  ( i\omega - D q^2)}+ \ldots\,,\\
	\label{eq:abelianCorrelators2}
		G_{03}^{a b} &= G_{3 0}^{a b}  = - \frac{ N^2 T \omega q\, \delta^{a b}}{ 16 \pi  ( i\omega - D q^2)}+ \cdots\,,\\
		G_{33}^{a b} &=  \frac{ N^2 T \omega^2\, \delta^{ab}}{ 16 \pi  ( i\omega - D q^2)}+ \ldots\,,  
	\end{align}  
	where $D=1/(2\pi T)$ .  

\subsection{Correlation functions}
\label{sec:correlationFunctions}
	
	In this section we list the real and imaginary parts of the flavor currents
	in the first two flavor-directions~$a=1,\,2$ and in the third
	Lorentz-direction coupling to the supergravity-fields $X_3$ and $\widetilde
	X_3$ (as defined in \eqref{eq:flavorTrafo}).
	\begin{alignat}{3}
	\label{eq:reGX3X3t>=}
		\Re G_{3 \widetilde 3}(\omega\ge 0)  & =
		\Re G_{\widetilde 3 3}(\omega < 0)  && =
		-\frac{N\, q^2\, (\omega^2+\mu\left|\omega\right|)}{16\pi^2 \left[
		2\mu\left|\omega\right|+q^4/(2\pi T)^2 \right]}, \\[\medskipamount]
	\label{eq:imGX3X3t>=}
		 \Im G_{3 \widetilde 3}(\omega\ge 0) &= 
		 -\Im G_{\widetilde 3 3}(\omega < 0)  &&= 
		-\frac{N T\, \sqrt{2\mu\left|\omega\right|}\,
		 (\omega^2+\mu\left|\omega\right|)}
		 {8\pi \left[
		2\mu\left|\omega\right|+q^4/(2\pi T)^2 \right]}\, ,\\[\medskipamount]
	\label{eq:reGX3X3t<}
		\Re G_{3 \widetilde 3}(\omega < 0)   & =
		\Re G_{\widetilde 3 3}(\omega \ge 0) && = 
		-\frac{N T\, (\omega^2-\mu\left|\omega\right|)}{8\pi \left[
		\sqrt{2\mu\left|\omega\right|}+q^2/(2\pi T) \right]}\,,\\[\medskipamount]
	\label{eq:imGX3X3t<}
		\Im G_{3 \widetilde 3}(\omega  <  0)  &= 0,
		\Im G_{\widetilde 3 3}(\omega \ge 0)  &&= 0.
	\end{alignat}

} 

\section{Coupling constant for vector meson interaction}
\label{sec:mesoncoupling}

In this section we show how the coupling constant for the interaction of vector
mesons can be computed, extending the ideas presented in
\cite{Kruczenski:2003be}. This coupling constant in the effective
four-dimensional meson theory can be determined by redefinition of the gauge
fields such that the kinetic term has canonical form. This coupling constant
depends on the geometry of the extra dimensions.

First we consider the eight-dimensional theory determined by the DBI action
$S^{(2)}_{\text{DBI}}$ expanded to second order in the fluctuations $A$,
\begin{equation}
	S^{(2)}_{\text{DBI}}=\frac{T_7(2\pi\alpha')^2}{4}\int\!\dd^8\xi\sqrt{-\mathcal G}\;\mathcal G^{\mu\alpha}\mathcal G^{\nu\beta}\hat F_{\alpha\nu}\hat F_{\beta\mu}\,,
\end{equation}
where $\mathcal G$ contains the background fields and we simplify the analysis
by considering only Abelian gauge fields. Defining the dimensionless coordinate
$\bar\rho=\varrho/R$ and integrating out the contribution of the $\S^3$, we
obtain
\begin{equation}
	S^{(2)}_{\text{DBI}}=\frac{T_7(2\pi\alpha')^2\mathrm{vol}(\S_3)R^4}{4}\int\!\dd^4 x\int\!\dd\bar\rho\sqrt{-\mathcal G}\;\mathcal G^{\mu\alpha}\mathcal G^{\mu\beta}\hat F_{\alpha\nu}\hat F_{\beta\mu}\,.
\end{equation}
To obtain a four-dimensional effective theory we have to integrate over the
coordinate $\bar\rho$. This contribution depends on the geometry induced by the
$\bar\rho$ dependence of the metric factors. However, we expect that it is
independent of the 't~Hooft coupling $\lambda$. We parametrize this contribution
by $c_{\scriptscriptstyle A}'$. The kinetic term of the effective theory is then
given by
\begin{equation}
  S^{(2)}_{\text{DBI}}=\frac{T_7(2\pi\alpha')^2\mathrm{vol}(\S_3)R^4 c_{\scriptscriptstyle A}'}{4}\int\!\dd^4 x \;\hat F_{\mu\nu}\hat F^{\mu\nu}\,,
\end{equation}
where the prefactor may be written as
\begin{equation}
  \frac{T_7(2\pi\alpha')^2\mathrm{vol}(\S_3)R^4 c_{\scriptscriptstyle A}'}{4}=\frac{\lambda}{\gym^2 c_{\scriptscriptstyle A}^2}\,,
\end{equation}
where the numerical values independent of the 't~Hooft coupling are grouped into
the coefficient $c_{\scriptscriptstyle A}$. From this we can read off that a
rescaling of the form
\begin{equation}
  \hat A\mapsto\frac{c_{\scriptscriptstyle A}}{\sqrt{\lambda}}\hat A
\end{equation}
casts the Lagrangian into canonical form with a prefactor of $1/\gym^2$.

\section[Chemical potentials in field theories and BEC]{Chemical potentials in field theories: Runaway potential and Bose-Einstein condensation}
\label{sec:scalarPotential}

In our setup we consider a field theory which is supersymmetric in the far UV.
Its fundamental matter consists of complex scalars (squarks) and fermionic
fields (quarks). In this section we describe the effect of the chemical
potential on the field theory Lagrangian and on the vacuum as e.\,g.\  in
ref.~\cite{Harnik:2003ke}. We consider a theory with one complex scalar $\phi$ and
one fermionic field $\psi$ with the same mass $m_q$ coupled to an $\U(1)$ gauge
field $A_\nu$. The time component of the $\U(1)$ gauge field has a non-zero vev
which induces the chemical potential $\mu$,
\begin{equation}
	A_\nu=\mu\delta_{\nu 0}.
\end{equation}
The Lagrangian is given by
\begin{equation}
	\L=-\left(D_\mu\phi\right)^* D^\mu\phi-m_q^2\phi^*\phi-\bar\psi(D\!\llap/+m_q)\psi-\frac{1}{4}F_{\mu\nu}F^{\mu\nu}\,,
\end{equation}
where $D_\mu=\del_\mu-i A_\mu$ is the covariant derivative and
$F_{\mu\nu}=\del_\mu A_\nu-\del_\nu A_\mu$ the field strength tensor. Expanding
the Lagrangian around the non-zero vev of the gauge field, it becomes
\begin{equation}
\label{eq:fieldl}
	\L=-\del_\mu\phi^*\del^\mu\phi-(m_q^2-\mu^2)\phi^*\phi+\mu J^S_0-\bar\psi(\del\llap/+m_q)\psi+\mu J_0^F\,
\end{equation}
where $J^S_\mu=i\left((\del_\mu\phi^*)\phi-\phi^*(\del_\mu\phi)\right)$, and
$\left(J^\mu\right)^F=-i\bar\psi\gamma^\mu\psi$ are conserved currents. These
conserved currents are the population densities $N_S$ for the scalar field and
$N_F$ for the fermionic field, such that the linear terms in the Lagrangian are
$\mu N_S$ and $\mu N_F$.\par The mass term $-(m_q^2-\mu^2)\phi^2$ of the
Lagrangian (\ref{eq:fieldl}) introduces an instability if $\mu>m_q$ since the
corresponding potential $V=(m_q^2-\mu^2)\phi^2+\cdots$ is not bounded from
below. In some systems this runaway potential is stabilized by higher
interactions and becomes a Mexican hat potential such that the scalar condenses
and the scalar density becomes non-zero. This condensation is known as
Bose-Einstein condensation (BEC).

\providecommand{\href}[2]{#2}\begingroup\raggedright\endgroup

\end{document}